\documentclass[aps,prd,twocolumn,superscriptaddress,nofootinbib]{revtex4-1}

\usepackage{xspace}
\usepackage{xcolor}
\usepackage{graphicx}
\usepackage{amssymb}
\usepackage{amsmath}
\usepackage{bm}
\usepackage{epsfig}
\usepackage{subfigure}
\usepackage[section]{placeins}
\usepackage[export]{adjustbox}
\usepackage[utf8]{inputenc}
\usepackage[english]{babel}
\usepackage{blindtext}
\usepackage{enumitem}

\newtheorem{jigsaw}{Jigsaw Rule}[section]
\newtheorem{subjigsaw}{JR case}[jigsaw]

\newcommand{\minMtsq}{min $\Sigma M_{\rm top}^2~$}
\newcommand{\minDMt}{min $\Delta M_{\rm top}~$}
\newcommand{\minMt}{$M_{\rm top}^a = M_{\rm top}^b~$}
\newcommand{\minMW}{$M_{W}^a = M_{W}^b~$}

\newcommand{\bea}{\begin{eqnarray}}
\newcommand{\eea}{\end{eqnarray}}
\def\be{\begin{eqnarray}}
\def\ee{\end{eqnarray}}

\def\pslash{\not{\hbox{\kern-2pt p}}}

\definecolor{mygreen}{rgb}{0.12, 0.3, 0.17}
\definecolor{myred}{rgb}{0.45, 0.0, 0.09}
\definecolor{myblue}{rgb}{0.1, 0.1, 0.44}
\newcommand{\I}[1]{\ensuremath{\mathrm{\textcolor{mygreen}{\textbf{#1}}}}}
\newcommand{\V}[1]{\ensuremath{\mathrm{\textcolor{myblue}{\textbf{#1}}}}}
\newcommand{\D}[1]{\ensuremath{\mathrm{\textcolor{myred}{\textbf{#1}}}}}
\newcommand{\lab}{\ensuremath{\mathrm{\textbf{lab}}}}

\newcommand{\pfour}[2]{\ensuremath{\mathbf{p}_{\footnotesize #1}^{\footnotesize~#2}}}
\newcommand{\pthree}[2]{\ensuremath{\vec{p}_{\footnotesize#1}^{\footnotesize~#2}}}
\newcommand{\pone}[2]{\ensuremath{p_{\footnotesize#1}^{\footnotesize~#2}}}
\newcommand{\phat}[2]{\ensuremath{\hat{p}_{\footnotesize#1}^{\footnotesize~#2}}}
\newcommand{\E}[2]{\ensuremath{E_{\footnotesize#1}^{\footnotesize~#2}}}
\newcommand{\vbeta}[2]{\ensuremath{\vec{\beta}_{\footnotesize#1}^{\footnotesize~#2}}}
\newcommand{\sbeta}[2]{\ensuremath{\beta_{\footnotesize#1}^{\footnotesize~#2}}}
\newcommand{\hbeta}[2]{\ensuremath{\hat{\beta}_{\footnotesize#1}^{\footnotesize~#2}}}
\newcommand{\sgamma}[2]{\ensuremath{\gamma_{\footnotesize#1}^{\footnotesize~#2}}}
\newcommand{\mass}[2]{\ensuremath{m_{\footnotesize#1}^{\footnotesize~#2}}}
\newcommand{\Mass}[2]{\ensuremath{M_{\footnotesize#1}^{\footnotesize~#2}}}
\newcommand{\boost}[2]{\ensuremath{\Lambda_{\footnotesize#1}^{\footnotesize~#2}}}

\newcommand{\Hvar}[3]{\ensuremath{H_{\V{#1},\I{#2}}^{\footnotesize~\D{#3}}}}

\newcommand{\met}{\ensuremath{\vec{E}_{T}^{\rm  miss}}}

\begin{document}


\title{Recursive Jigsaw Reconstruction: HEP event analysis in the presence of kinematic and combinatoric ambiguities}

\author{Paul Jackson}
\email{p.jackson@adelaide.edu.au}
\affiliation{University of Adelaide, Department of Physics, Adelaide, SA 5005, Australia}
\affiliation{ARC Centre of Excellence for Particle Physics at the Tera-scale, University of Adelaide}
\author{Christopher Rogan}
\email{crogan@fas.harvard.edu}
\affiliation{Harvard University, Department of Physics, 17 Oxford Street, Cambridge, MA 02138}

\date{\today}

\begin{abstract}
We introduce \emph{Recursive Jigsaw Reconstruction}, a technique for analyzing reconstructed particle interactions in the presence of kinematic and combinatoric unknowns associated with unmeasured and indistinguishable particles, respectively. By factorizing missing information according to decays and rest frames of intermediate particles, an interchangeable and configurable set of {\it Jigsaw Rules}, algorithms for resolving these unknowns, are applied to approximately reconstruct decays with arbitrarily many particles, in their entirety. 

That the Recursive Jigsaw Reconstruction approach can be used to analyze {\it any} event topology of interest, with any number of ambiguities, is demonstrated through a twelve different simulated LHC physics examples. These include the production and decay of $W$, $Z$, Higgs bosons, and supersymmetric particles including gluinos, stop quarks, charginos, and neutralinos.

\end{abstract}

\pacs{Pacs numbers}

\maketitle

\section{Introduction}
\label{sec:intro}

Using a growing dataset of high energy collisions, experimentalists have many choices in strategy for searching for evidence of phenomena beyond the Standard Model (SM) of particle physics. Motivated by a lack of explanation in the SM for anomalies like the masses of neutrinos, matter/anti-matter asymmetry, and the identity of Dark Matter, they are looking for signs of new sub-atomic particle states appearing in the debris of collisions, whose presence is inferred kinematically from the particles reconstructed in dedicated detectors.  

Extensions to the SM, like supersymmetry (SUSY)~\cite{Borschensky:2014cia, Golfand:1971iw, Neveu:1971rx, Ramond:1971gb, Volkov:1973ix, Wess:1974tw, Salam:1974ig, Wess:1974jb, Ferrara:1974pu, Martin:1997ns}, introduce many additional, undiscovered particles, including dark matter candidates like the lightest weak-scale SUSY particle. Massive dark matter particles are often prevented from decaying to lighter SM particles by a conserved symmetry, such as $R$-parity in SUSY, which results in them being stable and having a low probability of interacting with particle detectors if they are produced in collisions. 

The key to observing the production and decay of new particles in HEP experiments, and performing precision studies of existing ones to infer deviations from SM predictions, is being able to precisely resolve their presence in collision events through the identification and measurement of their decay products in particle detectors. If the four vectors of all the final state particles are measured accurately, they can be combined to estimate the properties of any intermediate particles appearing in these events. Unfortunately, there are complications in studying many processes of interest beyond how well these particles are measured. If there are neutrinos, or dark matter particles, produced in collisions, they will escape undetected, taking with them crucial information about not only their properties, but also those of any intermediate particles which appeared in their production. Similarly, many of the reconstructed final state particles appearing in detectors can be indistinguishable, with a loss of information as to how they should be correctly combined into the actual intermediate resonances that realized them. 

These challenges are examples of kinematic and combinatoric ambiguities appearing in the analysis of particle interactions. In this paper we introduce \emph{Recursive Jigsaw Reconstruction} (RJR), a technique that can be applied to resolve such unknowns in any production or decay topology of interest, irrespective of their number. The RJR approach analyzes events by factorizing any unknowns according to the different decay steps of intermediate particles expected to appear in an event, making choices for the those associated with each particle decay using a combination of one or more {\it Jigsaw Rules} (JR's), interchangeable and configurable algorithms for resolving individual decays. Each event is then analyzed recursively, iteratively moving from the laboratory frame, where particle four vectors are measured, through each expected intermediate decay frame.

An overview of the RJR approach is provided in Section~\ref{sec:rjr}, where it is compared to existing HEP event reconstruction strategies. What sets apart the RJR technique is its general applicability achieved through its comprehensive library of JR's, which includes algorithms for resolving any combination of unknowns that can be encountered. Each of these JR's are motivated and described in this paper through a series of increasingly complex examples involving a collection of SM and new physics processes at a hadron collider. The notation used throughout this paper is defined in Section~\ref{sec:notation}, while the event generator used for simulating the physics processes discussed in each example is described in Section~\ref{sec:restframes}.

Twelve different physics examples are considered, organized according to the types of JR's that are being studied. Section~\ref{sec:Part1} introduces JR's for final states with a single invisible particle, including examples with the production and decay of $W$ bosons, top quarks, and charged Higgs bosons, as may appear in some extensions of the SM. JR's for decays with two invisible particles are described in Section~\ref{sec:Part2}, with their use demonstrated in examples involving Higgs bosons decaying to $W$'s, and both stop quark and neutralino pair production in SUSY scenarios, with decays through top quarks, Higgs, and $Z$ bosons.  To demonstrate how the RJR approach can independently measure the masses of many particles in a single event, examples of resonant and non-resonant top pair production are described in Section~\ref{sec:Part3}, with final states containing $b$-quarks, leptons and neutrinos with intermediate $W$ bosons. The SUSY analogue of this process, $\tilde{t}\tilde{t} \rightarrow b\tilde{\chi}^{\pm}(\ell\tilde{\nu})b\tilde{\chi}^{\mp}(\ell\tilde{\nu})$, is also discussed. 

JR's for decays to an arbitrary number of invisible particles are described through the example of non-resonant $N \geq 2~W(\ell\nu)$ production in Section~\ref{sec:Part4}, followed by a demonstration of how many JR's can be combined, recursively, to analyze events with both many kinematic and combinatoric unknowns through the examples $H \rightarrow hh \rightarrow 4W(\ell\nu)$ and $\tilde{g}\tilde{g} \rightarrow bb\tilde{\chi}^{0}_{1} bb\tilde{\chi}^{0}_{1}$.

\section{Recursive Jigsaw Reconstruction}
\label{sec:rjr}

The RJR approach constitutes both a methodology for analyzing reconstructed particle interactions, and a collection of techniques to resolve combinatorial and kinematic ambiguities, event-by-event, for a given sample. 

As used in this paper, {\it event} refers to a collection of measurements corresponding to a particle collision, including the three momenta of reconstructed particles and, potentially, additional measurements associated with their masses, missing momentum in one or more directions, or the center-of-mass energy of the interaction. In the RJR approach, a ``particle'' view of the event is used - in the form of a decay tree diagram - where intermediate heavy states are introduced, decaying to the indivisible ``visible'' objects, whose four vectors follow from detector particle reconstruction and identification,
and ``invisible'' objects, which correspond to weakly interacting particles hypothesized to have escaped detection.

Each decay tree not only describes how different intermediate states decay to the final state particles in an event, but also implicitly introduces a {\it kinematic basis} for analyzing the event; just as the collection of four vectors of all the individual final state particles fully describes an event, so too does the set of all of the masses and decay angles of the particles appearing in a decay tree. The latter is a natural kinematic basis for studying an event, in that the masses of intermediate particles of interest are included, along with decay angles sensitive to their spin and quantum numbers, while other uninformative degrees of freedom (ex. rotational symmetry of an entire event) are clearly isolated. 

The number of degrees of freedom associated with a decay tree need not be restricted to just the number of kinematic measurements made in the reconstruction of each event. If there are invisible particles appearing in a decay tree, one or more elements of their kinematic description may be under-constrained. Similarly, some reconstructed particles may be indistinguishable, leading to combinatoric ambiguities related to where each should appear in a decay tree. The masses and decay angles of a decay tree can be thought of as functions of these unknowns, such that the choice of how to analyze an event amounts to specifying how to resolve these under-constrained degrees of freedom.

There have been many strategies proposed for how to resolve kinematic ambiguities on an event-by-event basis. For combinatoric ambiguities, one generally chooses a particular metric to 
minimize, considering all possible combinatoric assignments, such as intermediate masses, distance metrics like $\Delta R$, and those used in jet clustering. The RJR algorithm uses many of the same concepts, with JR's designed to make combinatoric assignments which can be combined recursively to treat any number of unknowns.

For kinematic ambiguities associated with invisible or undetected particles, there is a large literature of strategies for measuring the properties of event kinematics despite missing information~
\cite{Lester:2007fq, Ross:2007rm, Cho:2007qv, Cho:2007dh, Gripaios:2007is, Barr:2007hy, Cheng:2008hk, Cho:2008cu, Tovey:2008ui, Nojiri:2008vq, Nojiri:2008hy, Barr:2008ba, Barr:2008hv, Cho:2008tj, Burns:2008va, Konar:2008ei, Kim:2009nq, Barr:2009wu, Alwall:2009zu, Barr:2009jv, Konar:2009wn, Han:2009ss, Cho:2009ve,  Barr:2010zj, Han:2012nm, Han:2012nr}. In these cases, the momentum and masses of invisible particles are unmeasured, and the functional dependence of kinematic quantities of interest on these unknowns must be mitigated. 

Some approaches advocate imposing mass constraints on events, solving the associated system of equations for any kinematic unknowns. While potentially useful, such approaches generally involve high-order polynomials, implying many solutions, sometimes with none guaranteed to be real~\cite{Gripaios:2011jm}. Aside from the practical difficulties involved in analyzing data when there is an ensemble of (potentially complex) solutions for each event, there is the additional complication
of using such an approach in the context of searches for new particle states, where it can be difficult to guess the masses to include in constraints, or impracticle to include interpretations 
for many ``test'' constraints. 

Alternatively, under-constrained degrees of freedom that are expected to be small can be ignored, constraining multiple unknowns simultaneously. The efficacy of observables obtained from such an approach depends on the accuracy of the approximation and, as for mass constraints, the existence of real solutions is not guaranteed~\cite{Rogan:2010kb}.

A collection of approaches involve expressing masses of interest as a function of kinematic unknowns, choosing them to minimize or maximize those masses, subject to desired constraints. Examples are $M_{T2}$~\cite{Lester:1999tx, Barr:2003rg}, $M_{CT}$~\cite{Nojiri:2007pq, Polesello:2009rn}, and a whole collection of similarly constructed ``singularity variables'' for estimating masses in final states with two invisible particles~\cite{Kim:2009si}. The key property of these observables is that, if the applied constraints hold, their distribution should bound the true mass from above or below (depending on the minimization or maximization 
being imposed). Many of the JR's in the RJR approach use this same strategy, with the important distinction that only a subset of unknowns are considered in each minimization/maximization, with potentially more than one used in a self-consistent reconstruction of a single event.

The observation that motivates the RJR technique is that there is not necessarily only one quantity of interest in each event, but rather, potentially many observables which are useful in concert, and depend on many of the same under-constrained degrees of freedom. This means that in order to ensure that the resulting basis of observables are maximally uncorrelated, care must be taken with how multiple unknowns are parameterized and resolved. In the RJR approach, this is achieved by considering how under-constrained degrees of freedom effect our determination of the {\it velocities} relating the different reference frames corresponding to the rest frames of the intermediate states in a given decay tree. We exploit the fact that the measured quantities in an event correspond to a known reference frame, the laboratory frame, and identify the subset of unknowns necessary only for determining the velocity relating it to the next reference frame appearing in the decay tree of interest. Any additional unknowns related to the velocities of subsequent decay frames are considered separately, with the factorization of unknowns repeated recursively through the entirety of a decay tree. 

In the RJR framework, an algorithm for resolving the unknowns in a single decay step is called a JR, or {\it jigsaw}. Each event is analyzed through the recursive application of a series of JR's, moving through a decay tree from the lab frame to the rest frame of each intermediate particle appearing in the event. The factorized approach to their use means JR's are also {\it interchangeable}; different JR's can be chosen to resolve the same unknowns, resulting in different behavior of the derived observables. By considering a subset of unknowns at a time, each JR isolates their effects to only a few observables. The RJR algorithm can be summarized as the application of the following steps:
\begin{enumerate}[noitemsep]
\item Choose a decay tree to impose on the event, including any intermediate particle states of interest, with measured and invisible particles appearing in the final decay steps.
\item Express the velocity relating the current reference frame to the next one(s) in the decay tree as a function of unknown and measured quantities and choose a JR to resolve these this missing information.
\item Proceed to the subsequent reference frame(s) in the decay tree and repeat (2) until the ends of the tree are reached.
\end{enumerate}
The applied algorithm for analyzing an event is determined by the JR's that are chosen. Each is based only on an abstraction of the event evaluated in a single reference frame, with at least one JR for each type of decay topology and set of unknowns that can be encountered. The recursive application of JR's ensures that observables corresponding to different reference frames (masses and decay angles) are maximally uncorrelated from those associated with the frames that precede and follow in the tree.

The RJR framework is simply a library of JR's which, like puzzle pieces, can be assembled to analyze events according to a chosen decay tree. Each JR resolves unknowns using different constraints and assumptions, with a customizable combination available for studying any process of interest.

\section{Notation}
\label{sec:notation}

Throughout this paper, four vectors are denoted in bold, with subscripts indicating the object, or group of objects, to which the four vector corresponds, and superscripts the reference frame the four vector is being evaluated in, such that \pfour{\V{a}}{\D{b}} represents the four vector of object \V{a}, evaluated in reference frame \D{b}. The energy and momentum components of this four vector are denoted by $\pfour{\V{a}}{\D{b}} \equiv \{ \E{\V{a}}{\D{b}}, \pthree{\V{a}}{\D{b}} \}$, with individual momentum components written \pone{\V{a},x}{\D{b}}.

For readers viewing this paper through a color-sensitive medium, the type of different objects, or collections of objects, is indicated in color: Blue, for objects like \V{a}, implies that they are ``visible'', with measurements of their four vectors in the lab frame assumed to come from detector reconstruction. ``Invisible'' particles, corresponding to those that escape detection, are green, while intermediate particle states are shown in red. These labels may also indicate groups of particles. If \V{a} represents many individual particles ($\V{a} \equiv \{ \V{a$_{i}$},\cdots,\V{a$_{n}$} \}$) then it can be used interchangeably to indicate the set of elements in the group or their four vector sum, with $\pfour{\V{a}}{\D{b}} = \sum^{\footnotesize\V{a}}_{i}\pfour{\V{a$_{i}$}}{\D{b}}$. Similarly, the label of an intermediate particle state is used to represent both the set of invisible and visible particles which follow from the decay of this particle and its four vector. If an intermediate particle \D{b} decays to a visible particle \V{a}, and invisible particle \I{c}, ``\D{b}'' may be used like $\D{b} \equiv \{ \V{a},\I{c} \}$ or $\pfour{\D{b}}{\D{d}} = \pfour{\V{a}}{\D{d}}+\pfour{\I{c}}{\D{d}}$. Additionally, \D{b} can also refer to the rest frame of the object \D{b}, with $\pfour{\D{b}}{\D{b}} = \{ \mass{\D{b}}{},\vec{0} \} $.

Velocities representing the relative motion of different references frames are expressed as \vbeta{\D{d}}{\D{b}}, indicating the velocity, in units $c$, of reference frame \D{b} in the rest frame of \D{d}. $\sbeta{\D{d}}{\D{b}} = |\vbeta{\D{d}}{\D{b}}|$ and $\sgamma{\D{d}}{\D{b}} = [1-(\sbeta{\D{d}}{\D{b}})^2]^{-1/2}$. The associated Lorentz transformation is \boost{\D{d}}{\D{b}} or \boost{\vbeta{\D{d}}{\D{b}}}{}, with $\boost{\D{d}}{\D{b}}\pfour{\V{a}}{\D{d}} = \pfour{\V{a}}{\D{b}}$.

Kinematic observables corresponding to physical quantities, like particle masses or decay angles, always refer to those calculated in the RJR approximate event reconstruction, unless otherwise noted. The ``true'' values of particles' masses, on an event-by-event basis (not pole masses), are written in lower case (\mass{\V{a}}{}), while masses calculated to approximate these correct masses are expressed in capital letters (\Mass{\V{a}}{}). Until the relevant JR's are defined, some observables may depend on unknown quantities which have yet to be specified. These dependencies may be shown explicitly, like $\Mass{\D{b}}{}(\pone{\I{c},z}{\lab})$, or omitted and discussed in the text. Once JR's have been chosen and applied, and all unknowns associated with a decay tree have been resolved, these observables represent those calculated with these choices. 

Throughout, particle notation is often simplified by referring to particles and anti-particles using the same symbol, or by omitting the charge of particles. 

\section{Event Generation}
\label{sec:restframes}

Each of the examples described in Sections~\ref{sec:Part1}-\ref{sec:Part4} corresponds to a specific production and decay tree, and the \texttt{RestFrames}~\cite{RestFrames} code package is used to generate and analyze Monte Carlo events for each process. Specific code for reproducing each of the examples described in this paper, including all of the figures, is included in the software distribution.

The~\texttt{RestFrames} package provides the ability to define fully-configurable decay trees to both generate and analyze reconstructed particle interactions. The event generator includes flat (isotropic) decay kinematics, with options for including phase-space effects from parton distribution functions and particle propagators. This is accomplished through a Markov Chain Monte Carlo (MCMC) scheme, with interchangeable modules available for different types of decays in the tree. No effects from hadronization or imperfect detector resolution are considered; in general, these shortcomings have small effects on the distributions of kinematic observables, as their resolution in correspondence to ``true'' quantities most strongly depends on the effects of kinematic unknowns. Of greater importance are realistic acceptance requirements on the final state particles assumed to have come from detector reconstruction, which can have large effects on kinematic distributions. These are included in the event generation. The examples in this paper are simulated assuming a $q\bar{q}$ initial state at the 13 TeV LHC, using numerical parameterizations of parton distribution functions~\cite{Martin:2009iq}.

Analysis of these events is done using the implementation of the RJR algorithm and JR's within~\texttt{RestFrames}. A configurable decay tree is defined for each {\it view} of the event, with corresponding JR's chosen to resolve any unknowns. \texttt{RestFrames} automates the process of determining whether a decay tree is valid, whether a sufficient number and type of JR's are specified to resolve all the included unknowns, and the order in which each of the JR's should be applied to the event. Each particle appearing at any stage in a decay tree is associated with a \texttt{RestFrames} object which, after an event has been analyzed, can be queried about the value of its mass, decay angles, and momentum, evaluated in any reference frame in the reconstructed event.

\section{Jigsaws for an invisible particle}
\label{sec:Part1}

Whenever a weakly interacting particle is produced in a collider event
and escapes undetected, it carries with it irrecoverable kinematic
information. In collider experiments, the masses of escaping particles are not directly
measured, and their momenta can only be inferred in dimensions where
the total momentum is constrained. In this section, we introduce the
RJR algorithm through simple examples of event topologies containing a
single invisible particle in the final state. The application
of JR's to resolve missing information associated with this particle
allows for the accurate extraction of a number of other useful pieces of
information from these events. 

\subsection{$W\rightarrow\ell\nu$ at a hadron collider}
\label{subsec:Part1_exampleA}

\begin{figure}[!htbp]
\centering
\includegraphics[width=.28\textwidth]{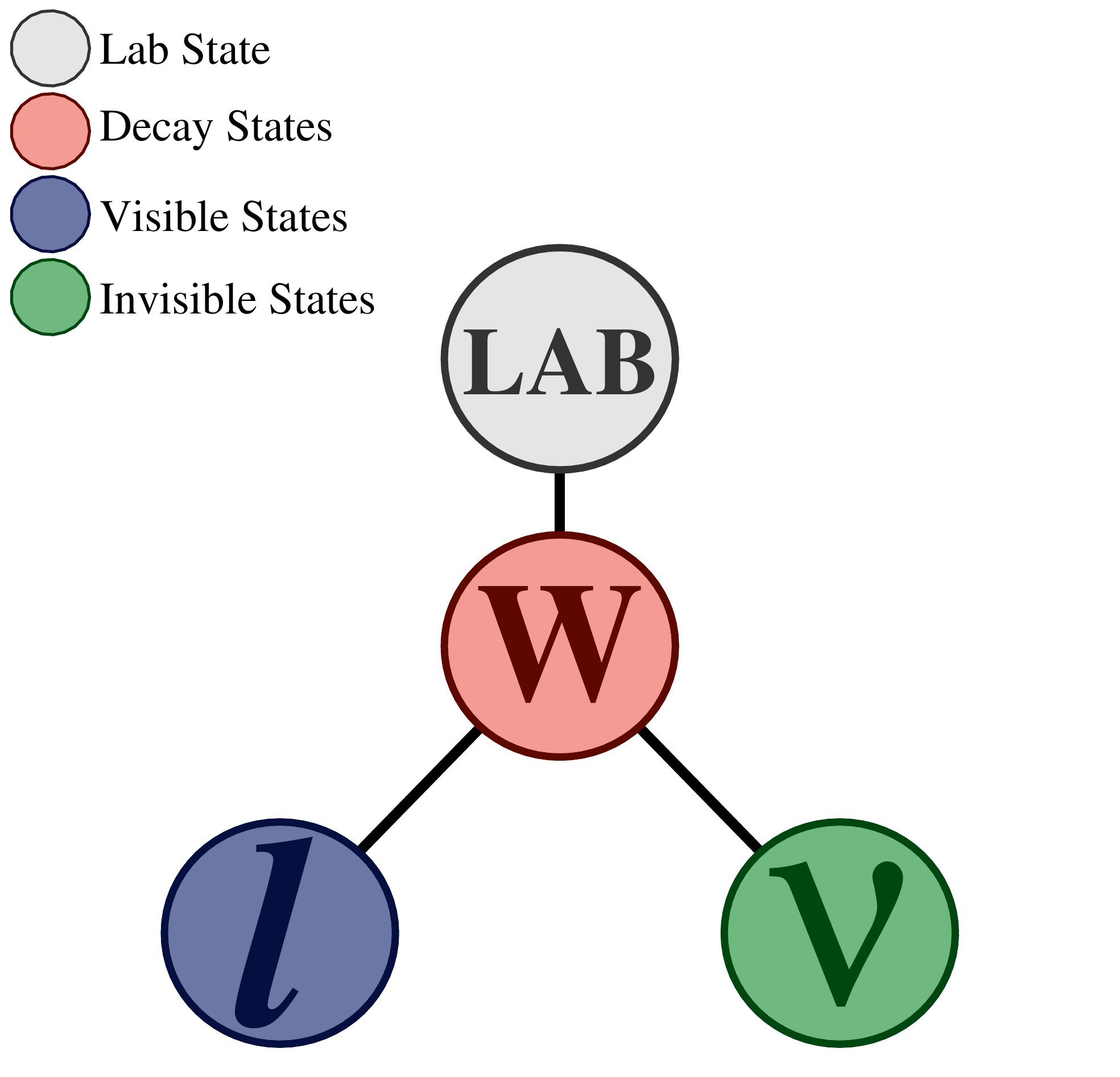}
\vspace{-0.3cm}
\caption{\label{fig:decayTree_Wlnu} A decay tree diagram for single
  $W$ production and decay. The $W$ boson is moving in the lab frame
  and decays to a lepton and neutrino.}
\end{figure}

A simple case involving a single invisible particle is the production
of a $W$ boson with decay $W\rightarrow\ell\nu$, with the decay tree
for this process shown in Fig.~\ref{fig:decayTree_Wlnu}. 

\FloatBarrier

We assume that the lepton is identified and reconstructed by a
detector, and that its four vector, \pfour{\V{$\ell$}}{\lab}, is
  measured in the lab frame. The neutrino escapes undetected,
  preventing a direct measurement of its energy and
  momentum. Exploiting conservation of momentum in the plane
  transverse to the beam axis, we interpret the measurement of the event
  missing transverse momentum, \met, as the transverse momentum of
  the neutrino, with \pthree{\I{$\nu$},T}{\lab} = \met.

The decay tree shown in Fig.~\ref{fig:decayTree_Wlnu} not only
describes the event, but also implicitly defines a kinematic basis of
useful observables based on this view of the event: the mass, decay
angles and lab frame momentum of the $W$ boson. Unfortunately, we are
unable to calculate any of these quantities from our measurements
\pfour{\V{$\ell$}}{\lab} and \pthree{\I{$\nu$},T}{\lab}, as they all
also require knowledge of the momentum of the neutrino along the beam
axis, \pone{\I{$\nu$},z}{\lab}, and its mass, \mass{\I{$\nu$}}{},
information which is carried with the exiting neutrino.

The shortcomings in our knowledge of the neutrino's kinematics leave
two options: we can either abandon our desired observable basis and restrict our analysis to the quantities we have measured or attempt to resolve, or guess, these unknown quantities using additional constraints. One could, for example, assume that there was an on-shell $W$ boson in the event, and require that the invariant mass of the lepton and neutrino be equal to its pole mass. Assuming the neutrino mass is zero, this approach will always result in two solutions for \pone{\I{$\nu$},z}{\lab}, with both potentially complex. Apart from the inconvenience of dealing with multiple and complex solutions, such an approach effectively trades any information about the mass of the $W$ boson for its momentum and decay kinematics, and is certainly only appropriate as far as its assumptions are valid. The RJR approach also resolves these same unknowns, but does so without any prior assumptions about the masses or decay kinematics of the event, instead using the decay tree to decompose each event into a basis of observables closely approximating those we are interested in.

This is accomplished by considering the under-constrained energy/momentum components of the neutrino instead as unknown components of the velocity relating the lab frame to the $W$ rest frame, \vbeta{\D{$W$}}{\lab}. Specifying this three vector is equivalent to choosing values \pone{\I{$\nu$},z}{\lab} and \mass{\I{$\nu$}}{}, and sufficient for calculating an approximation of any quantity of interest in the event:  It can be used to boost the measured lepton to the \D{$W$} frame, and to calculate the $W$ decay angle $(\mathrm{cos}~\theta_{\D{\it W}} = \hbeta{\D{\it W}}{\lab} \cdot \phat{\V{$\ell$}}{\D{\it W}})$, compare the azimuthal orientation of the $W$ decay plane to the $W$ momentum and beam axis plane ($\Delta \phi_{\D{\it W}} = \phat{\V{$\ell$}}{\D{\it W}} \times \hbeta{\D{\it W}}{\lab}~\angle~\hat{n}_{\mathrm{z}} \times \hbeta{\D{\it W}}{\lab}$), and measure the $W$ mass $(\E{\V{$\ell$}}{\D{\it W}} = \frac{\Mass{\D{\it W}}{2} - \mass{\I{$\nu$}}{2} + \mass{\V{$\ell$}}{2}}{2\Mass{\D{\it W}}{}})$. This is true in general, where the measured four vectors of visible particles and the velocities relating adjacent reference frames fully specify the kinematics of a decay tree.

When choosing the unknown components of \vbeta{\D{$W$}}{\lab} we decompose this velocity into two pieces: \sbeta{\D{$W$},z}{\lab}, the longitudinal boost to a reference frame, $\lab,z$, where $\pone{\D{$W$},z}{\lab,z} = 0$, and \vbeta{\D{$W$},T}{\lab,z}, the transverse velocity relating that intermediate frame to the $W$ rest frame. For the first of these velocities, rather than trying to make the most accurate guess (by, for example, using likelihoods sensitive to parton distribution functions) we choose a value that ensures that any quantities we calculate using our guess are {\it independent} of the true value. This is ensured by choosing \sbeta{\D{$W$},z}{\lab} to satisfy
\begin{equation}
\frac{\partial \E{\V{$\ell$}}{\lab,z}(\sbeta{\D{$W$},z}{\lab})}{\partial \sbeta{\D{$W$},z}{\lab}} = 0~,
\label{eqn:betaz}
\end{equation}
such that $\sbeta{\D{\it W},z}{\lab} = \pone{\V{$\ell$},z}{\lab}/\E{\V{$\ell$}}{\lab}$. This choice also sets \pone{\I{$\nu$},z}{\lab}, with $\pone{\I{$\nu$},z}{\lab,z} = \pone{\V{$\ell$},z}{\lab,z} = \pone{\D{$W$},z}{\lab,z} = 0$, and is equivalent to setting the rapidity of the neutrino equal to that of the lepton. This algorithmic determination of \sbeta{\D{\it W},z}{\lab} is an example of a JR, where a guess for an unknown quantity is made based on the four momenta of visible particles, evaluated in a particular reference frame: 

\begin{jigsaw}[Invisible Rapidity]
\label{jr:rapidity}
If the momentum of an invisible particle, \I{I}, in a reference frame, \D{F}, is unknown along an axis $\hat{n}_{\parallel}$, it can be chosen such that the rapidity of \I{I} along $\hat{n}_{\parallel}$ is set equal to that of a visible system of particles \V{V} according to:
\begin{equation}
\label{eqn:rapidity_jigsaw}
\pone{\I{I},\parallel}{\D{F}} = \pone{\V{V},\parallel}{\D{F}} \frac{\sqrt{|\pthree{\I{I},\perp}{\D{F}}|^{2} + \mass{\I{I}}{2}}}{\sqrt{|\pthree{\V{V},\perp}{\D{F}}|^{2} + \mass{\V{V}}{2}}}~,
\end{equation}
where ``$\perp$'' indicates the plane normal to $\hat{n}_{\parallel}$. This choice is equivalent to minimizing \Mass{\V{V}\I{I}}{} w.r.t. \pone{\I{I},\parallel}{\D{F}}. 
\end{jigsaw}

That the application of JR~\ref{jr:rapidity} ensures that all observables depending on $\sbeta{\D{\it W},z}{\lab}$ are independent of its true value (or, alternatively, that they are invariant under longitudinal boosts in the lab frame) can be understood intuitively by noting that our definition of the {\it \lab,z} reference frame is itself longitudinally boost invariant, in that we will always arrive at the same $\lab,z$ irrespective of the $W$ boson's velocity along the beam axis. As all subsequent estimators (and choices for other unknowns) follow from the choices made in this frame, in a sense observables associated with reference frames appearing below $\lab,z$ in the decay tree {\it inherit} this invariance property. 

With $\sbeta{\D{\it W},z}{\lab}$ specified, we need to guess the remaining velocity \vbeta{\D{$W$},T}{\lab,z}, whose magnitude depends on \mass{\I{$\nu$}}{}. We could attempt to specify this unknown using a similar approach to $\sbeta{\D{\it W},z}{\lab}$, for example choosing \sbeta{\D{$W$},T}{\lab,z} to minimize \Mass{\D{$W$}}{}. Unfortunately, this can result in unphysical values of \Mass{\I{$\nu$}}{}, with a tachyonic reconstructed neutrino. This is true in general for the individual masses of invisible particles, in that partial derivatives of derived quantities w.r.t. these masses are never guaranteed to be physically viable. In this case, we set $\Mass{\I{$\nu$}}{} = 0$ both because the neutrino mass is negligible on the scale of this event and because it is the smallest Lorentz invariant choice which guarantees a viable interpretation of the event. The importance of this latter distinction will become clear in later examples with multiple invisible particles in the final state. 

\begin{figure}[htb]
\centering 
\subfigure[]{\includegraphics[width=.28\textwidth]{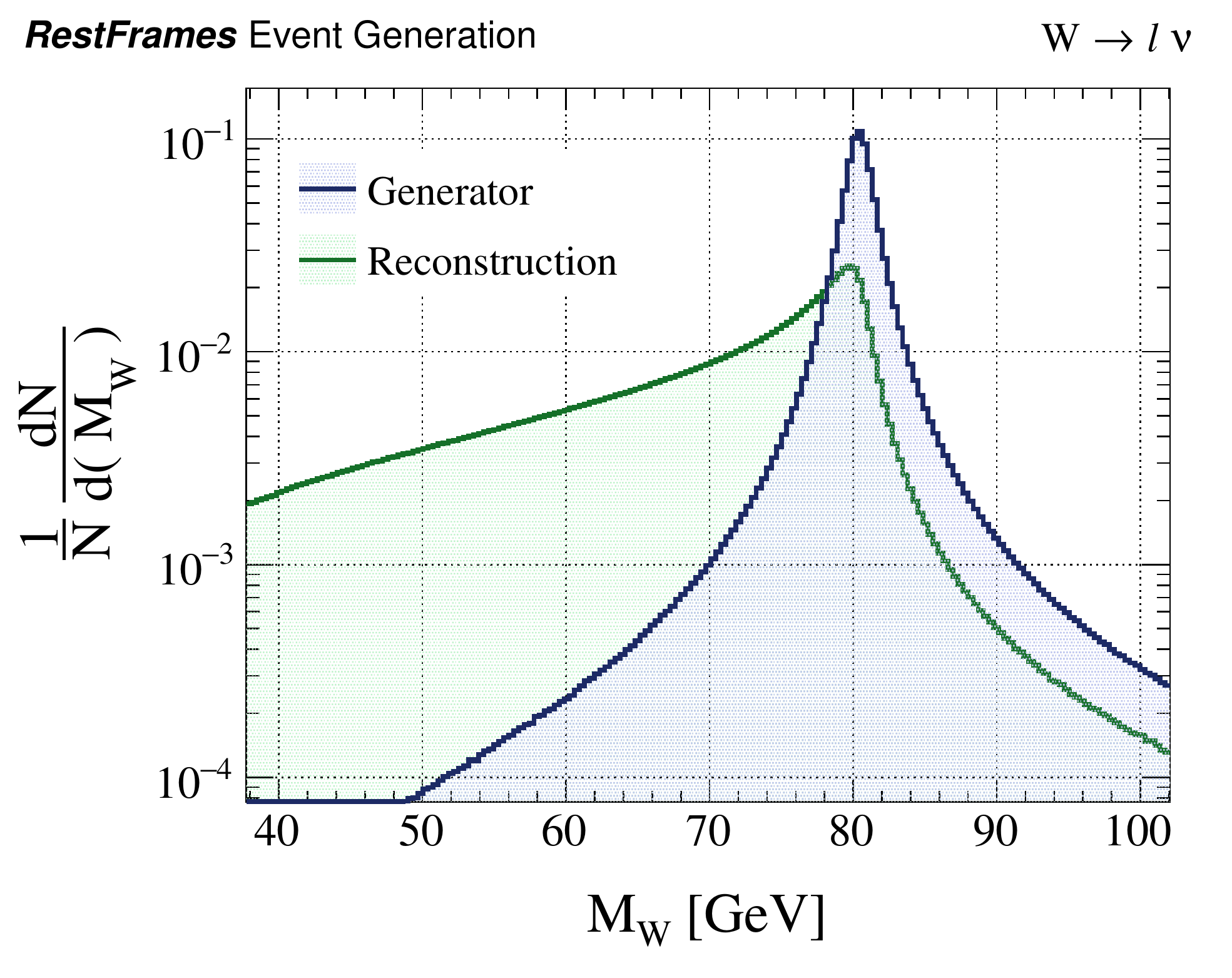}} \hspace{0.5cm}
\subfigure[]{\includegraphics[width=.28\textwidth]{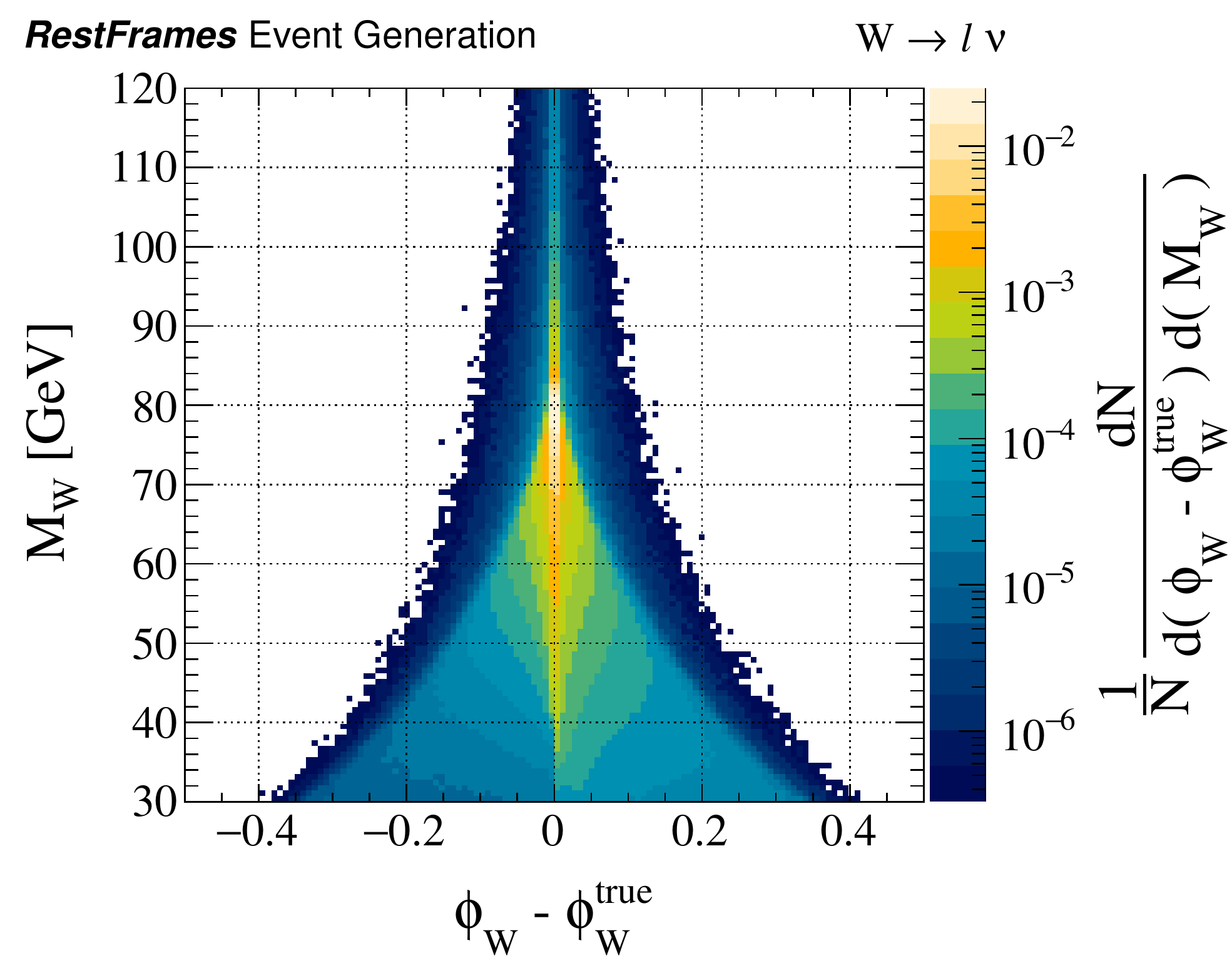}} \hspace{0.5cm}
\subfigure[]{\includegraphics[width=.28\textwidth]{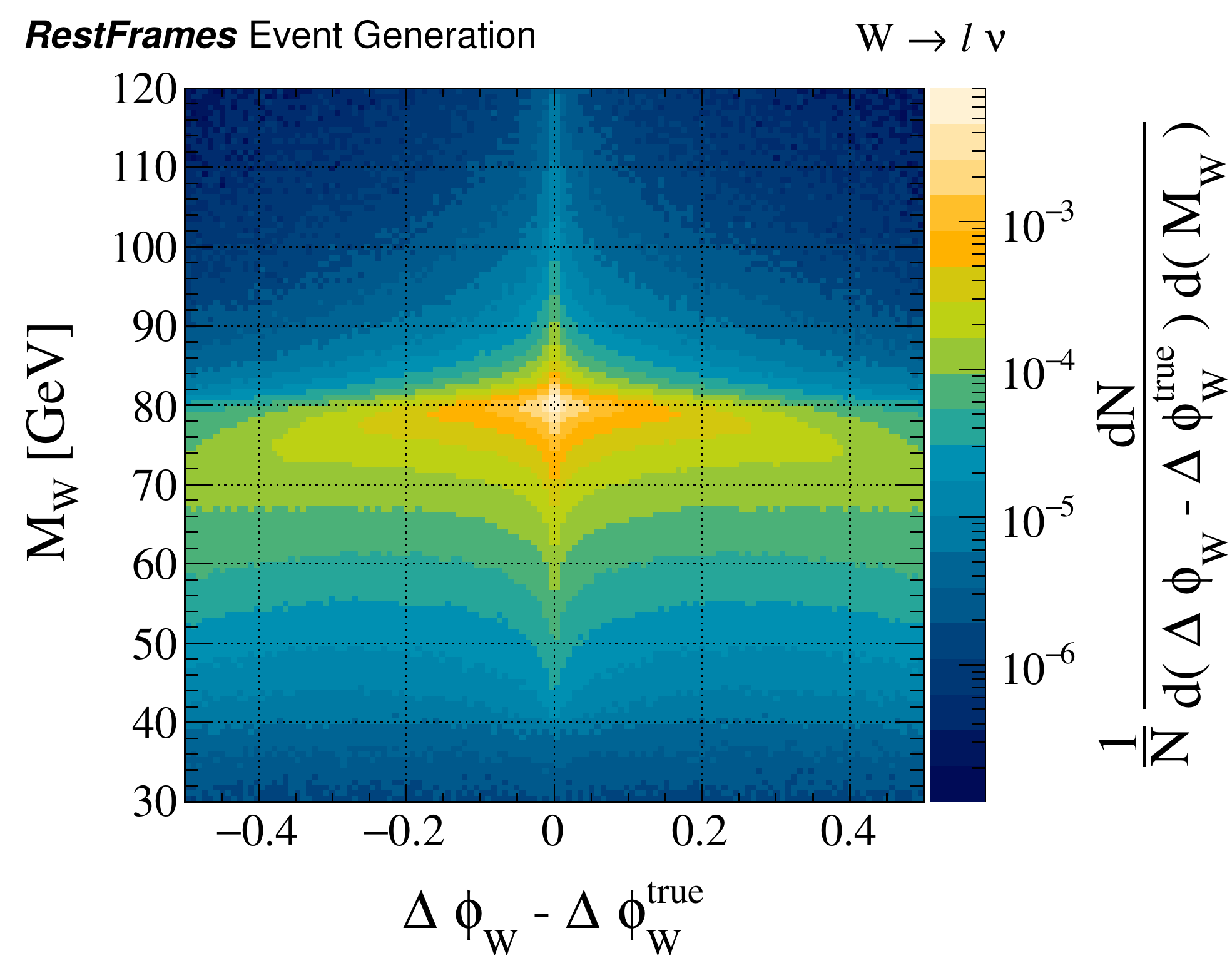}}
\vspace{-0.3cm}
\caption{\label{fig:example_Wlnu-MW} (a) Distribution of \Mass{\D{\it W}}{} for simulated $W \rightarrow \ell \nu$ events at both generator level and after reconstruction. Each distribution is normalized to unit area. Distributions of (b) azimuthal component of the $W$ decay angle, $\phi_{\D{\it W}}$, and (c) $\Delta \phi_{\D{\it W}}$, relative to their true values, compared with the reconstructed $W$ mass. Angles are shown in units radian.}
\end{figure}

Once values for the unknowns associated with the neutrino have been specified, we can calculate any observable of interest in our approximate view of the event. Our estimator for the $W$ mass can be expressed in terms of measurable quantities in the lab frame:
\bea
\label{eqn:MTW}
\Mass{\D{\it W}}{2} &=& \nonumber \\ \mass{\V{$\ell$}}{2} &+& 2\left( |\pthree{\I{$\nu$},T}{\lab}|\sqrt{\mass{\V{$\ell$}}{2} + |\pthree{\V{$\ell$},T}{\lab}|^{2}} - \pthree{\V{$\ell$},T}{\lab}\cdot\pthree{\I{$\nu$},T}{\lab} \right)~,
\eea
where we note that we have re-derived the transverse mass. The distribution of \Mass{\D{\it W}}{} is shown in Fig.~\ref{fig:example_Wlnu-MW}(a), where we observe $\Mass{\D{\it W}}{} \leq \mass{\D{\it W}}{}$, due to our implicit minimization in our choice of \pone{\I{$\nu$},z}{\lab}.\\

\onecolumngrid

\twocolumngrid 

In addition to \Mass{\D{\it W}}{}, we can also calculate estimators of other quantities of interest in the event. Distributions for the azimuthal component of the $W$ decay angle, $\phi_{\D{$W$}}$, and $\Delta \phi_{\D{\it W}}$ are shown in Fig.~\ref{fig:example_Wlnu-MW}(b,c), where the differences between the reconstructed and true quantities for these observables are examined as a function of \Mass{\D{\it W}}{}. We observe that both $\phi_{\D{$W$}}$ and $\Delta \phi_{\D{\it W}}$ can be reconstructed with excellent resolution, and largely independently of \Mass{\D{\it W}}{}, despite the missing information associated with the neutrino.

\begin{figure}[!htb]
\centering 
\subfigure[]{\includegraphics[width=.28\textwidth]{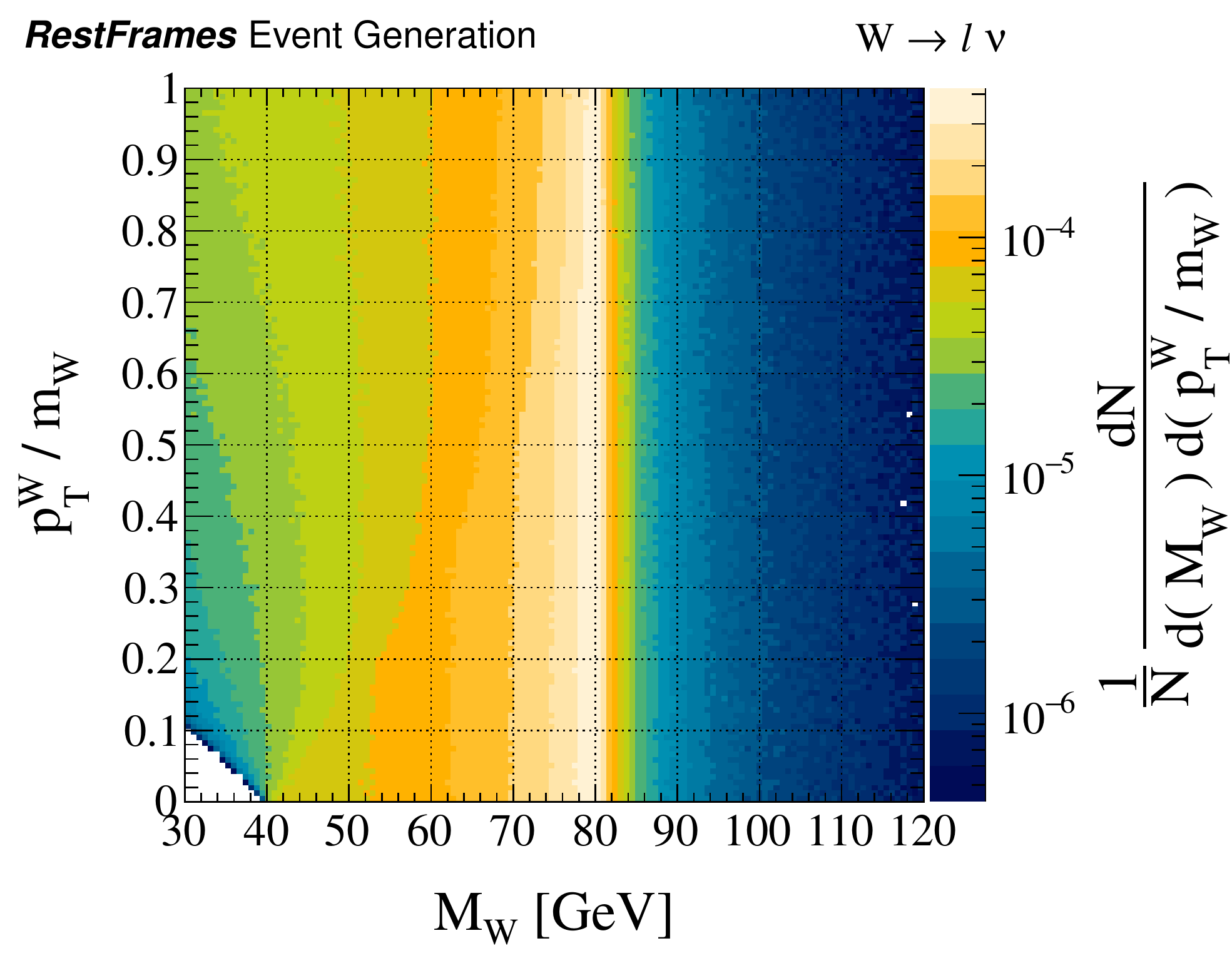}} \hspace{0.5cm}
\subfigure[]{\includegraphics[width=.28\textwidth]{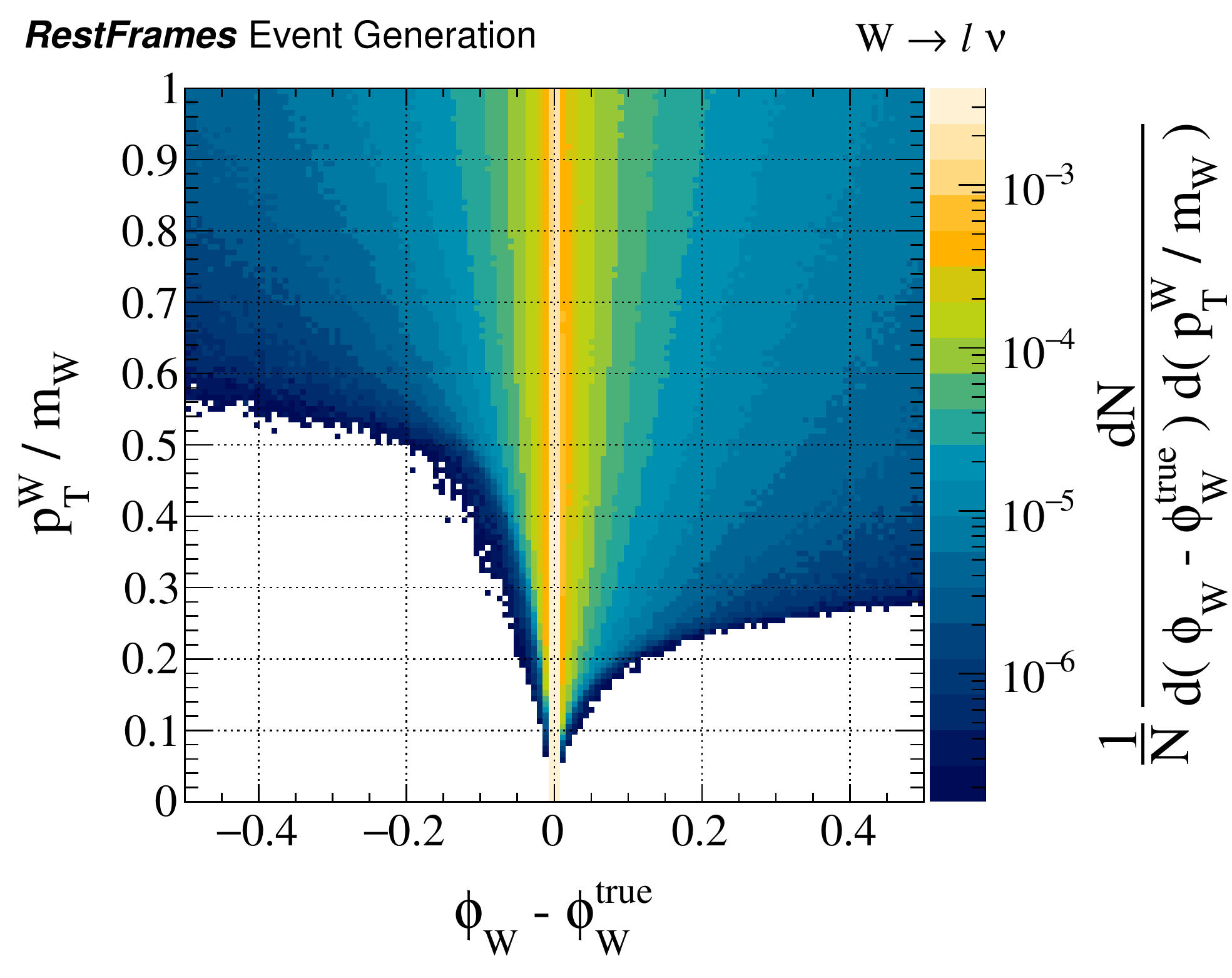}} \hspace{0.5cm}
\subfigure[]{\includegraphics[width=.28\textwidth]{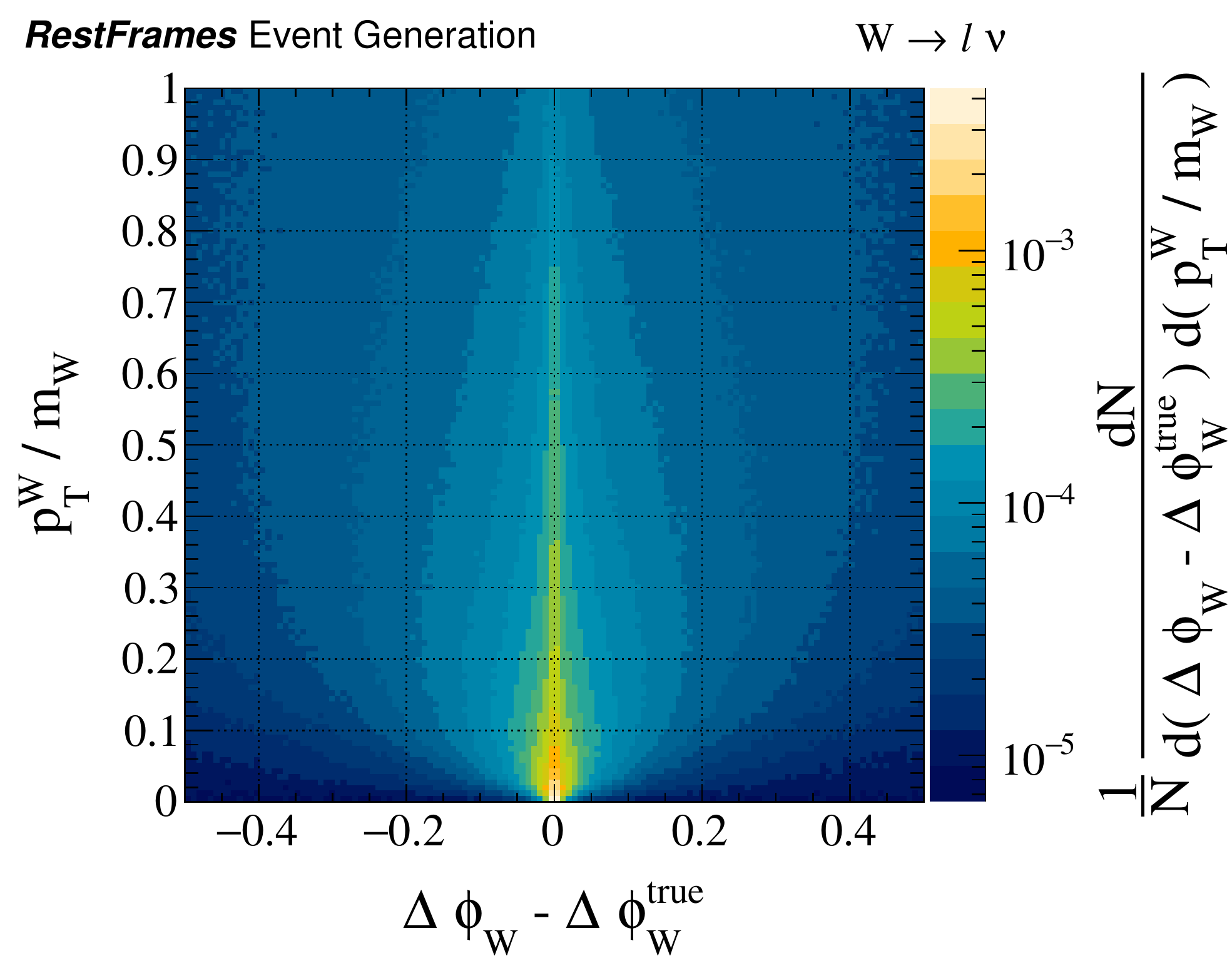}}
\vspace{-0.3cm}
\caption{\label{fig:example_Wlnu-PTW_v_phiW} Distributions of (a) the reconstructed $W$ mass, \Mass{\D{\it W}}{}, (b) the azimuthal component of the $W$ decay angle, $\phi_{\D{\it W}}$, and (c) $\Delta \phi_{\D{\it W}}$, shown as a function of $\pone{\D{\it W},T}{\lab}/\mass{\D{\it W}}{}$, for simulated $W\rightarrow \ell \nu$ events. Angles are shown in units radian.}
\end{figure}

Not only are these estimators nearly entirely independent from each other, but they are almost entirely insensitive to the true momentum of the $W$ boson in the lab frame. In addition to being manifestly invariant under longitudinal boosts, the reconstructed quantities \Mass{\D{\it W}}{}, $\phi_{\D{$W$}}$, and $\Delta \phi_{\D{\it W}}$ have almost no dependence on the $W$ transverse momentum, as demonstrated in Fig.~\ref{fig:example_Wlnu-PTW_v_phiW}. The RJR analysis prescription for these events results in a basis of observables, each with a strong correspondence to the true value of the quantity it is estimating, and little correlation between them introduced by the reconstruction algorithm. 
\onecolumngrid

\twocolumngrid 

\subsection{Single-top production with $t\rightarrow bW(\ell\nu)$}
\label{subsec:Part1_exampleB}
\begin{figure}[hb]
\centering 
\includegraphics[width=.28\textwidth]{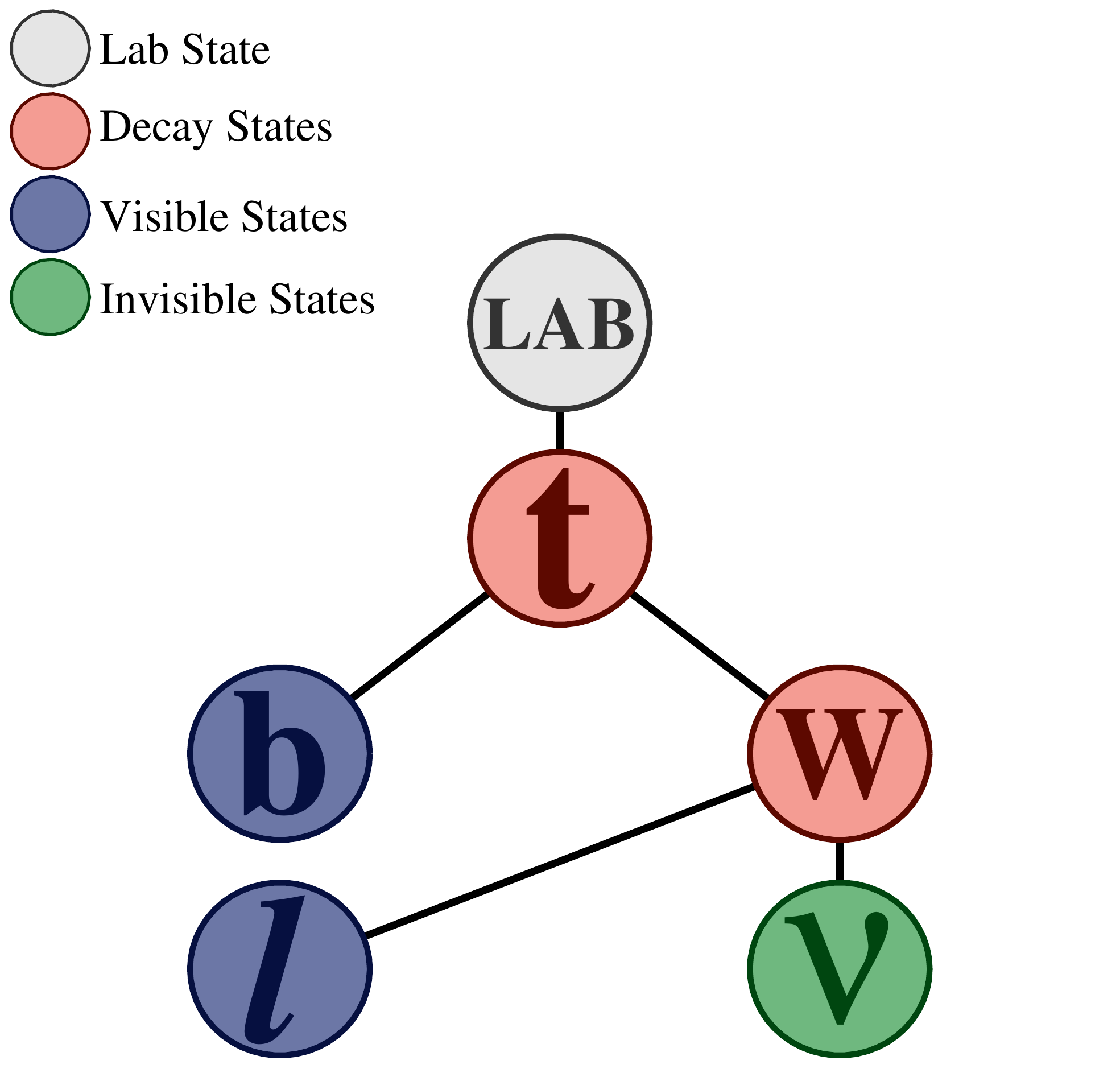}
\vspace{-0.3cm}
\caption{\label{fig:decayTree_tblnu} A decay tree diagram for a single top quark decaying to a $b$-quark, lepton, and neutrino, through an intermediate $W$ boson.}
\end{figure}

We expand on the previous example by considering the case of single-top production at a hadron collider, with subsequent decay to a $b$-quark, lepton, and neutrino, via an intermediate $W$ boson. The decay tree for analysis of this final state is shown in Fig.~\ref{fig:decayTree_tblnu}. As in Section~\ref{subsec:Part1_exampleA}, we assume that the lepton four vector, \pfour{\V{$\ell$}}{\lab}, and transverse momentum of the neutrino, \pthree{\I{$\nu$},T}{\lab}, are measured in the lab frame, with an additional $b$-tagged jet associated with the final state $b$-quark also reconstructed, with four vector \pfour{\V{$b$}}{\lab}.

\FloatBarrier

\onecolumngrid

\begin{figure}[htbp]
\centering 
\subfigure[]{\includegraphics[width=.238\textwidth]{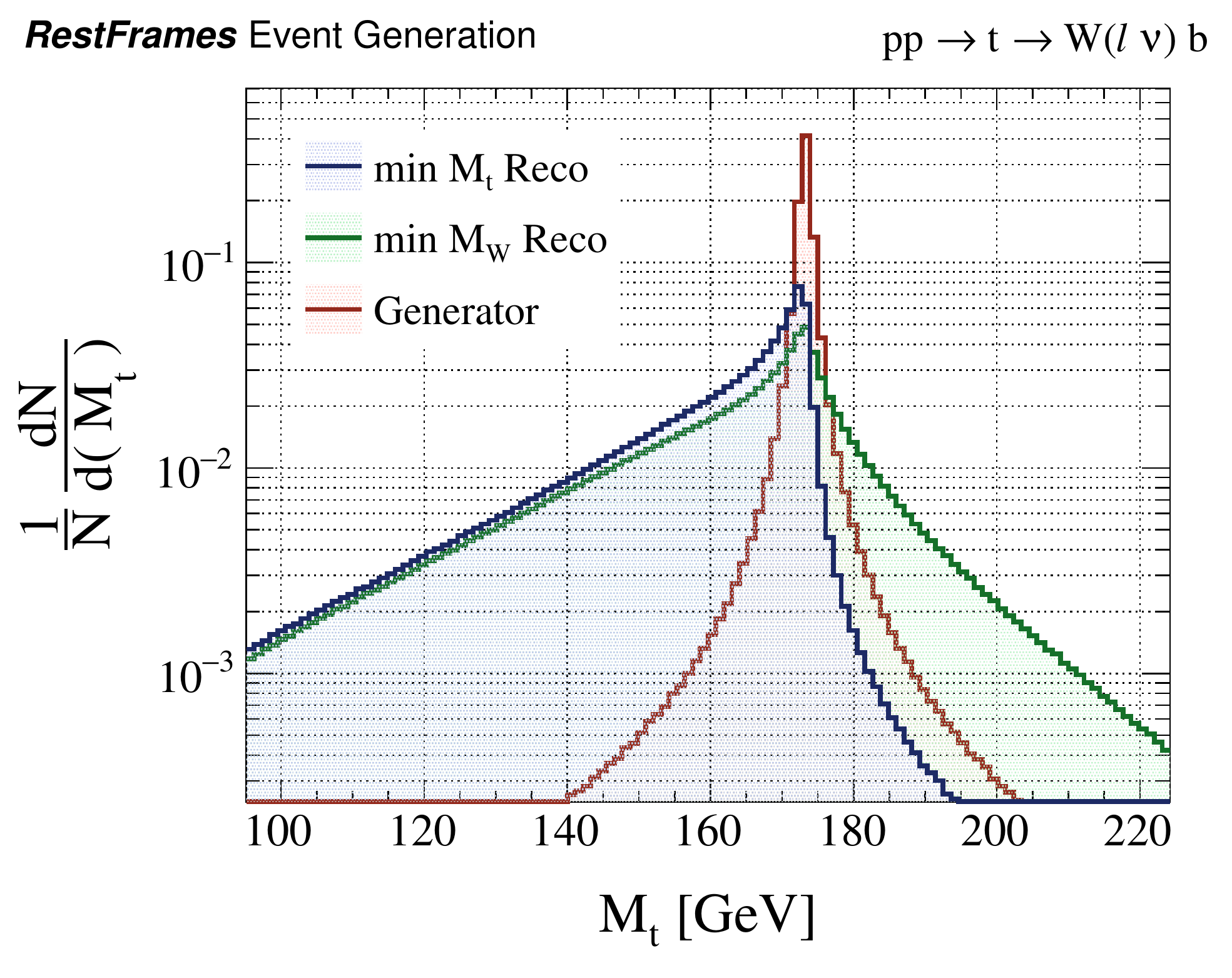}}
\subfigure[]{\includegraphics[width=.238\textwidth]{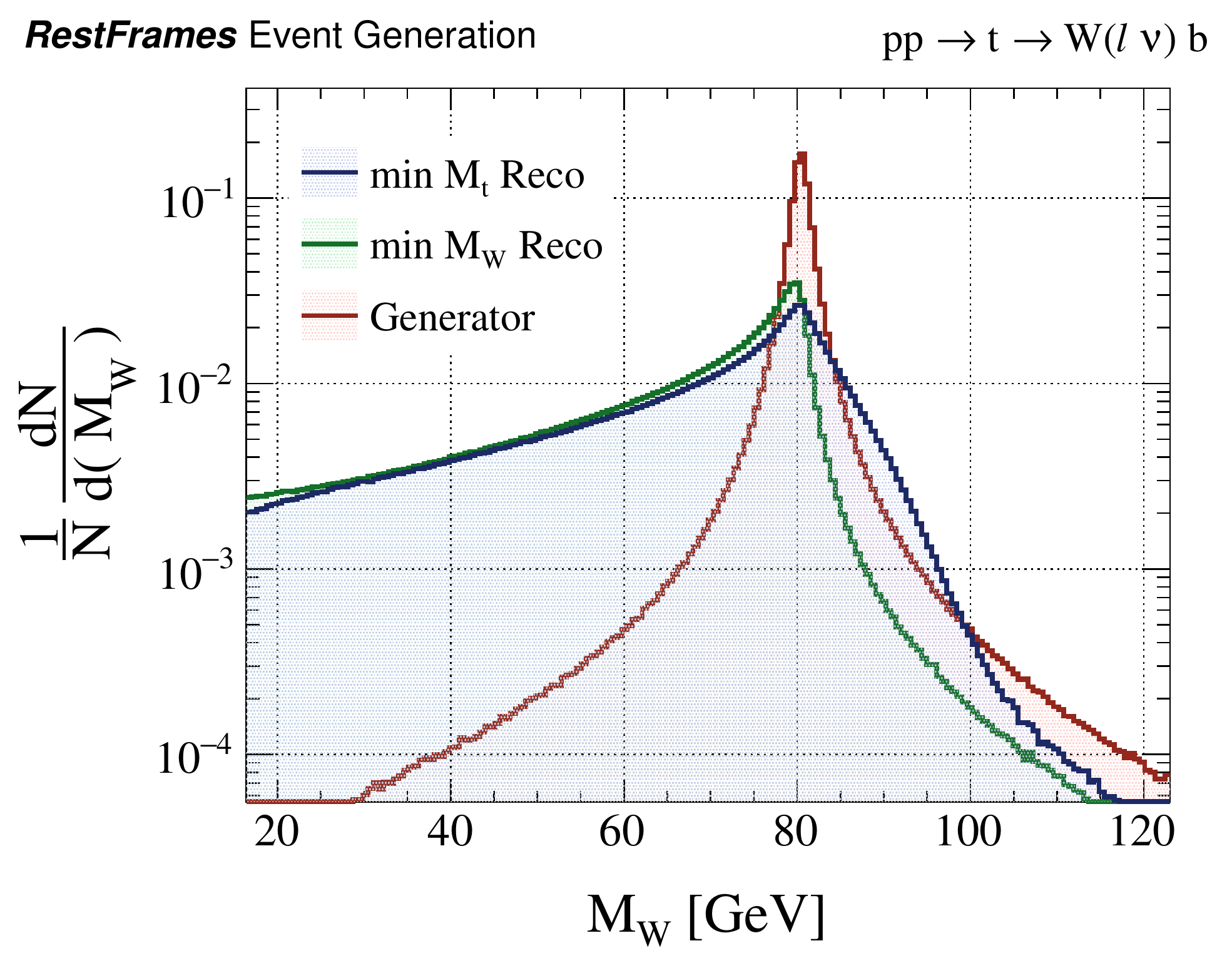}}
\subfigure[]{\includegraphics[width=.238\textwidth]{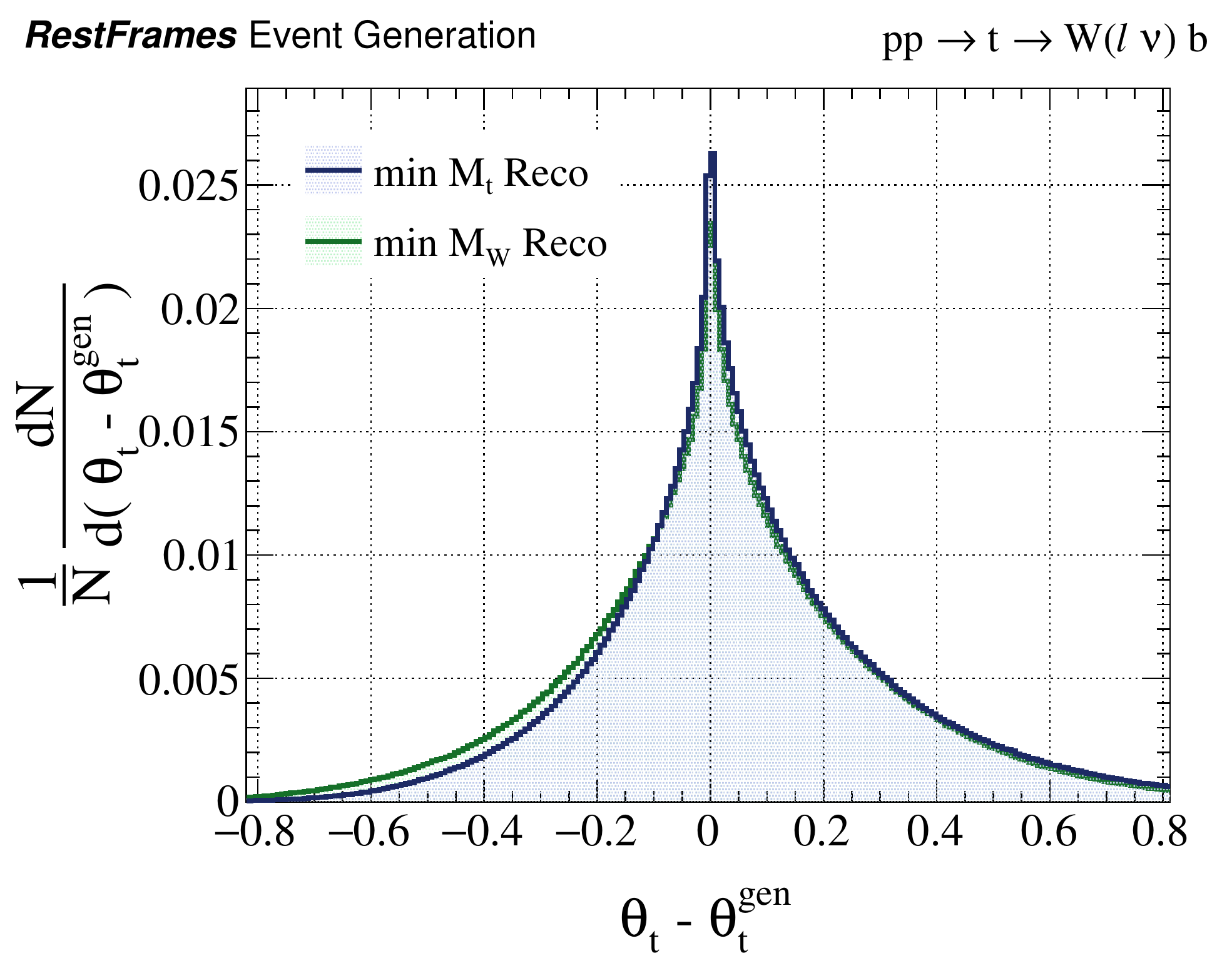}}
\subfigure[]{\includegraphics[width=.238\textwidth]{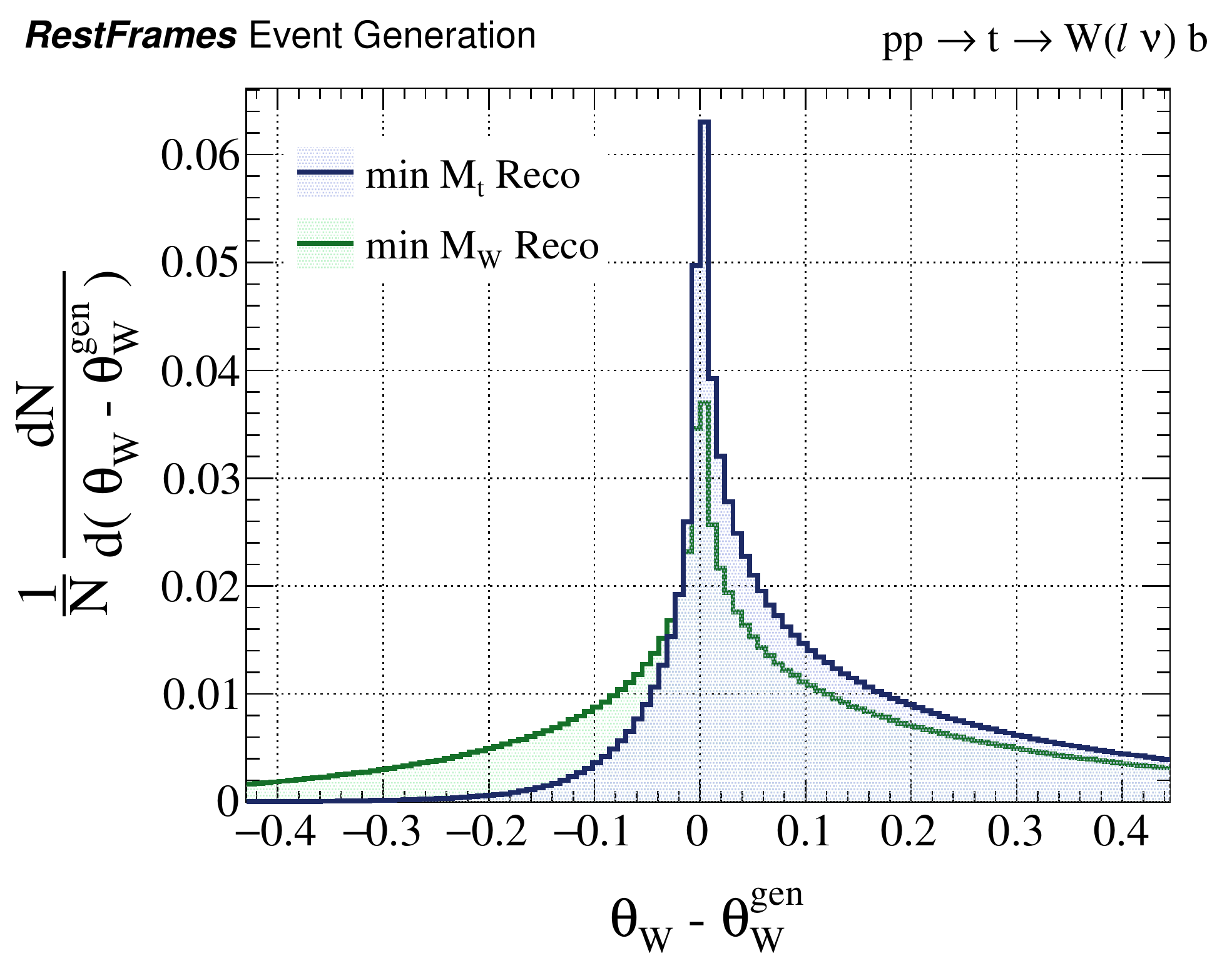}}
\vspace{-0.3cm}
\caption{\label{fig:example_top_to_bWlnu-masses} Distributions of reconstructed (a) top quark mass, \Mass{\D{$t$}}{},  (b) $W$ boson mass, \Mass{\D{$W$}}{}, (c) top quark decay angle, $\theta_{\D{$t$}}$, and (d) $W$ boson decay angle, $\theta_{\D{$W$}}$, for simulated $t\rightarrow bW(\ell\nu)$ events. Each mass is reconstructed in both the min $M_{W}$ and min $M_{t}$ schemes and compared with the true, generated values of these masses. Decay angles are shown relative to their true values in units radian.}
\end{figure}
\twocolumngrid

We are unable to calculate the masses and decay angles of the intermediate particles in the event due to missing information associated with the neutrino. With $\Mass{\I{$\nu$}}{} = 0$, we can set the neutrino's longitudinal momentum through an application of the invisible rapidity JR~\ref{jr:rapidity} except now, with two visible particles in the final state, we have a choice as to which combination to use in the JR. Setting the neutrino's rapidity equal to the lepton's implicitly chooses \pone{\I{$\nu$},z}{\lab} to minimize \Mass{\D{$W$}}{}, an approach which we will denote ``min $M_{W}$ reconstruction''. Alternatively, the set of both visible objects, $\V{V} = \{\V{$\ell$},\V{$b$}\}$, can be used in the JR, effectively minimizing \Mass{\D{$t$}}{} according to ``min $M_{t}$ reconstruction''. The distributions of the top and $W$ mass estimators, \Mass{\D{$t$}}{} and \Mass{\D{$W$}}{}, respectively, are shown in Fig.~\ref{fig:example_top_to_bWlnu-masses}(a,b) for these two reconstruction schemes. 

\onecolumngrid

\begin{figure}[!htbp]
\centering 
\subfigure[]{\includegraphics[width=.28\textwidth]{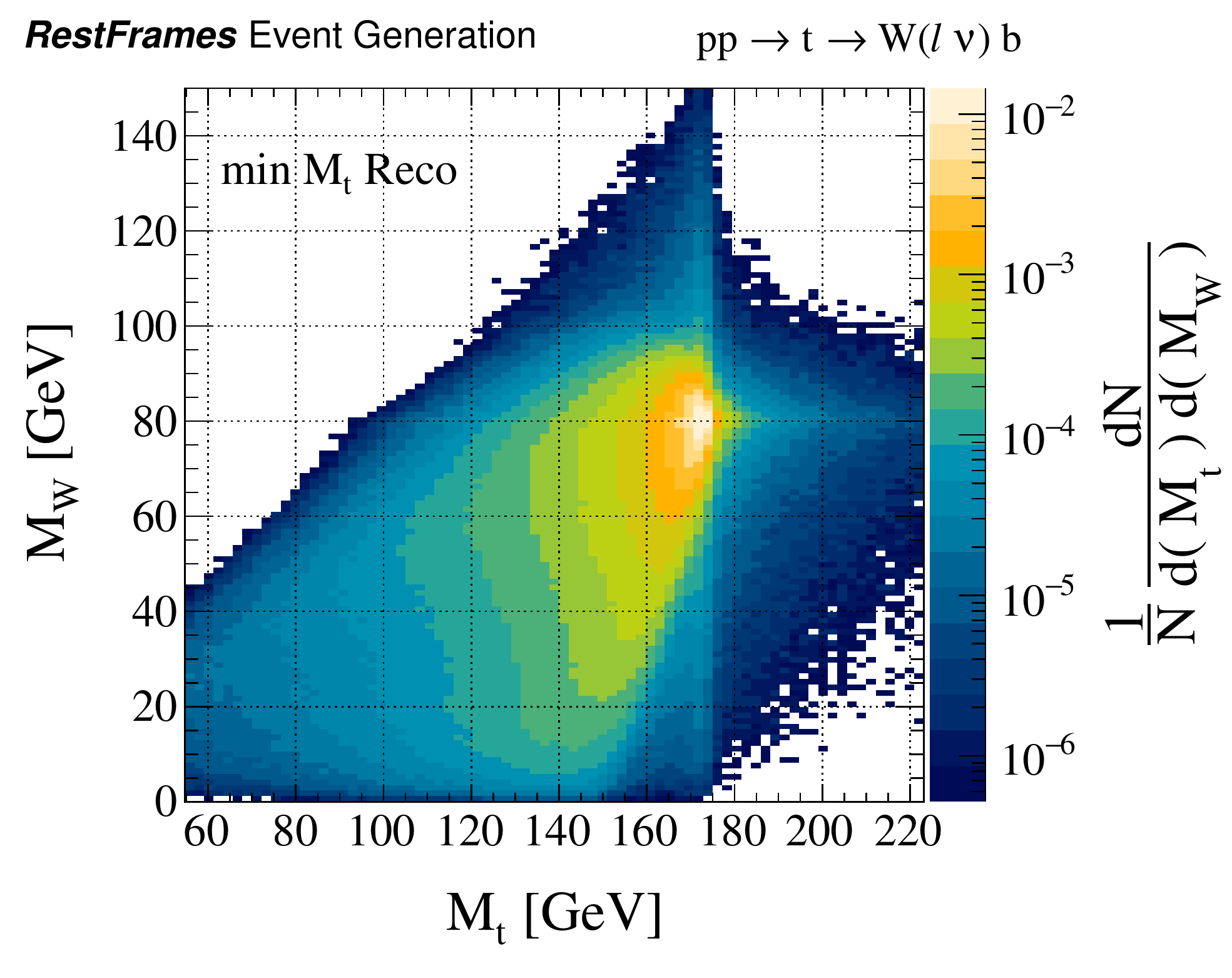}} \hspace{0.5cm}
\subfigure[]{\includegraphics[width=.28\textwidth]{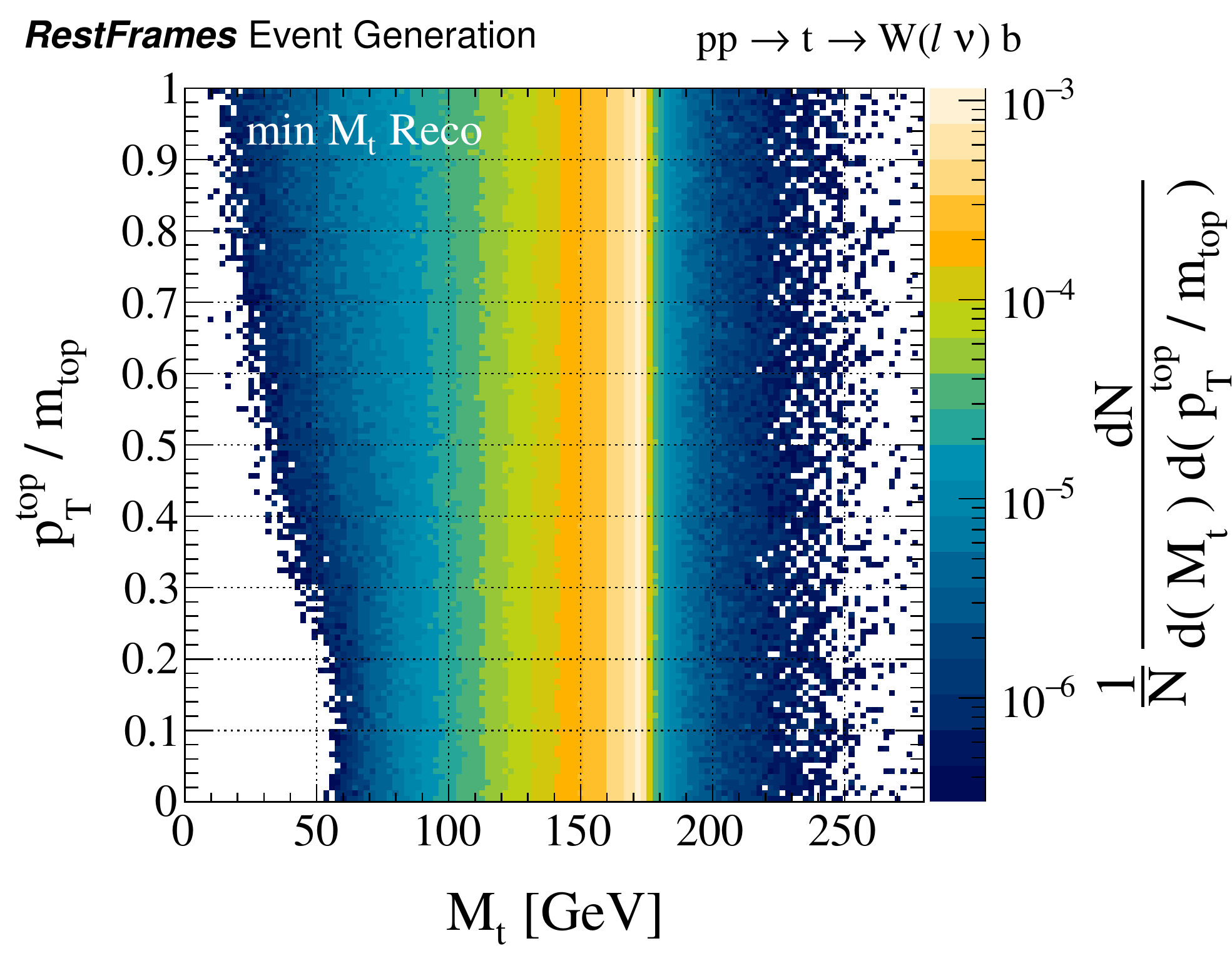}} \hspace{0.5cm}
\subfigure[]{\includegraphics[width=.28\textwidth]{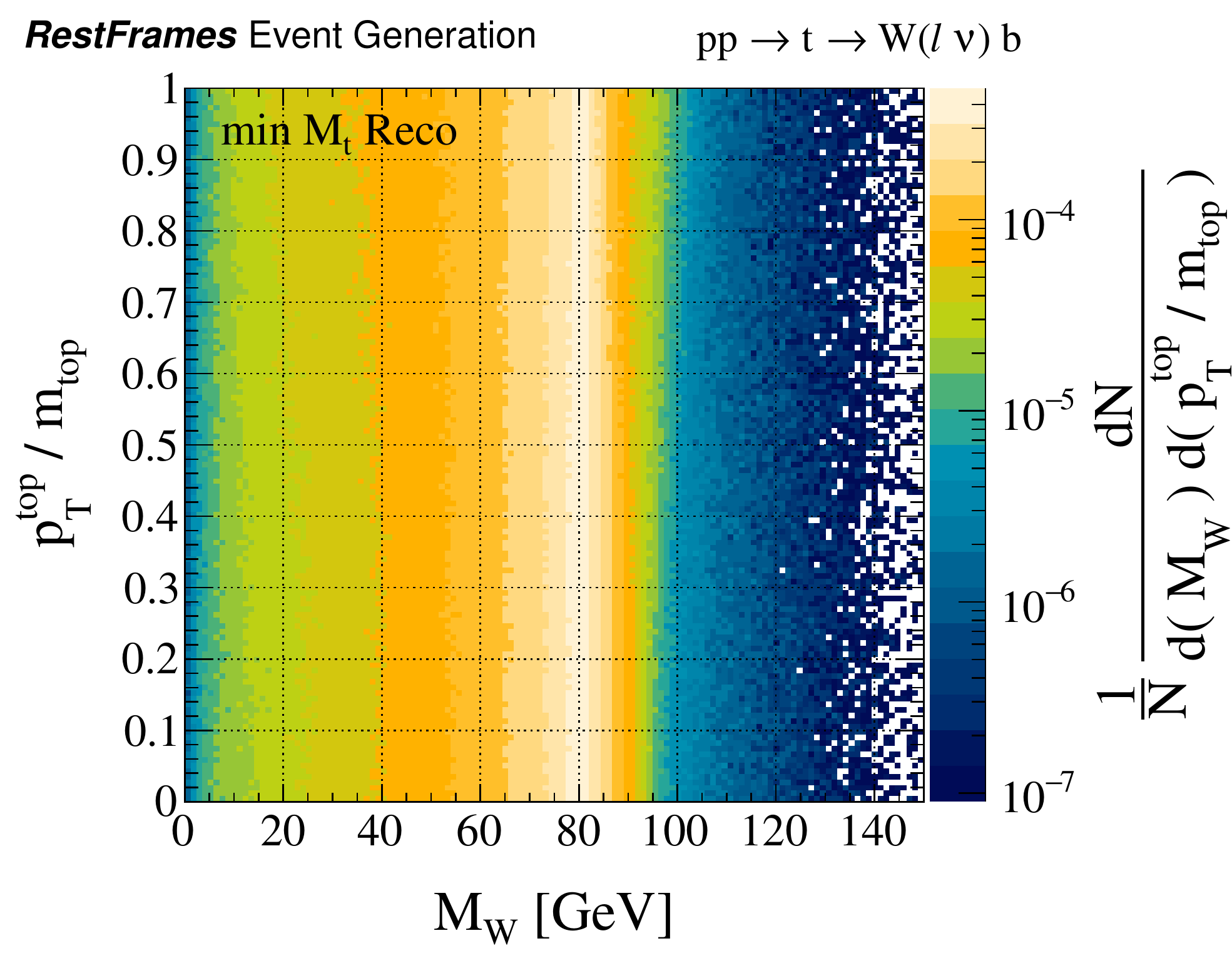}}
\vspace{-0.3cm}
\caption{\label{fig:example_top_to_bWlnu-2D} Distributions of (a) the $W$ boson mass estimator, \Mass{\D{\it W}}{}, as a function of the top mass estimator, \Mass{\D{\it t}}{}, (b) \Mass{\D{\it t}}{}, and (c) \Mass{\D{\it W}}{} as a function of $\pone{\D{\it t},T}{\lab}/\mass{\D{\it t}}{}$, for simulated $t\rightarrow bW(\ell\nu)$ events. The min $M_{t}$ reconstruction scheme is used for all observables.}
\end{figure}
\twocolumngrid 

In both reconstruction approaches, the mass estimator which is effectively minimized through the application of the invisible rapidity jigsaw corresponds to a transverse mass, with the characteristic Jacobian edges appearing at the true masses in the distributions of Fig.~\ref{fig:example_top_to_bWlnu-masses}(a,b). The distribution of the observable \Mass{\D{$t$}}{}, when calculated in the min $M_{W}$ scheme, exhibits a more pronounced tail at high values, an undesirable feature in the context of searches looking for new particles, with potentially larger masses than their SM counterparts, and similar decays. Other observables calculated using each approach exhibit noticeable differences in behavior, such as the top and $W$ boson decay angles, as seen in Fig.~\ref{fig:example_top_to_bWlnu-masses}(c,d). While the resolution of the reconstructed $\theta_{\D{$t$}}$ is similar between the two schemes, the $\theta_{\D{$t$}}$ resolution is better for the min $M_{t}$ approach, albeit with a larger bias. The improved resolution indicates that using both visible particles yields a generally more accurate estimation of \pone{\I{$\nu$},z}{\lab}.

When using the min $M_{t}$ strategy, the resulting \Mass{\D{$t$}}{} estimator takes the form of the transverse mass of the neutrino and visible particles:
\bea
\label{eqn:Mt_minMt}
\Mass{\D{$t$}}{2} &=& \\ \mass{\V{V}}{2} &+& 2\left( \sqrt{\mass{\V{V}}{2} + |\pthree{\V{V},T}{\lab}|^{2}}|\pthree{\I{$\nu$},T}{\lab}| - \pthree{\V{V},T}{\lab}\cdot\pthree{\I{$\nu$},T}{\lab} \right)~. \nonumber
\eea
Since we have not chosen to minimize \Mass{\D{\it W}}{} w.r.t \pone{\I{$\nu$},z}{\lab}, the expression for its estimator takes a different form:
\bea
\label{eqn:MW_minMt}
\Mass{\D{\it W}}{2} &=& 2\left( \E{\V{$\ell$}}{\lab}\E{\V{V}}{\lab} - \pone{\V{$\ell$},z}{\lab}\pone{\V{V},z}{\lab} \right) \frac{|\pthree{\I{$\nu$},T}{\lab}|}{\sqrt{\mass{\V{V}}{2}+|\pthree{\V{V},T}{\lab}|^{2}}} \nonumber \\
&+&\mass{\V{$\ell$}}{2} - 2\pthree{\V{$\ell$},T}{\lab}\cdot\pthree{\I{$\nu$},T}{\lab}~.
\eea
This expression is longitudinally boost invariant despite using information about the lepton and $b$-jet's longitudinal momentum in the lab frame. The advantage to using a single choice for \pone{\I{$\nu$},z}{\lab} when calculating \Mass{\D{\it t}}{} and \Mass{\D{\it W}}{}, rather than simply using the \Mass{\D{\it W}}{} transverse mass estimator of Eq.~\ref{eqn:MTW}, is that we can estimate the two masses with only small correlation, as demonstrated in Fig.~\ref{fig:example_top_to_bWlnu-2D}(a). This is despite their common dependence on \pone{\I{$\nu$},z}{\lab}, the source of residual correlation between the observables.

While the invariance to longitudinal boosts of the \Mass{\D{\it t}}{} and \Mass{\D{\it W}}{} observables is exact, they also exhibit little sensitivity to transverse boosts, as can be seen in Fig.~\ref{fig:example_top_to_bWlnu-2D}(b,c). This behavior is indicative of the fact that observables calculated in reference frames below the lab frame in the decay tree (intermediate particle masses, decay angles) are almost entirely insensitive to the top's velocity in the lab frame. Furthermore, the observables calculated in a particular frame are largely independent of those calculated in {\it all} other reference frames. This property makes the variables calculated in the RJR approach an excellent {\it basis} for studying processes, as it not only allows for relatively accurate estimations of many quantities of interest, but also independently. 

\FloatBarrier


\subsection{Heavy Charged Higgs production with $\boldmath{H^{+}\rightarrow W^{+}(\ell\nu)h^{0}(\gamma\gamma)}$ }
\label{subsec:Part1_exampleC}

\begin{figure}[htbp]
\centering 
\vspace{-0.3cm}
\includegraphics[width=.32\textwidth]{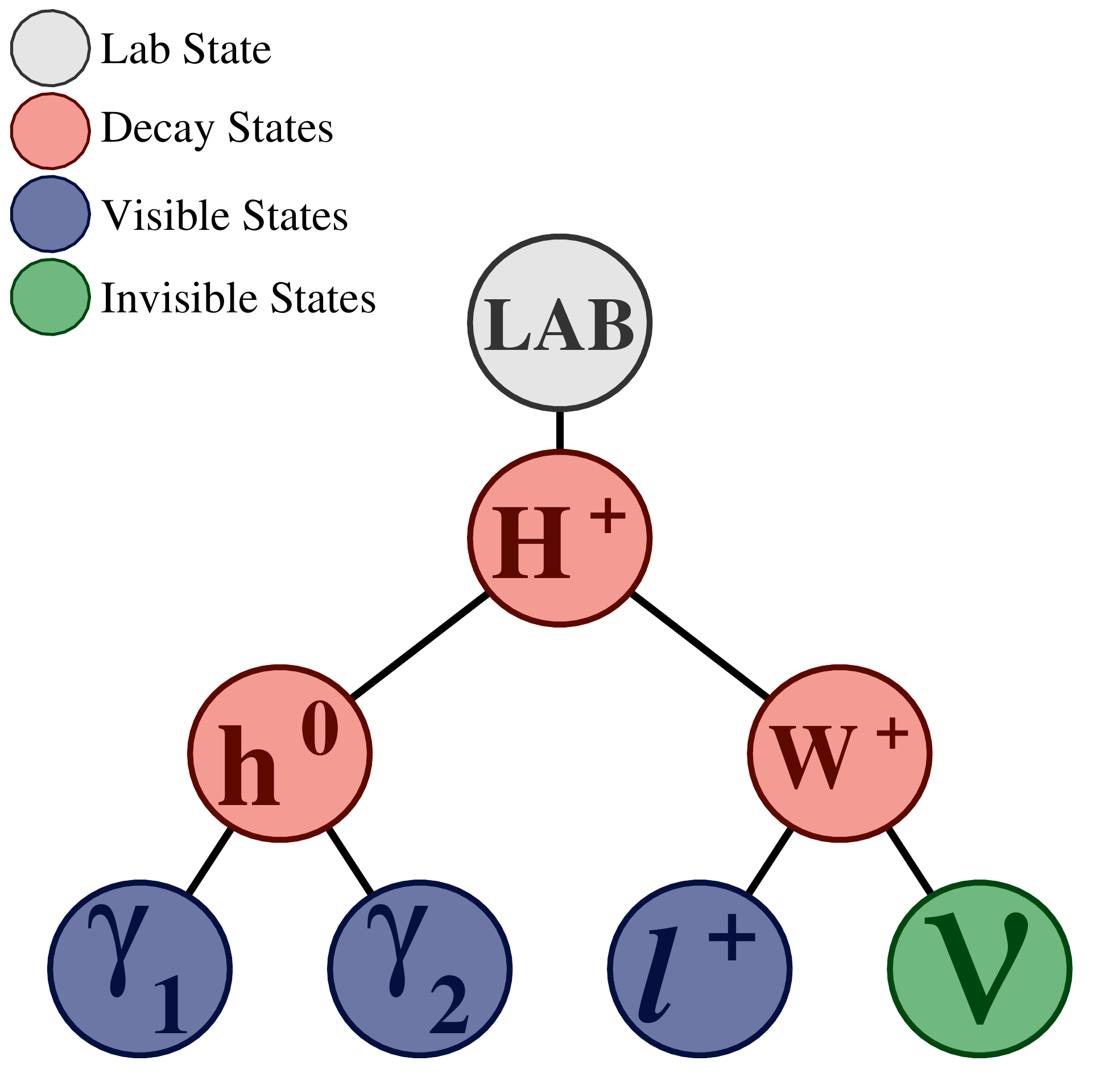}
\vspace{-0.3cm}
\caption{\label{fig:decayTree_HW} A decay tree diagram of a heavy charged Higgs boson, \D{$H^{+}$}, decaying to a neutral, SM-like Higgs boson, $h^{0}$, and a $W$ boson. The neutral Higgs decays to two photons while the $W$ decays to a lepton and neutrino. }
\end{figure}

Finally, we conclude our discussion of events with a single invisible particle by adding a small embellishment to the previous example. In this case, we consider the production of a heavy, charged Higgs boson (\D{$H^{+}$}) at a hadron collider, with the charged Higgs decaying to a neutral, SM-like Higgs (\D{$h^{0}$}) and a $W$ boson. The neutral Higgs decays to two photons, whose four vectors \pfour{\V{$\gamma_{1}$}}{\lab} and \pfour{\V{$\gamma_{2}$}}{\lab} are assumed to have been measured in the detector, while the $W$ decays to a lepton, with four vector \pfour{\V{$\ell$}}{\lab}, and a neutrino, whose transverse momentum is estimated from the \met. The decay tree for interpreting this final state is shown in Fig.~\ref{fig:decayTree_HW}.

\begin{figure}[htbp]
\centering 
\includegraphics[width=.35\textwidth]{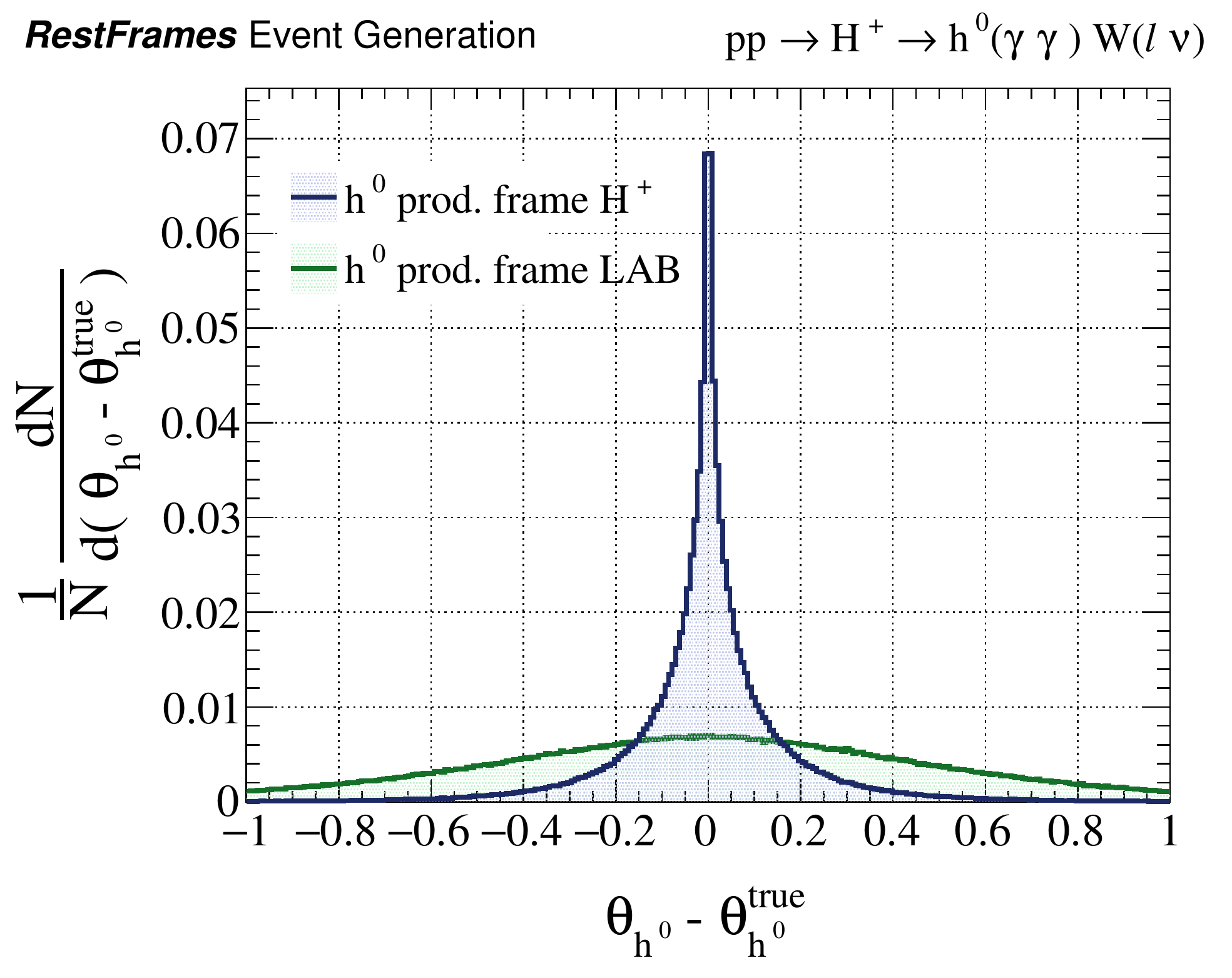}
\caption{\label{fig:example_Hp_to_HggWlnu-thetah} The decay angle of the neutral Higgs, cos$~\theta_{\D{$h^{0}$}}$, in simulated $\boldmath{H^{+}\rightarrow W^{+}(\ell\nu)h^{0}(\gamma\gamma)}$ events. Two different approaches to reconstructing cos$~\theta_{\D{$h^{0}$}}$ are used, with one using the RJR approximation of the \D{$h^{0}$} production frame and the other treating the lab frame as that frame. Angles are shown in units radian and the mass of the charged Higgs is chosen $\mass{\D{$H^{+}$}}{} = 750$ GeV.}
\end{figure}

As the missing information associated with the neutrino is identical to the previous two examples, we can resolve it by choosing the neutrino mass, $\Mass{\I{$\nu$}}{} = 0$, and applying the invisible rapidity JR~\ref{jr:rapidity} using the collection of all the visible particles, $\V{V} = \{ \V{$\ell$},  \V{$\gamma_{1}$}, \V{$\gamma_{2}$} \}$. With these choices, we are able to calculate estimators of each of the intermediate particle masses in each event, along with their decay angles. While the reconstructed neutral Higgs mass has no dependence on the neutrino kinematics, its decay angle does, as its calculation requires knowledge of the \D{$H^{+}$} rest frame. The RJR approach not only allows for accurate estimation of quantities related to invisible particles, but also angles like this. Fig.~\ref{fig:example_Hp_to_HggWlnu-thetah} demonstrates the resolution that can be achieved in the estimation of this angle for simulated events, comparing the RJR observable with a calculation of cos$~\theta_{\D{$h^{0}$}}$ using the lab frame as an approximation to the \D{$h^{0}$} production frame. The improvement in resolution when using the RJR approximations is dramatic.

\section{Jigsaws for two invisible particles}
\label{sec:Part2}
To this point, we have only considered cases with a single invisible particle in the final state, and only two relevant missing pieces of information associated with its momentum and mass. When there are two invisible particles, the amount of information lost with their escape from the detector increases, as we must now guess how momentum is shared between these particles in order to approximately reconstruct events. As for Section~\ref{sec:Part1}, we introduce new JR's for these cases through three examples, each with increasing complexity. 

\subsection{$H^{0}\rightarrow W^{+}(\ell\nu)W^{-}(\ell\nu)$ at a hadron collider}
\label{subsec:Part2_exampleA}

\begin{figure}[htbp]
\centering 
\includegraphics[width=.32\textwidth]{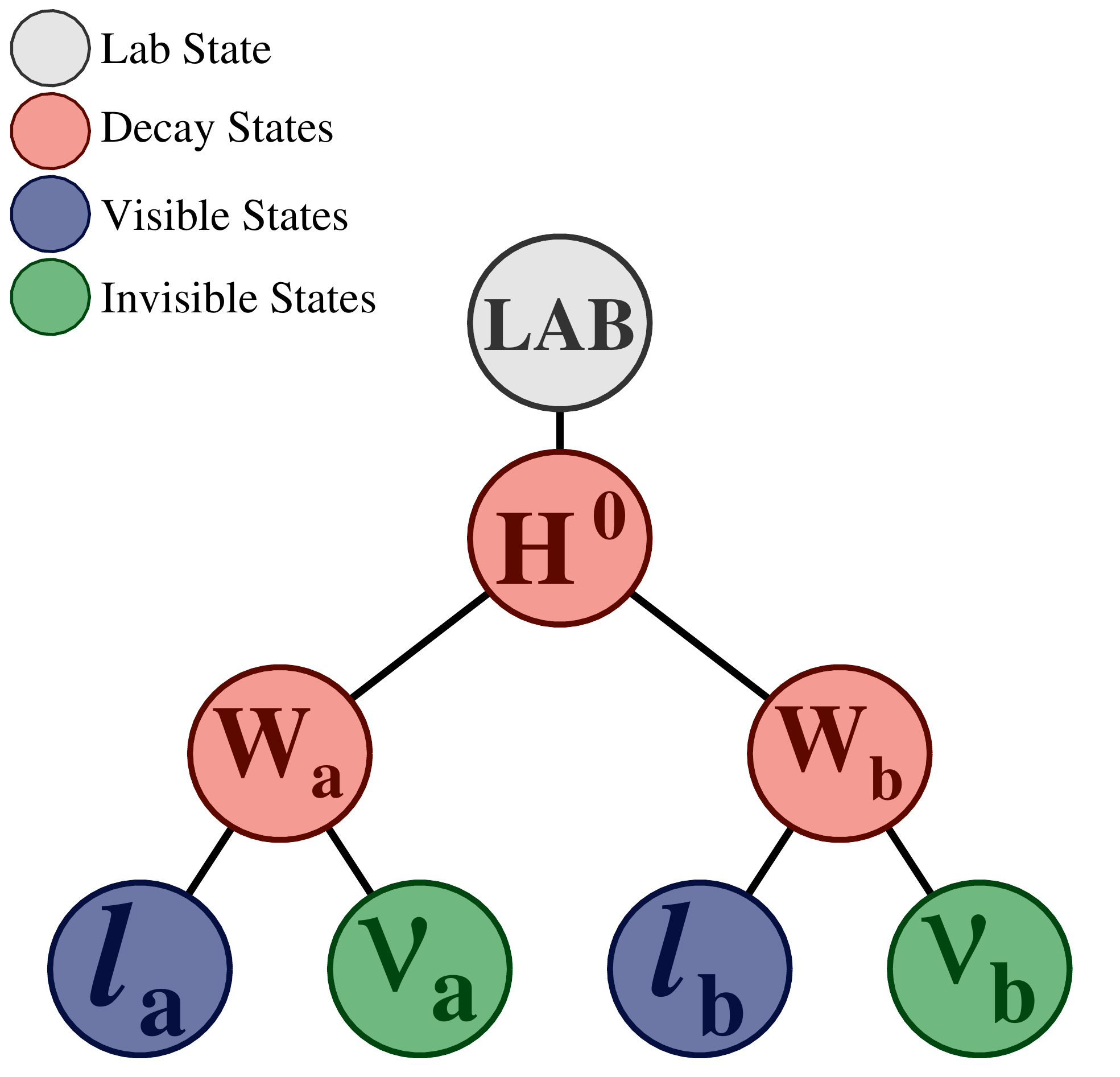}
\vspace{-0.3cm}
\caption{\label{fig:decayTree_HWW} The decay tree for analyzing $H^{0}\rightarrow W^{+}(\ell\nu)W^{-}(\ell\nu)$ events. The two sides of the event are labelled ``a" and ``b",  each including a lepton and neutrino from their respective $W$ decay.}
\end{figure}

We consider an example with production of a single, neutral Higgs boson (\D{$H^{0}$}) at a hadron collider, with decays to two $W$ bosons which, in-turn, each decay to a lepton and neutrino. The decay tree imposed on this final state is shown in Figure~\ref{fig:decayTree_HWW}. The two lepton four vectors, \pfour{\V{$\ell_{a}$}}{\lab} and \pfour{\V{$\ell_{b}$}}{\lab}, are assumed to have been measured and, working at a hadron collider, the measured \met~ is interpreted as the sum transverse momentum of the two neutrinos, such that
\begin{equation}
\pthree{\I{I},T}{\lab} = \pthree{\I{$\nu_{a}$},T}{\lab}+\pthree{\I{$\nu_{b}$},T}{\lab} = \met~,
\end{equation} 
where $\I{I} = \{ \nu_{a}, \nu_{b} \}$ is the set of all invisible particles in the event.

In the previous examples with only one invisible particle in the final state we were only missing two associated pieces of information: its momentum along the beam axis and its mass. Now, with two missing particles there are only two measured constraints on the eight degrees of freedom associated with the two neutrinos' four vectors. The total longitudinal momentum of the di-neutrino system, \pone{\I{I},z}{\lab}, can be chosen using the invisible rapidity JR~\ref{jr:rapidity} with the set of visible particles in the event, $\V{V} = \{ \V{$\ell_{a}$},\V{$\ell_{b}$} \}$, leaving only the mass of the neutrino system and how it is shared between the momentum of the two neutrinos to specify. We would like the JR which resolves these quantities to result in observables which, as much as possible, are independent of those we can already calculate related to the Higgs momentum in the lab frame. This is achieved by basing the determination of these additional neutrino unknowns on the four vectors of the visible leptons evaluated in the hypothetical, and yet-to-be determined, \D{$H$} rest frame.

Working in this reference frame, the momentum of the \I{I} system must be equal and opposite to that of the \V{V} system, as it is the center-of-mass frame of all the final state objects. Similarly, the momentum of the $W$ bosons must also be equal and opposite. With only these constraints, there are many different ways to choose the unknown individual momenta of the two neutrinos. The RJR approach is to consider these unknowns as the components of the velocities relating the two $W$ rest frames to this \D{$H$} frame, \vbeta{\D{$W_{a}$}}{\D{$H$}} and \vbeta{\D{$W_{b}$}}{\D{$H$}}. Ideally, we would like our derived estimators to be independent of the true values of these velocities, in a manner similar to longitudinal boost invariance through the invisible rapidity JR~\ref{jr:rapidity}. To achieve this, we introduce the additional constraint that the two $W$ bosons have the same mass. While this assumption may not be unreasonable in this particular case (although not when $\mass{\D{$H$}}{} < 2 \mass{\D{$W$}}{}$, and one of the $W$'s will likely be forced off-shell) this choice is primarily motivated by it allowing the two $W$'s velocities to be written in terms of a single vector:
\begin{equation}
\vec{\beta}_{c} = \vbeta{\D{$W_{a}$}}{\D{$H$}} = -\vbeta{\D{$W_{b}$}}{\D{$H$}}~.
\end{equation}
In order to ensure that our common $W$ mass estimator, \Mass{\D{$W$}}{}, is independent of $\vec{\beta}_{c}$, we minimize it w.r.t. this unknown quantity. Neglecting the mass of the leptons and neutrinos, the dependence of \Mass{\D{$W$}}{} on $\vec{\beta}_{c}$ can be expressed through the relation 
\bea
\Mass{\D{$W$}}{} &=& \E{\V{$\ell_{a}$}}{\D{$W_{a}$}} + \E{\V{$\ell_{b}$}}{\D{$W_{b}$}} \nonumber \\
&=& \gamma_{c}\left(\E{\V{$\ell_{a}$}}{\D{$H$}} + \E{\V{$\ell_{b}$}}{\D{$H$}} - \vec{\beta}_{c} \cdot (\pthree{\V{$\ell_{a}$}}{\D{$H$}} - \pthree{\V{$\ell_{b}$}}{\D{$H$}})\right)~.
\eea
 We choose $\vec{\beta}_{c}$ to satisfy
 \begin{equation}
 \frac{\partial \left(\E{\V{$\ell_{a}$}}{\D{$W_{a}$}} + \E{\V{$\ell_{b}$}}{\D{$W_{b}$}}\right)}{\partial \vec{\beta}_{c}} = 0~,
 \end{equation}
 which results in 
 \begin{equation}
 \vec{\beta}_{c} = \frac{\pthree{\V{$\ell_{a}$}}{\D{$H$}} - \pthree{\V{$\ell_{b}$}}{\D{$H$}}}{\E{\V{$\ell_{a}$}}{\D{$H$}} - \E{\V{$\ell_{b}$}}{\D{$H$}}}~.
 \end{equation}

With this choice, a quantity that appears in expressions for derived estimators is the {\it contra-boost invariant}, or Euclidean mass:
\begin{equation}
M_{c}^2(\pfour{\V{$\ell_{a}$}}{\D{$H$}},\pfour{\V{$\ell_{b}$}}{\D{$H$}}) = 2\left( \E{\V{$\ell_{a}$}}{\D{$H$}} \E{\V{$\ell_{b}$}}{\D{$H$}} + \pthree{\V{$\ell_{a}$}}{\D{$H$}}\cdot \pthree{\V{$\ell_{b}$}}{\D{$H$}}\right)~.
\end{equation}
$M_{c}^2$, the inner product of two four vectors with a Euclidean metric, is unchanged under the application of any contra-boost $\vec{\beta}_{c}$ such that $M_{c}^2(\pfour{a}{},\pfour{b}{}) = M_{c}^2(\boost{\vec{\beta}_{c}}{}\pfour{a}{},\boost{-\vec{\beta}_{c}}{}\pfour{b}{})$ for any two four vectors. The energy of each of the leptons, evaluated in their respective $W$ production frames, can be expressed as:
\begin{equation}
\E{\V{$\ell_{i}$}}{\D{$W_{i}$}} = \frac{\mass{\V{$\ell_{i}$}}{2} + \frac{1}{2}M_{c}^{2}(\pfour{\V{$\ell_{a}$}}{\D{$H$}}, \pfour{\V{$\ell_{b}$}}{\D{$H$}})}{\sqrt{\mass{\V{$\ell_{a}$}}{2} + \mass{\V{$\ell_{b}$}}{2} + M_{c}^{2}(\pfour{\V{$\ell_{a}$}}{\D{$H$}}, \pfour{\V{$\ell_{b}$}}{\D{$H$}})}}~,
\end{equation}
which is manifestly contra-boost invariant, as desired. 

After choosing $\vec{\beta}_{c}$, there is still one unspecified degree of freedom associated with the individual neutrino, and di-neutrino, masses. Expressing this remaining parameter as $c$, we can write
\bea
\label{eqn:numasses}
\Mass{\I{$\nu_{a}$}}{2} &=& (c-1)^{2}\mass{\V{$\ell_{a}$}}{2} + c^{2}\mass{\V{$\ell_{b}$}}{2} + (c-1)cM_{c}^{2}(\pfour{\V{$\ell_{a}$}}{\D{$H$}}, \pfour{\V{$\ell_{b}$}}{\D{$H$}}) \nonumber \\
\Mass{\I{$\nu_{b}$}}{2} &=& (c-1)^{2}\mass{\V{$\ell_{b}$}}{2} + c^{2}\mass{\V{$\ell_{a}$}}{2} + (c-1)cM_{c}^{2}(\pfour{\V{$\ell_{a}$}}{\D{$H$}}, \pfour{\V{$\ell_{b}$}}{\D{$H$}}) \nonumber \\
\Mass{\I{I}}{2} &=& (2c - 1)^{2}(\E{\V{V}}{\D{$H$}})^{2} - |\pthree{\V{V}}{\D{$H$}}|^{2}~.
\eea
where different values of $c$ are seen to increase and decrease each of the neutrino-related masses coherently. Neglecting the individual lepton masses, a choice of $c = 1$ sets the neutrino masses to zero, and also implies $\Mass{\I{I}}{} = \mass{\V{V}}{}$. From Eq.~\ref{eqn:numasses} we also observe that a smaller choice for $c$ (or, alternatively, smaller choice for \Mass{\I{I}}{}) could lead to tachyonic individual neutrinos. This means that while a choice of \Mass{\I{I}}{} larger than \mass{\V{V}}{} would be consistent with the above prescription, smaller choices are not sufficiently large to use this approach. As the calculation of the velocity relating the lab frame to the \D{$H$} rest frame depends on \Mass{\I{I}}{}, in the RJR scheme this choice must be made appealing only to the four vectors of visible particles in the lab frame. Fortunately, the minimum value necessary to ensure our ultimate neutrino approximations are physically viable, \mass{\V{V}}{}, is a Lorentz-invariant function of these four vectors, meaning we can make this assignment only knowing our prescription for analyzing the event in the \D{$H$} rest frame, but not necessarily having enough information to evaluate any visible four vectors there. This choice corresponds to a JR:

\begin{jigsaw}[Invisible Mass]
\label{jr:mass}
If the mass of an invisible particle, \I{I}, is unknown it can be chosen to be the smallest Lorentz invariant function of visible four vectors that is sufficiently large to accommodate any other applied JR's which correspond to dividing \I{I} into other invisible particles.
\end{jigsaw}

Similarly, the above prescription for choosing $\vec{\beta}_{c}$ can be generalized as another JR:
\begin{jigsaw}[Contra-boost Invariant]
\label{jr:contra1}
If the internal degrees of freedom specifying how an invisible particle, $\I{I} = \{ \I{I$_{a}$}, \I{I$_{b}$} \}$, should split into two particles are unknown, they can be specified by choosing a corresponding pair of visible particles,  $\V{V} = \{ \V{V$_{a}$}, \V{V$_{b}$} \}$, and applying the constraint $\Mass{\V{V$_{a}$}\I{I$_{a}$}}{} = \Mass{\V{V$_{b}$}\I{I$_{b}$}}{}$. It is assumed that the four vectors of the visible particles are known in the center-of-mass frame, $\D{F} = \{ \V{V}, \I{I}~\}$, as is the four vector of the total \I{I} system, \pfour{\I{I}}{\D{F}}. 
The four vectors of the invisible particles can be chosen in the \D{F} frame as:
\bea
\pthree{\I{I$_{a}$}}{\D{F}} &=& (c-1) \pthree{\V{V$_{a}$}}{\D{F}} - c \pthree{\V{V$_{b}$}}{\D{F}} \nonumber \\
\pthree{\I{I$_{b}$}}{\D{F}} &=& (c-1) \pthree{\V{V$_{b}$}}{\D{F}} - c \pthree{\V{V$_{a}$}}{\D{F}}  \\
\E{\I{I$_{a}$}}{\D{F}} &=& (c-1)\E{\V{V$_{a}$}}{\D{F}} + c\E{\V{V$_{b}$}}{\D{F}}  \nonumber \\
\E{\I{I$_{b}$}}{\D{F}} &=& (c-1)\E{\V{V$_{b}$}}{\D{F}} + c\E{\V{V$_{a}$}}{\D{F}}~, \nonumber
\eea
where
\bea
c = \frac{1}{2}\left[ 1 +   \frac{\sqrt{(\E{\V{V}}{\D{F}})^{2} - \mass{\V{V}}{2} +\Mass{\I{I}}{2}}}{\E{\V{V}}{\D{F}}}\right ]~.
\eea

If the visible particles \V{V$_{a}$} and \V{V$_{b}$} are massless, the minimum value of \Mass{\I{I}}{} required to guarantee that the individual invisible particles will not be tachyonic is \mass{\V{V}}{}.
\end{jigsaw}

With these JR's defined, we can summarize the RJR approach for analyzing $H^{0}\rightarrow W^{+}(\ell^{+}\nu)W^{-}(\ell^{-}\nu)$ events:
\begin{enumerate}[noitemsep]
\item Apply the invisible mass JR~\ref{jr:mass}, \\ choosing $\Mass{\I{I}}{} = \mass{\V{V}}{}$ 
\item Apply the invisible rapidity JR~\ref{jr:rapidity}, \\ choosing \pone{\I{I},z}{\lab} using the leptons \V{V}
\item Apply the contra-boost invariant JR~\ref{jr:contra1}, \\ specifying the neutrino four vectors \\ using the constraint $\Mass{\D{$W_{a}$}}{} = \Mass{\D{$W_{b}$}}{}$
\end{enumerate}
After the application of these JR's, values for all of the unknowns in the event are specified and any kinematic quantity of interest can be estimated. 

\begin{figure}[htbp]
\centering 
\includegraphics[width=.35\textwidth]{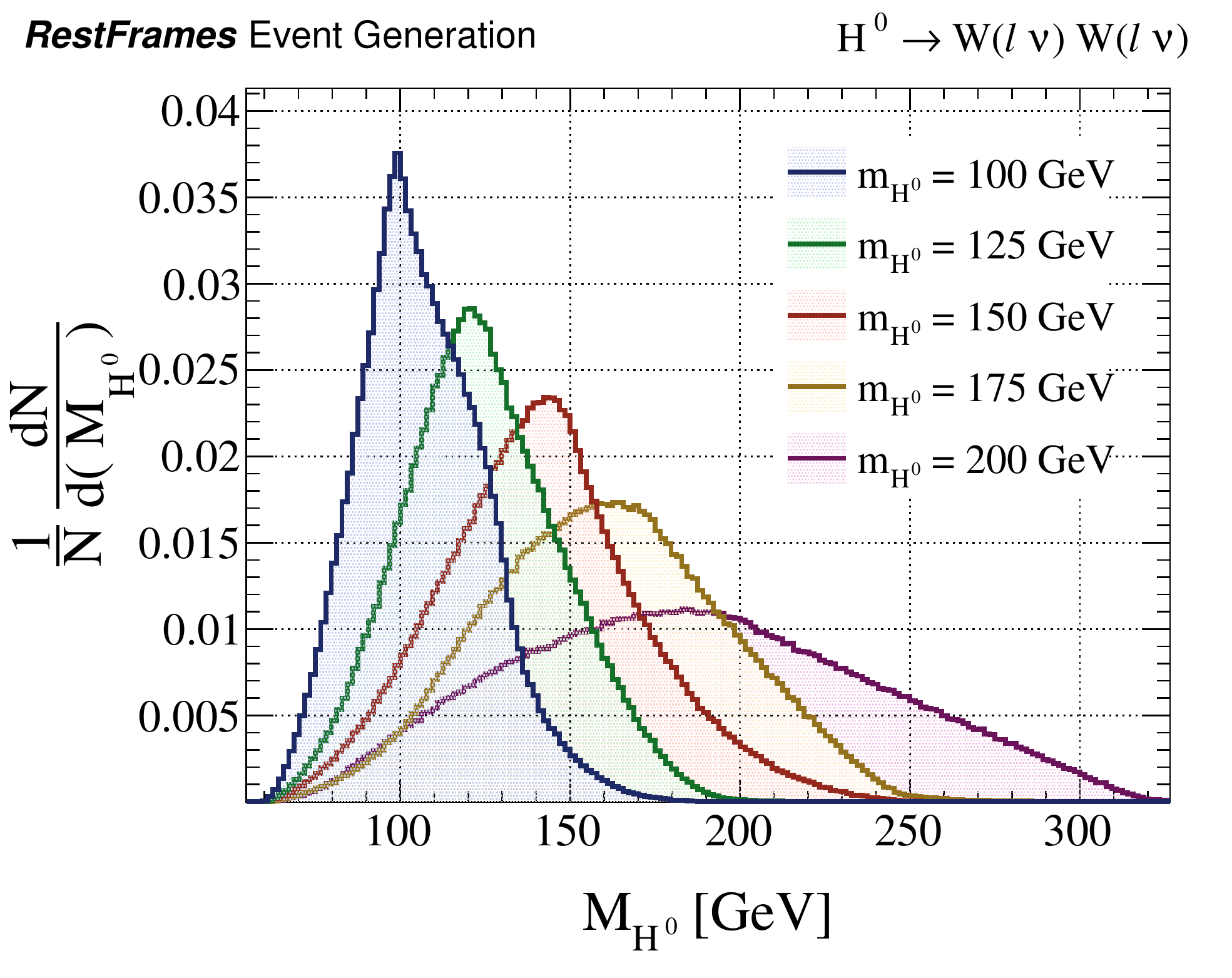}
\vspace{-0.3cm}
\caption{\label{fig:example_H_to_WlnuWlnu-MH} Distribution of \Mass{\D{$H$}}{} for simulated $H^{0}\rightarrow W^{+}(\ell\nu)W^{-}(\ell\nu)$ events with various values for \mass{\D{$H$}}{}.}
\end{figure}

One natural quantity of interest is the mass of the heavy, neutral Higgs, shown in Fig.~\ref{fig:example_H_to_WlnuWlnu-MH} for different values of \mass{\D{$H$}}{}. The relative resolution of the Higgs mass estimator, \Mass{\D{$H$}}{}, is approximately the same for $\mass{\D{$H$}}{} > 2m_{W}$, with the peak of each distribution scaling with the true value, and a slight underestimation visible due to the implicit minimization in the invisible rapidity JR. 

\begin{figure}[htbp]
\centering 
\includegraphics[width=.35\textwidth]{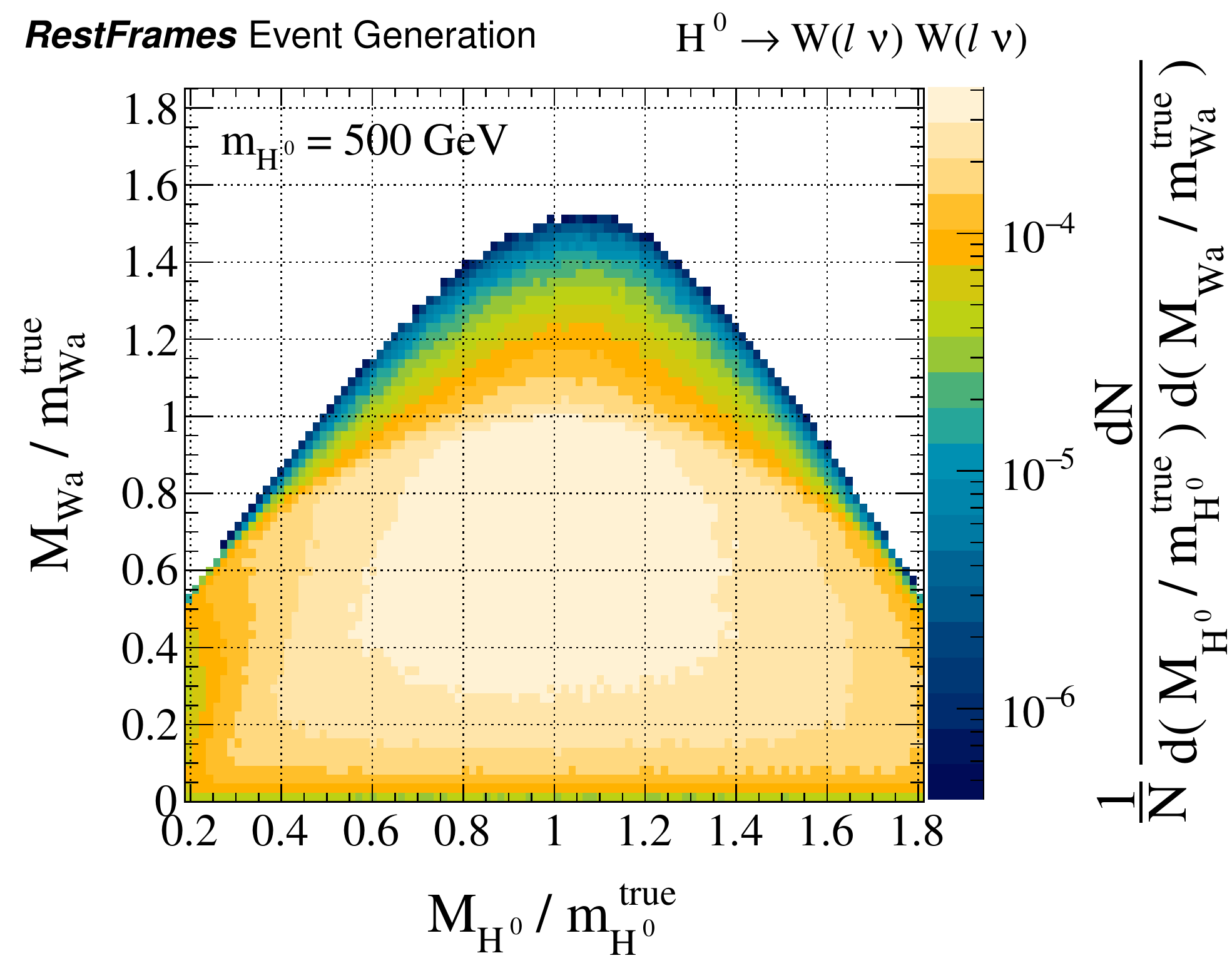}
\vspace{-0.3cm}
\caption{\label{fig:example_H_to_WlnuWlnu-MW_v_MH} Distribution of \Mass{\D{$W$}}{}, as a function of \Mass{\D{$H$}}{}, for simulated $H^{0}\rightarrow W^{+}(\ell\nu)W^{-}(\ell\nu)$ events with $\mass{\D{$H$}}{} = 500$ GeV. Each mass estimator is normalized by the true value. }
\end{figure}

Like \Mass{\D{$H$}}{}, the $W$ mass estimator, \Mass{\D{$W$}}{}, also provides sensitivity to the true value, exhibiting a kinematic edge as can be observed in Fig.~\ref{fig:example_H_to_WlnuWlnu-MW_v_MH}. The shape of the  \Mass{\D{$W$}}{} distribution is largely independent of the value of \mass{\D{$H$}}{}, and event-by-event it is estimated largely independently of \Mass{\D{$H$}}{}, with the resolution of \Mass{\D{$H$}}{} correlated with how close \Mass{\D{$W$}}{} is to its approximate kinematic boundary. This lack of correlation is a result of the JR rules applied. The contra-boost invariant JR ensures that \Mass{\D{$W$}}{} is largely insensitive to the $W$ boson velocity in its production frame, which is roughly proportional to the mass of \D{$H$}. 

\begin{figure}[htbp]
\centering 
\subfigure[]{\includegraphics[width=.238\textwidth]{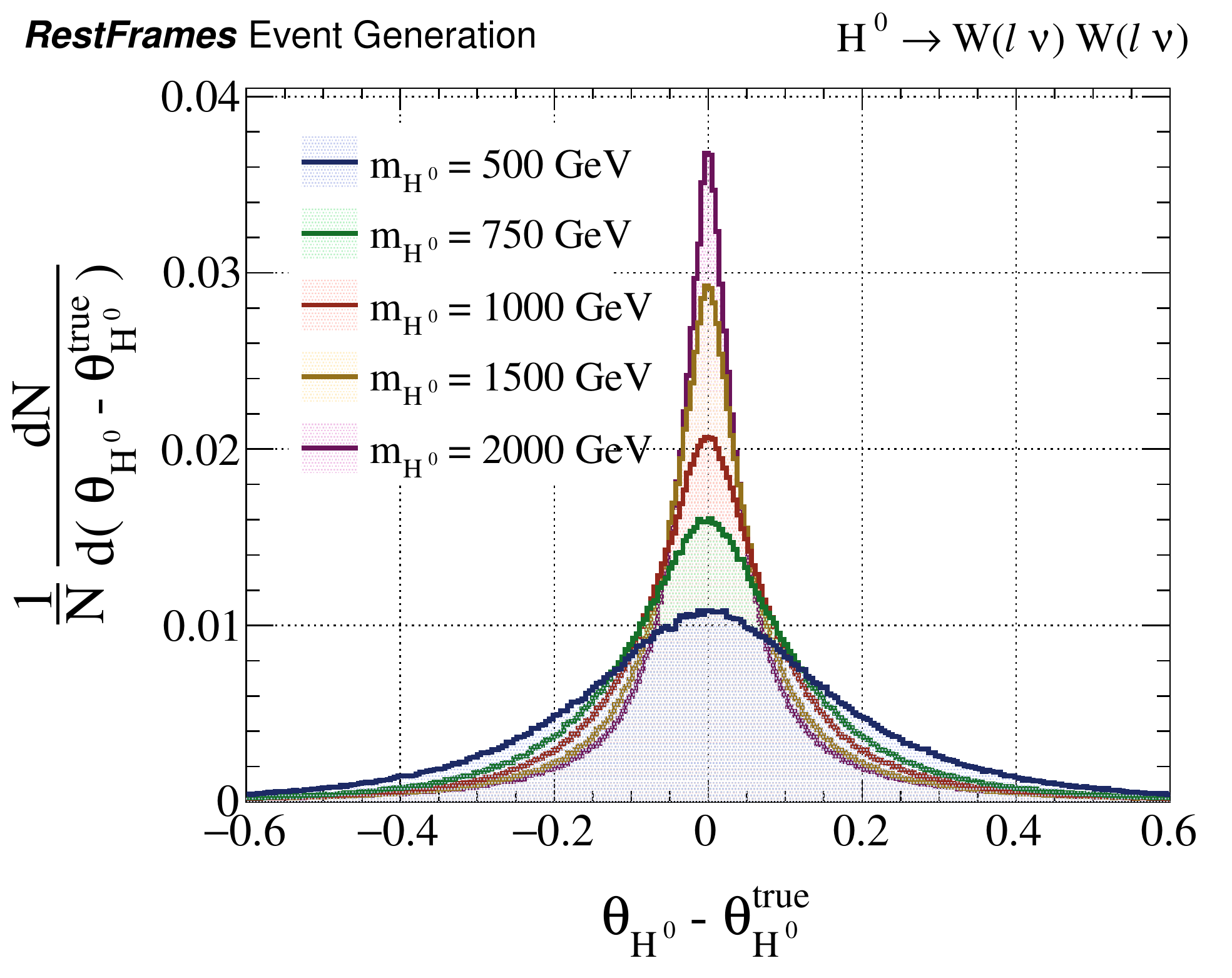}}
\subfigure[]{\includegraphics[width=.238\textwidth]{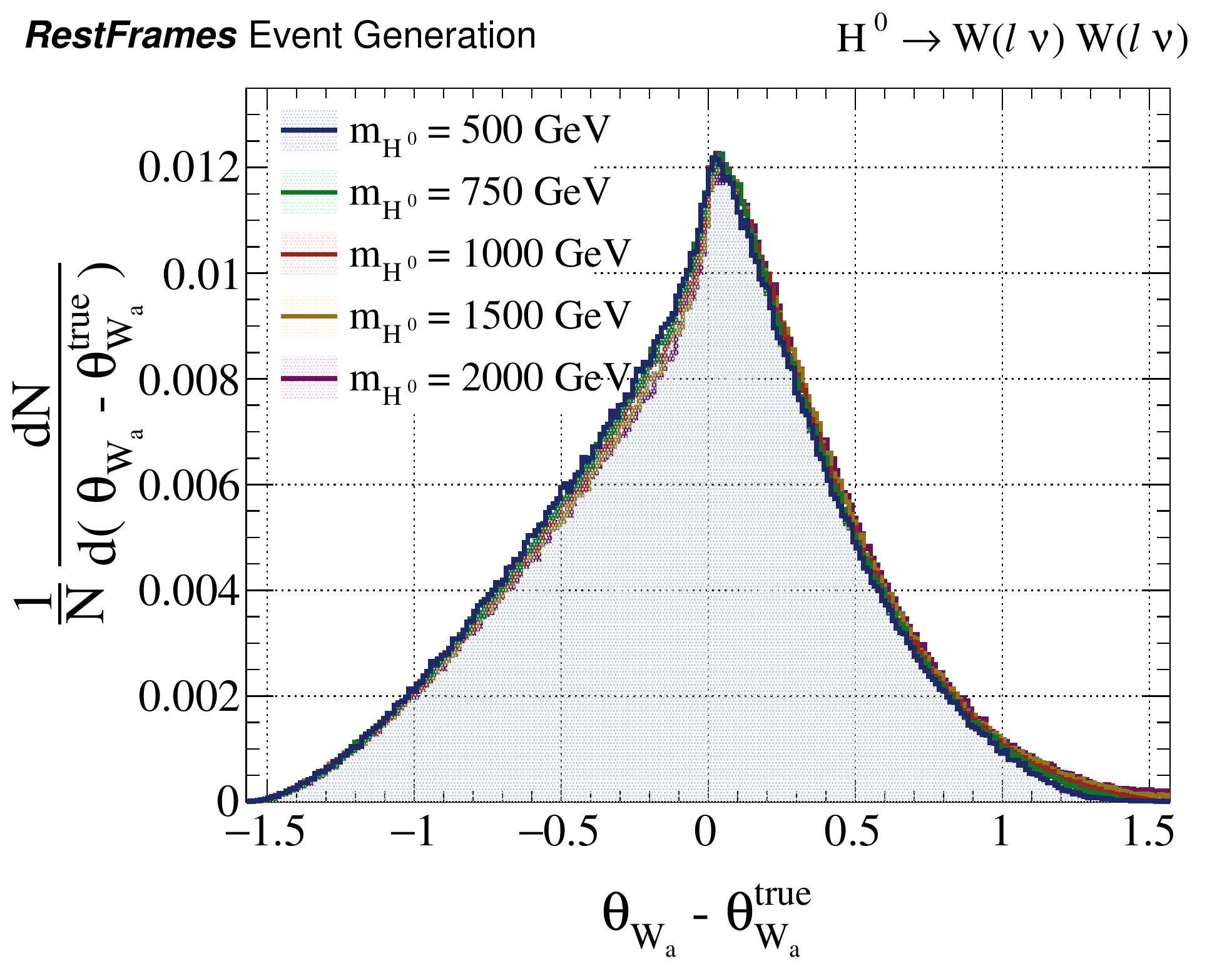}}
\subfigure[]{\includegraphics[width=.238\textwidth]{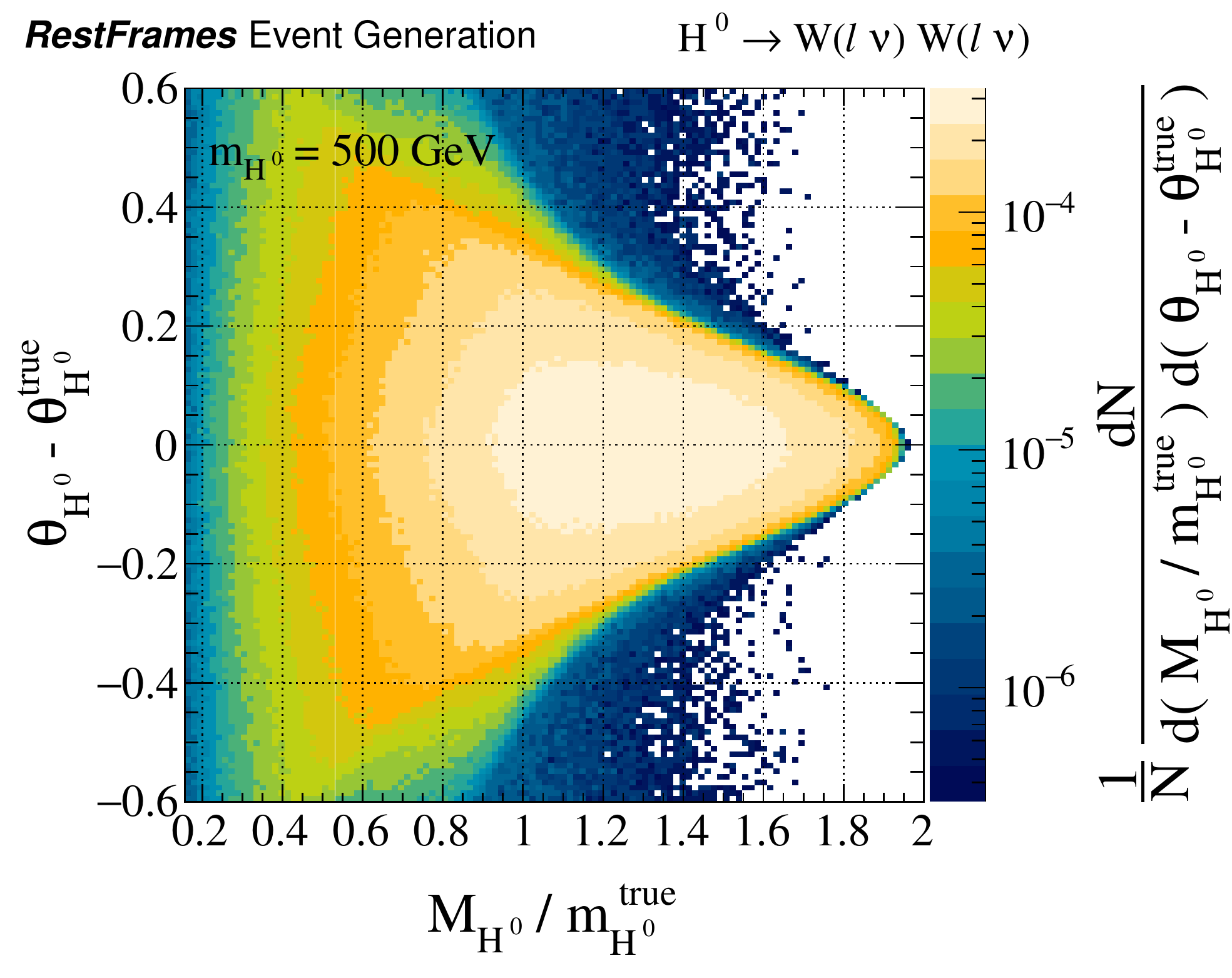}}
\subfigure[]{\includegraphics[width=.238\textwidth]{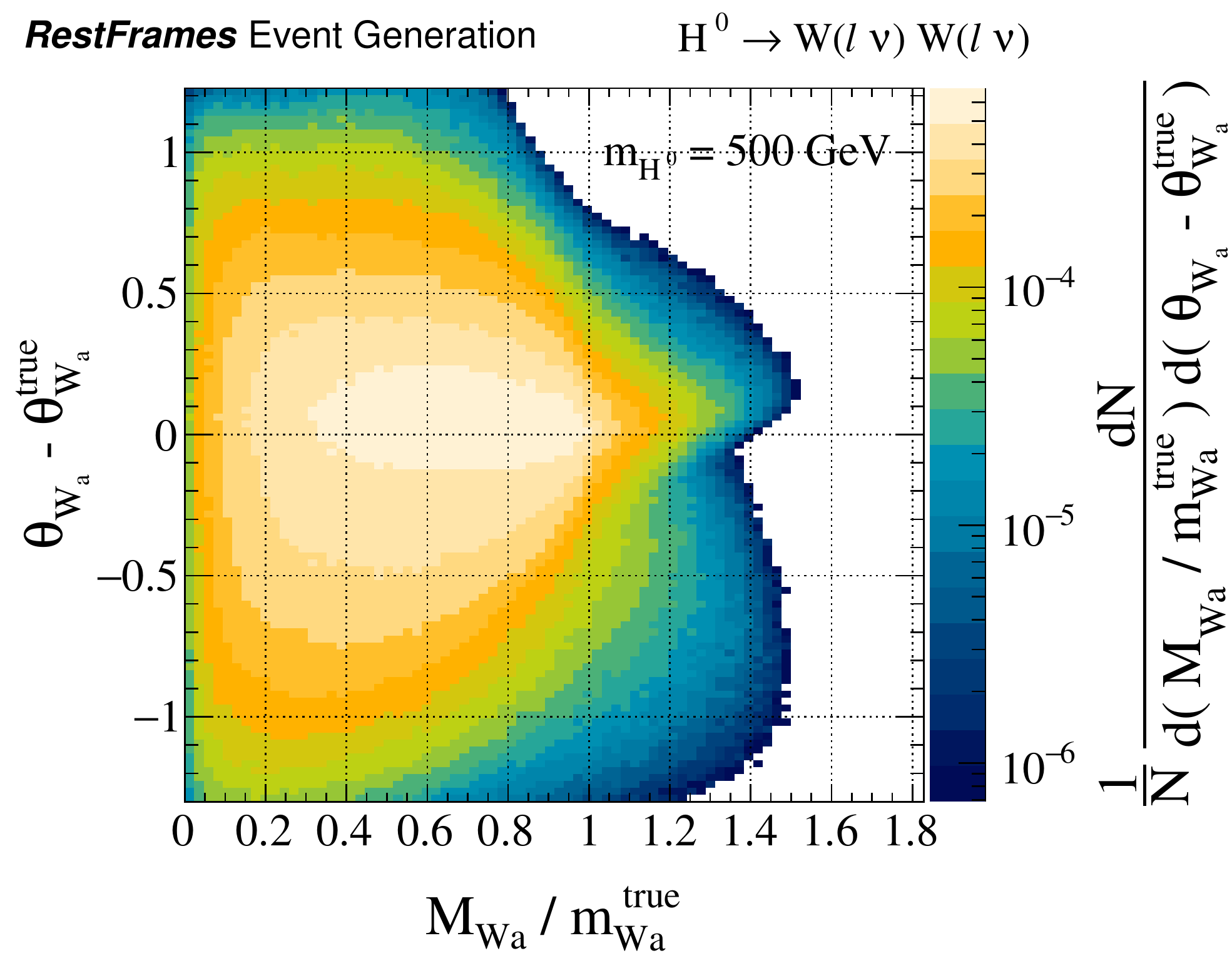}}
\vspace{-0.3cm}
\caption{\label{fig:example_H_to_WlnuWlnu-thetas} Distributions of (a) Higgs decay angle, $\theta_{\D{$H$}}$ and (b) $W$ decay angle, $\theta_{\D{$W$}}$, estimated using the RJR scheme for simulated 
 $H^{0}\rightarrow W^{+}(\ell\nu)W^{-}(\ell\nu)$ events and various values for \mass{\D{$H$}}{}. Figures (c,d) show these estimators as a function of their corresponding mass estimators, \Mass{\D{$H$}}{} and \Mass{\D{$W$}}{}, respectively. Each observable is normalized appropriately by the true value of the quantity it is estimating, with angles in units radian.}
\end{figure}

Other observables can be estimated with some accuracy, like the Higgs and $W$ decay angles, shown in Fig.~\ref{fig:example_H_to_WlnuWlnu-thetas}. The resolution of $\theta_{\D{$H$}}$ improves with increasing \mass{\D{$H$}}{}, as it is easier to resolve the Higgs decay axis as $\pone{\D{$W$}}{\D{$H$}} / \mass{\D{$W$}}{}$ grows larger, while the resolution of $\theta_{\D{$H$}}$ is insensitive to the Higgs mass. The poorer accuracy of the $\theta_{\D{$W$}}$ estimator is to be expected; with the constraints from the applied JR's the $W$ decay angle estimators are set equal and opposite (cos$~\theta_{\D{$W_{a}$}} = -$cos$~\theta_{\D{$W_{b}$}}$), so the single estimator corresponds to a combination of the two. Regardless, it is still sensitive to these quantities and the information about spin correlations they represent.

In addition to being estimated accurately, it is noteworthy that these observables are measured largely independently of each other, as demonstrated in Fig.~\ref{fig:example_H_to_WlnuWlnu-thetas}(c,d) when comparing the reconstructed decay angles with their corresponding mass estimators. While the resolution of the decay angles can vary with estimated mass, no significant biases are observed. 

The  kinematic observables resulting from the RJR approximate reconstruction of these events are also known as {\it super-razor variables}, with a thorough discussion of their phenomenology in the corresponding reference~\cite{Buckley:2013kua}.


\subsection{$\tilde{t}\tilde{t} \rightarrow (t\chi_{1}^{0})(t\chi_{1}^{0})$ at a hadron collider}
\label{subsec:Part2_exampleB}

\begin{figure}[htbp]
\centering 
\includegraphics[width=.32\textwidth]{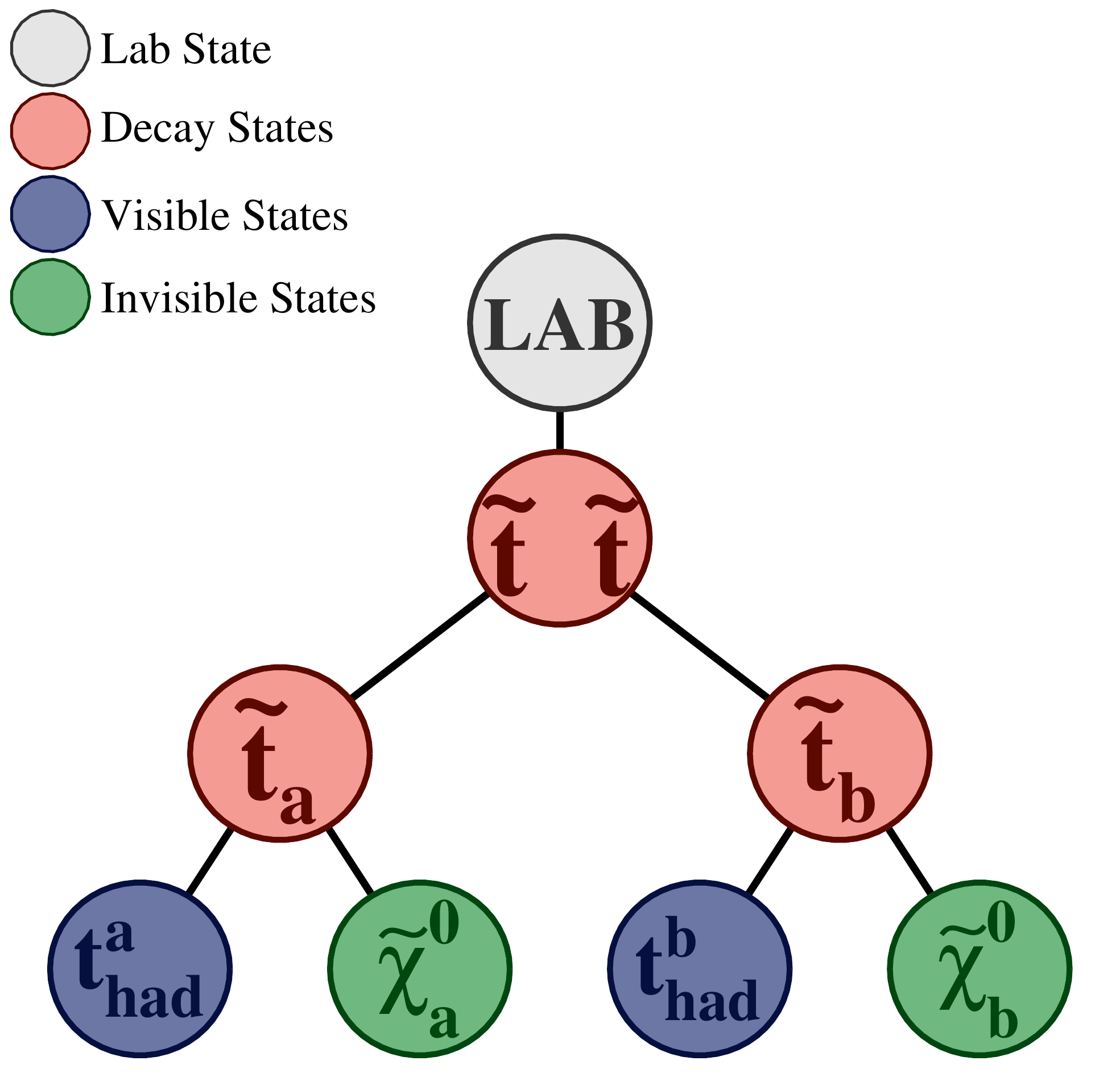}
\vspace{-0.3cm}
\caption{\label{fig:decayTree_DiStop_to_hadtopXhadtopX} The decay tree for analyzing $\tilde{t}\tilde{t} \rightarrow (t\chi_{1}^{0})(t\chi_{1}^{0})$ events. The two sides of the event are labelled ``a" and ``b",  each including a hadronic top and neutralino from their respective stop decays.}
\end{figure}

To further generalize the JR's for final states with two invisible particles, we consider the example process of stop quark pair production at a hadron collider, with each stop decaying to a top quark and undetected neutralino. We assume that the top quarks decay hadronically, and that each is identified and reconstructed in the detector, with measured four vectors \pfour{\V{$t_{a}$}}{\lab} and \pfour{\V{$t_{b}$}}{\lab}. In a real experiment, there can be significant kinematic dependencies on the efficiency for reconstructing and identifying hadronically decaying top quarks, along with imperfect resolution in estimating the top's momentum and mass. We neglect these effects in this example, focusing only on shortcomings in event reconstruction due to missing information associated with the escaping neutralinos. Similarly, we assume that the \met~ provides a reliable estimate of the transverse mass of the di-neutralino system, $\I{I} = \{ \I{$\tilde{\chi}_{a}$}, \I{$\tilde{\chi}_{b}$} \}$, with $\pthree{\I{I},T}{\lab} = \met$. The decay tree used in analyzing this final state is shown in Fig.~\ref{fig:decayTree_DiStop_to_hadtopXhadtopX}.

This decay topology is nearly identical to the $H^{0}\rightarrow W^{+}(\ell\nu)W^{-}(\ell\nu)$ process studied in Section~\ref{subsec:Part2_exampleA}, with the two intermediate stop quarks playing the role of the $W$ bosons. and the di-stop system the neutral Higgs boson. One important difference is that the stop quarks are produced non-resonantly, with \mass{\D{$\tilde{t}\tilde{t}$}}{} having a much larger variance than \mass{\D{$H$}}{}. But this distinction does nothing to prohibit the use of the same event analysis approach described in the previous example. The more pernicious complications follow from the masses of the individual visible tops and neutralinos, which now can have non-negligible values relative to the scale of the process. 

These masses have notable implications for the application of a contra-boost invariant JR, like that described in Section~\ref{subsec:Part2_exampleA}. Taking the same approach, we constrain $\Mass{\D{$\tilde{t}_{a}$}}{} = \Mass{\D{$\tilde{t}_{b}$}}{}$, and use the visible particle four vectors, evaluated in the putative di-stop rest frame, to make guesses for the momentum and energy of the two neutralinos. In the application of the contra-boost invariant JR~\ref{jr:contra1} we estimate the velocities relating the di-stop rest frame to the two stop rest frames as
\bea
\label{eqn:betatop}
 \vbeta{\D{$\tilde{t}_{a}$}}{\D{$\tilde{t}\tilde{t}$}} = -\vbeta{\D{$\tilde{t}_{b}$}}{\D{$\tilde{t}\tilde{t}$}} = \frac{\pthree{\V{$t_{a}$}}{\D{$\tilde{t}\tilde{t}$}} - \pthree{\V{$t_{b}$}}{\D{$\tilde{t}\tilde{t}$}}}{\E{\V{$t_{a}$}}{\D{$\tilde{t}\tilde{t}$}} + \E{\V{$t_{b}$}}{\D{$\tilde{t}\tilde{t}$}}}~.
\eea
This choice constrains the estimators for the neutralino and di-neutralino masses, which can be expressed by an additional free parameter, $c$, in equations analogous to Eq.~\ref{eqn:numasses}:
\bea
\label{eqn:chimasses}
\Mass{\I{$\tilde{\chi}_{a}$}}{2} &=& (c-1)^{2}\mass{\V{$t_{a}$}}{2} + c^{2}\mass{\V{$t_{b}$}}{2} + (c-1)cM_{c}^{2}(\pfour{\V{$t_{a}$}}{\D{$\tilde{t}\tilde{t}$}}, \pfour{\V{$t_{b}$}}{\D{$\tilde{t}\tilde{t}$}}) \nonumber \\
\Mass{\I{$\tilde{\chi}_{b}$}}{2} &=& (c-1)^{2}\mass{\V{$t_{b}$}}{2} + c^{2}\mass{\V{$t_{a}$}}{2} + (c-1)cM_{c}^{2}(\pfour{\V{$t_{a}$}}{\D{$\tilde{t}\tilde{t}$}}, \pfour{\V{$t_{b}$}}{\D{$\tilde{t}\tilde{t}$}}) \nonumber \\
\Mass{\I{I}}{2} &=& (2c - 1)^{2}(\E{\V{V}}{\D{$\tilde{t}\tilde{t}$}})^{2} - |\pthree{\V{V}}{\D{$\tilde{t}\tilde{t}$}}|^{2}~,
\eea
where $\V{V} = \{ \V{$t_{a}$}, \V{$t_{b}$} \}$ is the collection of all visible particles resulting from the interaction of interest. While appropriate in the previous example, the simple choice $\Mass{\I{I}}{} = \mass{\V{V}}{}$ would result in each of the neutralinos taking nonzero masses, with $\Mass{\I{$\tilde{\chi}_{a/b}$}}{} = \mass{\V{$t_{b/a}$}}{}$. For the moment, we assume that $\mass{\V{$t_{a}$}}{} = \mass{\V{$t_{b}$}}{} = \mass{\V{$t$}}{}$ to simplify the discussion, generalizing later. If we instead attempt to constrain this final degree of freedom by requiring $\Mass{\I{$\tilde{\chi}_{a}$}}{} = \Mass{\I{$\tilde{\chi}_{b}$}}{} = 0$, for lack of a better choice, this implies an expression for \Mass{\I{I}}{} of
\bea
\label{eqn:MI0}
\Mass{\I{I}}{2}|_{\Mass{\I{$\tilde{\chi}$}}{}=0} = \frac{4\left(|\pthree{\V{$t_{a}$}}{\D{$\tilde{t}\tilde{t}$}}|^{2}|\pthree{\V{$t_{b}$}}{\D{$\tilde{t}\tilde{t}$}}|^{2} - (\pthree{\V{$t_{a}$}}{\D{$\tilde{t}\tilde{t}$}} \cdot \pthree{\V{$t_{b}$}}{\D{$\tilde{t}\tilde{t}$}})^{2}  \right)}{2\mass{\V{$t$}}{2} + M_{c}^{2}(\pfour{\V{$t_{a}$}}{\D{$\tilde{t}\tilde{t}$}}, \pfour{\V{$t_{b}$}}{\D{$\tilde{t}\tilde{t}$}})}~.
\eea
Unfortunately, this is not a viable choice for \Mass{\I{I}}{}, as its expression in Eq.~\ref{eqn:MI0} is not a Lorentz invariant function of the visible four vectors, which violates one of the requirements of the invisible mass JR~\ref{jr:mass} we must apply in order to calculate the velocity relating the lab frame to the di-stop rest frame, \vbeta{\D{$\tilde{t}\tilde{t}$}}{\lab}. One could imagine trying to remedy this situation through a brute-force approach, writing Eq.~\ref{eqn:MI0} as a function of this unknown velocity by boosting the top four vectors from the lab frame to the di-stop rest frame, and further constraining \Mass{\I{I}}{} and \vbeta{\D{$\tilde{t}\tilde{t}$}}{\lab} through the relation
\bea
\label{eq:hard}
\vbeta{\D{$\tilde{t}\tilde{t}$}}{\lab} = \frac{\pthree{\D{$\tilde{t}\tilde{t}$}}{\lab}}{\E{\V{V}}{\lab}+ \sqrt{|\pthree{\I{I}}{\lab}|^{2} + \Mass{\I{I}}{2}(\vbeta{\D{$\tilde{t}\tilde{t}$}}{\lab})}}~.
\eea
Although numerically viable, solving Eq.~\ref{eq:hard} for \vbeta{\D{$\tilde{t}\tilde{t}$}}{\lab}, a high order polynomial equation, will lead to multiple solutions, none of which are guaranteed to be real. 

We choose to take a different approach, effectively factorizing the JR's applied at different stages in the decay tree by choosing \Mass{\I{I}}{} as the smallest Lorentz invariant expression that is strictly greater or equal to that in Eq.~\ref{eqn:MI0}:
\bea
\label{eqn:MI}
\Mass{\I{I}}{2} = \mass{\V{V}}{2} - 4\mass{\V{$t$}}{2} \geq \Mass{\I{I}}{2}|_{\Mass{\I{$\tilde{\chi}$}}{}=0}~.
\eea
That this expression is Lorentz invariant means it can be evaluated using only the four vectors of visible particles measured in the lab frame, and doesn't require knowledge of any approximate reference frames in the event. But this choice also requires a concession, in that our estimators for the neutrinalino masses, \Mass{\I{$\tilde{\chi}_{a}$}}{} and \Mass{\I{$\tilde{\chi}_{b}$}}{}, which we previously tried to constrain to zero, will now take non-zero values, with 
\bea
\label{eqn:Mchi}
\Mass{\I{$\tilde{\chi}_{a/b}$}}{} = \mass{\V{$t$}}{} \frac{|\pthree{\V{$t_{a}$}}{\D{$\tilde{t}\tilde{t}$}} - \pthree{\V{$t_{b}$}}{\D{$\tilde{t}\tilde{t}$}}|}{\E{\V{$t_{a}$}}{\D{$\tilde{t}\tilde{t}$}} + \E{\V{$t_{b}$}}{\D{$\tilde{t}\tilde{t}$}}}~.
\eea
A portion of the ``extra'' mass we assigned to \Mass{\I{I}}{} in Eq.~\ref{eqn:MI} in order to make it Lorentz invariant has been absorbed by \Mass{\I{$\tilde{\chi}_{a/b}$}}{}. The expression \label{eqn:Mchi} does not contain new information, in the sense that it is a combination of \mass{\V{$t$}}{} and the estimator for \vbeta{\D{$\tilde{t}_{a/b}$}}{\D{$\tilde{t}\tilde{t}$}} from Eq.~\ref{eqn:betatop}, and is not sensitive to the true values \mass{\I{$\tilde{\chi}_{a/b}$}}{}. We also note that our estimators for the top energies in their respective production frames, \E{\V{$t_{a}$}}{\D{$\tilde{t}_{a}$}} and \E{\V{$t_{b}$}}{\D{$\tilde{t}_{b}$}}, do not depend on this choice of \Mass{\I{I}}{}, with
\bea
\E{\V{$t_{a/b}$}}{\D{$\tilde{t}_{a/b}$}} = \frac{\sqrt{2\mass{\V{$t$}}{2} + M_{c}^{2}(\pfour{\V{$t_{a}$}}{\D{$\tilde{t}\tilde{t}$}}, \pfour{\V{$t_{b}$}}{\D{$\tilde{t}\tilde{t}$}})}}{2}~.
\eea

Recalling that $\E{\V{$t_{a/b}$}}{\D{$\tilde{t}_{a/b}$}} = (\mass{\D{$\tilde{t}_{a/b}$}}{2} - \mass{\I{$\tilde{\chi}_{a/b}$}}{2} + \mass{\V{$t_{a/b}$}}{2})/2\mass{\D{$\tilde{t}_{a/b}$}}{}$, we are reminded that we are only sensitive to the mass {\it difference} between stops and neutralinos in these events, and our estimator of this quantity, \E{\V{$t_{a/b}$}}{\D{$\tilde{t}_{a/b}$}}, can be extracted with contra-boost invariance independent of our choices for the individual neutralino masses. In practice, allowing our estimators \Mass{\I{$\tilde{\chi}_{a/b}$}}{} to acquire mass in our approximate view of each event is a book-keeping device to account for the lack of commutation between the boosts (\vbeta{\D{$\tilde{t}\tilde{t}$}}{\lab}) and contra-boosts (\vbeta{\D{$\tilde{t}_{a/b}$}}{\D{$\tilde{t}\tilde{t}$}}) which describe the visible tops' path from their production frames to the lab frame. While providing no information about the true mass values of the invisible particles, this approach minimizes the effect this lack of knowledge has on our ability to extract other information from the event, like the mass splittings of sparticles. It also allows for the JR's applied in analyzing the event to be factorized and, as we will see in later examples, inter-changed, further decoupling the observables measured in each approximate reference frame.

In service of this last consideration, we imagine a case where one may want to ensure that the estimators \Mass{\I{$\tilde{\chi}_{a/b}$}}{}, and the corresponding quantities in our reconstructed view of each event, are greater than or equal to some non-trivial value, with $\Mass{\I{$\tilde{\chi}_{a/b}$}}{} \geq \mu$. This need may occur if there is prior knowledge, outside of quantities measured in the event, about these masses, or if one wants to examine the dependence of observables in different test masses $M_{\rm min}$. Allowing for a minimum mass also permits this contra-boost invariant JR to accommodate future invisible JR's appearing later in the decay tree, where \I{$\tilde{\chi}_{a}$} and \I{$\tilde{\chi}_{b}$} may be sub-divided into more invisible particles.

Returning to Eq.~\ref{eqn:chimasses}, requiring $\Mass{\I{$\tilde{\chi}_{a/b}$}}{} = \mu$ implies that \Mass{\I{I}}{} can be expressed as
\bea
\Mass{\I{I}}{2}|_{\Mass{\I{$\tilde{\chi}$}}{}=\mu} &=&  \nonumber \\
\mass{\V{V}}{2} &+& \frac{ 4(\mu^2-\mass{\V{$t$}}{2})}{2\mass{\V{$t$}}{2} + M_{c}^{2}(\pfour{\V{$t_{a}$}}{\D{$\tilde{t}\tilde{t}$}}, \pfour{\V{$t_{b}$}}{\D{$\tilde{t}\tilde{t}$}})}(\E{\V{V}}{\D{$\tilde{t}\tilde{t}$}})^{2}~.
\eea
As was the case for the expression for $\Mass{\I{I}}{}|_{\Mass{\I{$\tilde{\chi}$}}{}=0}$ from Eq.~\ref{eqn:MI0}, $\Mass{\I{I}}{}|_{\Mass{\I{$\tilde{\chi}$}}{}=\mu}$ is not a Lorentz invariant function, so we instead choose \Mass{\I{I}}{} to correspond to the smallest Lorentz invariant expression that is guaranteed to be greater than or equal to $\Mass{\I{I}}{}|_{\Mass{\I{$\tilde{\chi}$}}{}=\mu}$:
\bea
\label{eqn:MImin}
\Mass{\I{I}}{2} =
  \begin{cases}
    \mass{\V{V}}{2} + 4(\mu^2-  \mass{\V{$t$}}{2})     & \quad \mu \leq \mass{\V{$t$}}{} \\\
    \frac{\mu^2}{ \mass{\V{$t$}}{2}}\mass{\V{V}}{2}  & \quad \mu> \mass{\V{$t$}}{}~.\\
  \end{cases}
\eea
These choices result in the mass estimators \Mass{\I{$\tilde{\chi}_{a/b}$}}{} taking values larger than $\mu$, with
\bea
\label{eqn:Mchimin}
\Mass{\I{$\tilde{\chi}_{a/b}$}}{2} =
  \begin{cases}
     \mu^2 + (\mass{\V{$t$}}{2} - \mu^2)\frac{|\pthree{\V{$t_{a}$}}{\D{$\tilde{t}\tilde{t}$}} - \pthree{\V{$t_{b}$}}{\D{$\tilde{t}\tilde{t}$}}|^{2}}{(\E{\V{$t_{a}$}}{\D{$\tilde{t}\tilde{t}$}} + \E{\V{$t_{b}$}}{\D{$\tilde{t}\tilde{t}$}})^2}     &  \mu \leq \mass{\V{$t$}}{} \\\
     \mu^2 + \frac{(\mu^2-\mass{\V{$t$}}{2})}{\mass{\V{$t$}}{2}}\frac{|\pthree{\V{$t_{a}$}}{\D{$\tilde{t}\tilde{t}$}} \times \pthree{\V{$t_{b}$}}{\D{$\tilde{t}\tilde{t}$}}|^2}{(\E{\V{$t_{a}$}}{\D{$\tilde{t}\tilde{t}$}} + \E{\V{$t_{b}$}}{\D{$\tilde{t}\tilde{t}$}})^2}  & \mu > \mass{\V{$t$}}{}~.\\
  \end{cases}
\eea

An important feature of the mass expressions in Eq.~\ref{eqn:MImin} and Eq.~\ref{eqn:Mchimin} is that they are divergent for $\mass{\V{$t$}}{} \rightarrow 0$ when $\mu > 0$. When either of two visible particles used in a contra-boost invariant JR are massless, there is no finite, Lorentz invariant expression for the total invisible system mass that can guarantee the masses of the individual neutralinos will remain larger than any non-zero value. In these cases, alternative JR's can be used, as described in later examples.

Returning to the more general case where $\mass{\V{$t_{a}$}}{} \neq \mass{\V{$t_{b}$}}{}$, we can generalize the contra-boost invariant JR~\ref{jr:contra1} to cases with non-trivial visible and invisible particle masses:
\begin{jigsaw}[Contra-boost Invariant]
\label{jr:contra2}
If the internal degrees of freedom specifying how an invisible particle, $\I{I} = \{ \I{I$_{a}$}, \I{I$_{b}$} \}$, should split into two particles are unknown, they can be specified by choosing a corresponding pair of visible particles,  $\V{V} = \{ \V{V$_{a}$}, \V{V$_{b}$} \}$, and applying the constraint $\Mass{\V{V$_{a}$}\I{I$_{a}$}}{} = \Mass{\V{V$_{b}$}\I{I$_{b}$}}{}$. It is assumed that the four vectors of the visible particles are known in the center-of-mass frame, $\D{F} = \{ \V{V}, \I{I}~\}$, as is the four vector of the total \I{I} system, \pfour{\I{I}}{\D{F}}. 
The four vectors of the invisible particles can be chosen in the \D{F} frame as:
\bea
\pthree{\I{I$_{a}$}}{\D{F}} &=& (c-1) \pthree{\V{V$_{a}$}}{\D{F}} - c \pthree{\V{V$_{b}$}}{\D{F}} \nonumber \\
\pthree{\I{I$_{b}$}}{\D{F}} &=& (c-1) \pthree{\V{V$_{b}$}}{\D{F}} - c \pthree{\V{V$_{a}$}}{\D{F}}  \\
\E{\I{I$_{a}$}}{\D{F}} &=& (c-1)\E{\V{V$_{a}$}}{\D{F}} + c\E{\V{V$_{b}$}}{\D{F}}  \nonumber \\
\E{\I{I$_{b}$}}{\D{F}} &=& (c-1)\E{\V{V$_{b}$}}{\D{F}} + c\E{\V{V$_{a}$}}{\D{F}}~, \nonumber
\eea
where
\bea
c = \frac{1}{2}\left[ 1 +   \frac{\sqrt{(\E{\V{V}}{\D{F}})^{2} - \mass{\V{V}}{2} +\Mass{\I{I}}{2}}}{\E{\V{V}}{\D{F}}}.\right ]~.
\eea

Assuming $\mass{\V{V$_{a}$}}{} \geq \mass{\V{V$_{b}$}}{}$, in order for the individual invisible particle masses to be guaranteed to be greater than some known value, $\Mass{\I{I$_{a/b}$}}{} \geq \mu \geq 0$, the mass estimator \Mass{\I{I}}{} must be chosen to be at least as large as 
\bea
\Mass{\I{I}}{2} \geq
  \begin{cases}
    \mass{\V{V}}{2} + 4(\mu^2-  \mass{\V{V$_{b}$}}{2})     & \quad \mu \leq \mass{\V{V$_{b}$}}{} \\\
    \frac{\mu^2}{ \mass{\V{V$_{b}$}}{2}}\mass{\V{V}}{2}  & \quad \mu> \mass{\V{V$_{b}$}}{}~.\\
  \end{cases}
\eea
\end{jigsaw}
 
The RJR approach to analyzing $\tilde{t}\tilde{t} \rightarrow (t\chi_{1}^{0})(t\chi_{1}^{0})$ events at a hadron collider can be summarized as: 
\begin{enumerate}[noitemsep]
\item Apply the invisible mass JR~\ref{jr:mass}, choosing \Mass{\I{I}}{} as the smallest possible quantity consistent with JR~\ref{jr:contra2} and non-tachyonic neutralinos.
\item Apply the invisible rapidity JR~\ref{jr:rapidity}, \\ choosing \pone{\I{I},z}{\lab} using the tops \V{V}.
\item Apply the contra-boost invariant JR~\ref{jr:contra2}, \\ specifying the neutralino four vectors \\ using the constraint $\Mass{\D{$t_{a}$}}{} = \Mass{\D{$t_{b}$}}{}$.
\end{enumerate}

\begin{figure}[htbp]
\centering
\includegraphics[width=.35\textwidth]{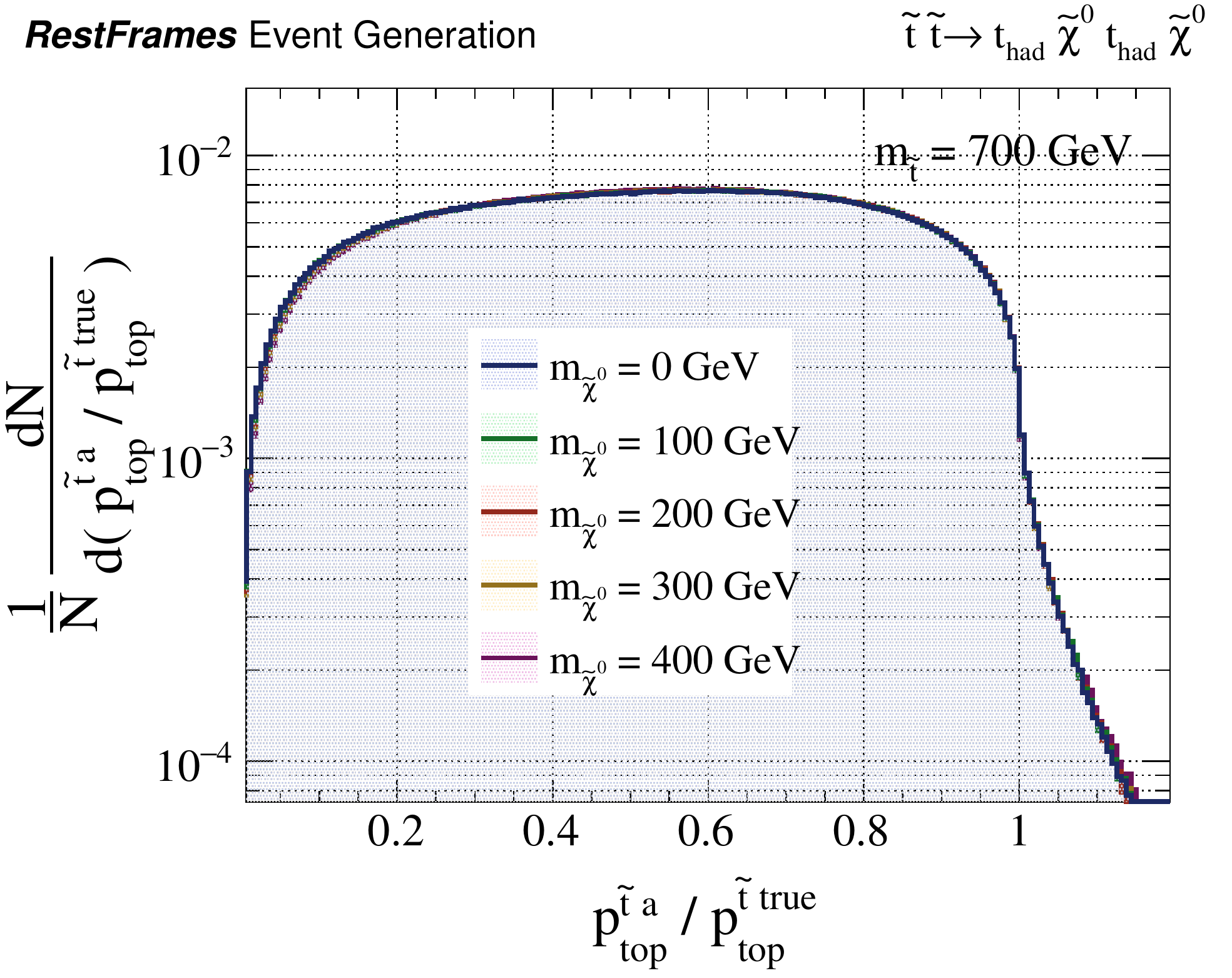}
\vspace{-0.3cm}
\caption{\label{fig:example_DiStop_to_hadtopXhadtopX-Ptop} The distribution of the reconstructed top momentum in the approximation of its production frame, normalized by the true value, in simulated  $\tilde{t}\tilde{t} \rightarrow (t\chi_{1}^{0})(t\chi_{1}^{0})$ events with varying choices of \mass{\I{$\tilde{\chi}^0$}}{}.}
\end{figure}

With these choices for resolving the neutralino-related unknowns in the event, the phenomenology of the resulting estimators is studied in simulated $\tilde{t}\tilde{t} \rightarrow (t\chi_{1}^{0})(t\chi_{1}^{0})$ events, for various values of \mass{\I{$\tilde{\chi}^0$}}{}. The distribution of the reconstructed top's momentum in the approximation of its production frame, \pone{\V{$t_{a}$}}{\D{$\tilde{t}_{a}$}}, is shown in Fig.~\ref{fig:example_DiStop_to_hadtopXhadtopX-Ptop} for simulated events. The estimator for \pone{\V{$t_{a}$}}{\D{$\tilde{t}_{a}$}} exhibits no visible dependence on \mass{\I{$\tilde{\chi}^0$}}{}, reliably providing sensitivity to the true value with a kinematic edge. While the estimators for the individual neutralino masses may take non-zero values, it has a negligible effect on the determination of the velocities relating the reconstructed reference frames.

\begin{figure}[htbp]
\centering 
\subfigure[]{\includegraphics[width=.238\textwidth]{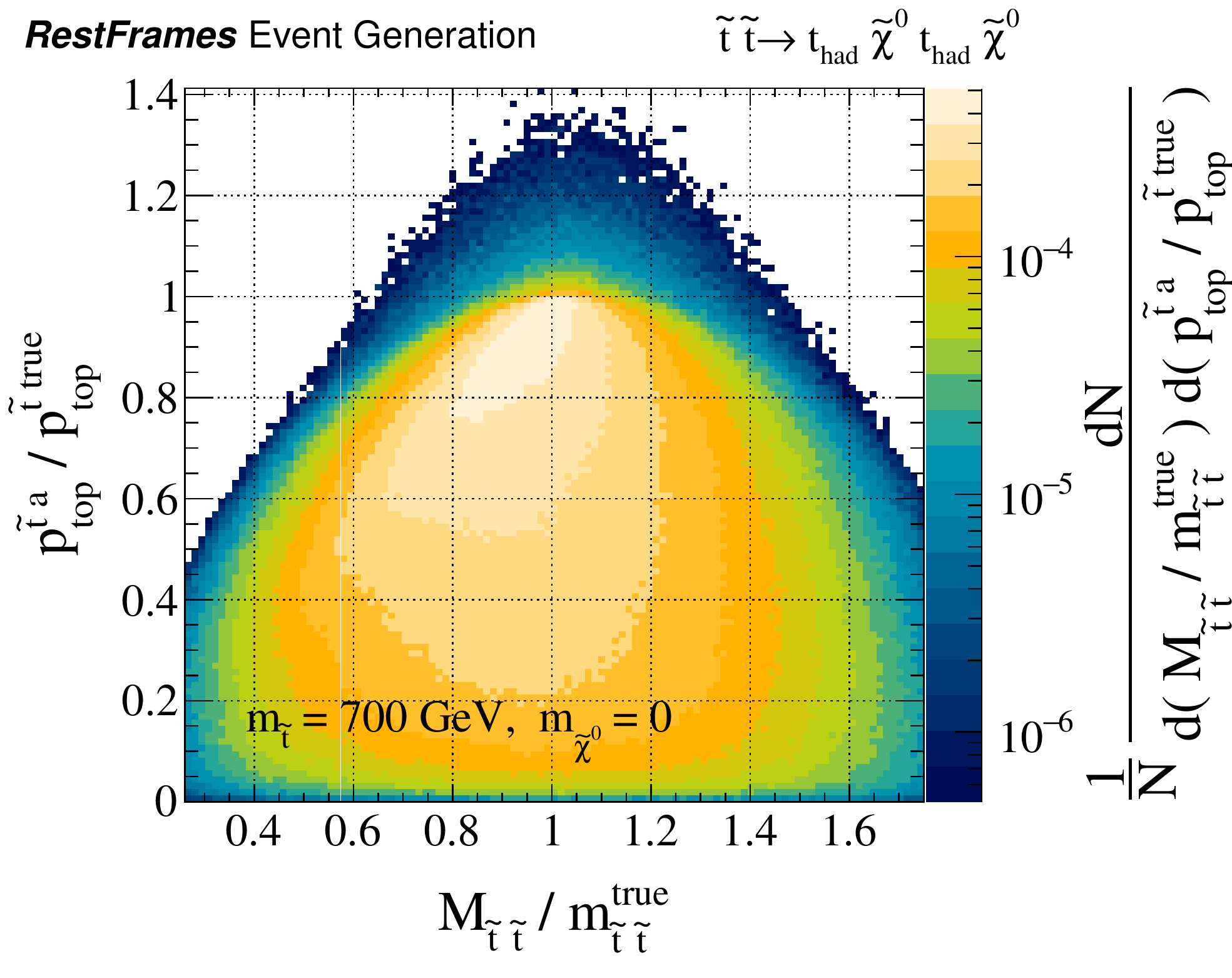}}
\subfigure[]{\includegraphics[width=.238\textwidth]{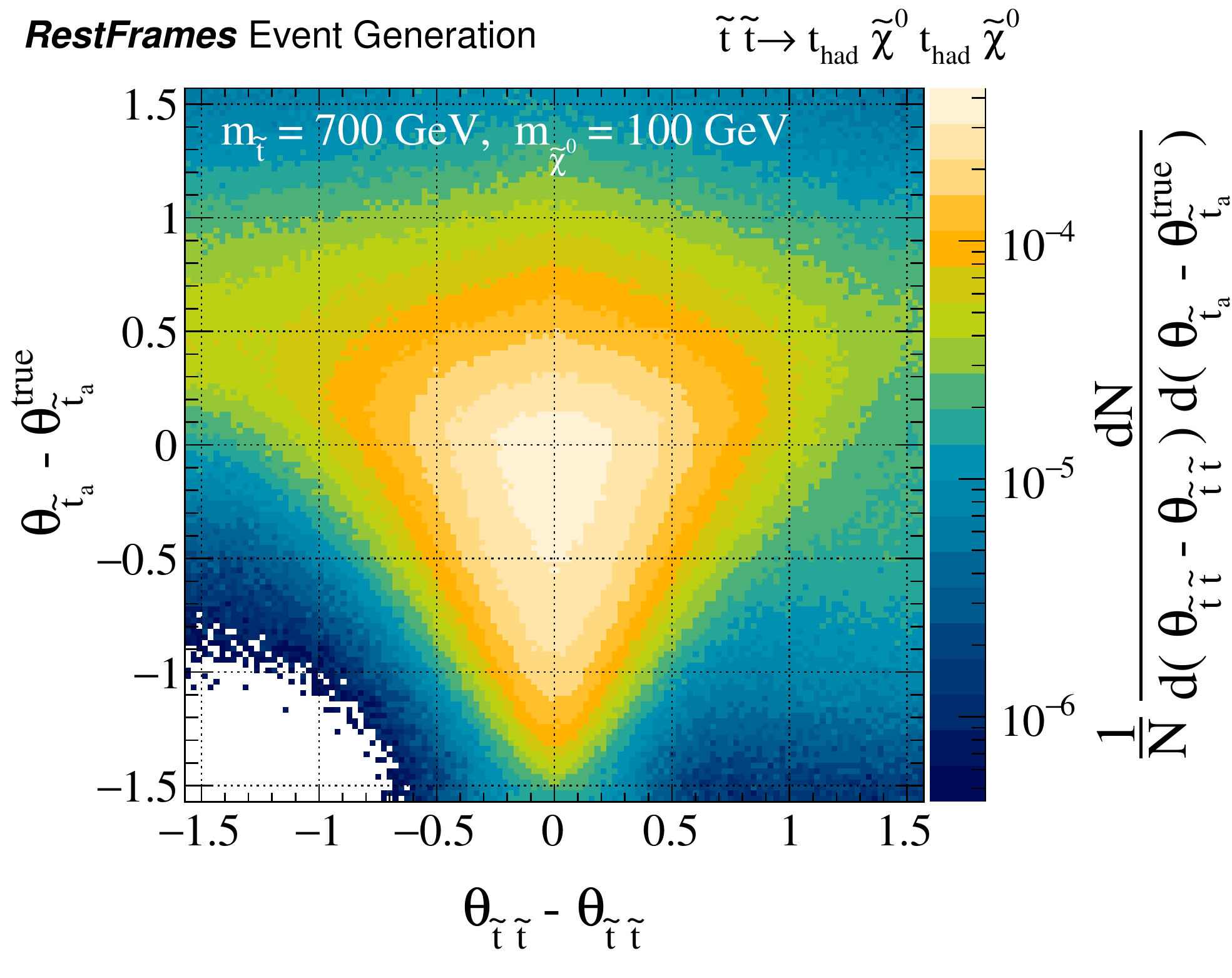}}
\vspace{-0.3cm}
\caption{\label{fig:example_DiStop_to_hadtopXhadtopX-Ptop_v_MTT} (a) Distribution of the reconstructed top momentum in the approximation of its production frame as a function of the estimated di-stop invariant mass. (b) Distribution of the di-stop decay angle, $\theta_{\D{$\tilde{t}\tilde{t}$}}$, as a function of the stop decay angle, $\theta_{\D{$\tilde{t}_{a}$}}$. All observables are normalized appropriately by the true values of the quantities they are estimating, with angles in units radian. }
\end{figure}

As was the case for the $W$ mass estimator in the previous example, \pone{\V{$t_{a}$}}{\D{$\tilde{t}_{a}$}} (which is effectively an estimator of the stop/neutralino mass difference) can be estimated independently of $\Mass{\D{$\tilde{t}\tilde{t}$}}{}$, as demonstrated in Fig.~\ref{fig:example_DiStop_to_hadtopXhadtopX-Ptop_v_MTT}(a). Similarly, the decay angles of the intermediate particle states can be measured, with their distributions for reconstructed events shown in Fig.~\ref{fig:example_DiStop_to_hadtopXhadtopX-Ptop_v_MTT}(b). In particular, the decay angle of the di-stop system can be measured with excellent resolution. 


\subsection{$\tilde{\chi}^{0}_{2}\tilde{\chi}^{0}_{2} \rightarrow Z(\ell\ell) \tilde{\chi}^{0}_{1} h(\gamma\gamma) \tilde{\chi}^{0}_{1}$  at a hadron collider}
\label{subsec:Part2_exampleC}

\begin{figure}[htbp]
\centering
\includegraphics[width=.35\textwidth]{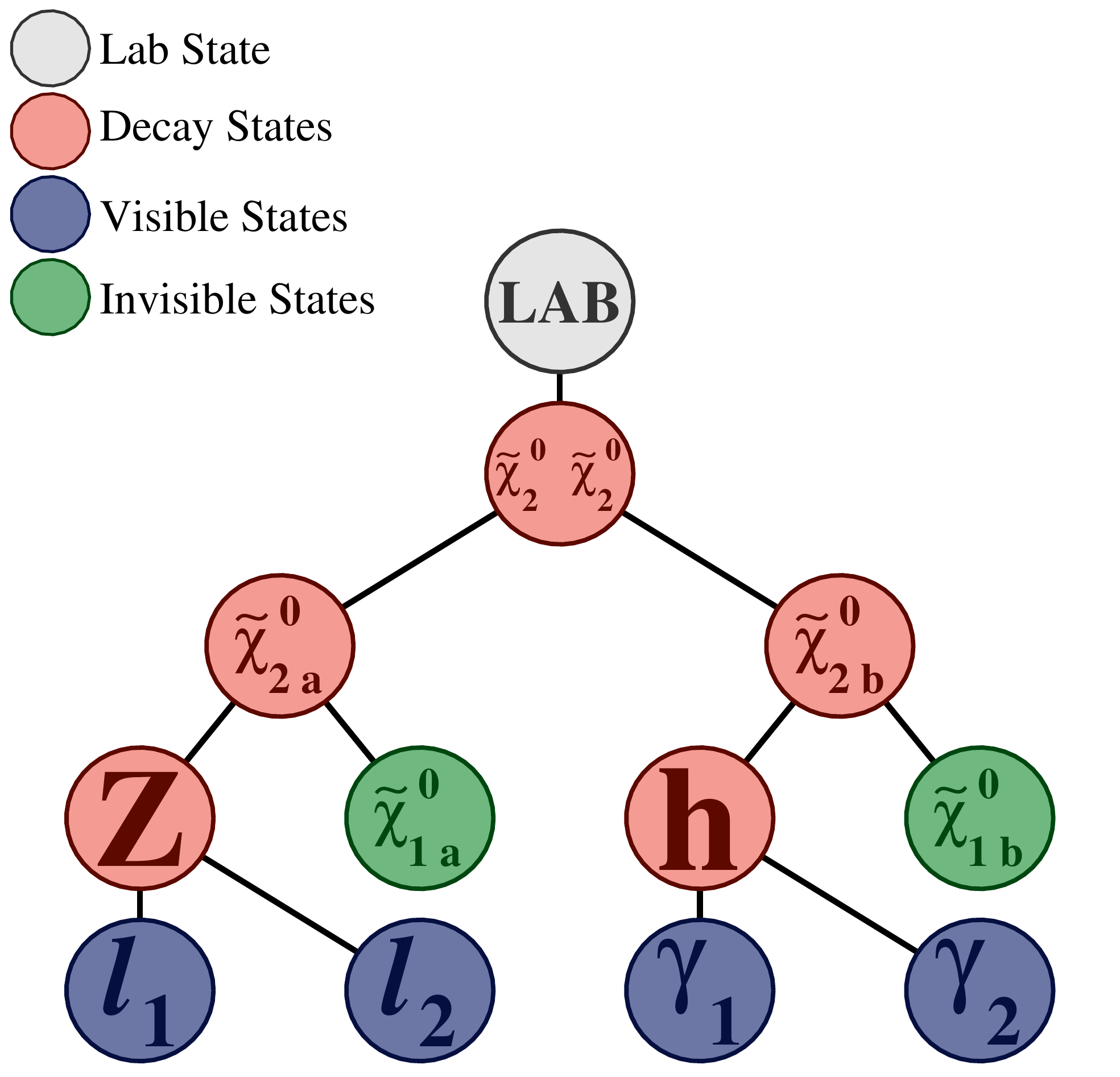}
\vspace{-0.1cm}
\caption{\label{fig:decayTree_Chi2Chi2} The decay tree for analyzing $\tilde{\chi}_{2}^{0}\tilde{\chi}_{2}^{0}\rightarrow Z(\ell^{+}\ell^{-})\tilde{\chi}_{1}^{0} h(\rightarrow \gamma\gamma)\tilde{\chi}_{1}^{0}$ events. The two $\tilde{\chi}_{2}^{0}$'s are expected to be produced non-resonantly and each proceeds through two intermediate resonances of differing mass with different decay products. The two final state $\tilde{\chi}_{1}^{0}$ particles are potentially massive.}
\end{figure}

We conclude our discussion of decay topologies with two visible and two invisible particles in the final state by considering the example of neutralino, $\tilde{\chi}^{0}_{2}$, pair production at a hadron collider, where the second neutralino mass eigenstates each decay to a lighter neutralino, $\tilde{\chi}^{0}_{1}$, and a boson, either a $Z$ or Higgs (with $m_{h} \sim 125$ GeV). The two final state bosons each decay to a pair of visible particles which we assume have been identified and reconstructed in the detector, with $Z\rightarrow \ell^+\ell^-$ and $h\rightarrow \gamma\gamma$. The decay tree for analyzing this final state is shown in Fig.~\ref{fig:decayTree_Chi2Chi2}.

The similarity of this decay topology to the previous two examples (Fig.~\ref{fig:decayTree_HWW} and~\ref{fig:decayTree_DiStop_to_hadtopXhadtopX}) is clear; a pair of massive particles, produced resonantly or non-resonantly, each decay to a visible system of particles and an invisible system. The fact that our two visible particles, the Higgs and $Z$ bosons, themselves decay to a pair of measured particles is immaterial to the strategy for choosing missing information associated with the invisible particles in the event, as we can analyze the event as a function of the measured four vectors, $\pfour{\V{$Z$}}{\lab} = \pfour{\V{$\ell_{1}$}}{\lab}+\pfour{\V{$\ell_{2}$}}{\lab}$ and $\pfour{\V{$h$}}{\lab} = \pfour{\V{$\gamma_{1}$}}{\lab}+\pfour{\V{$\gamma_{2}$}}{\lab}$, using the same JR's as in the previous example from Section~\ref{subsec:Part2_exampleB}.

The important distinction in this example is that not only are our visible states, $Z$ and $h$, massive, but their masses are distinctly different by a non-negligible amount, adding an asymmetry to the kinematics of the event. Defining $\I{I} = \{ \I{$\tilde{\chi}_{1a}$}, \I{$\tilde{\chi}_{1b}$} \}$ to be the collection of the invisible final state particles in the event, we interpret the measured \met~ as the transverse momentum of the $\I{I}$ system, and further define $\V{V} = \{ \V{$\ell_{1}$}, \V{$\ell_{2}$}, \V{$\gamma_{1}$}, \V{$\gamma_{2}$} \}$ to be the collection of all the visible particles in the final state. Choosing the unknown degrees of freedom describing how the total $\I{I}$ system momentum is shared between the two neutralinos using the contra-boost invariant JR~\ref{jr:contra2}, we find that our estimators for \Mass{\I{$\tilde{\chi}_{1a/b}$}}{} are non-zero, even if we prefer they are always zero. If we choose the parameters of the JR to make these masses as small as possible, while ensuring they remain non-negative, we find that
\bea
\label{eqn:Mchimax}
\Mass{\I{$\tilde{\chi}_{1a}$}}{2} &=& \Mass{\I{$\tilde{\chi}_{1b}$}}{2} + (\mass{\V{$h$}}{2}-\mass{\V{$Z$}}{2})\sqrt{1- \frac{4\mass{\V{$Z$}}{2}}{( \E{\V{V}}{\D{cm}})^2}} \nonumber \\
\Mass{\I{$\tilde{\chi}_{1b}$}}{2} &=&
  \mass{\V{$Z$}}{2}\frac{|\pthree{\V{$Z$}}{\D{cm}} - \pthree{\V{$h$}}{\D{cm}}|^{2}}{(\E{\V{$Z$}}{\D{cm}} + \E{\V{$h$}}{\D{cm}})^2}  \\
  &+&   \frac{1}{2}(\mass{\V{$h$}}{2}-\mass{\V{$Z$}}{2})\left[1-\sqrt{1- \frac{4\mass{\V{$Z$}}{2}}{( \E{\V{V}}{\D{cm}})^2}} \right]~, \nonumber
\eea
where $\D{cm} = \{\V{V},\I{I}~\}$ denotes the center-of-mass frame of the interaction and we have assumed $\mass{\V{$h$}}{} \geq \mass{\V{$Z$}}{}$. Compared to Eq.~\ref{eqn:Mchimin} in the previous example, the expression for \Mass{\I{$\tilde{\chi}_{1b}$}}{} in Eq.~\ref{eqn:Mchimax} has an additional term proportional to  $\mass{\V{$h$}}{2} - \mass{\V{$Z$}}{2}$, implying that differences in the visible particle masses are additionally absorbed into the invisible particle masses. More concerning is that \Mass{\I{$\tilde{\chi}_{1a}$}}{} is systematically larger than \Mass{\I{$\tilde{\chi}_{1b}$}}{}, where Eq.~\ref{eqn:Mchimax} implies
\bea
\Mass{\I{$\tilde{\chi}_{1a}$}}{2}-\Mass{\I{$\tilde{\chi}_{1b}$}}{2} \geq (\mass{\V{$h$}}{}-\mass{\V{$Z$}}{})^{\frac{3}{2}}(\mass{\V{$h$}}{}+3\mass{\V{$Z$}}{})^{\frac{1}{2}} ~.
\eea
The constraint $\Mass{\D{$\tilde{\chi}_{2a}$}}{} = \Mass{\D{$\tilde{\chi}_{2b}$}}{}$ associated with the contra-boost invariant JR has caused our \Mass{\I{$\tilde{\chi}_{1a/b}$}}{} estimators to develop a large systematic difference, a kinematic feature that is not present in the process we are hoping to study. In our case, the two lightest neutralinos have the same mass, meaning that the one associated with the $Z$ decay should systematically receive more momentum from its parent's decay rather than increase in mass itself. More generally, for lack of a better choice, a practical application would desire that both the mass estimators \Mass{\I{$\tilde{\chi}_{1a/b}$}}{} take values as close to zero as possible. To achieve this behavior, we reconsider how the contra-boost invariant JR we apply is constructed. 

With the assumption $\Mass{\D{$\tilde{\chi}_{2a}$}}{} = \Mass{\D{$\tilde{\chi}_{2b}$}}{}$, we can again relate the velocities of these two particles in their mutual center-of-mass frame by a single contra-boost, $\vec{\beta}_{c}$, and choose its value to constrain the momentum of the invisible neutralinos. Instead of setting $\vec{\beta}_{c}$ according to an explicit minimization as we did in previous examples, we consider an {\it ad-hoc} generalization that maintains the contra-boost invariance of the choice:
\bea
\label{eqn:betac_gen}
\vec{\beta}_{c} = \vbeta{\D{$\tilde{\chi}_{2a}$}}{\D{cm}} = - \vbeta{\D{$\tilde{\chi}_{2b}$}}{\D{cm}} = \frac{ c_a \pthree{\V{$Z$}}{\D{cm}} - c_b \pthree{\V{$h$}}{\D{cm}}}{c_a \E{\V{$Z$}}{\D{cm}} + c_b \E{\V{$h$}}{\D{cm}}}~,
\eea
where $c_a$ and $c_b$ are unspecified functions of the visible four vectors in the interaction center-of-mass frame, \D{cm}. With this choice, the reconstructed energies of the visible particles in their respective production frames can be expressed as
\bea
\label{eqn:EZH}
\E{\V{$Z$}}{\D{$\tilde{\chi}_{2a}$}} &=& \frac{ c_a \mass{\V{$Z$}}{2} + (c_b/2) M_{c}^{2}(\pfour{\V{$Z$}}{\D{cm}},\pfour{\V{$h$}}{\D{cm}}) }{ 
\sqrt{c_a^2 \mass{\V{$Z$}}{2} +    c_b^2 \mass{\V{$h$}}{2} + c_a c_b M_{c}^{2}(\pfour{\V{$Z$}}{\D{cm}},\pfour{\V{$h$}}{\D{cm}})}} \\
\E{\V{$h$}}{\D{$\tilde{\chi}_{2b}$}} &=& \frac{ c_b \mass{\V{$h$}}{2} + (c_a/2) M_{c}^{2}(\pfour{\V{$Z$}}{\D{cm}},\pfour{\V{$h$}}{\D{cm}}) }{ 
\sqrt{c_a^2 \mass{\V{$Z$}}{2} +    c_b^2 \mass{\V{$h$}}{2} + c_a c_b M_{c}^{2}(\pfour{\V{$Z$}}{\D{cm}},\pfour{\V{$h$}}{\D{cm}})}}~. \nonumber
\eea
As there are only contra-boost invariant quantities appearing in Eq.~\ref{eqn:EZH} this means that, as long as there are also only contra-boost invariant quantities appearing in $c_a$ and $c_b$, our estimators \E{\V{$Z$}}{\D{$\tilde{\chi}_{2a}$}} and \E{\V{$h$}}{\D{$\tilde{\chi}_{2b}$}} will exhibit this property. 

Hence, there is a family of contra-boost invariant choices for $\beta_{c}$, defined by the different contra-boost invariant choices for the factors $c_a$ and $c_b$, which can be constructed from factors like \mass{\V{$Z$}}{2}, \mass{\V{$h$}}{2}, and $M_{c}^{2}(\pfour{\V{$Z$}}{\D{cm}},\pfour{\V{$h$}}{\D{cm}})$ which have this invariance property. We would like to use the additional degree of freedom associated with this choice to mitigate the large values of the estimators of invisible particle masses seen in Eq.~\ref{eqn:Mchimax}.

Defining
\bea
c_a &=& \frac{1}{2}\left(1+\hat{k}k_{a}\right) \\
c_b &=& \frac{1}{2}\left(1+\hat{k}k_{b}\right)~, \nonumber
\eea
the mass-squared difference of our estimators \Mass{\I{$\tilde{\chi}_{1a/b}$}}{}, using the more general contra-boost of Eq.~\ref{eqn:betac_gen}, can be expressed as
\bea
&& \Delta \Mass{\I{$\tilde{\chi}_{1}$}}{2} \equiv \Mass{\I{$\tilde{\chi}_{1a}$}}{2} - \Mass{\I{$\tilde{\chi}_{1b}$}}{2} = \\
&& \hat{k} \left[ (k_a - k_b)M_{c}^{2}(\pfour{\V{$Z$}}{\D{cm}},\pfour{\V{$h$}}{\D{cm}})/2 - (k_a \mass{\V{$Z$}}{2} - k_b \mass{\V{$h$}}{2}) \right]~. \nonumber
\eea
We use our choice for the factors $k_a$ and  $k_b$ to minimize $\Delta \Mass{\I{$\tilde{\chi}_{1}$}}{2}$, in particular ensuring that $\lim_{M_{c}^{2} \rightarrow \infty} \Delta \Mass{\I{$\tilde{\chi}_{1}$}}{2} = 0$, with
\bea
k_a &=&  \mass{\V{$Z$}}{2} - \mass{\V{$h$}}{2} + M_{c}^{2}(\pfour{\V{$Z$}}{\D{cm}},\pfour{\V{$h$}}{\D{cm}}) - 2 \mass{\V{$Z$}}{2}\mass{\V{$h$}}{2} \\
k_b &=&  \mass{\V{$h$}}{2} - \mass{\V{$Z$}}{2} + M_{c}^{2}(\pfour{\V{$Z$}}{\D{cm}},\pfour{\V{$h$}}{\D{cm}}) - 2 \mass{\V{$Z$}}{2}\mass{\V{$h$}}{2}~, \nonumber
\eea
which results in
\bea
\Delta \Mass{\I{$\tilde{\chi}_{1}$}}{2} = \hat{k} (\mass{\V{$h$}}{}+\mass{\V{$Z$}}{})(\mass{\V{$h$}}{}-\mass{\V{$Z$}}{})^{3} ~.
\eea
The factor $\hat{k}$ effectively normalizes $k_a$ and  $k_b$ to be dimensionless, and with $\lim_{M_{c}^{2} \rightarrow \infty} \hat{k}^{-1} = M_{c}^{2}(\pfour{\V{$Z$}}{\D{cm}},\pfour{\V{$h$}}{\D{cm}})$ the invisible particle mass difference will approach zero when the mass splitting between parent $\tilde{\chi}_{2}^{0}$ and invisible $\tilde{\chi}_{1}^{0}$ is large, relative to the $Z$ and Higgs boson masses. 

To see that our ultimate choice for $\hat{k}$ has the expected asymptotic behavior, we choose it's value by setting $\Mass{\I{$\tilde{\chi}_{1b}$}}{} = 0$ and solving for $\hat{k}$:
\bea
&& \hat{k} = \\
&& \frac{ k_b \mass{\V{$h$}}{2} - k_a \mass{\V{$Z$}}{2} + \frac{k_a-k_b}{2}M_{c}^{2} + \frac{k_a+k_b}{2}\sqrt{M_{c}^{4}-4 \mass{\V{$Z$}}{2} \mass{\V{$h$}}{2} }}{ k_a^2 \mass{\V{$Z$}}{2} + k_b^2 \mass{\V{$h$}}{2} +k_a k_b M_{c}^{2}}~, \nonumber
\eea
where $M_{c}^{2} \equiv M_{c}^{2}(\pfour{\V{$Z$}}{\D{cm}},\pfour{\V{$h$}}{\D{cm}})$.  As $\lim_{M_{c}^{2} \rightarrow \infty} k_{a} = \lim_{M_{c}^{2} \rightarrow \infty} k_{a} = M_{c}^{2}$, our choices for $k_a$, $k_b$, and $\hat{k}$ ensure that $\Delta \Mass{\I{$\tilde{\chi}_{1}$}}{2}$, and the individual neutralino masses, are reconstructed to be as small as possible.

As was the case in the previous example, our expression for \Mass{\I{I}}{} with these choices is not Lorentz invariant, which means we must identify the smallest Lorentz invariant function of the visible particles' four vectors which is at least as large as this current estimator, which ensures that the invisible particle mass estimators remain non-negative. In this case, we find
\bea
\Mass{\I{I}}{2} = \Mass{\V{V}}{2} - 4 \mass{\V{$Z$}}{}\mass{\V{$h$}}{} \geq \Mass{\I{I}}{2} |_{\Mass{\I{$\tilde{\chi}_{1b}$}}{}=0}~,
\eea
which is smaller than the value $\Mass{\V{V}}{2} - 4 \mass{\V{$Z$}}{2}$ which would be required when using the previous contra-boost JR~\ref{jr:contra2}. 

The details of how the invisible neutralino four vectors are chosen with this approach can be summarized as a more general contra-boost JR:
\begin{jigsaw}[Contra-boost Invariant]
\label{jr:contra}
If the internal degrees of freedom specifying how an invisible particle, $\I{I} = \{ \I{I$_{a}$}, \I{I$_{b}$} \}$, should split into two particles are unknown, they can be specified by choosing a corresponding pair of visible particles,  $\V{V} = \{ \V{V$_{a}$}, \V{V$_{b}$} \}$, and applying the constraint $\Mass{\V{V$_{a}$}\I{I$_{a}$}}{} = \Mass{\V{V$_{b}$}\I{I$_{b}$}}{}$. It is assumed that the four vectors of the visible particles are known in the center-of-mass frame, $\D{F} = \{ \V{V}, \I{I}~\}$, as is the four vector of the total \I{I} system, \pfour{\I{I}}{\D{F}}. 
The four vectors of the invisible particles can be chosen in the \D{F} frame as:
\bea
\pthree{\I{I$_{a}$}}{\D{F}} &=& (\hat{c}c_a-1) \pthree{\V{V$_{a}$}}{\D{F}} - \hat{c}c_b \pthree{\V{V$_{b}$}}{\D{F}} \nonumber \\
\pthree{\I{I$_{b}$}}{\D{F}} &=& (\hat{c}c_b-1) \pthree{\V{V$_{b}$}}{\D{F}} - \hat{c}c_a \pthree{\V{V$_{a}$}}{\D{F}}  \\
\E{\I{I$_{a}$}}{\D{F}} &=& (\hat{c}c_a-1)\E{\V{V$_{a}$}}{\D{F}} + \hat{c}c_b\E{\V{V$_{b}$}}{\D{F}}  \nonumber \\
\E{\I{I$_{b}$}}{\D{F}} &=& (\hat{c}c_b-1)\E{\V{V$_{b}$}}{\D{F}} + \hat{c}c_a\E{\V{V$_{a}$}}{\D{F}}~, \nonumber
\eea
where $\hat{c}$, $c_a$, and $c_b$ are factors whose functional forms depend on the masses of the individual visible and invisible particles. We assume, without loss of generality, that $\mass{\V{V$_a$}}{} \geq \mass{\V{V$_b$}}{}$ and that the masses of the individual invisible particles are required to satisfy $\Mass{\I{I$_a$}}{} \geq \mu_a$ and $\Mass{\I{I$_b$}}{} \geq \mu_b$, with $\mu = \max (\mu_a, \mu_b)$.
\end{jigsaw}
\begin{subjigsaw}[$\mu \leq \mass{\V{V$_b$}}{}$] 
\label{jr:contra_a}
The factors $c_a$ and $c_b$ are defined in terms of parameters $\hat{k}$, $k_a$, and $k_b$, with
\bea
c_a = \frac{1}{2}\left(1+\hat{k}k_{a}\right) ~,~c_b = \frac{1}{2}\left(1+\hat{k}k_{b}\right)~, 
\eea
with $\hat{k}$, $k_a$, and $k_b$ chosen as
\bea
k_a &=&  \mass{\V{V$_a$}}{2} - \mass{\V{V$_b$}}{2} +M_{c}^{2} - 2 \mass{\V{V$_a$}}{2}\mass{\V{V$_b$}}{2} \nonumber \\
k_b &=&  (\mass{\V{V$_a$}}{2} - \mass{\V{V$_b$}}{2})(\frac{2\mu}{\mass{\V{V$_b$}}{}}-1) + M_{c}^{2} - 2 \mass{\V{V$_a$}}{2}\mass{\V{V$_b$}}{2} \nonumber \\
\hat{k} &=& \frac{\hat{k}_n}{\hat{k}_d^2}~,~with,\\
\hat{k}_d^2 &=& k_a^2 \mass{\V{V$_a$}}{2} + k_b^2 \mass{\V{V$_b$}}{2} +k_a k_b M_{c}^{2} \nonumber \\
\hat{k}_n &=& k_a \mass{\V{V$_a$}}{2} - k_b \mass{\V{V$_b$}}{2} + \frac{k_b-k_a}{2}M_{c}^{2} \nonumber \\
&+& \frac{1}{2}\sqrt{(k_a+k_b)^2(M_{c}^{4}-4 \mass{\V{V$_a$}}{2} \mass{\V{V$_b$}}{2}) + 16 \mu^2 \hat{k}_d^2 } \nonumber 
\eea
where $M_{c}^{2} \equiv M_{c}^{2}(\pfour{\V{V$_a$}}{\D{F}},\pfour{\V{V$_b$}}{\D{F}})$. $\hat{c}$ is given by
\bea
\hat{c} = \frac{1}{2}\left[ \frac{\E{\V{V}}{\D{F}} +   \sqrt{(\E{\V{V}}{\D{F}})^{2} +\Mass{\I{I}}{2} - \mass{\V{V}}{2}}}{c_a\E{\V{V$_a$}}{\D{F}}+c_b\E{\V{V$_b$}}{\D{F}}} \right]~.
\eea
In order to guarantee that $\Mass{\I{I$_b$}}{} \geq \Mass{\I{I$_a$}}{} \geq \mu$, \Mass{\I{I}}{} must be chosen to be at least as large as
\bea
\Mass{\I{I}}{2} \geq \mass{\V{V}}{2} + 4(\mu + \mass{\V{V$_{a}$}}{})(\mu-\mass{\V{V$_{b}$}}{})~.
\eea
\end{subjigsaw}
\vspace{0.2cm}
\begin{subjigsaw}[$\mu > \mass{\V{V$_b$}}{}$] 
\label{jr:contra_b}
The factors $\hat{c}$, $c_a$ and $c_b$ are defined as
\bea
c_a &=& c_b = 1 \\
\hat{c} &=& \frac{1}{2}\left[1 +   \frac{\sqrt{(\E{\V{V}}{\D{F}})^{2} +\Mass{\I{I}}{2} - \mass{\V{V}}{2}}}{\E{\V{V}}{\D{F}}} \right]~. \nonumber
\eea
In order to guarantee that $\Mass{\I{I$_a$}}{} \geq \mu_a$ and $\Mass{\I{I$_b$}}{} \geq \mu_b$, \Mass{\I{I}}{} must be chosen to be at least as large as
\bea
\Mass{\I{I}}{2} \geq \frac{\max (\mu_a^2-\mass{\V{V$_a$}}{2},\mu_b^2-\mass{\V{V$_b$}}{2})}{\mass{\V{V$_b$}}{2}}\mass{\V{V}}{2}~.
\eea
\end{subjigsaw}
 
The RJR steps to analyzing $\tilde{\chi}^{0}_{2}\tilde{\chi}^{0}_{2} \rightarrow Z(\ell\ell)\tilde{\chi}^{0}_{1} h(\gamma\gamma) \tilde{\chi}^{0}_{1}$ events at a hadron collider can be summarized as: 
\begin{enumerate}[noitemsep]
\item Apply the invisible mass JR~\ref{jr:mass}, choosing \Mass{\I{I}}{} as the smallest possible quantity consistent with JR~\ref{jr:contra_a} and non-tachyonic neutralinos.
\item Apply the invisible rapidity JR~\ref{jr:rapidity}, \\ choosing \pone{\I{I},z}{\lab} using all the visible particles \V{V}.
\item Apply the contra-boost invariant JR~\ref{jr:contra_a}, \\ specifying the neutralino four vectors \\ using the constraint $\Mass{\D{$\tilde{\chi}_{2a}$}}{} = \Mass{\D{$\tilde{\chi}_{2b}$}}{}$.
\end{enumerate}

\begin{figure}[htbp]
\centering 
\includegraphics[width=.35\textwidth]{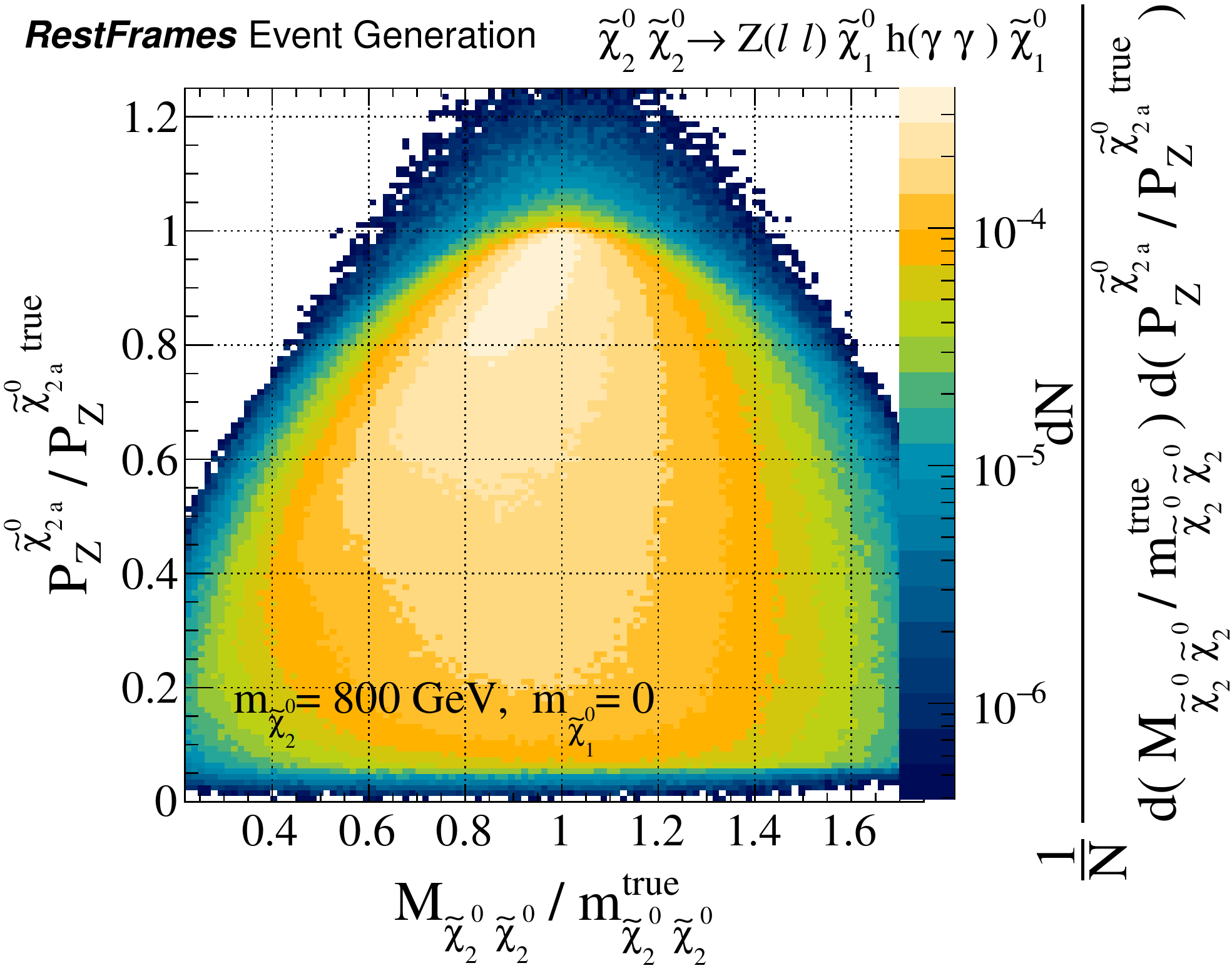}
\vspace{-0.3cm}
\caption{\label{fig:example_X2X2_to_ZllXHggX-PZ_v_MCM} The distribution of the reconstructed $Z$ boson momentum in the approximation of its production frame, as a function of the estimator \Mass{\D{cm}}{}, for simulated $\tilde{\chi}_{2}^{0}\tilde{\chi}_{2}^{0}\rightarrow Z(\ell^{+}\ell^{-})\tilde{\chi}_{1}^{0} h(\gamma\gamma)\tilde{\chi}_{1}^{0}$ events. Each of the estimated quantities is normalized by the true value event-by-event. }
\end{figure}

Despite the complications involved with the non-trivial $Z$ and Higgs masses, the contra-boost invariant jigsaw allows the kinematics of these visible systems in their production frames to be estimated with little bias. The reconstructed $Z$ boson momentum in the approximation of its production frame, \pone{\V{$Z$}}{\D{$\tilde{\chi}_{2a}$}}, is shown in Fig.~\ref{fig:example_X2X2_to_ZllXHggX-PZ_v_MCM}, where the similarities to Fig.~\ref{fig:example_H_to_WlnuWlnu-MW_v_MH} and~\ref{fig:example_DiStop_to_hadtopXhadtopX-Ptop_v_MTT} showing analogous observables from previous examples is striking. \pone{\V{$Z$}}{\D{$\tilde{\chi}_{2a}$}}, which is sensitive to the mass splitting between the two neutralino states, is estimated nearly independently of the total di-$\tilde{\chi}_{2}^{0}$ mass, \Mass{\D{cm}}{}, with a kinematic endpoint reliably falling at the true value. Relative to the previous examples, no additional distortion in the \pone{\V{$Z$}}{\D{$\tilde{\chi}_{2a}$}} distribution due to the $Z$ and Higgs masses is visible.

\begin{figure}[htbp]
\centering 
\includegraphics[width=.35\textwidth]{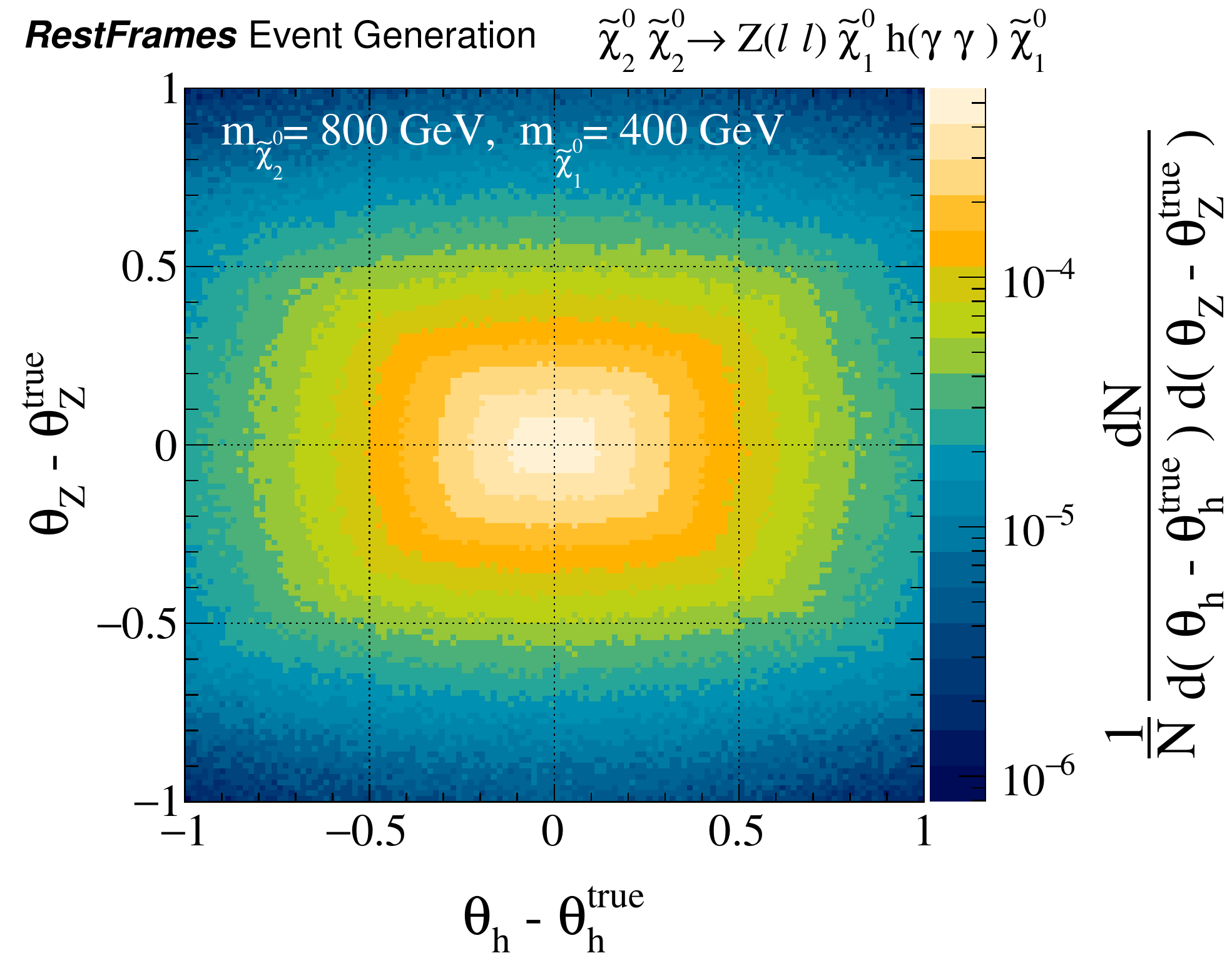}
\vspace{-0.3cm}
\caption{\label{fig:example_X2X2_to_ZllXHggX-thetas} The distribution of the reconstructed $Z$ decay angle, $\theta_{\V{$Z$}}$, as a function of the Higgs decay angle, $\theta_{\V{$h$}}$, for simulated $\tilde{\chi}_{2}^{0}\tilde{\chi}_{2}^{0}\rightarrow Z(\ell^{+}\ell^{-})\tilde{\chi}_{1}^{0} h(\gamma\gamma)\tilde{\chi}_{1}^{0}$ events. Both observables are shown relative to the true decay angles they are estimating, with angles in units radian.}
\end{figure}

That the approximations of the neutralino rest frames, \D{$\tilde{\chi}_{2a}$} and \D{$\tilde{\chi}_{2b}$}, have a strong correspondence to the true ones can be seen by considering the reconstructed decay angles of the $Z$ and Higgs bosons, which require knowledge of these reference frames as they are where the bosons are produced. The estimators $\theta_{\V{$Z$}}$ and $\theta_{\V{$h$}}$ can resolve the true quantities quite well, as demonstrated in Fig.~\ref{fig:example_X2X2_to_ZllXHggX-thetas}, and are insensitive to the momentum of the center-of-mass system in the lab frame and the masses of the sparticles in the event. This is possible because the approximate \D{$\tilde{\chi}_{2a}$} and \D{$\tilde{\chi}_{2b}$} rest frames have effectively {\it inherited} the invariance properties of the reference frames that proceed them in the decay tree, with longitudinal and contra-boost invariant definitions. As was the case for the $h \rightarrow \gamma\gamma$ decays in Section~\ref{subsec:Part1_exampleC}, the approximate reconstruction of the intermediate decay frames in the event allows the decay angles of these visible systems to be estimated with excellent precision.

\FloatBarrier

\section{More Jigsaws for Two Invisible Particles}
\label{sec:Part3}

In this section, we expand the library of JR's designed to study final states with two invisible particles by considering a series of examples motivated by top pair production. In fully-leptonic $t\bar{t}$ events, two top quarks each decay to a $b$-quark and a leptonically decaying $W$ boson, resulting in a final state with four visible, reconstructable particles, and two neutrinos. The additional two visible particles relative to the examples of Section~\ref{sec:Part2} provide both new challenges and opportunities. While a combinatoric ambiguity must be resolved as to which $b$-tagged jet to associate with each lepton, more visible particles allow for better resolution of the under-constrained neutrino kinematics. 

We consider three examples with this same final state, including non-resonant top pair-production (Section~\ref{subsec:Part3_exampleA}), resonant $t\bar{t}$ production through a heavy Higgs boson (Section~\ref{subsec:Part3_exampleB}), and stop quark pair-production with decays through charginos and sneutrinos (Section~\ref{subsec:Part3_exampleC}).  Additional JR's for studying this final state are described in Section~\ref{subsec:Part3_exampleA} with several approaches compared throughout the examples. 


\subsection{$t\bar{t} \rightarrow bW(\ell\nu)bW(\ell\nu)$ at a hadron collider}
\label{subsec:Part3_exampleA}

\begin{figure}[htbp]
\centering 
\includegraphics[width=.33\textwidth]{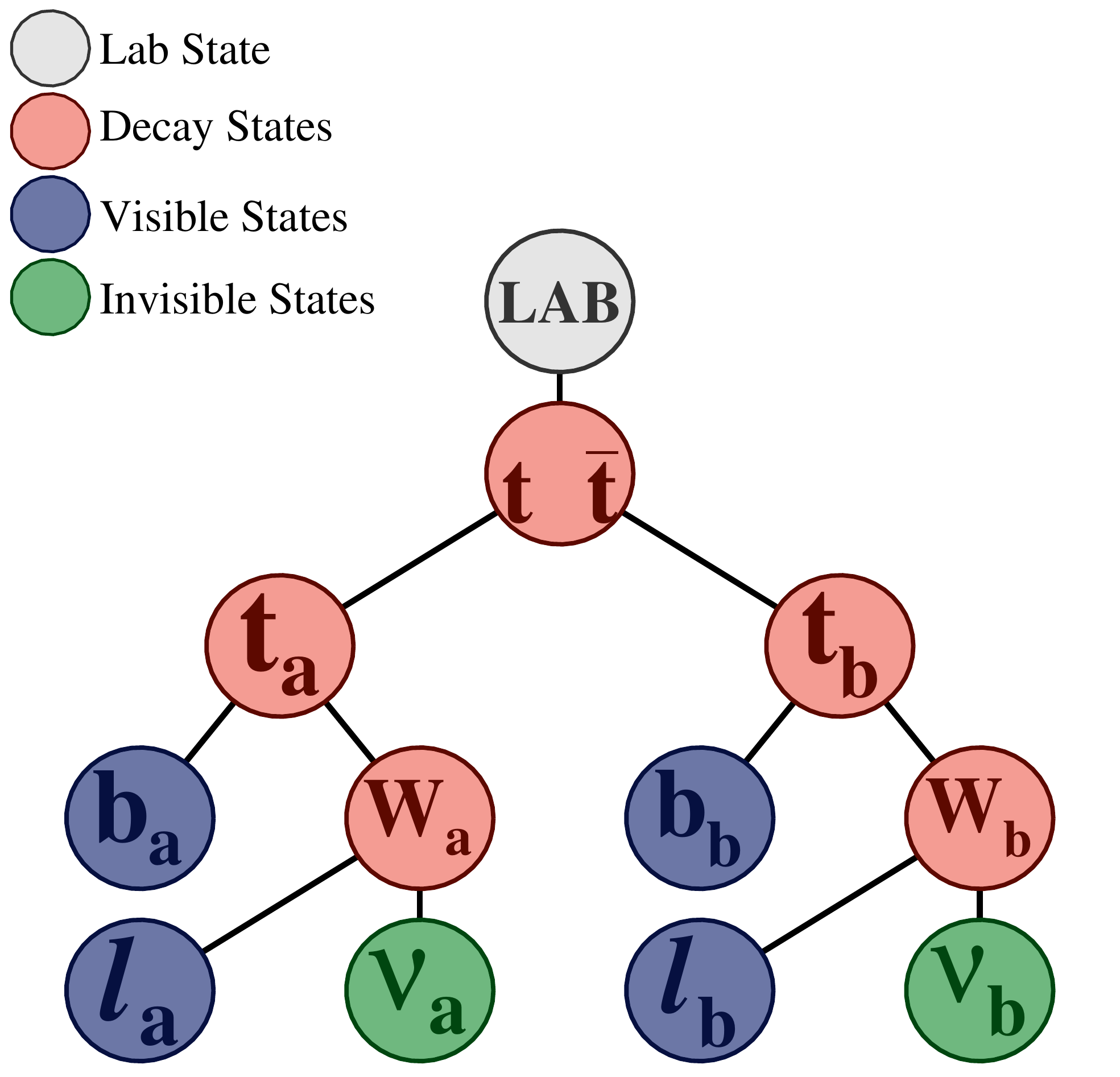}
\vspace{-0.3cm}
\caption{\label{fig:example_ttbar_to_bWlnubWlnu-decay} The decay tree for analyzing $t\bar{t} \rightarrow bW(\ell\nu)bW(\ell\nu)$ events. Four visible particles are reconstructed in the final state, along with two invisible particles which are constrained by the measured \met. There are several masses of interest in the event, with two intermediate top quarks and two $W$ bosons appearing in the decays.}
\end{figure}

The first case we consider with four visible and two invisible particles in the final state is top pair production at a hadron collider, with subsequent decays to $b$-quarks, leptons, and neutrinos via intermediate $W$ bosons. We assume in these events that two $b$-tagged jets are identified and reconstructed in the detector, with measured four vectors \pfour{\V{$b_a$}}{\lab} and \pfour{\V{$b_b$}}{\lab}, along with two leptons, with four vectors \pfour{\V{$\ell_a$}}{\lab} and \pfour{\V{$\ell_b$}}{\lab}. The measured \met~ is interpreted as the transverse momentum of the di-neutrino system, \pthree{\I{I},T}{\lab}, with $\I{I} = \{ \I{$\nu_a$}, \I{$\nu_{b}$} \}$. The decay tree used for analyzing this final state is shown in Fig.~\ref{fig:example_ttbar_to_bWlnubWlnu-decay}.

In addition to unknowns associated with invisible particles in the event, there is a {\it combinatoric ambiguity} in deciding which reconstructed $b$-tagged jet should be associated with each lepton. We assume that we are unable to reliably distinguish between $b$-tagged jets initiated by $b$-quarks and those from anti-particles, so we make this choice solely relying on the kinematics of each event. Defining $\V{V$_{a}$} = \{ \V{$b_a$}, \V{$\ell_a$} \}$ and $\V{V$_{b}$} = \{ \V{$b_b$}, \V{$\ell_b$} \}$ as the two sets of visible particles associated with each top decay, we define a JR to choose the $\V{$b$}/\V{$\ell$}$ pairing which minimizes the function $\Mass{\V{V$_{a}$}}{2} + \Mass{\V{V$_{b}$}}{2}$. This is a simple, and generally correct, prescription as it chooses combinations where the sum of four vector inner products is smallest, effectively pairing particles flying closest together as expected from a common decay source. More generally, we can define this combinatoric JR as follows:
\begin{jigsaw}[Combinatoric Minimization]
If there is a set of $n$ visible particles, $\V{V} = \{ \V{V$_{1}$},\cdots,\V{V$_{n}$} \}$, we can choose a partition of \V{V} into $m \leq n$ subsets, $P_{\V{V}} = \{ S_{\V{V$1$}}, \cdots,  S_{\V{V$m$}} \}$, by minimizing a chosen metric over the space of all valid partitions, $P_{\V{V}} \in \mathbb{P}_{\V{V}}$.

A partition, $P_{\V{V}}$, is valid if it satisfies the conditions
\bea
&(a)&~\bigcup_{S_{\V{V}} \in P_{\V{V}}} S_{\V{V}} = \V{V} \nonumber \\
&(b)&~\forall S_{\V{V$a$}},S_{\V{V$b$}} \in P_{\V{V}}~,  \\
&& ~S_{\V{V$a$}} \neq S_{\V{V$b$}} \implies S_{\V{V$a$}} \cap S_{\V{V$b$}} = \emptyset \nonumber \\
&(c)&~|S_{\V{V$i$}}| \geq or = n_{i} \geq 1~for~any~requirements~n_i, \nonumber \\
&(d)&~Q(S_{\V{V$i$}}) = q_{i}~for~any~charge~requirements~q_i~. \nonumber 
\eea
We assume there is at least one valid partition in each event, $\ni |\mathbb{P}_{\V{V}}| \geq 1$, and choose a partition, $\hat{P}_{\V{V}}$, from this set by minimizing a function of the event's kinematics, $f(P_{\V{V}})$, that is sensitive to this choice, with
\bea
\label{eqn:comb_min}
f(\hat{P}_{\V{V}}) = \min_{P_{\V{V}} \in \mathbb{P}_{\V{V}}} f(P_{\V{V}})~.
\eea
If the function $f(P_{\V{V}})$ depends on other unknown kinematic or combinatoric information that depends on the application of other JR's which, in-turn, depend on the choice $\hat{P}_{\V{V}}$, then these JR's are evaluated independently for each value $P_{\V{V}}$ in Eq.~\ref{eqn:comb_min}. 
\end{jigsaw}
\begin{subjigsaw}[Minimize Masses Squared]
\label{jr:minM2}
A convenient choice for $f(P_{\V{V}})$ is the sum of four vector inner products of the elements of the $m$ sets in a partition, potentially including other particles in the event. With $\V{O} = \{ \V{O$_{1}$},\cdots,\V{O$_{m}$} \}$ the set of other particles associated with each combinatoric subset in a partition $P_{\V{V}}$, we can define $f(P_{\V{V}})$ as
\bea
f(P_{\V{V}}) &=& \sum_{i=1}^{m} \Mass{S_{\V{V$i$}}\V{O$_i$}}{2} = \sum_{i=1}^{m}(\pfour{S_{\V{V$i$}}}{}+\pfour{\V{O$_i$}}{})^{2}~.  
\eea
The appeal of this functional form is that it can be linearly factorized, in the sense that if $P_{\V{V}}^{'} = \{ S_{\V{V$1$}}^{'}, \cdots,  S_{\V{V$m$}^{'}}^{'} \}$ is a partition of \V{V} where $P_{\V{V}}$ is its refinement, such that $|P_{\V{V}}| \leq |P_{\V{V}}^{'}|$ and $\forall S_{\V{V}} \in  P_{\V{V}},~ \exists~S_{\V{V}}^{'} \in P_{\V{V}}^{'}~with~S_{\V{V}} \subseteq S_{\V{V}}^{'}$, then $f(P_{\V{V}})$ can be expressed as
\bea
f(P_{\V{V}}) = \sum_{S_{\V{V}}^{'} \in P_{\V{V}}^{'}} f(S_{\V{V}}^{'}) =  \sum_{S_{\V{V}}^{'} \in P_{\V{V}}^{'}} \sum_{S_{\V{V}} \in S_{\V{V}}^{'}} f(S_{\V{V}})~.
\eea
This implies one can use recursive applications of this JR corresponding to a sequence of progressively fine partitions of \V{V}, effectively minimizing the same function as a single application, but in factorized steps, potentially improving the resolution of intermediate particle structure in an event if chosen correspondingly.
\end{subjigsaw}

With this choice for the $\V{$b$}/\V{$\ell$}$ pairing in each event, the remaining unknowns associated with the neutrinos are the same as in each of the examples of Section~\ref{sec:Part2}, and can be represented as the mass of the total invisible system, \Mass{\I{I}}{}, its momentum along the beam axis, the orientation of the ``decay'' of \I{I} into two neutrinos, and the two individual neutrino masses. The increased number of visible particles means that there are choices in strategy when applying the contra-boost invariant JR~\ref{jr:contra} to resolve these unknowns. For example, the two leptons could be paired with the two neutrinos, and a contra-boost invariant JR imposing $\Mass{\D{$W_a$}}{} = \Mass{\D{$W_b$}}{}$ applied. Alternatively, the two $\V{$b$}/\V{$\ell$}$ pairs, \V{V$_{a}$} and \V{V$_{b}$}, can be used, imposing the constraint $\Mass{\D{$t_a$}}{} = \Mass{\D{$t_b$}}{}$. While both approaches are perfectly applicable to this final state, there are cases where these mass equality constraints may not be appropriate. Furthermore, there may be cases where insisting on contra-boost invariance at the cost of the estimators of invisible particle masses acquiring additional mass (as described thoroughly in Section~\ref{sec:Part2}) may not be desirable. To address these cases, we introduce additional JR's for choosing the degrees of freedom associated with splitting apart a di-invisible system.

We take the same approach as Section~\ref{sec:Part2}, imagining that we are able to evaluate each of the visible four vectors in the event in the total center-of-mass frame, $\D{cm} = \D{$t\bar{t}$} = \{\V{V},\I{I} \}$, and the four vector associated with the total invisible system, \pfour{\I{I}}{\D{cm}}. Additionally, we now impose exact constraints on the individual neutrino masses, with $\Mass{\I{$\nu_a$}}{} = \Mass{\I{$\nu_b$}}{} = 0$. The remaining unknowns are associated with how the momentum of these two individual neutrinos are chosen in this reference frame, subject to the constraints
\bea
\pthree{\I{$\nu_a$}}{\D{cm}} + \pthree{\I{$\nu_b$}}{\D{cm}} + \pthree{\V{V}}{\D{cm}} = 0 \\
(\pfour{\I{$\nu_a$}}{\D{cm}} + \pfour{\I{$\nu_b$}}{\D{cm}})^2 = \Mass{\I{I}}{2}~, \nonumber
\eea
corresponding to two under-constrained degrees of freedom. 

As the individual neutrino masses are fixed, and the masses \mass{\V{V$_a$}}{} and \mass{\V{V$_b$}}{} will assume unequal, non-trivial values in these events, we are unable set our approximations of the two top masses exactly equal, as such a constraint could lead to tachyonic approximations of the neutrino four vectors. If we want to effectively minimize these masses w.r.t. our choices for the neutrino momentum, we must choose a new mass-sensitive metric.

A suitable choice for this metric can be seen more clearly if we instead work in the rest frame of the di-neutrino system, \D{I}. With our assumed knowledge of \pfour{\I{I}}{\D{cm}}, we can calculate the velocity relating \D{cm} to \D{I} as
\bea
\vbeta{\I{I}}{\D{cm}} = \frac{\pthree{\I{I}}{\D{cm}}}{\E{\I{I}}{\D{cm}}}~,
\eea
and, as we are in the di-neutrino rest frame, the two neutrinos must have equal and opposite momentum, with magnitude determined by \Mass{\I{I}}{}:
\bea
\pthree{\I{$\nu_a$}}{\D{I}} = -\pthree{\I{$\nu_b$}}{\D{I}} = \frac{\Mass{\I{I}}{}}{2}  \phat{\I{$\nu_a$}}{\D{I}}~.
\eea
We see that choosing the remaining degrees of freedom associated with the neutrinos amounts to choosing \phat{\I{$\nu_a$}}{\D{I}}. The momentum of the two visible systems of particles in this reference frame, \pthree{\V{V$_a$}}{\D{I}} and \pthree{\V{V$_b$}}{\D{I}}, define a plane, with normal vector $\hat{n}_{\V{V}} \propto \pthree{\V{V$_a$}}{\D{I}} \times \pthree{\V{V$_b$}}{\D{I}}$. As there is no visible momentum along the $\hat{n}_{\V{V}}$ direction, there is little information for choosing $\phat{\I{$\nu_a$}}{\D{I}} \cdot \hat{n}_{\V{V}}$. In fact, any mass estimators that depend only on these visible four vectors (without resolving the individual $b$-tagged jets and leptons) are completely insensitive to the sign of this inner product, indicating that a minimization of any of these masses will yield $\phat{\I{$\nu_a$}}{\D{I}} \cdot \hat{n}_{\V{V}} = 0$. We adopt this choice, leaving only one angle, describing the direction of \phat{\I{$\nu_a$}}{\D{I}} in the plane defined by $\hat{n}_{\V{V}}$, to choose.

There are many ways to determine this final angle. We consider two different approaches, corresponding to two different JR's, each resulting in observables with distinctively different behavior. The first follows by considering the sum of top mass estimators squared, evaluated in the \D{I} frame:
\bea
\label{eqn:invminM2}
f_1(\phat{\I{$\nu_a$}}{\D{I}}) &=& \Mass{\D{$t_{a}$}}{2} + \Mass{\D{$t_{b}$}}{2} = \nonumber \\
&=& \mass{\V{V$_a$}}{2} + \mass{\V{V$_b$}}{2} + 2\E{\V{V$_{a}$}}{\D{I}} \E{\I{$\nu_{a}$}}{\D{I}} +  2\E{\V{V$_{b}$}}{\D{I}} \E{\I{$\nu_{b}$}}{\D{I}} \nonumber \\
&& - 2\pthree{\I{$\nu_a$}}{\D{I}} \cdot \pthree{\V{V$_a$}}{\D{I}} - 2\pthree{\I{$\nu_b$}}{\D{I}} \cdot \pthree{\V{V$_b$}}{\D{I}}  \\
&=&  \mass{\V{V$_a$}}{2} + \mass{\V{V$_b$}}{2} + \Mass{\I{I}}{}(\E{\V{V$_{a}$}}{\D{I}} + \E{\V{V$_{b}$}}{\D{I}}) \nonumber \\
&& - \Mass{\I{I}}{} \phat{\I{$\nu_a$}}{\D{I}} \cdot (\pthree{\V{V$_a$}}{\D{I}} - \pthree{\V{V$_b$}}{\D{I}})~, \nonumber
\eea
where only the final term, $- \Mass{\I{I}}{} \phat{\I{$\nu_a$}}{\D{I}} \cdot (\pthree{\V{V$_a$}}{\D{I}} - \pthree{\V{V$_b$}}{\D{I}})$, depends on the unknown \phat{\I{$\nu_a$}}{\D{I}}. Choosing \phat{\I{$\nu_a$}}{\D{I}} to minimize $f_1(\phat{\I{$\nu_a$}}{\D{I}})$ from Eq.~\ref{eqn:invminM2}, we find
\bea
\phat{\I{$\nu_a$}}{\D{I}} \propto \pthree{\V{V$_a$}}{\D{I}} - \pthree{\V{V$_b$}}{\D{I}}~.
\eea
This choice effectively minimizes the two top mass estimators simultaneously, even if they have different values. 

Alternatively, we can make a different choice for \phat{\I{$\nu_a$}}{\D{I}} by considering another metric to minimize. Even though the two top mass estimators cannot be guaranteed to be equal, we can attempt to make them as similar as possible using this degree of freedom, by defining
\bea
f_2(\phat{\I{$\nu_a$}}{\D{I}}) = | \Mass{\D{$t_{a}$}}{} - \Mass{\D{$t_{b}$}}{} |~,
\eea
and choosing \phat{\I{$\nu_a$}}{\D{I}} to minimize $f_2(\phat{\I{$\nu_a$}}{\D{I}})$. As this function has units mass, this approach minimizes the two top masses while also minimizing their difference. Unlike the other JR prescriptions described to this point, this minimization does not have an analytic solution and is performed numerically.

We can define these JR's more generally as follows:
\begin{jigsaw}[Invisible Minimize Masses$^2$]
\label{jr:invminM2_1}
If the internal degrees of freedom specifying how an invisible particle, $\I{I} = \{ \I{I$_{a}$}, \I{I$_{b}$} \}$, should split into two particles are unknown, they can be specified by choosing a corresponding pair of visible particles,  $\V{V} = \{ \V{V$_{a}$}, \V{V$_{b}$} \}$, and minimizing the quantity $\Mass{\V{V$_{a}$}\I{I$_{a}$}}{2} + \Mass{\V{V$_{b}$}\I{I$_{b}$}}{2}$. It is assumed that the four vectors of the visible particles are known in the center-of-mass frame $\D{F} = \{ \V{V}, \I{I} \}$, as is the four vector of the total \I{I} system, \pfour{\I{I}}{\D{F}}. Furthermore, we assume that the individual invisible particle masses, \Mass{\I{I$_a$}}{} and \Mass{\I{I$_b$}}{}, are specified.
The four vectors of the invisible particles can be chosen in the di-invisible rest frame, \D{I}, as:
\bea
\pthree{\I{I$_{a}$}}{\D{I}} &=& \pone{\I{I$_{a}$}}{\D{I}} \frac{\pthree{\V{V$_{a}$}}{\D{I}} -  \pthree{\V{V$_{b}$}}{\D{I}}}{|\pthree{\V{V$_{a}$}}{\D{I}} -  \pthree{\V{V$_{b}$}}{\D{I}}|}  \\
\pthree{\I{I$_{b}$}}{\D{I}} &=& \pone{\I{I$_{b}$}}{\D{I}} \frac{\pthree{\V{V$_{b}$}}{\D{I}} -  \pthree{\V{V$_{a}$}}{\D{I}}}{|\pthree{\V{V$_{a}$}}{\D{I}} -  \pthree{\V{V$_{b}$}}{\D{I}}|}~,  \nonumber 
\eea
with
\bea
\pone{\I{I$_{a}$}}{\D{I}} &=& \pone{\I{I$_{b}$}}{\D{I}} \nonumber \\
&=& \frac{\sqrt{\left(\Mass{\I{I}}{2}-(\Mass{\I{I$_a$}}{} + \Mass{\I{I$_b$}}{})^2\right)\left(\Mass{\I{I}}{2}-(\Mass{\I{I$_a$}}{} - \Mass{\I{I$_b$}}{})^2\right)}}{2\Mass{\I{I}}{}} \nonumber \\
\E{\I{I$_{a}$}}{\D{I}} &=& \sqrt{\Mass{\I{I$_a$}}{2} +  (\pone{\I{I$_{a}$}}{\D{I}})^2}  \\
\E{\I{I$_{b}$}}{\D{I}} &=& \sqrt{\Mass{\I{I$_b$}}{2} +  (\pone{\I{I$_{b}$}}{\D{I}})^2}~. \nonumber
\eea
The mass \Mass{\I{I}}{} must be chosen to be at least as large as the sum of individual particle masses.
\end{jigsaw}
\begin{jigsaw}[Invisible Minimize $\Delta$Masses]
\label{jr:invminDeltaM}
If the internal degrees of freedom specifying how an invisible particle, $\I{I} = \{ \I{I$_{a}$}, \I{I$_{b}$} \}$, should split into two particles are unknown, they can be specified by choosing a corresponding pair of visible particles,  $\V{V} = \{ \V{V$_{a}$}, \V{V$_{b}$} \}$, and minimizing the quantity $|\Mass{\V{V$_{a}$}\I{I$_{a}$}}{} - \Mass{\V{V$_{b}$}\I{I$_{b}$}}{}|$, subject to constraints. It is assumed that the four vectors of the visible particles are known in the center-of-mass frame, $\D{F} = \{ \V{V}, \I{I}~\}$, as is the four vector of the total \I{I} system, \pfour{\I{I}}{\D{F}}. Furthermore, we assume that the individual invisible particle masses, \Mass{\I{I$_a$}}{} and \Mass{\I{I$_b$}}{}, are specified.
The four vectors of the invisible particles can be chosen in the di-invisible rest frame, \D{I}, as:
\bea
\pthree{\I{I$_{a}$}}{\D{I}} &=& \pone{\I{I$_{a}$}}{\D{I}}\hat{n}   \\
\pthree{\I{I$_{b}$}}{\D{I}} &=& -\pone{\I{I$_{b}$}}{\D{I}}\hat{n} ~,  \nonumber 
\eea
with
\bea
\pone{\I{I$_{a}$}}{\D{I}} &=& \pone{\I{I$_{b}$}}{\D{I}} \nonumber \\
&=& \frac{\sqrt{\left(\Mass{\I{I}}{2}-(\Mass{\I{I$_a$}}{} + \Mass{\I{I$_b$}}{})^2\right)\left(\Mass{\I{I}}{2}-(\Mass{\I{I$_a$}}{} - \Mass{\I{I$_b$}}{})^2\right)}}{2\Mass{\I{I}}{}} \nonumber \\
\E{\I{I$_{a}$}}{\D{I}} &=& \sqrt{\Mass{\I{I$_a$}}{2} +  (\pone{\I{I$_{a}$}}{\D{I}})^2}  \\
\E{\I{I$_{b}$}}{\D{I}} &=& \sqrt{\Mass{\I{I$_b$}}{2} +  (\pone{\I{I$_{b}$}}{\D{I}})^2}~. \nonumber
\eea
and $\hat{n}$ chosen to correspond to the minimum
\bea
\min_{\hat{n},~\hat{n} \cdot (\pthree{\V{V$_a$}}{\D{I}} \times \pthree{\V{V$_b$}}{\D{I}})=0} |\Mass{\V{V$_{a}$}\I{I$_{a}$}}{} - \Mass{\V{V$_{b}$}\I{I$_{b}$}}{}|~.
\eea
The mass \Mass{\I{I}}{} must be chosen to be at least as large as the sum of individual particle masses.
\end{jigsaw}

With these additional JR's, we now have several choices in how to analyze events in the $t\bar{t} \rightarrow bW(\ell\nu)bW(\ell\nu)$ final state at a hadron collider. We consider four different strategies, described below, in order to compare the relative merits of the approaches:

\begin{enumerate}[noitemsep]
\item ``$M_{\rm top}^a = M_{\rm top}^b$ reconstruction''
{\footnotesize 
\begin{enumerate}[noitemsep]
\item Apply the combinatoric JR~\ref{jr:minM2}, choosing \\ the \V{$b$}/\V{$\ell$} pairing which minimizes $\mass{\V{V$_a$}}{2}+\mass{\V{V$_b$}}{2}$.
\item Apply the invisible mass JR~\ref{jr:mass}, choosing \\ $\Mass{\I{I}}{2}= \mass{\V{V}}{2} - 4\mass{\V{V$_a$}}{}\mass{\V{V$_b$}}{}$.
\item Apply the invisible rapidity JR~\ref{jr:rapidity}, choosing \\ \pone{\I{I},z}{\lab} using all the visible particles \V{V}.
\item Apply the contra-boost invariant JR~\ref{jr:contra}, \\ specifying the neutrino four vectors using \\ the constraint $\Mass{\D{$t_a$}}{} = \Mass{\D{$t_b$}}{}$.
\end{enumerate}
}
\item ``$M_{W}^a = M_{W}^b$ reconstruction''
{\footnotesize 
\begin{enumerate}[noitemsep]
\item Apply the combinatoric JR~\ref{jr:minM2}, choosing \\ the \V{$b$}/\V{$\ell$} pairing which minimizes $\mass{\V{V$_a$}}{2}+\mass{\V{V$_b$}}{2}$.
\item Apply the invisible mass JR~\ref{jr:mass}, choosing \\ $\Mass{\I{I}}{2}= \mass{\V{$\ell_a$}\V{$\ell_b$}}{2} - 4\mass{\V{$\ell_a$}}{}\mass{\V{$\ell_b$}}{}$.
\item Apply the invisible rapidity JR~\ref{jr:rapidity}, choosing \\ \pone{\I{I},z}{\lab} using all the visible particles \V{V}.
\item Apply the contra-boost invariant JR~\ref{jr:contra}, \\ specifying the neutrino four vectors using \\ the constraint $\Mass{\D{$W_a$}}{} = \Mass{\D{$W_b$}}{}$.
\end{enumerate}
}
\item ``min $\Sigma M_{\rm top}^2$ reconstruction''
{\footnotesize 
\begin{enumerate}[noitemsep]
\item Apply the combinatoric JR~\ref{jr:minM2}, choosing \\ the \V{$b$}/\V{$\ell$} pairing which minimizes $\mass{\V{V$_a$}}{2}+\mass{\V{V$_b$}}{2}$.
\item Apply the invisible mass JR~\ref{jr:mass}, choosing \\ $\Mass{\I{I}}{}= 2|\pthree{\V{$\ell_{a/b}$}}{\V{$\ell\ell$}}|$, $\Mass{\I{$\nu_{a/b}$}}{} = 0$.
\item Apply the invisible rapidity JR~\ref{jr:rapidity}, choosing \\ \pone{\I{I},z}{\lab} using all the visible particles \V{V}.
\item Apply the JR~\ref{jr:invminM2_1}, specifying the neutrino \\ four vectors by minimizing $\sum_i \Mass{\D{$t_i$}}{2}$.
\end{enumerate}
}
\item ``min $\Delta M_{\rm top}^2$ reconstruction''
{\footnotesize 
\begin{enumerate}[noitemsep]
\item Apply the combinatoric JR~\ref{jr:minM2}, choosing \\ the \V{$b$}/\V{$\ell$} pairing which minimizes $\mass{\V{V$_a$}}{2}+\mass{\V{V$_b$}}{2}$.
\item Apply the invisible mass JR~\ref{jr:mass}, choosing \\ $\Mass{\I{I}}{}= 2|\pthree{\V{$\ell_{a/b}$}}{\V{$\ell\ell$}}|$, $\Mass{\I{$\nu_{a/b}$}}{} = 0$.
\item Apply the invisible rapidity JR~\ref{jr:rapidity}, choosing  \\ \pone{\I{I},z}{\lab} using all the visible particles \V{V}.
\item Apply the JR~\ref{jr:invminDeltaM}, specifying the neutrino \\ four vectors by minimizing $\Delta \Mass{\D{$t_i$}}{2}$.
\end{enumerate}
}
\end{enumerate}
The factorization and interchangeability of the different JR's appearing in the four different approaches is clear in their descriptions, and extends to the resulting estimators each produces.  The invisible rapidity JR~\ref{jr:rapidity} ensures that all of the observables in each approach are invariant under longitudinal boosts, while they are approximately independent of the lab frame momentum of the center-of-mass di-top system. 

\begin{figure}[htbp]
\centering 
\subfigure[]{\includegraphics[width=.238\textwidth]{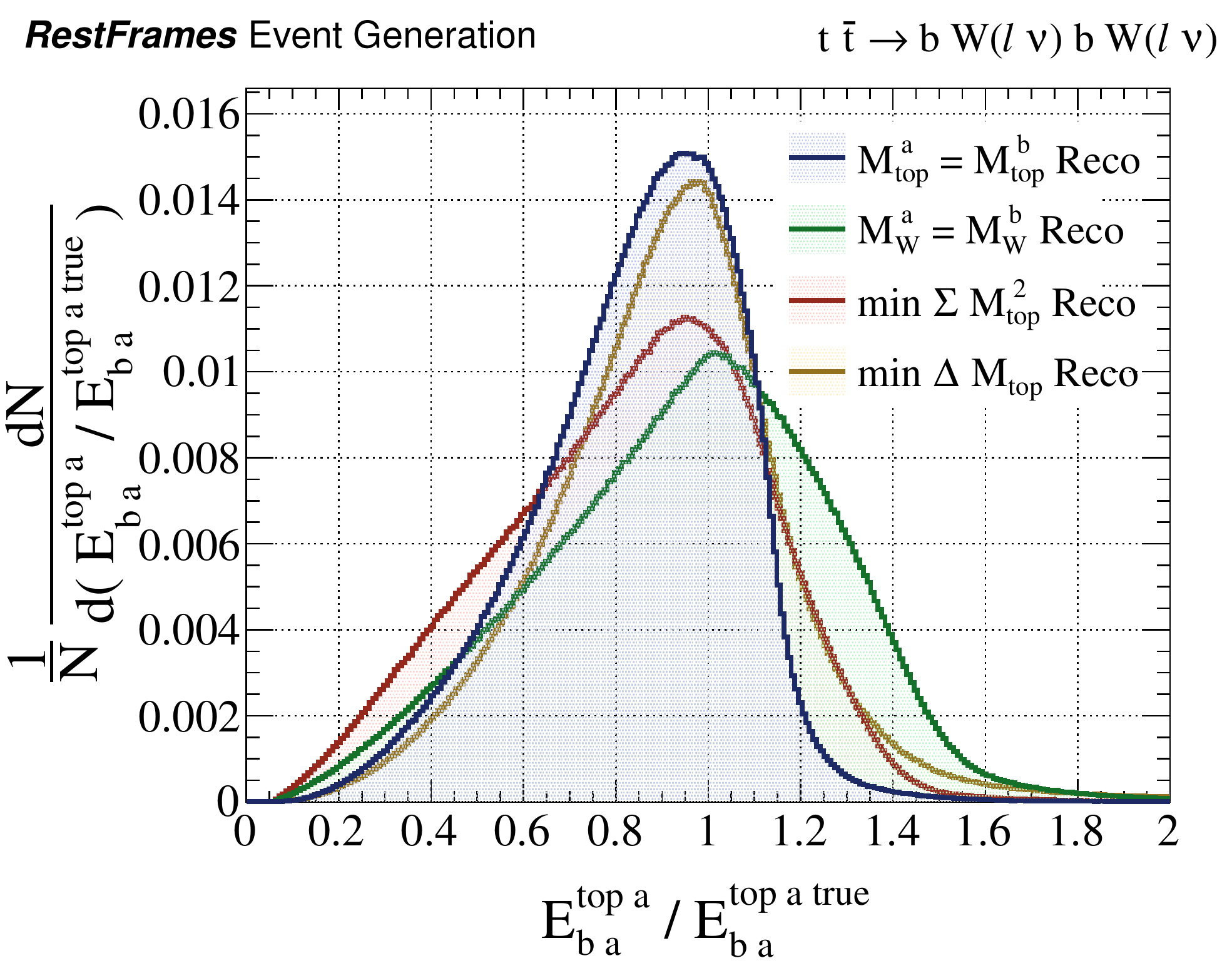}}
\subfigure[]{\includegraphics[width=.238\textwidth]{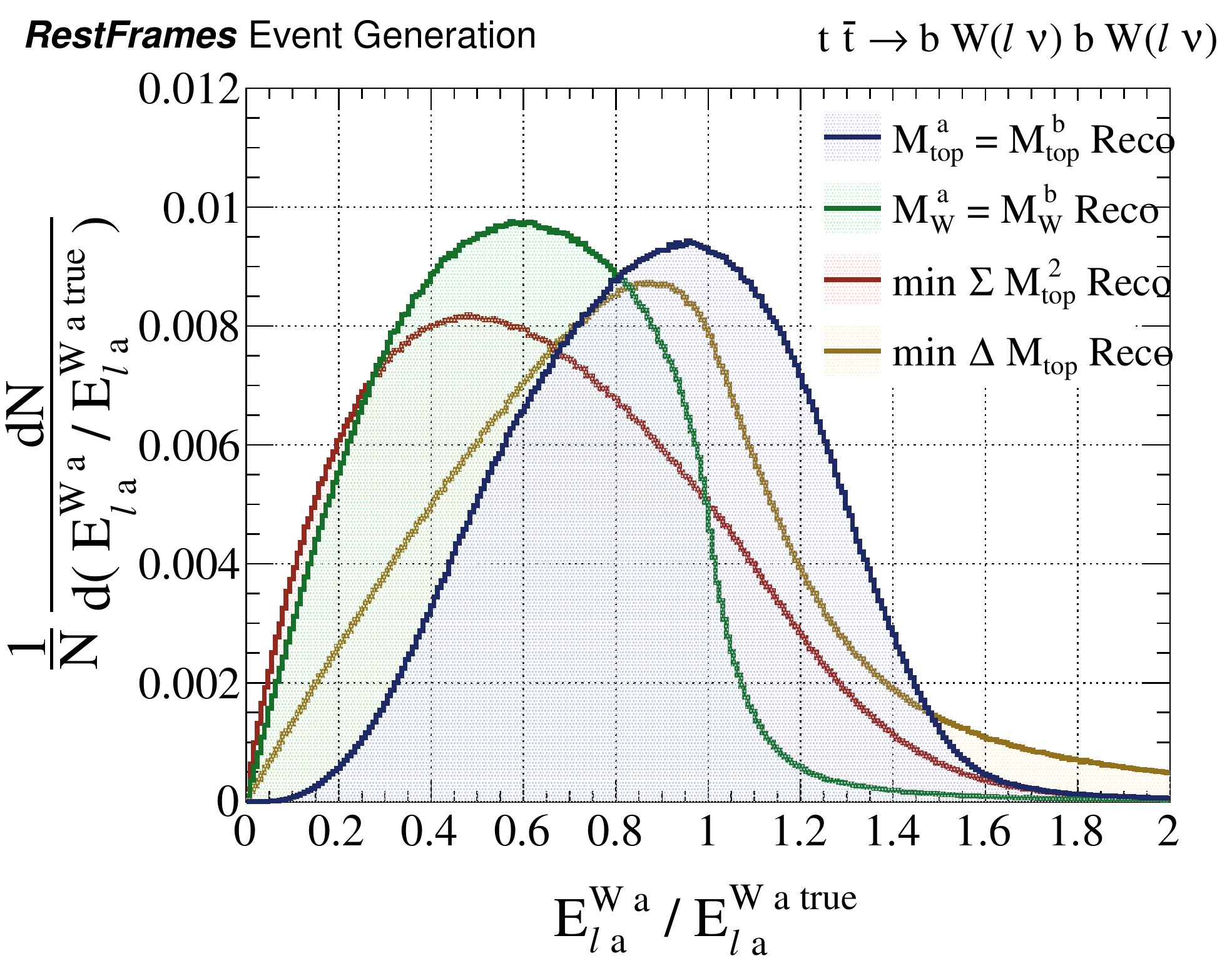}}
\vspace{-0.3cm}
\caption{\label{fig:example_ttbar_to_bWlnubWlnu-EbEl} Distributions of (a) the energy of a reconstructed $b$-tagged jet in its approximate production frame, \E{\V{$b_a$}}{\D{$t_a$}}, and (b) \E{\V{$\ell_a$}}{\D{$W_a$}}
  in simulated $t\bar{t} \rightarrow bW(\ell\nu)bW(\ell\nu)$ events at a hadron collider. These estimators are compared in four different reconstruction schemes. Each observable is normalized by the true value of the quantity it is estimating.}
\end{figure}

As the \minMt approach will result in non-trivial neutrino mass estimators, biasing any quantity that directly depends on them, we consider the energies of the different visible particles in their respective production frames, quantities sensitive to the mass splittings of the intermediate particle states. The estimators \E{\V{$b_a$}}{\D{$t_a$}} and \E{\V{$\ell_a$}}{\D{$W_a$}} are shown in Fig.~\ref{fig:example_ttbar_to_bWlnubWlnu-EbEl} for simulated events reconstructed in each of the four different ways. The \minMt and \minDMt approaches yield the most accurate estimates for \E{\V{$b_a$}}{\D{$t_a$}}, with small biases relative to the true value due to the minimizations in their application. While the \minMW approach introduces a smaller bias in \E{\V{$b_a$}}{\D{$t_a$}}, it exhibits much worse resolution, similar to \minMtsq. 

In Fig.~\ref{fig:example_ttbar_to_bWlnubWlnu-EbEl}(b), the different \E{\V{$\ell_a$}}{\D{$W_a$}} observables exhibit quite different behavior, with the \minMt approach providing the most accurate, unbiased, estimate. Using \minDMt results in similar behavior, with a slightly larger bias.  The \E{\V{$\ell_a$}}{\D{$W_a$}} distribution exhibits a kinematic edge at the true value in the \minMW approach, as was the case for two $W(\ell\nu)$ final states in Section~\ref{subsec:Part2_exampleA}. The worst estimate is provided by \minMtsq reconstruction, where the minimization of a sum of masses squared allows for longer tails in individual estimators' distributions.

\begin{figure}[htbp]
\centering 
\subfigure[]{\includegraphics[width=.238\textwidth]{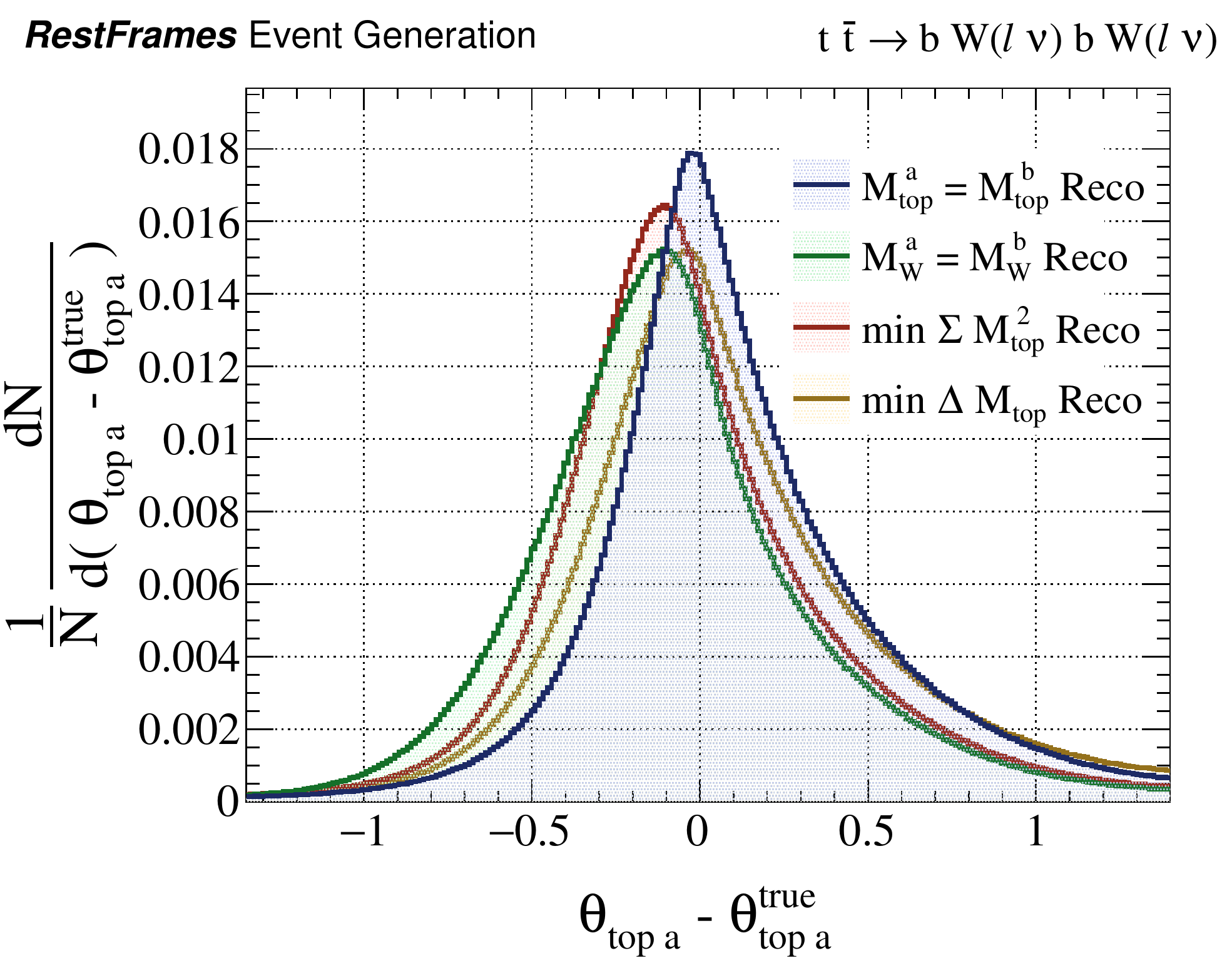}}
\subfigure[]{\includegraphics[width=.238\textwidth]{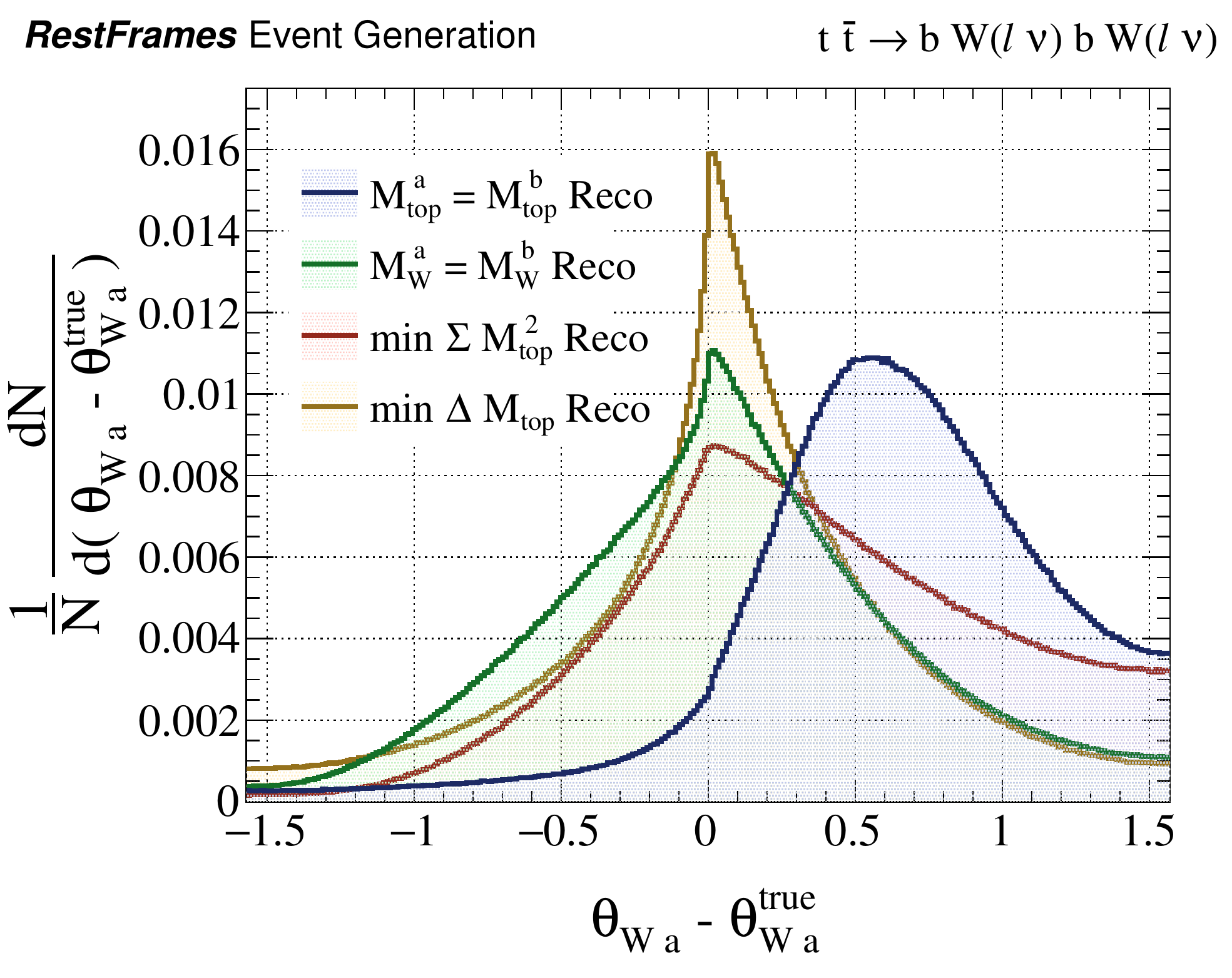}}
\vspace{-0.3cm}
\caption{\label{fig:example_ttbar_to_bWlnubWlnu-thetas} Distributions of (a) reconstructed top decay angle, $\theta_{\D{$t_a$}}$, and (b) $\theta_{\D{$W_a$}}$
  in simulated $t\bar{t} \rightarrow bW(\ell\nu)bW(\ell\nu)$ events at a hadron collider. These estimators are compared in four different reconstruction schemes. The decay angles are normalized by their their true values and shown in units radian.}
\end{figure}

Other observables of interest include the decay angles of the top and $W$ bosons, with their estimators in the different reconstruction schemes shown in in Fig.~\ref{fig:example_ttbar_to_bWlnubWlnu-thetas}. Similarly to the \E{\V{$b_a$}}{\D{$t_a$}} observable, the \minMt approach yields the best estimate of $\theta_{\D{$t_a$}}$, with \minDMt resulting in the next best. For $\theta_{\D{$W_a$}}$, the relative accuracy of the approaches is quite different, with \minMt resulting in a systematic bias. The approach results in the most accurate $\theta_{\D{$W_a$}}$, with quite impressive resolution, similar to that of $\theta_{\D{$t_a$}}$ in the scheme.

We see that, in general, the \minMt approach yields the best mass-sensitive estimators, with the exception of the total di-top invariant mass, which is biased by non-zero neutrino masses. The \minDMt scheme provides a compromise between accuracy in these observables, and significant improvements in resolving decay angles and other quantities. As we will see in the following example, the ability to fix the neutrino masses at zero in the \minDMt approach allows for an unbiased estimate of \mass{\D{$t\bar{t}$}}{}. 

\begin{figure}[htbp]
\centering 
\subfigure[]{\includegraphics[width=.28\textwidth]{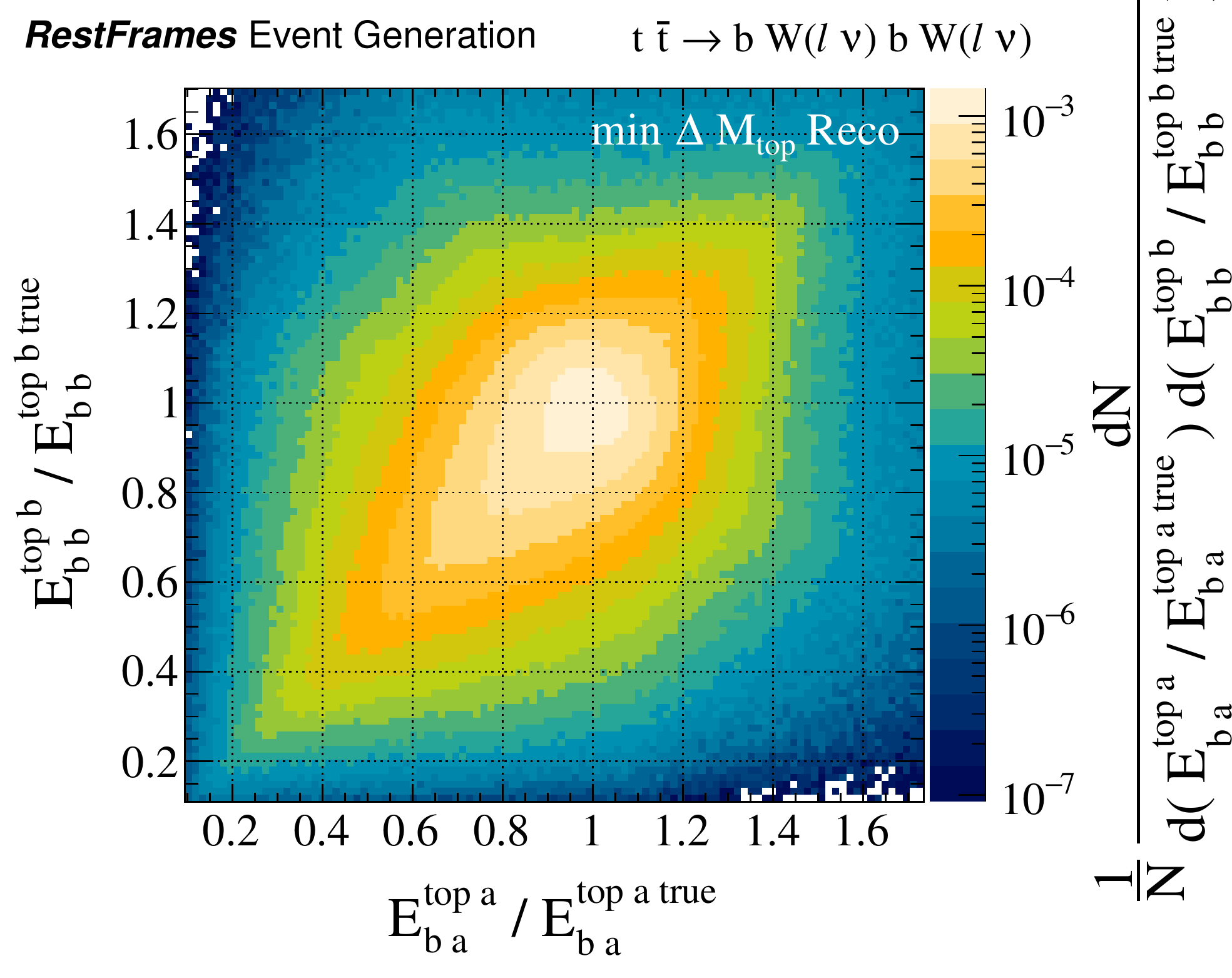}} \hspace{0.5cm}
\subfigure[]{\includegraphics[width=.28\textwidth]{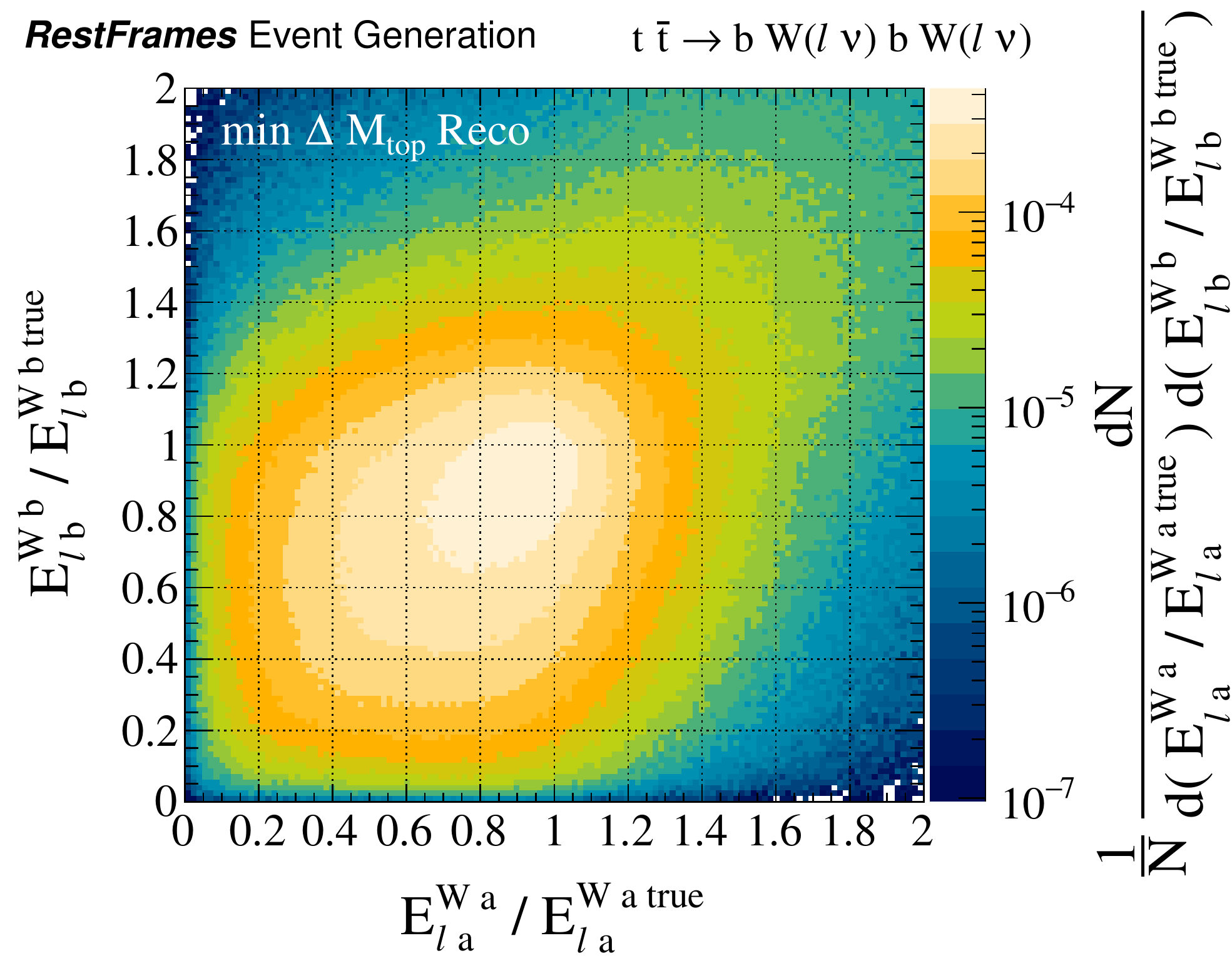}} \hspace{0.5cm}
\subfigure[]{\includegraphics[width=.28\textwidth]{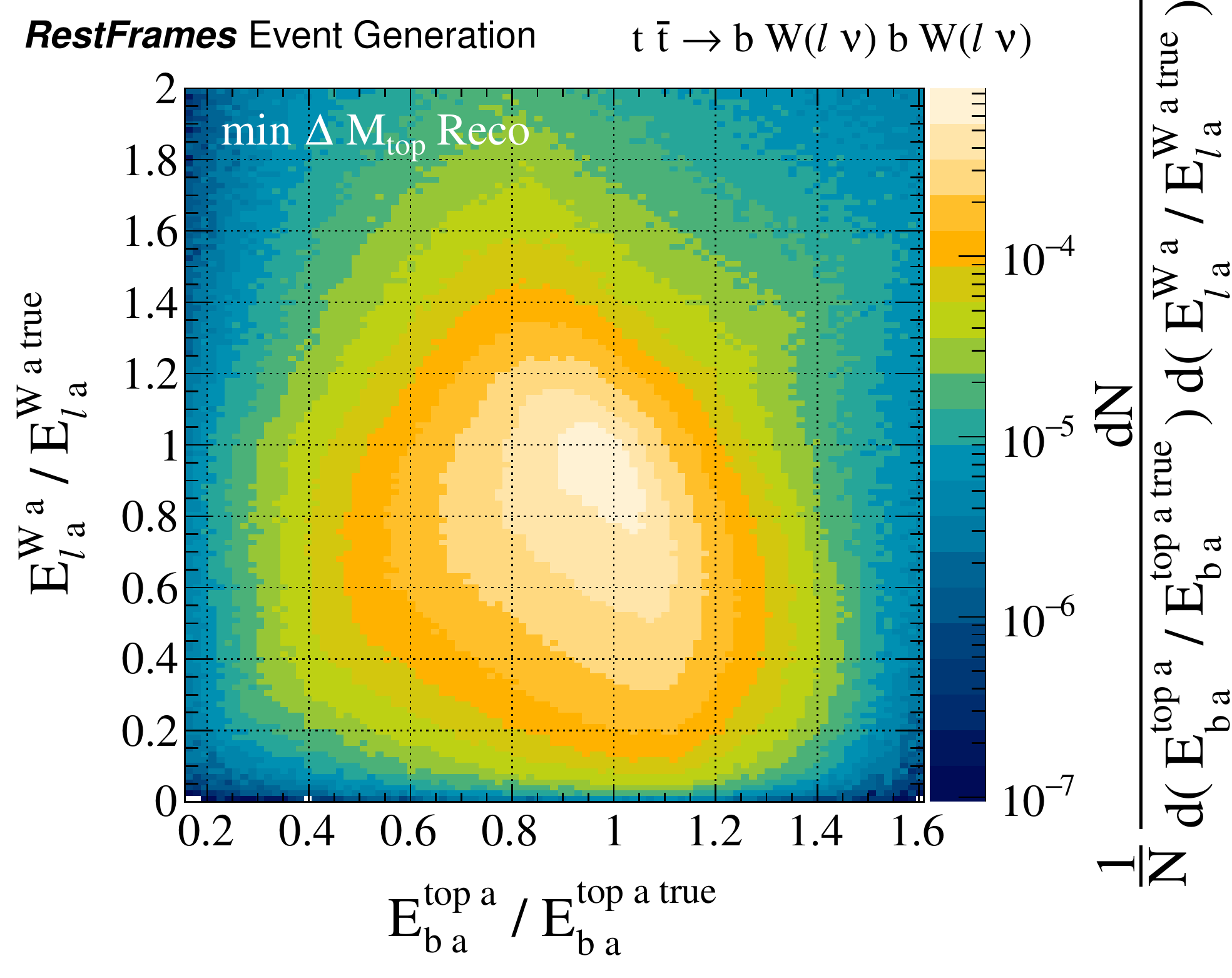}} 
\subfigure[]{\includegraphics[width=.28\textwidth]{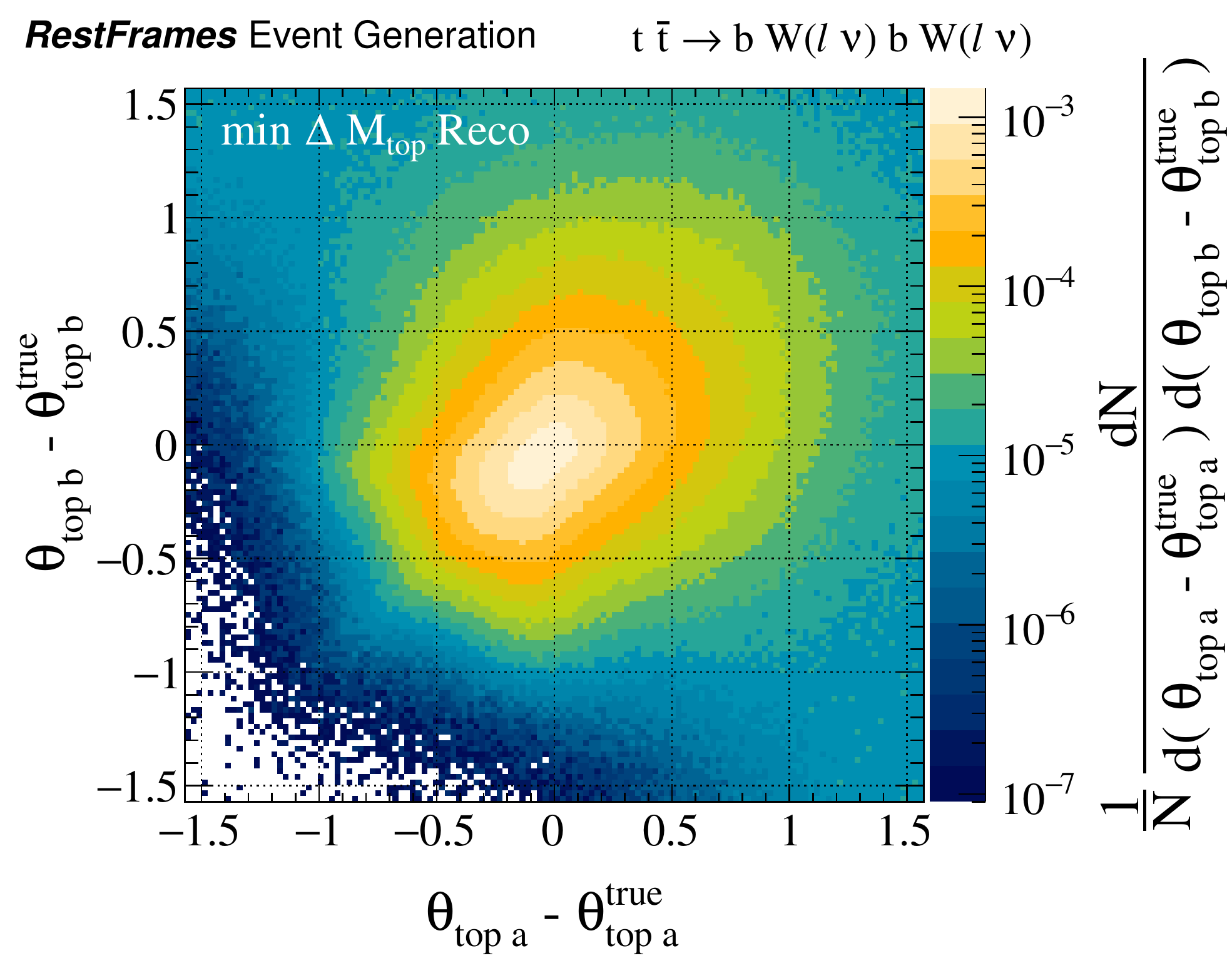}} \hspace{0.5cm}
\subfigure[]{\includegraphics[width=.28\textwidth]{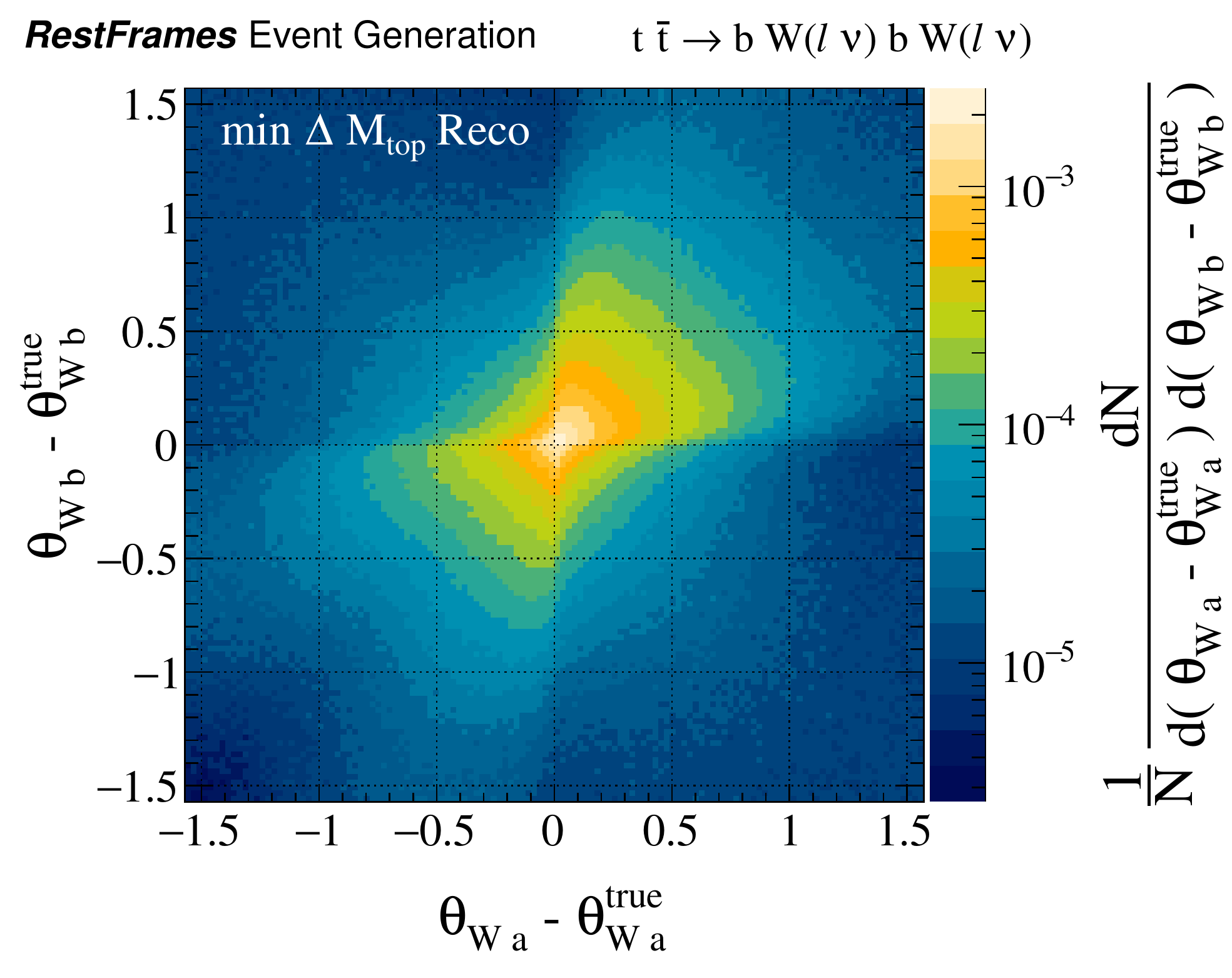}} \hspace{0.5cm}
\subfigure[]{\includegraphics[width=.28\textwidth]{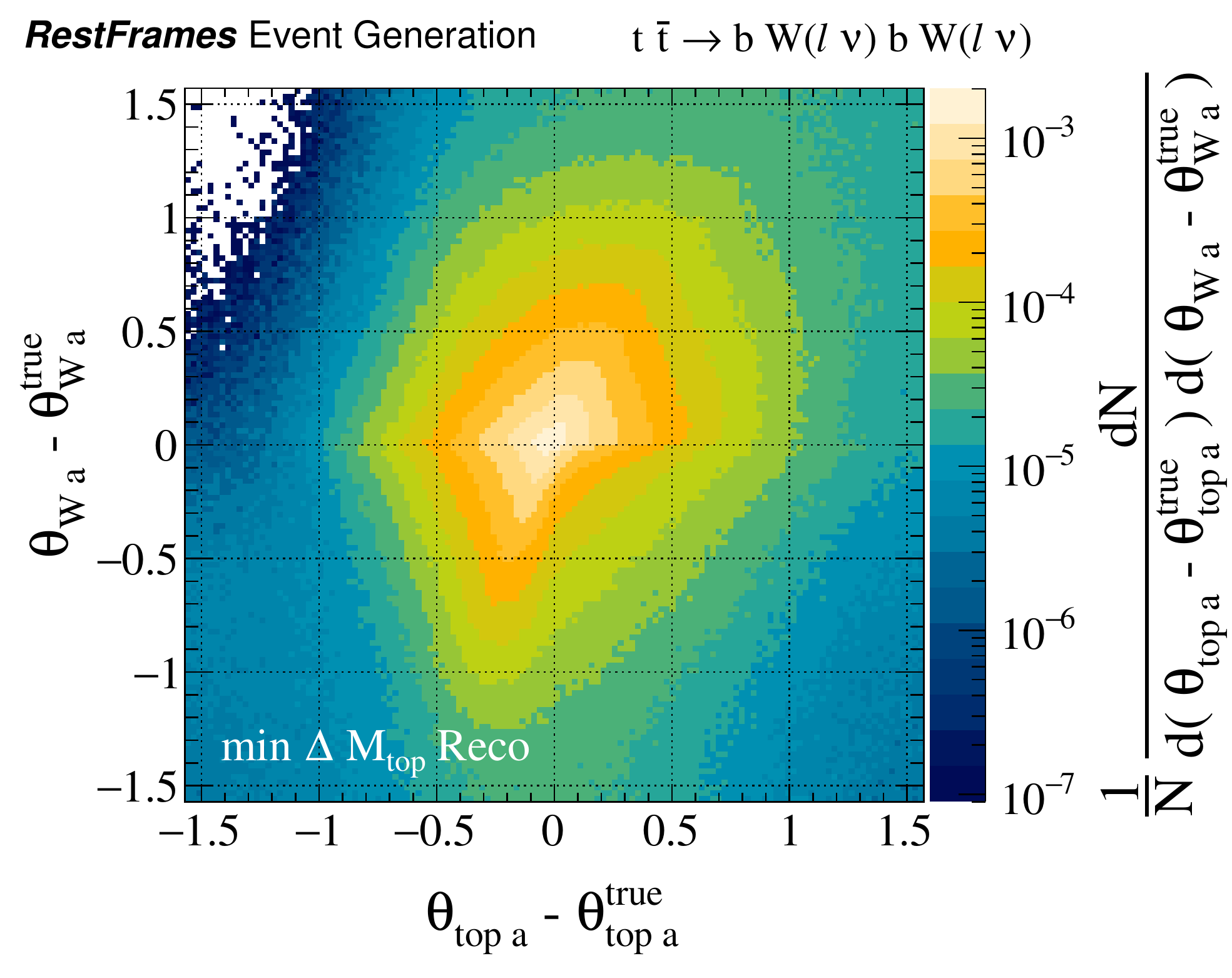}}
\vspace{-0.3cm}
\caption{\label{fig:example_ttbar_to_bWlnubWlnu-2D} Distributions of (a) \E{\V{$b_a$}}{\D{$t_a$}} vs. \E{\V{$b_b$}}{\D{$t_b$}}, (b) \E{\V{$\ell_a$}}{\D{$W_a$}} vs. \E{\V{$\ell_b$}}{\D{$W_b$}}, (c) \E{\V{$b_a$}}{\D{$t_a$}} vs. \E{\V{$\ell_a$}}{\D{$W_a$}}, (d) $\theta_{\D{$t_a$}}$ vs. $\theta_{\D{$t_b$}}$, (e) $\theta_{\D{$W_a$}}$ vs. $\theta_{\D{$W_b$}}$, and (f) $\theta_{\D{$t_a$}}$ vs. $\theta_{\D{$W_a$}}$ in simulated $t\bar{t} \rightarrow bW(\ell\nu)bW(\ell\nu)$ events at a hadron collider, calculated using the \minDMt approach. Each observable is normalized appropriately by the true value of the quantity it is estimating, with angles shown in units radian. }
\end{figure}

Another advantage to using the energies of visible particles evaluated in approximate reference frames, rather than explicit mass estimators, is that these quantities exhibit significantly smaller correlations between the two top quarks, even if the event was reconstructed with explicit constraints relating their masses. An event-by-event comparison of the \E{\V{$b_a$}}{\D{$t_a$}} and \E{\V{$b_b$}}{\D{$t_b$}} estimators using the \minDMt scheme, as seen in Fig.~\ref{fig:example_ttbar_to_bWlnubWlnu-2D}(a), indicates only modest correlation between the two quantities, with \E{\V{$\ell_a$}}{\D{$W_a$}} and \E{\V{$\ell_b$}}{\D{$W_b$}}, shown in Fig.~\ref{fig:example_ttbar_to_bWlnubWlnu-2D}(b) exhibiting similar behavior. It is seen in Fig.~\ref{fig:example_ttbar_to_bWlnubWlnu-2D}(c) that \E{\V{$b_{a/b}$}}{\D{$t_{a/b}$}} and \E{\V{$\ell_{a/b}$}}{\D{$W_{a/b}$}} can also be estimated largely independently, indicating that not only are the observables in each hemisphere largely decoupled, but so are those appearing at different stages of the decay chain. This is a consequence of the factorization of unknowns into different JR's, each describing how to choose only the information necessary to determine the kinematics in a particular reference frame. 
\onecolumngrid

\twocolumngrid 

Corresponding distributions of the correlations between the top and $W$ decay angles for each half of the event, shown in Fig.~\ref{fig:example_ttbar_to_bWlnubWlnu-2D}, confirm the near independence of the observables corresponding to different reference frames, with each estimating their respective true values accurately. In addition to the observables describing the mass and decay of the di-top system, which are studied in the following example, the total set of derived estimators in an RJR scheme like \minDMt constitute an excellent basis for studying these events, with only small correlations and uniformly good resolution of a collection of quantities, including masses and spin-sensitive decay angles. 


\subsection{$H^0 \rightarrow {t}\bar{t} \rightarrow bW(\ell\nu) bW(\ell\nu)$}
\label{subsec:Part3_exampleB}

\begin{figure}[hbp]
\centering 
\includegraphics[width=.33\textwidth]{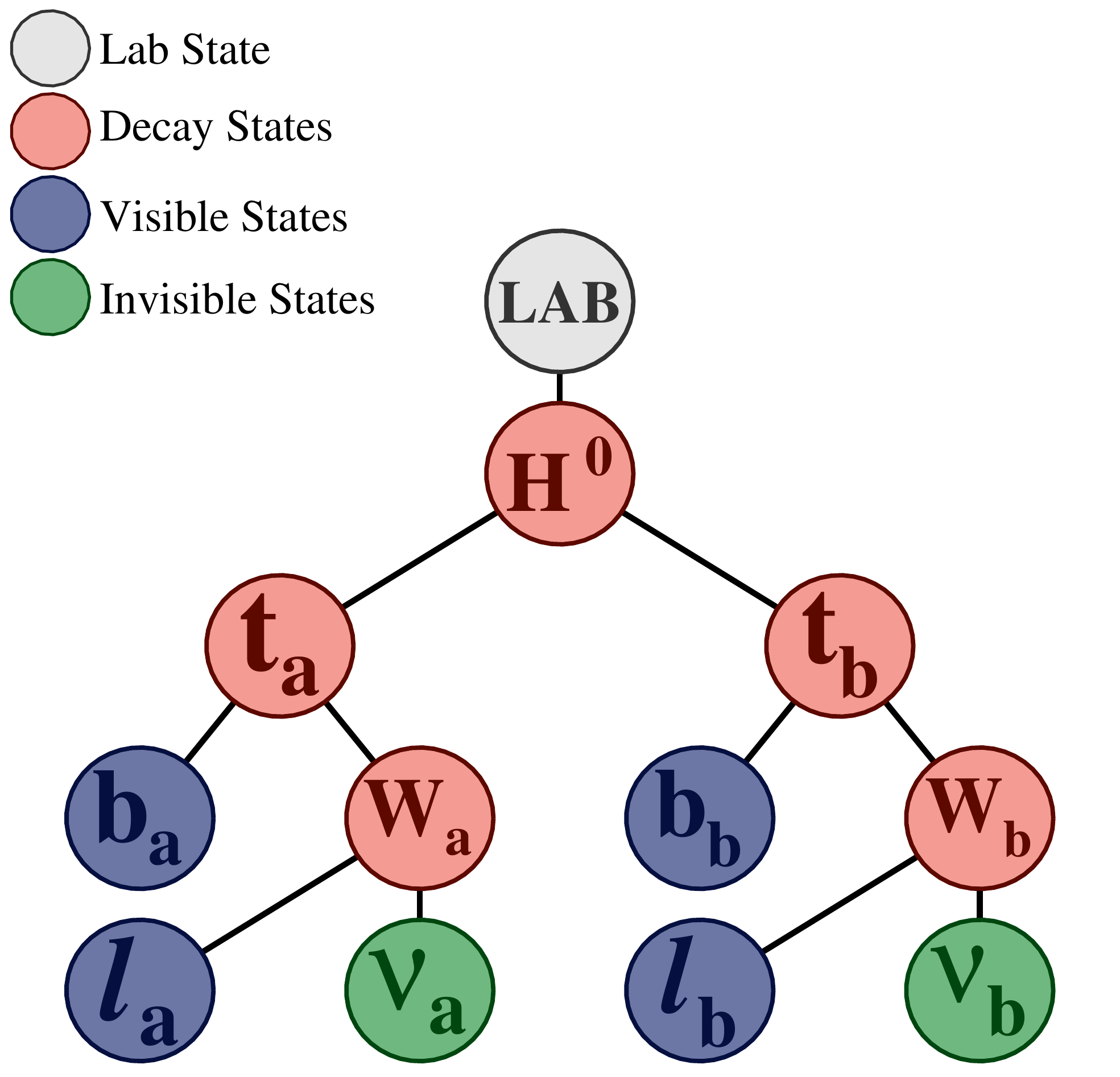}
\vspace{-0.3cm}
\caption{\label{fig:example_H_to_ttbar_to_bWlnubWlnu-decay} A decay tree for a heavy, neutral Higgs boson decaying to a $t\bar{t}$ pair, each of which decays to a $b$-quark and $W(\ell\nu)$.}
\end{figure}

We continue our discussion of di-leptonic $t\bar{t}$ final states by considering resonant top pair production, through a heavy, neutral Higgs boson, \D{$H^0$}. The kinematics of the decay tree describing this final state, shown in Fig.~\ref{fig:example_H_to_ttbar_to_bWlnubWlnu-decay}, is identical to that of the previous example, with four visible particles accompanied by two neutrinos. We will adopt the notation of Section~\ref{subsec:Part3_exampleA} throughout this example.

While for non-resonant top pair production we focused on the reconstruction of the top and $W$ rest frames, along with their associated estimators, here we are primarily interested in the approximation of the Higgs rest frame, its mass, and decay angles. In order to minimize any bias in the Higgs mass estimator, \Mass{\D{$H^0$}}{}, we adopt the \minDMt scheme, described in Section~\ref{subsec:Part3_exampleA}, for analyzing events in this example, which allows us to fix the individual neutrino masses to zero. This approach applies the Lorentz invariant choice for the total mass of the di-neutrino system $\Mass{\I{I}}{} = 2 |\pthree{\V{$\ell_{a/b}$}}{\V{$\ell\ell$}} |$, exploiting the symmetry of the neutrinos and leptons in their production.

\FloatBarrier

\begin{figure}[htbp]
\centering 
\includegraphics[width=.35\textwidth]{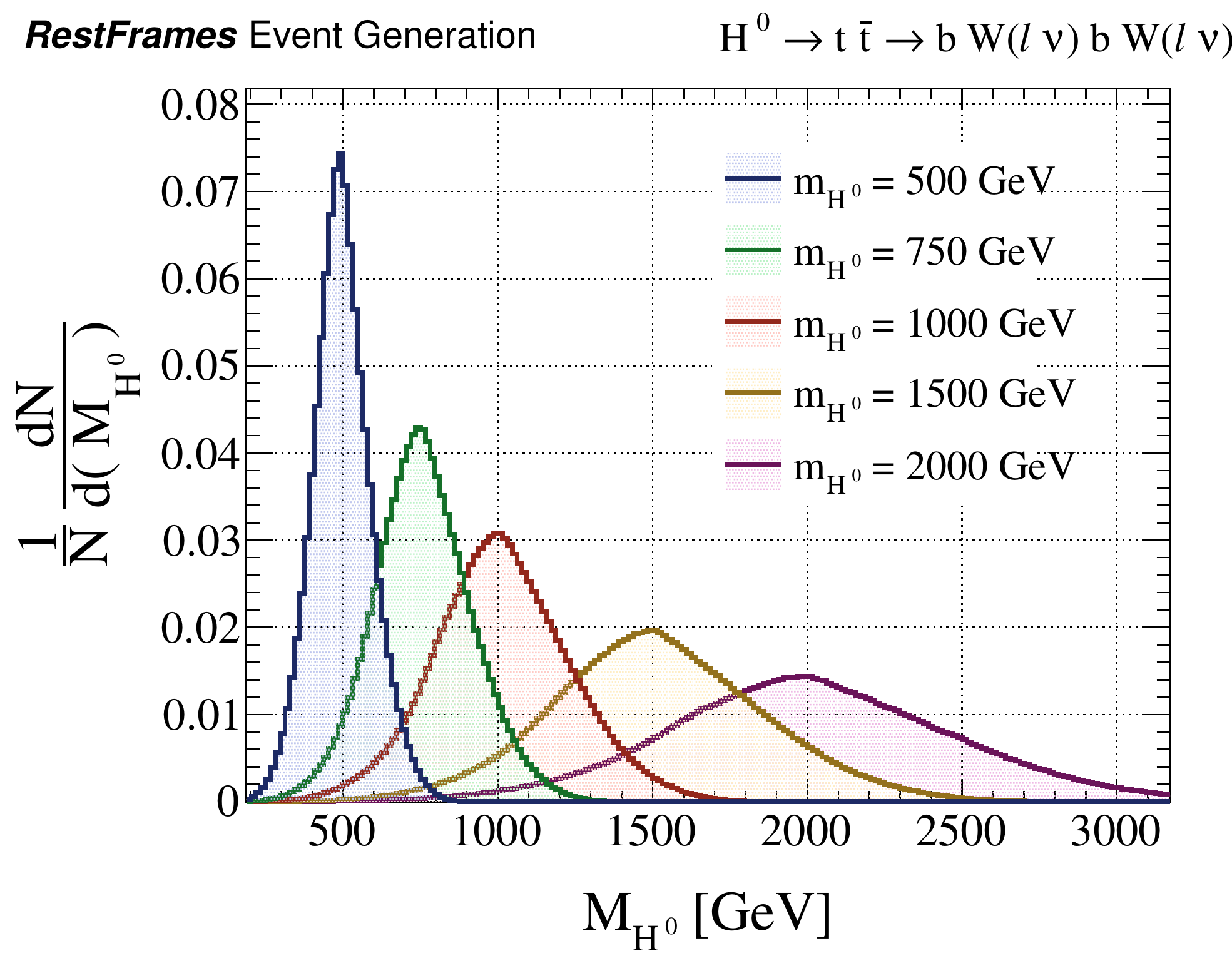}
\vspace{-0.3cm}
\caption{\label{fig:example_H_to_ttbar_to_bWlnubWlnu-MH} Distribution of the Higgs boson mass estimator, \Mass{\D{$H^0$}}{}, in simulated $H^0 \rightarrow {t}\bar{t} \rightarrow bW(\ell\nu) bW(\ell\nu)$ events with varying Higgs mass.}
\end{figure}

The distribution of \Mass{\D{$H^0$}}{}, for varying Higgs boson mass, is shown for simulated $H^0 \rightarrow {t}\bar{t} \rightarrow bW(\ell\nu) bW(\ell\nu)$ events in Fig.~\ref{fig:example_H_to_ttbar_to_bWlnubWlnu-MH}, where the observable peaks at the true value of the mass it is estimating with roughly constant relative resolution. This is indicative that, on average, our guess for the neutrino system's contribution to the Higgs mass is unbiased and not wholly inaccurate.

\begin{figure}[htbp]
\centering 
\subfigure[]{\includegraphics[width=.238\textwidth]{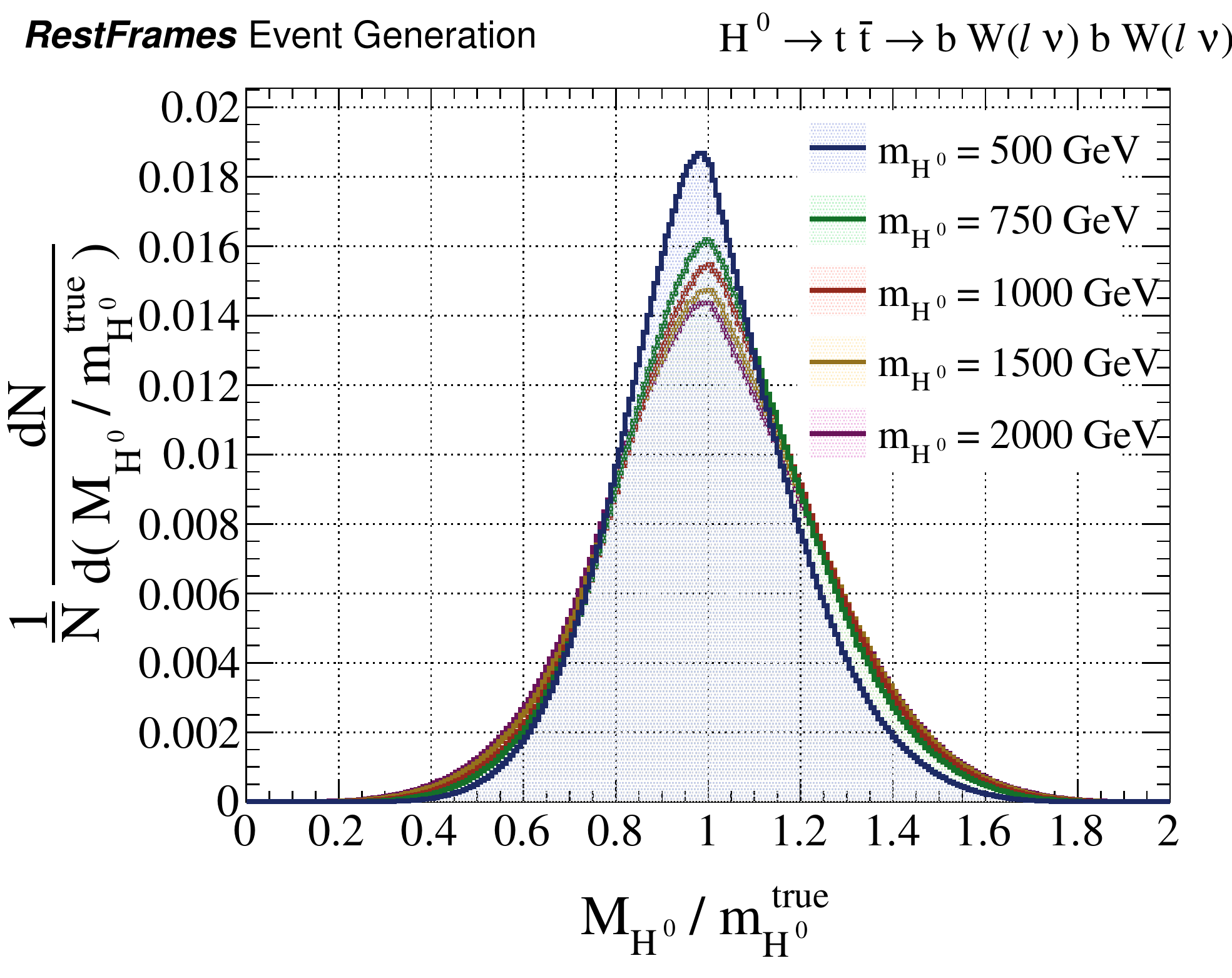}}
\subfigure[]{\includegraphics[width=.238\textwidth]{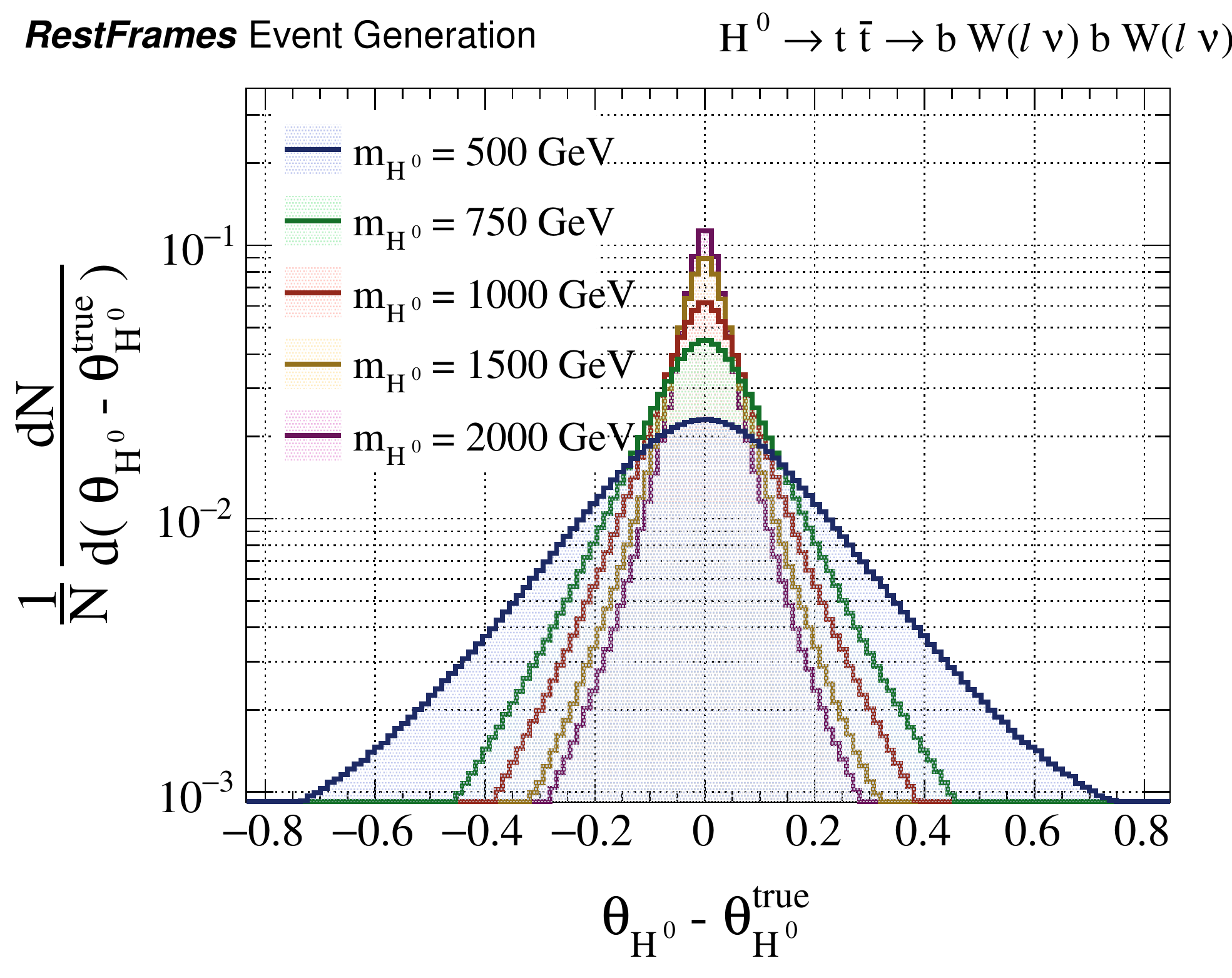}}
\vspace{-0.3cm}
\caption{\label{fig:example_H_to_ttbar_to_bWlnubWln-MHthetaH} Distributions of (a) the Higgs boson mass estimator, \Mass{\D{$H^0$}}{}, and (b) $\theta_{\D{$H^0$}}$ in simulated $H^0 \rightarrow {t}\bar{t} \rightarrow bW(\ell\nu) bW(\ell\nu)$ events with varying Higgs mass. Each observable is appropriately normalized by the true value of the quantity it is estimating, with angles in units radian.}
\end{figure}

 \Mass{\D{$H^0$}}{} estimates the Higgs boson mass with $\sim17-20\%$ resolution for those considered in this study, degrading with increasing mass, as demonstrated in Fig.~\ref{fig:example_H_to_ttbar_to_bWlnubWln-MHthetaH}(a). With this accuracy, the reconstructed  \Mass{\D{$H^0$}}{} has a width comparable to that in fully hadronic $t\bar{t}$ final states, with contributions from jet momentum and mass uncertainty resulting in a similar value. 

Conversely, the resolution of the Higgs decay angle estimator, $\theta_{\D{$H^{0}$}}$, improves with increasing Higgs mass, as can be seen in Fig.~\ref{fig:example_H_to_ttbar_to_bWlnubWln-MHthetaH}(b). This is indicative of the fact that as the velocity of the top quarks increases in the Higgs rest frame, the estimate of the magnitude of that velocity (related to \Mass{\D{$H^0$}}{}) becomes more uncertain while it's direction (related to $\theta_{\D{$H^{0}$}}$) is better resolved. 

\begin{figure}[htbp]
\centering 
\subfigure[]{\includegraphics[width=.238\textwidth]{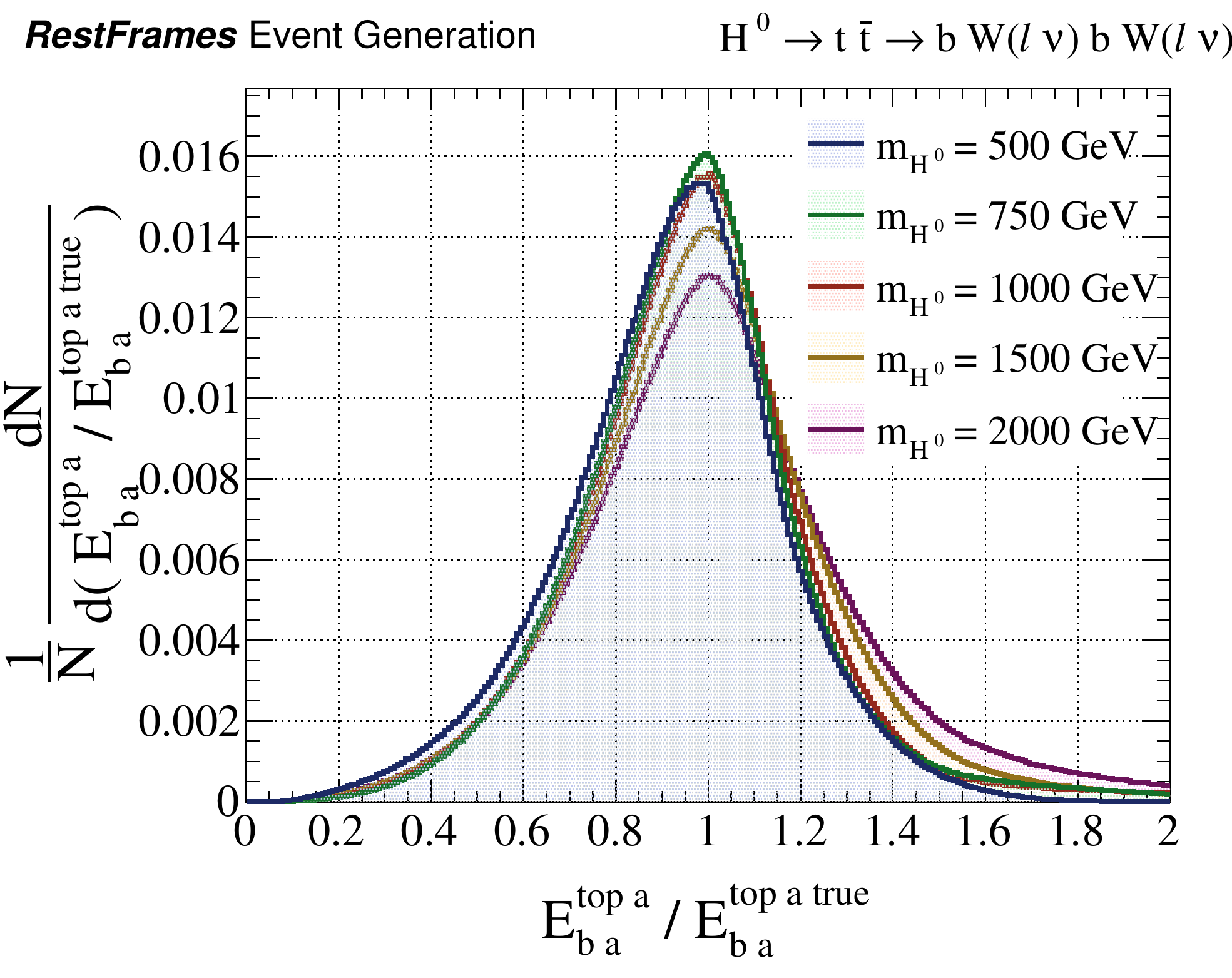}}
\subfigure[]{\includegraphics[width=.238\textwidth]{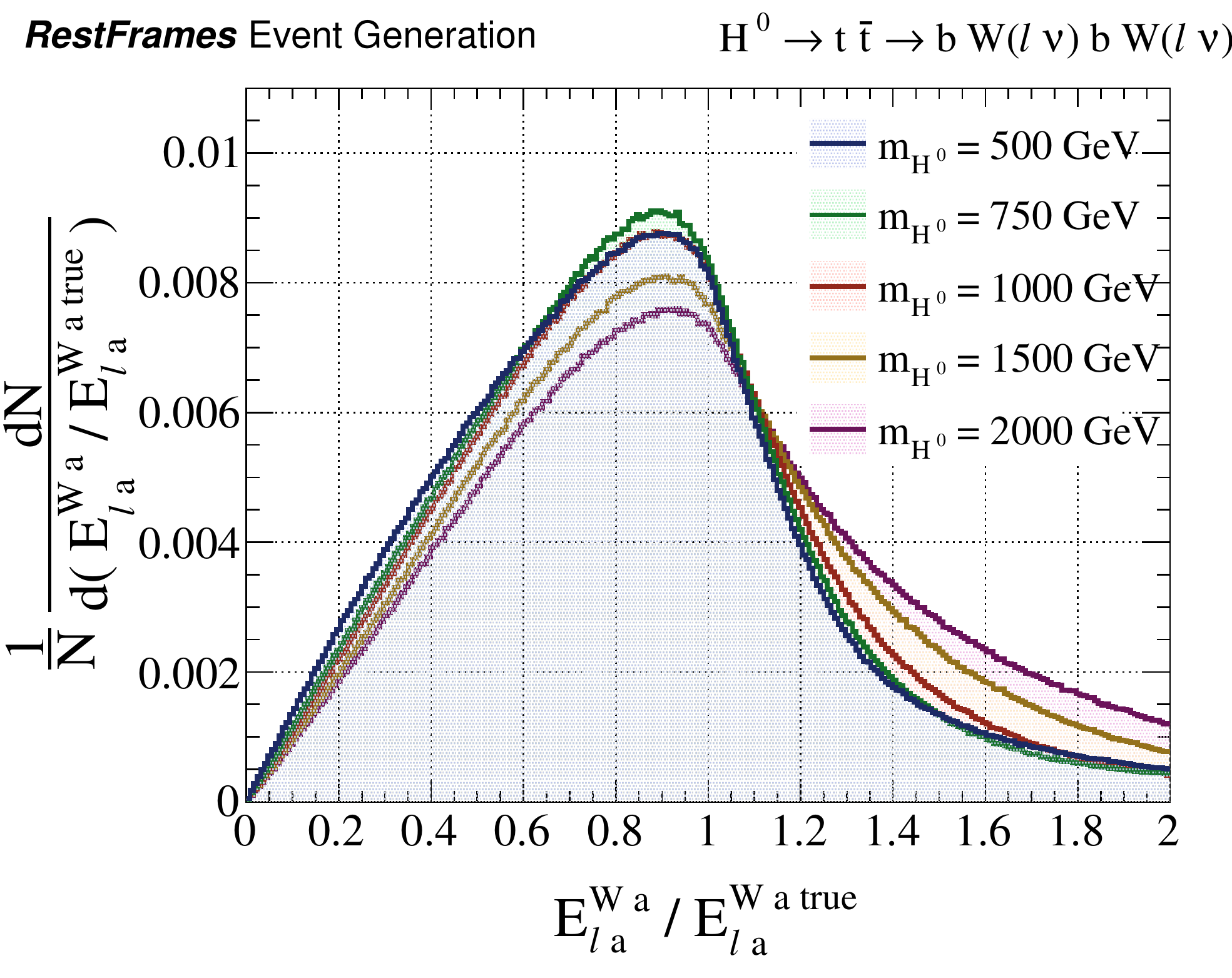}}
\vspace{-0.3cm}
\caption{\label{fig:example_H_to_ttbar_to_bWlnubWln-EbEl} Distributions of (a) the reconstructed energy of a $b$-tagged jet in the approximation of its production frame,  \E{\V{$b_a$}}{\D{$t_a$}}, and (b) \E{\V{$\ell_a$}}{\D{$W_a$}} in simulated $H^0 \rightarrow {t}\bar{t} \rightarrow bW(\ell\nu) bW(\ell\nu)$ events with varying Higgs mass. Each observable is appropriately normalized by the true value of the quantity it is estimating.}
\end{figure}

This mild dependency of the \vbeta{\D{$t_a$}}{\D{$H^0$}} estimate on the Higgs mass also has implications for the reconstruction of the top and $W$ rest frames. The distributions of the $b$-tagged jet and lepton energies in the approximations of their respective production frames are shown in Fig.~\ref{fig:example_H_to_ttbar_to_bWlnubWln-EbEl}. As the velocity relating the Higgs rest frame to the tops' respective rest frames becomes more uncertain, the resolution of these reference frames degrades, with corresponding effects on the observables associated with them. These higher order reconstruction effects, resulting from mis-measurements of the velocities relating different reference frames in a decay tree, introduce many of the small correlations observed between estimators in the RJR approach. 

\begin{figure}[!htbp]
\centering 
\subfigure[]{\includegraphics[width=.238\textwidth]{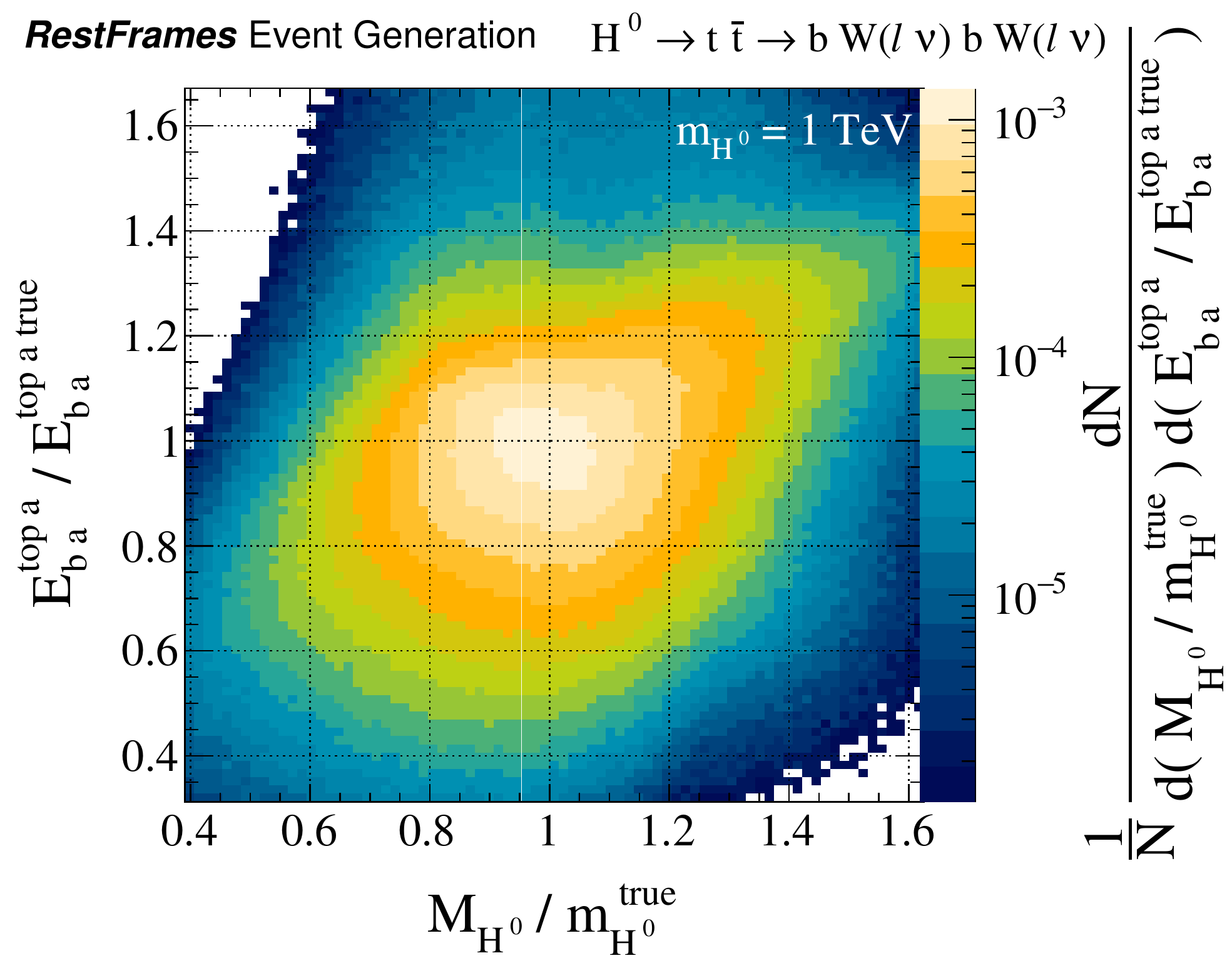}}
\subfigure[]{\includegraphics[width=.238\textwidth]{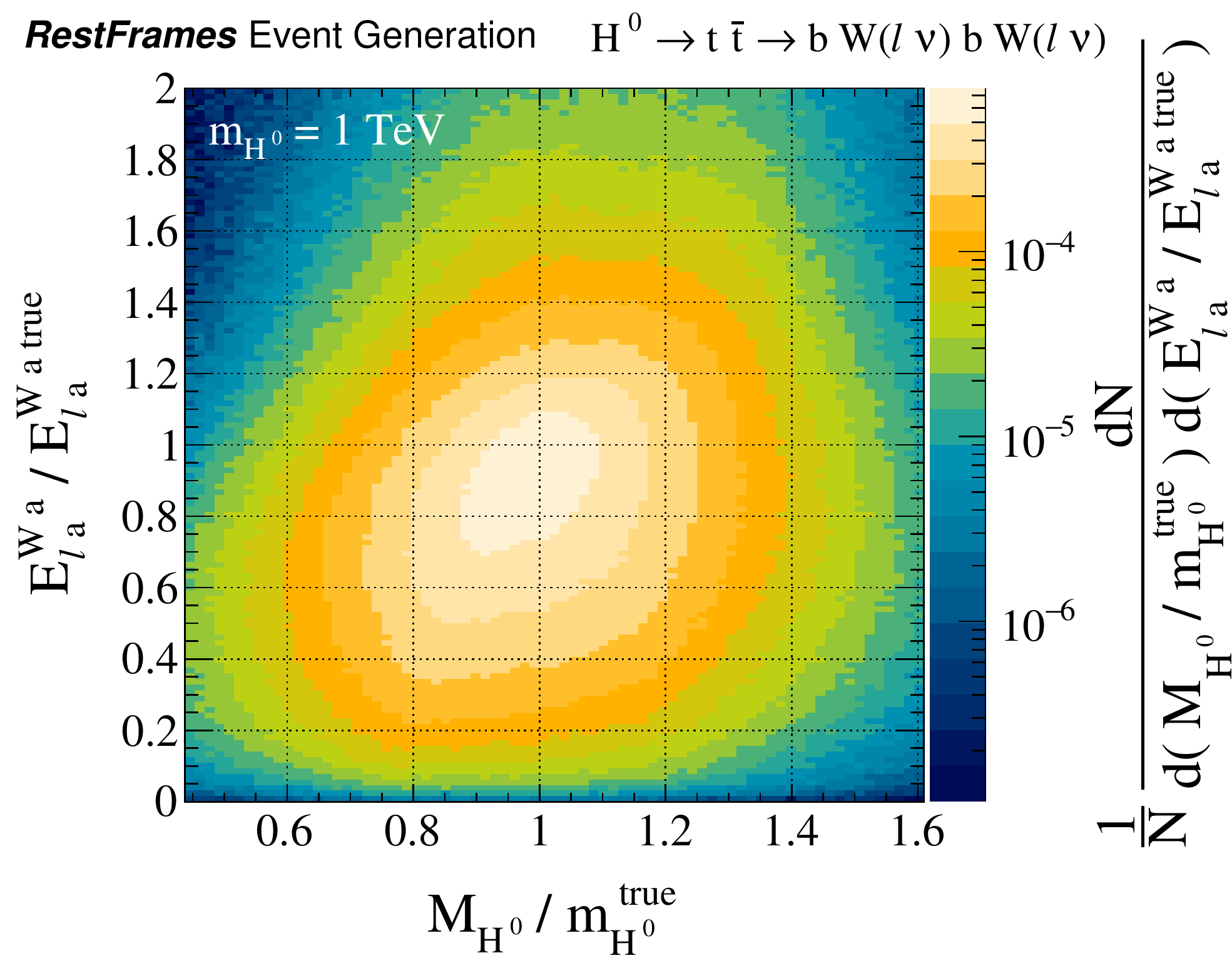}}
\subfigure[]{\includegraphics[width=.238\textwidth]{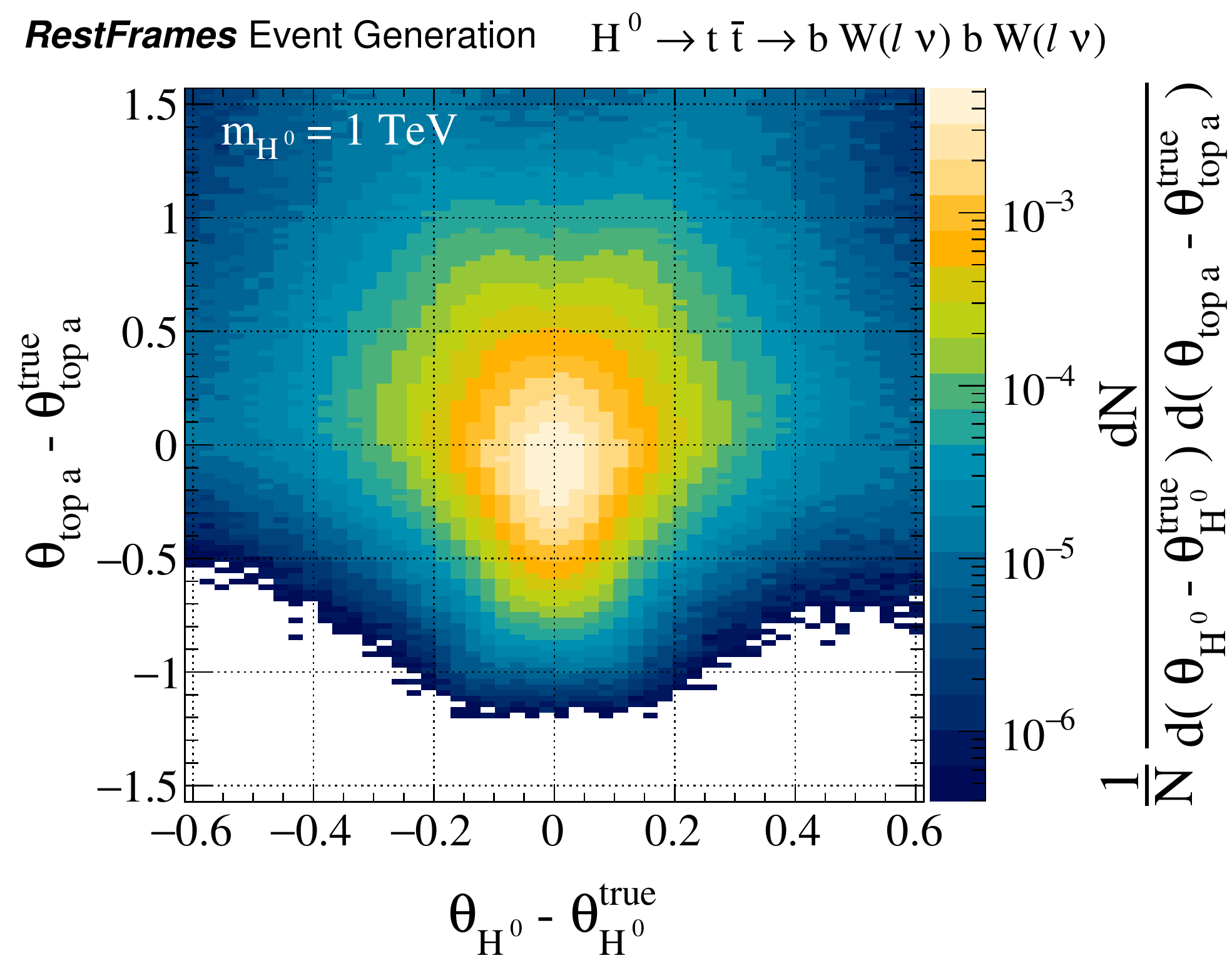}}
\subfigure[]{\includegraphics[width=.238\textwidth]{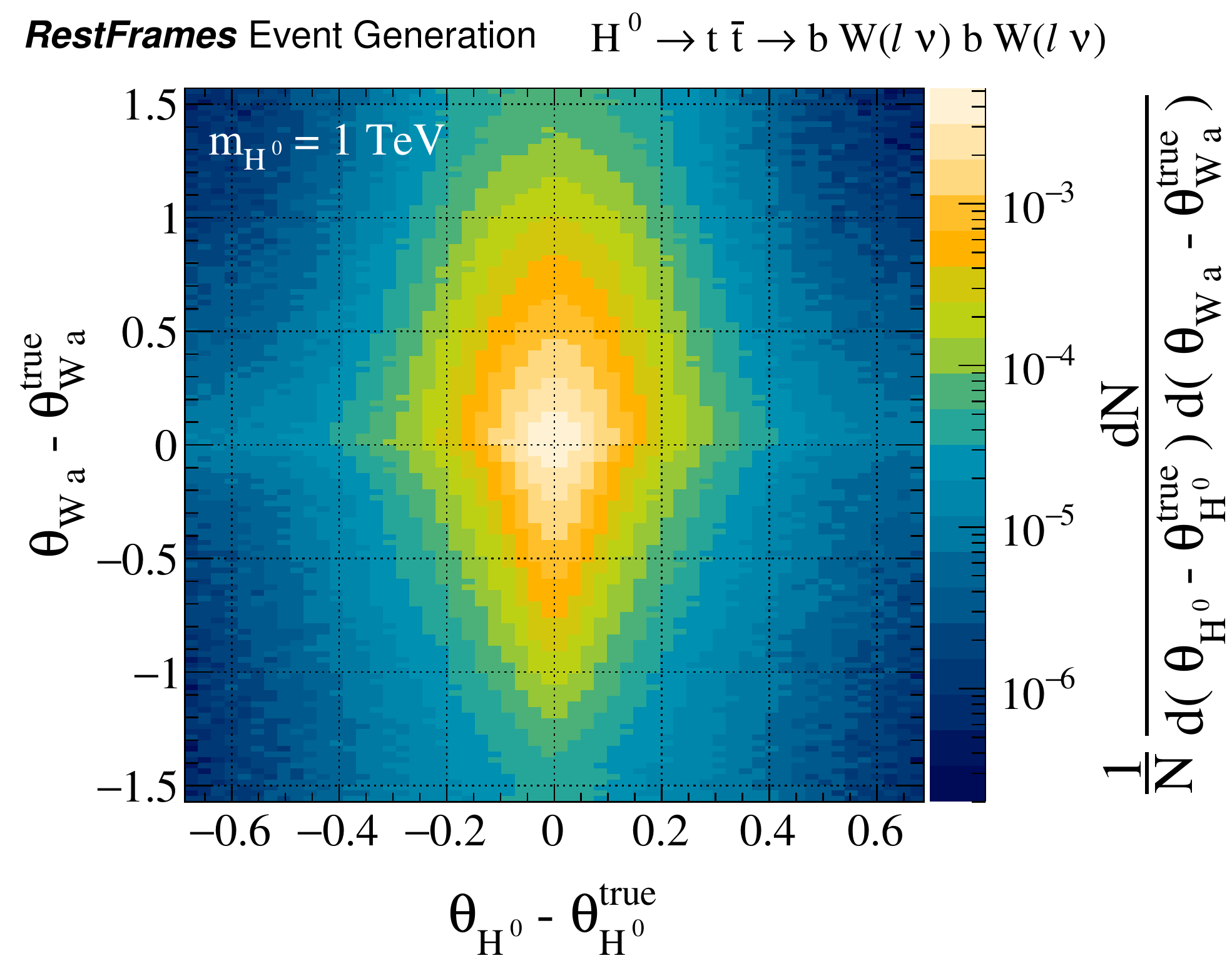}}
\vspace{-0.3cm}
\caption{\label{fig:example_H_to_ttbar_to_bWlnubWln-M2D} Distributions of (a) \E{\V{$b_a$}}{\D{$t_a$}}, and (b) \E{\V{$\ell_a$}}{\D{$W_a$}}, as a function of \Mass{\D{$H^0$}}{}, and (c) $\theta_{\D{$t_a$}}$, and (d) $\theta_{\D{$W_a$}}$, as a function of $\theta_{\D{$H$}}$, in simulated $H^0 \rightarrow {t}\bar{t} \rightarrow bW(\ell\nu) bW(\ell\nu)$ events. The true Higgs mass is set to $\mass{\D{$H^{0}$}}{} = 1$ TeV. Each observable is normalized by the true value, with angles in units radian.}
\end{figure}

The magnitude of these residual correlations on the estimates of the visible particle energies in their production frames can be observed when considering their dependence on the reconstructed Higgs boson mass, as can be seen in Fig.~\ref{fig:example_H_to_ttbar_to_bWlnubWln-M2D}(a,b) for simulated events with $\mass{\D{$H^{0}$}}{} = 1$ TeV.  The observables have small correlations, comparable to those between the visible energy estimators themselves as seen previously in Section~\ref{subsec:Part3_exampleA}. 

A similar conclusion holds when considering the dependence of the top and $W$ decay angle estimators on the analogous quantity for the Higgs, as demonstrated in Fig.~\ref{fig:example_H_to_ttbar_to_bWlnubWln-M2D}(c,d), where the observables exhibit negligible correlation. Taken with the conclusions from the previous example, it is clear that the estimators calculated in the RJR approach for this final state, in a particular reference frame, are almost entirely independent of those associated with different frames, with the accuracy of each of the estimators remaining stable - a crucial property of this basis of observables.  



\subsection{$\tilde{t}\tilde{t} \rightarrow b\tilde{\chi}_{1}^{\pm}(\ell\tilde{\nu})b\tilde{\chi}_{1}^{\mp}(\ell\tilde{\nu})$ at a hadron collider}
\label{subsec:Part3_exampleC}

\begin{figure}[htbp]
\centering 
\includegraphics[width=.34\textwidth]{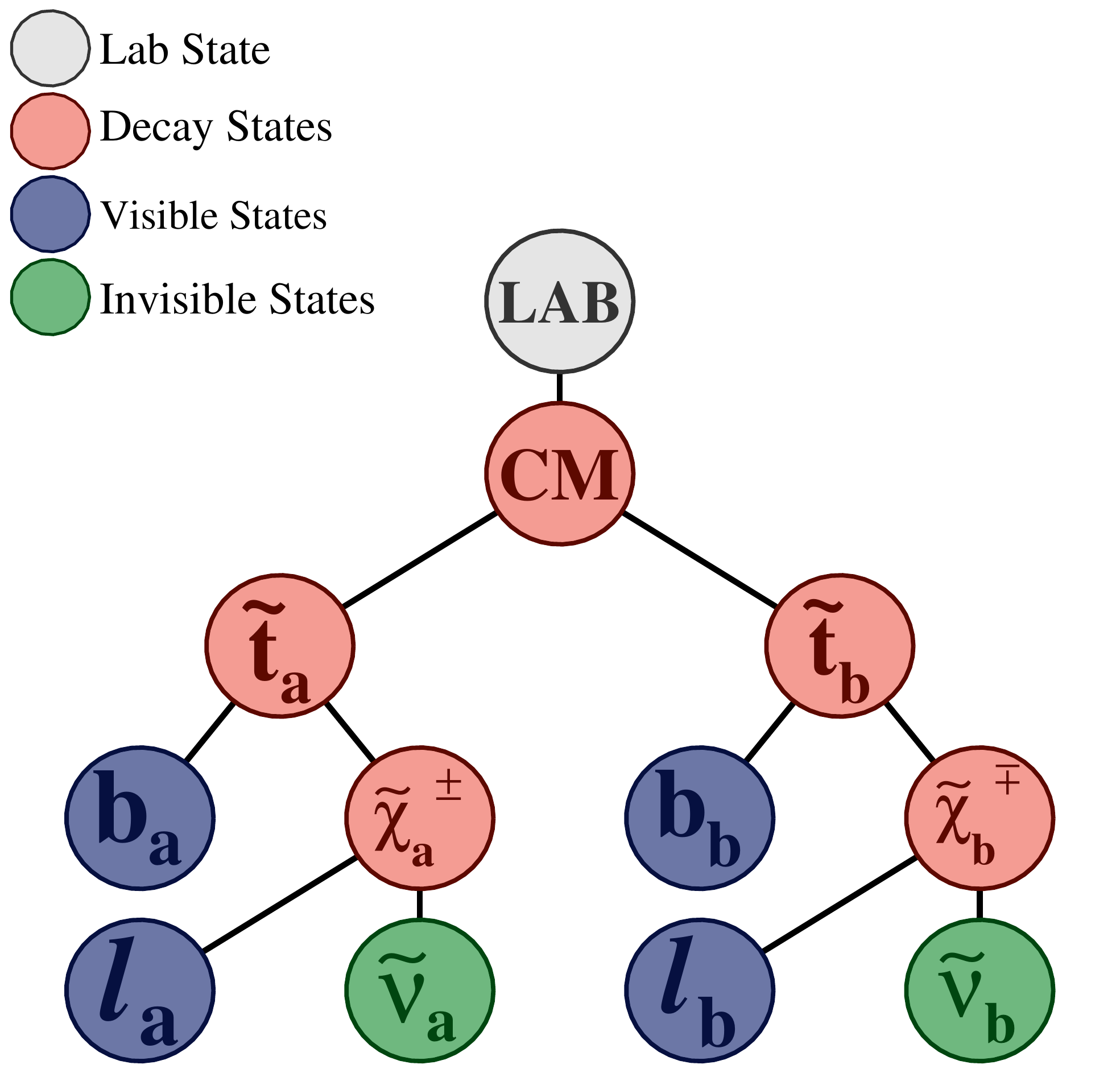}
\vspace{-0.3cm}
\caption{\label{fig:example_DiStop_to_bXp_bXm_to_blNblN-decay} Decay tree for stop pair production with decays to $b$-quarks, leptons, and sneutrinos, via a chargino. While the stop quarks and sneutrinos are assumed to have the same mass in each half of the event, the charginos may correspond to different mass eigenstates. }
\end{figure}

The final permutation of final states with two leptons, two $b$-tagged jets, and two invisible particles we consider is the case of stop quark pair production at a hadron collider, where each stop decays to a $b$-quark and chargino which, in turn, decays to a lepton and a sneutrino. We assume that the sneutrino is either the lightest supersymmetric particle or that it decays via $\tilde{\nu} \rightarrow \nu \tilde{\chi}^{0}_{1}$, such that it behaves as an individual invisible particle. The decay tree diagram for this final state, shown in Fig.~\ref{fig:example_DiStop_to_bXp_bXm_to_blNblN-decay}, is identical to that for fully-leptonic top pair production, with the intermediate particle states and invisible particles replaced by their supersymmetric counterparts.

The appeal of the RJR approach for studying events like these is that one can estimate the mass splittings of these particles and, in cases with many intermediate masses, do so largely independently of each other. The same strategy can be used whether the chargino appearing in these events has a mass degenerate with the stop, nearly as small as the sneutrino, or anywhere in between. In the context of a search for evidence of this phenomenon, this mass sensitivity is essential for distinguishing this process from a likely large $t\bar{t}$ background, where one or more mass splittings may be similar their SM counterparts, and difficult to distinguish from them.

If one were attempting to study the spin correlations of these stop decays, unbiased estimations of the total di-stop invariant mass and decay angles would be valuable, such that the \minDMt analysis scheme described in Section~\ref{subsec:Part3_exampleA} would be appropriate. In a discovery search, these observables are less relevant, with estimation of the intermediate particle mass splittings a higher priority. Recalling that the \minMt approach of Section~\ref{subsec:Part3_exampleA} provided the most accurate, and least biased, estimators of the relevant mass splittings, \E{\V{$b_{a/b}$}}{\D{$\tilde{t}_{a/b}$}} and \E{\V{$\ell_{a/b}$}}{\D{$\tilde{\chi}^{\pm}_{a/b}$}}, we adopt that strategy for the analysis of $\tilde{t}\tilde{t} \rightarrow b\tilde{\chi}_{1}^{\pm}(\ell\tilde{\nu})b\tilde{\chi}_{1}^{\mp}(\ell\tilde{\nu})$.

While the absolute values of the mass differences between the stop, chargino, and sneutrino may differ from their SM analogues, the behavior of the RJR observables, when compared with the true quantities we are attempting to estimate in these events, depends primarily on the ratio of the mass splittings in each decay step. We imagine in this example that each stop quark has $\mass{\D{$\tilde{t}$}}{} = 800$ GeV, and each sneutrino $\mass{\I{$\tilde{\nu}$}}{} = 100$ GeV. A range of intermediate chargino masses are considered, parameterized by the ratio of chargino/sneutrino and stop/sneutrino mass differences, $R_{\mass{\tilde{\chi}^{\pm}}{}}$:
\bea
R_{\mass{\tilde{\chi}^{\pm}}{}}= \frac{ \mass{\D{$\tilde{\chi}^{\pm}$}}{}-\mass{\I{$\tilde{\nu}$}}{} }{ \mass{\D{$\tilde{t}$}}{}-\mass{\I{$\tilde{\nu}$}}{} }~.
\eea

\begin{figure}[!htbp]
\centering 
\subfigure[]{\includegraphics[width=.28\textwidth]{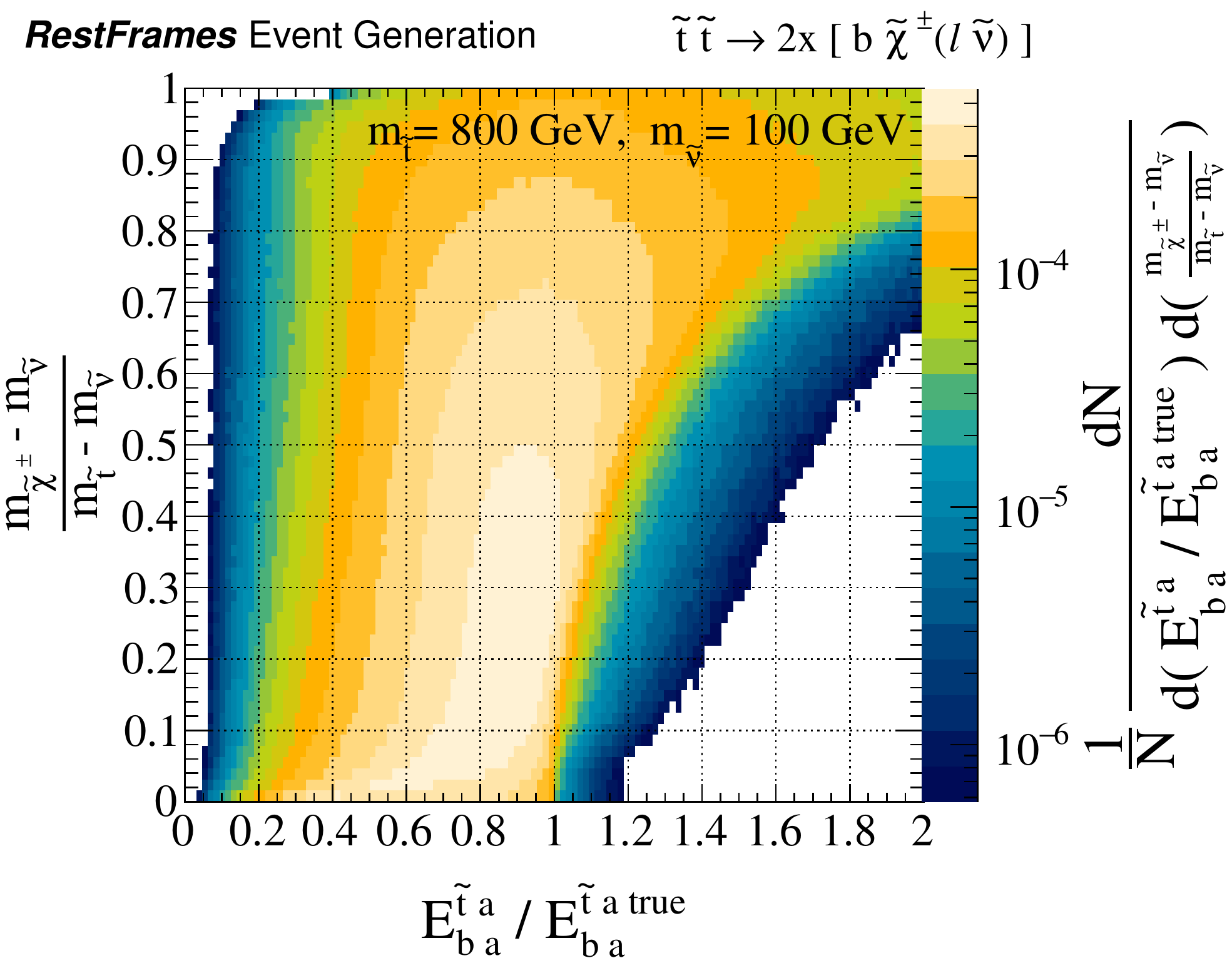}} \hspace{0.5cm}
\subfigure[]{\includegraphics[width=.28\textwidth]{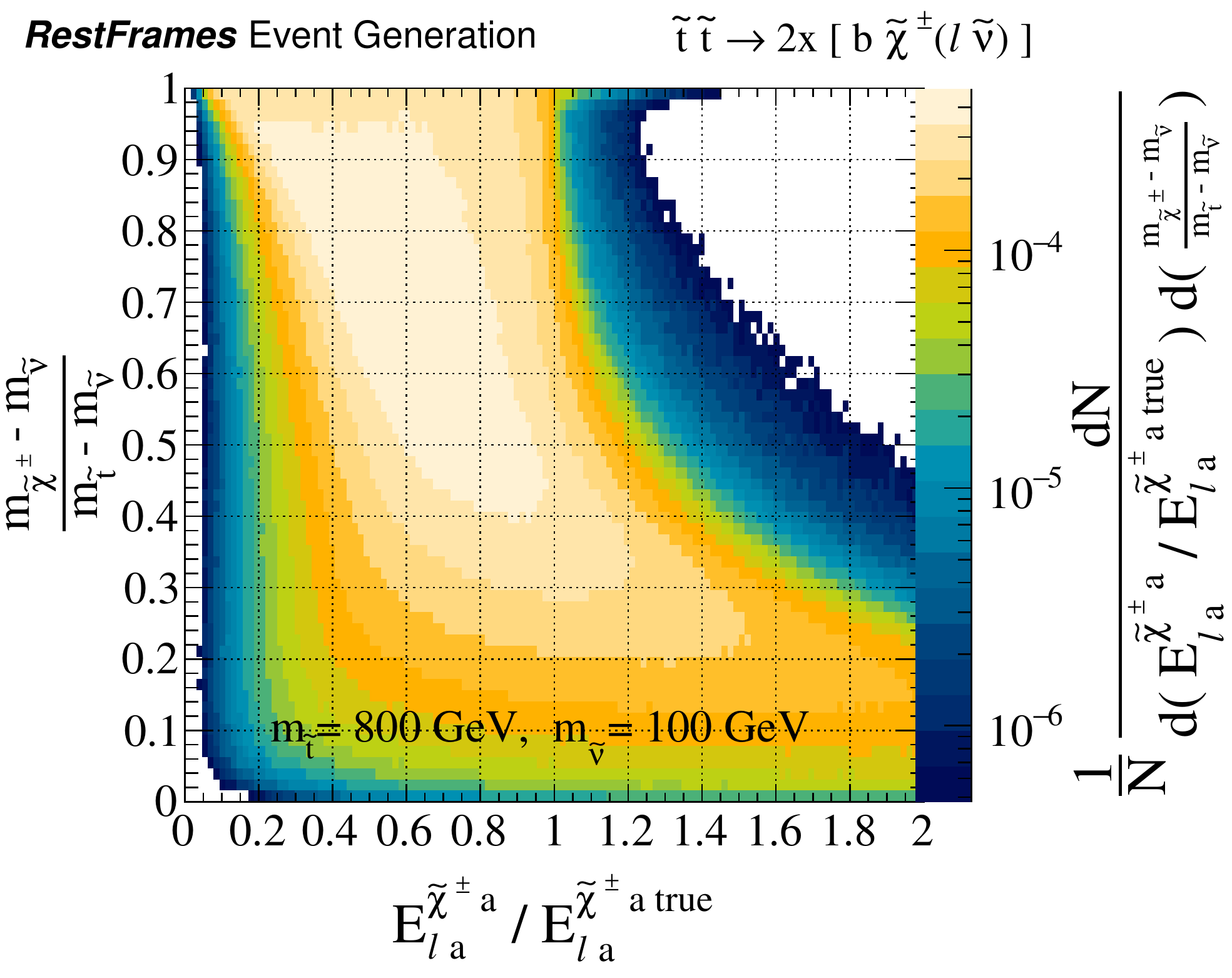}} \hspace{0.5cm}
\subfigure[]{\includegraphics[width=.28\textwidth]{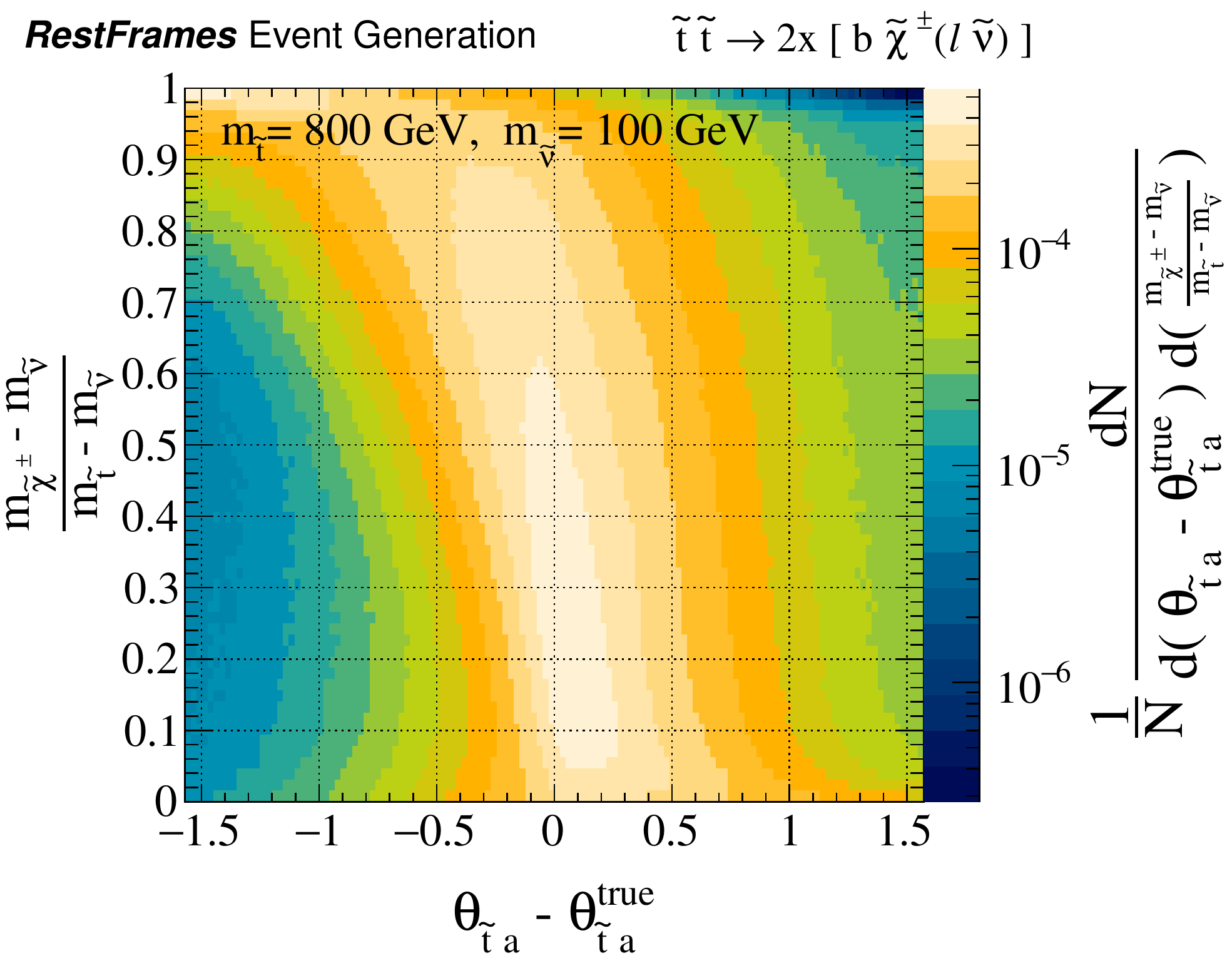}}
\subfigure[]{\includegraphics[width=.28\textwidth]{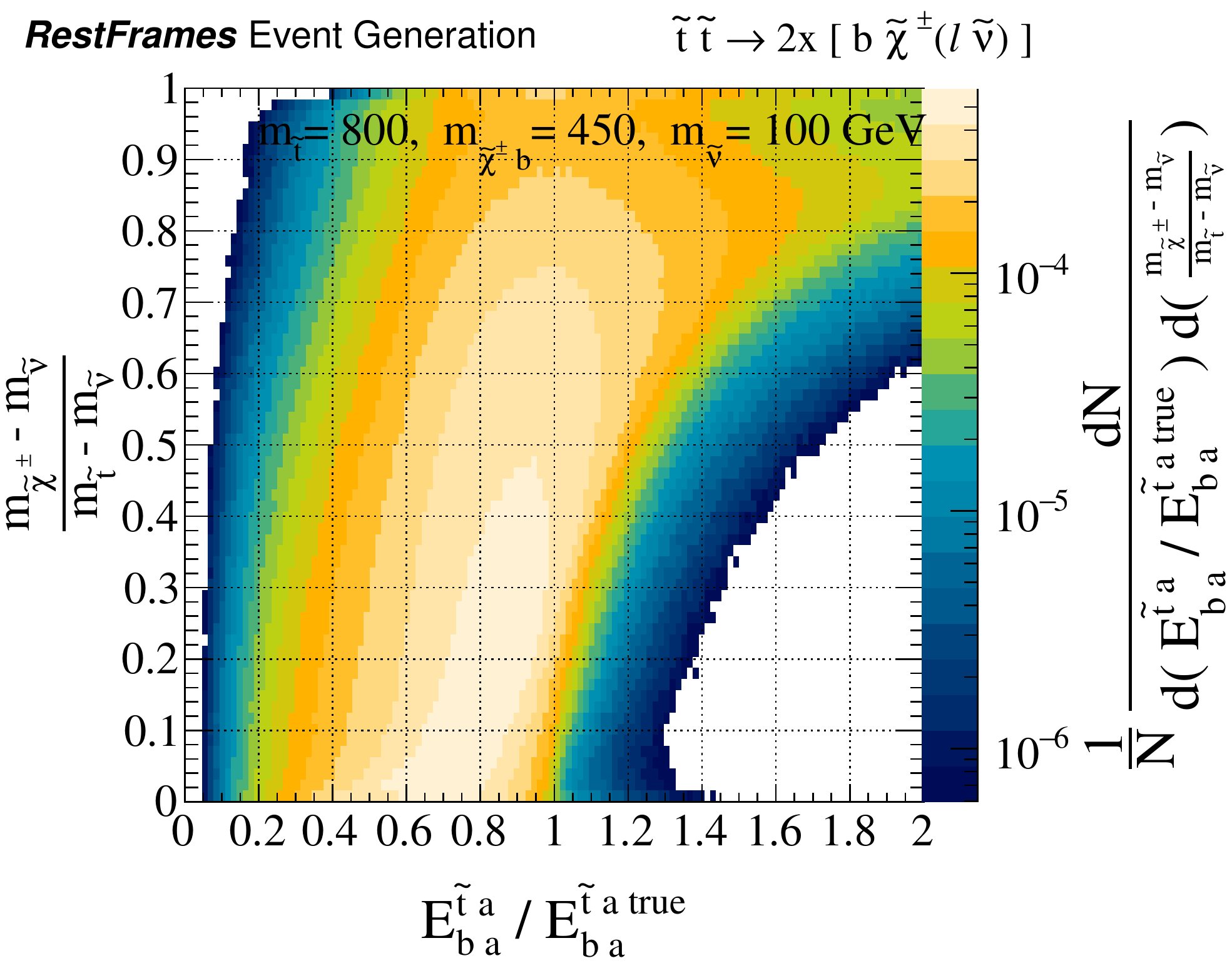}} \hspace{0.5cm}
\subfigure[]{\includegraphics[width=.28\textwidth]{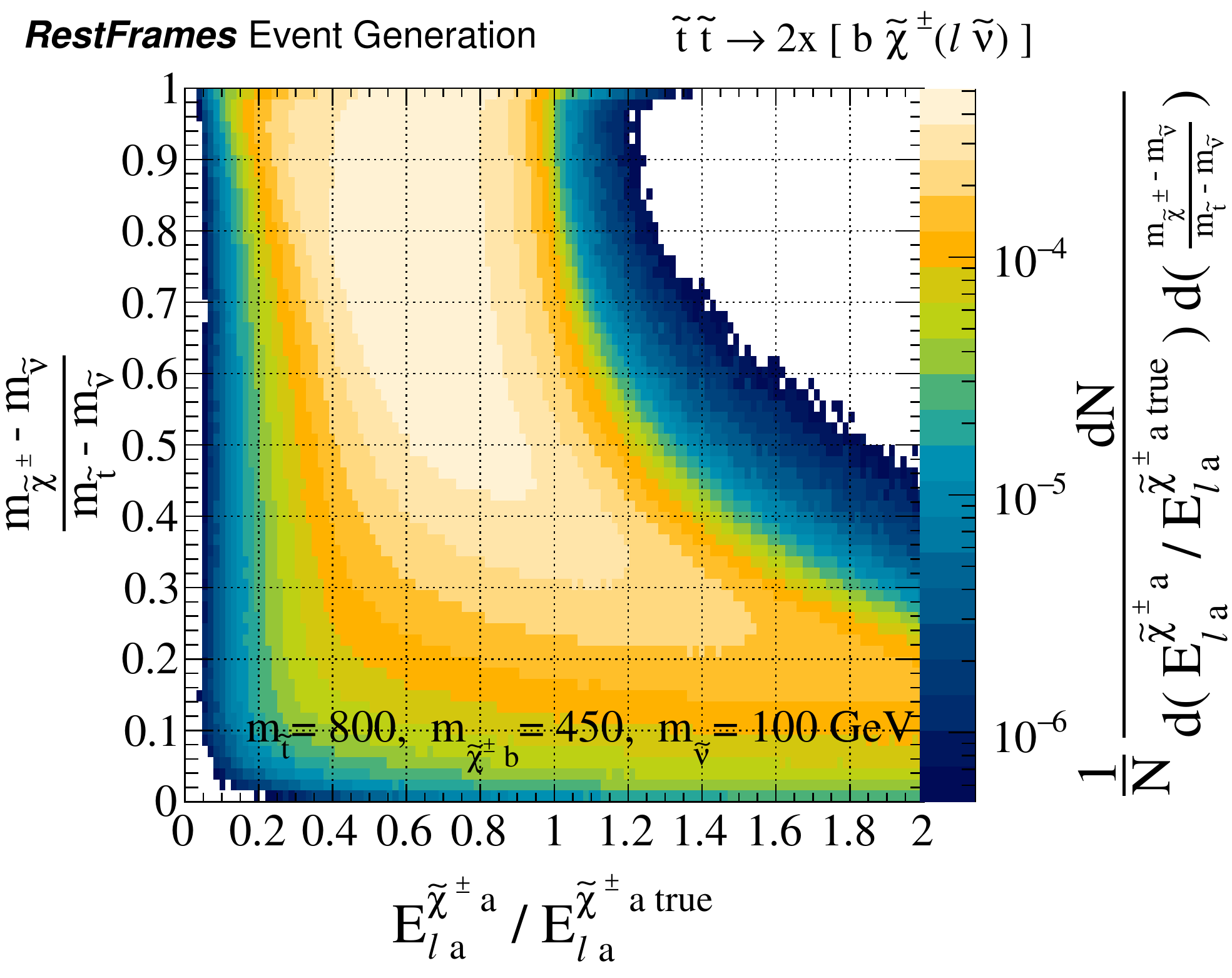}} \hspace{0.5cm}
\subfigure[]{\includegraphics[width=.28\textwidth]{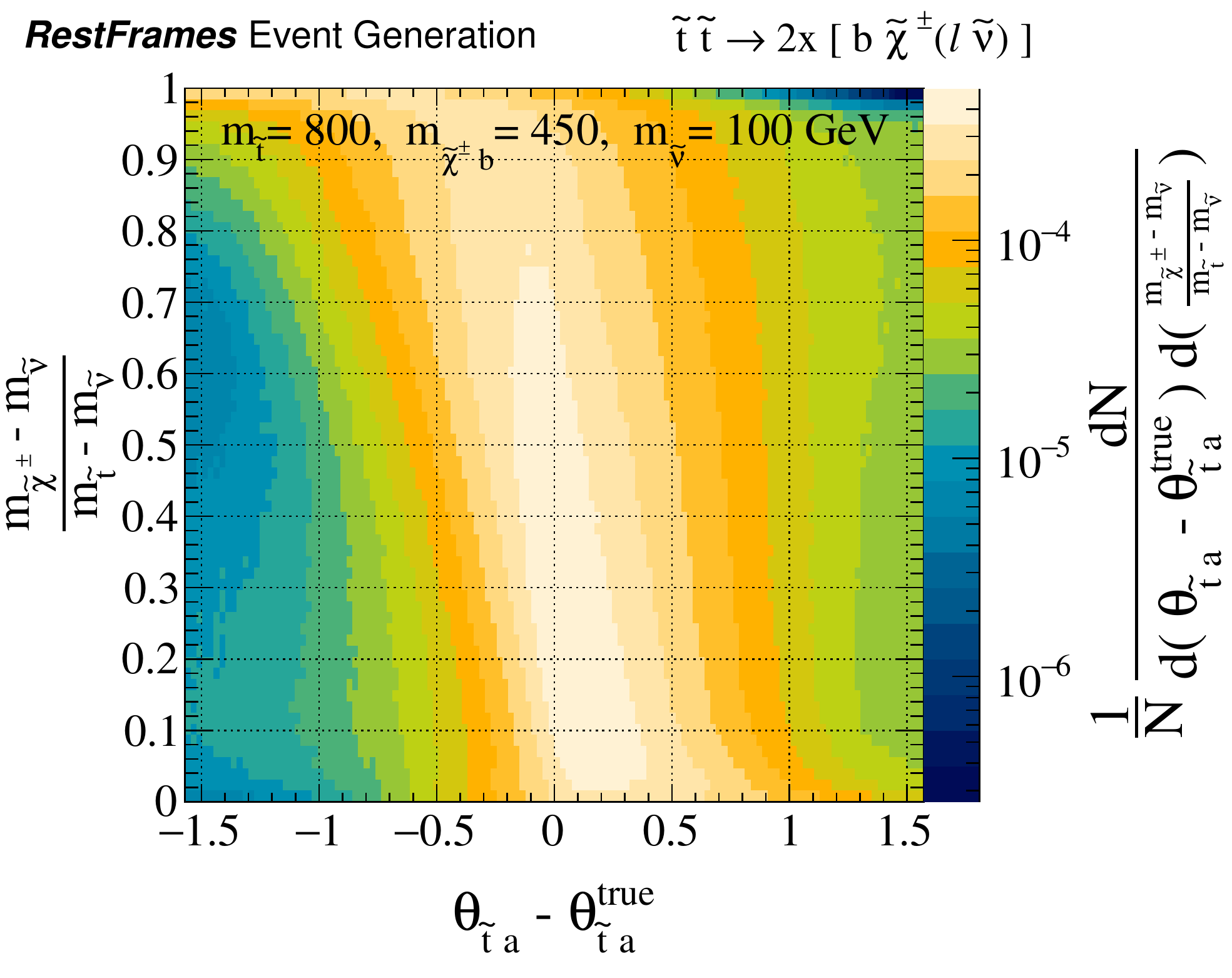}}
\vspace{-0.3cm}
\caption{\label{fig:example_DiStop_to_bXp_bXm_to_blNblN-RMX_v_Eb} Distributions of (a,d) \E{\V{$b_a$}}{\D{$\tilde{t}_a$}}, (b,e) \E{\V{$\ell_a$}}{\D{$\tilde{\chi}^{\pm}_a$}}, and (c,f) $\theta_{\D{$\tilde{t}_a$}}$, as a function of $R_{\mass{\tilde{\chi}^{\pm}}{}}$, for simulated $\tilde{t}\tilde{t} \rightarrow b\tilde{\chi}_{1}^{\pm}(\ell\tilde{\nu})b\tilde{\chi}_{1}^{\mp}(\ell\tilde{\nu})$ events. All observables are appropriately normalized by the true values they are estimating, with decay angles in units radian. The simulated stop and sneutrino masses are chosen as $\mass{\D{$\tilde{t}$}}{} = 800$ GeV and $\mass{\I{$\tilde{\nu}$}}{} = 100$ GeV, respectively. Figures (a,b,c) have $\mass{\D{$\tilde{\chi}^{\pm}_a$}}{} =   \mass{\D{$\tilde{\chi}^{\pm}_b$}}{}$, while figures (d,e,f) have one chargino mass fixed at $\mass{\D{$\tilde{\chi}^{\pm}_b$}}{} = 450$ GeV, and \mass{\D{$\tilde{\chi}^{\pm}_a$}}{} varying with $R_{\mass{\tilde{\chi}^{\pm}}{}}$.  }
\end{figure}

The distributions of the visible particle energy estimators, \E{\V{$b_a$}}{\D{$\tilde{t}$}} and \E{\V{$\ell_a$}}{\D{$\tilde{\chi}^{\pm}_a$}}, and the reconstructed stop decay angle are shown, as a function of $R_{\mass{\tilde{\chi}^{\pm}}{}}$, for simulated events in Fig.~\ref{fig:example_DiStop_to_bXp_bXm_to_blNblN-RMX_v_Eb}(a,b,c), where the true values of the chargino masses $\mass{\D{$\tilde{\chi}^{\pm}_a$}}{} =  \mass{\D{$\tilde{\chi}^{\pm}_b$}}{}$ are varied between their kinematically allowed values. As $R_{\mass{\tilde{\chi}^{\pm}}{}} \rightarrow 0$, the phase-space in the production of the lepton becomes negligible, meaning these events appear as if the stop quarks are decaying as $\tilde{t}\tilde{t}\rightarrow b\tilde{\nu}b\tilde{\nu}$.  This results in the \E{\V{$b_a$}}{\D{$\tilde{t}$}} distribution exhibiting a kinematic edge at the true value, similar to the analogous quantity shown in Fig.~\ref{fig:example_DiStop_to_hadtopXhadtopX-Ptop_v_MTT} for $\tilde{t}\tilde{t}\rightarrow t_{\rm had} \tilde{\chi}^{0}_{1}t_{\rm had}\tilde{\chi}^{0}_{1}$ events. Conversely, as $R_{\mass{\tilde{\chi}^{\pm}}{}} \rightarrow 1$ the events appear as if the stops are decaying $\tilde{t}\tilde{t}\rightarrow \ell \tilde{\nu}\ell\tilde{\nu}$, with the \E{\V{$\ell_a$}}{\D{$\tilde{\chi}^{\pm}_a$}} distribution in Fig.~\ref{fig:example_DiStop_to_bXp_bXm_to_blNblN-RMX_v_Eb}(b) taking a similar shape as for \E{\V{$\ell_a$}}{\D{$W_a$}} in $H \rightarrow W(\ell\nu)W(\ell\nu)$ events from Fig.~\ref{fig:example_H_to_WlnuWlnu-MW_v_MH}. As $R_{\mass{\tilde{\chi}^{\pm}}{}}$ approaches the opposite extremum for both energy estimators their distributions peak at the correct values, with increasingly degraded resolution as the associated decay phase-space shrinks. Similarly, the accuracy of the $\theta_{\D{$\tilde{t}_a$}}$ estimator becomes worse as the stop and chargino become degenerate in mass, developing a growing bias as this limit approaches, as can be seen in Fig.~\ref{fig:example_DiStop_to_bXp_bXm_to_blNblN-RMX_v_Eb}(c).

\onecolumngrid

\twocolumngrid

The RJR \minMt approach is able to retain sensitivity to the true mass splittings between the sparticles in these events over a broad phase space of decays, even in the limit of degeneracy between masses. But the most remarkable property of the derived estimators is the level of independence observables sensitive to the different stop kinematics exhibit. If we imagine a case where there are two chargino mass eigenstates in between the stop and sneutrino masses, a single event could contain one of each of these charginos. To study what happens to the RJR estimators when these masses are different, we fix one chargino mass, with $\mass{\D{$\tilde{\chi}^{\pm}_b$}}{} = 450$ GeV, while varying the other over the allowable on-shell phase-space between the stop and sneutrino. The same estimators, \E{\V{$b_a$}}{\D{$\tilde{t}$}}, \E{\V{$\ell_a$}}{\D{$\tilde{\chi}^{\pm}_a$}}, and $\theta_{\D{$\tilde{t}_a$}}$, are shown as a function of $R_{\mass{\tilde{\chi}^{\pm}}{}}$ for this differing chargino mass scenario in Fig.~\ref{fig:example_DiStop_to_bXp_bXm_to_blNblN-RMX_v_Eb}(d,e,f). The similarities between the these distributions and those for $\mass{\D{$\tilde{\chi}^{\pm}_a$}}{} =   \mass{\D{$\tilde{\chi}^{\pm}_b$}}{}$ in Fig.~\ref{fig:example_DiStop_to_bXp_bXm_to_blNblN-RMX_v_Eb}(a,b,c) are striking. With only small deformations, the estimators retain nearly identical dependence on $R_{\mass{\tilde{\chi}^{\pm}}{}}$ in the two cases, irrespective of the mass splittings of sparticles in the opposite half of the event. This independence between the two stops' kinematics is a result of the application of the contra-boost invariant JR~\ref{jr:contra} when analyzing the event, making the observables associated with each stop insensitive to the true velocity relating its rest frame to its production frame and, hence, also the other stop. This means that one can effectively analyze each part of the event in approximate isolation, allowing for the recursive application of even more JR's in events with additional kinematic structure, as is described in the following examples.

\section{Recursive Jigsaws For More Particles - Intermediate, Invisible, and Identical}
\label{sec:Part4}
The previous examples have introduced a large collection of configurable and interchangeable JR's, which can be chosen to analyze a variety of events. As the decays we hope to study grow further in complexity, with additional invisible particles in the final state and higher degrees of combinatoric ambiguity, there are two treatments available. We can introduce further generalizations of existing JR's to simultaneously choose more under-constrained degrees of freedom, an approach we discuss in Section~\ref{subsec:Part4_exampleA} for $N \geq 2~\times~W(\ell\nu)$ final states. Alternatively, we can combine existing JR's, recursively, into logical trees, iteratively sub-dividing each event kinematically while choosing the appropriate degrees of freedom. The examples in Sections~\ref{subsec:Part4_exampleB} and~\ref{subsec:Part4_exampleC} demonstrate this latter approach, considering the processes $H\rightarrow hh \rightarrow 4W(\ell\nu)$ and $\tilde{g}\tilde{g} \rightarrow bb\tilde{\chi}^{0}_{1}bb\tilde{\chi}^{0}_{1}$. The additional JR's defined in this section, combined with those described throughout this paper, constitute a sufficient collection for analyzing {\it any} final state, with an arbitrary degree of complexity and number unknowns.

\subsection{$\mathrm{pp} \rightarrow N \geq 2~\times~W(\ell\nu)$}
\label{subsec:Part4_exampleA}

As the number of invisible particles appearing in a final state increases, the number of unmeasured degrees of freedom grows quickly. If this increase is accompanied by additional mass constraints, for example between intermediate particles appearing in a decay, the additional unknowns can be mitigated with the recursive application of corresponding JR's, exploiting the expected structure in each event to better resolve quantities of interest.

\begin{figure}[!htbp]
\centering
\subfigure[]{\includegraphics[width=.28\textwidth]{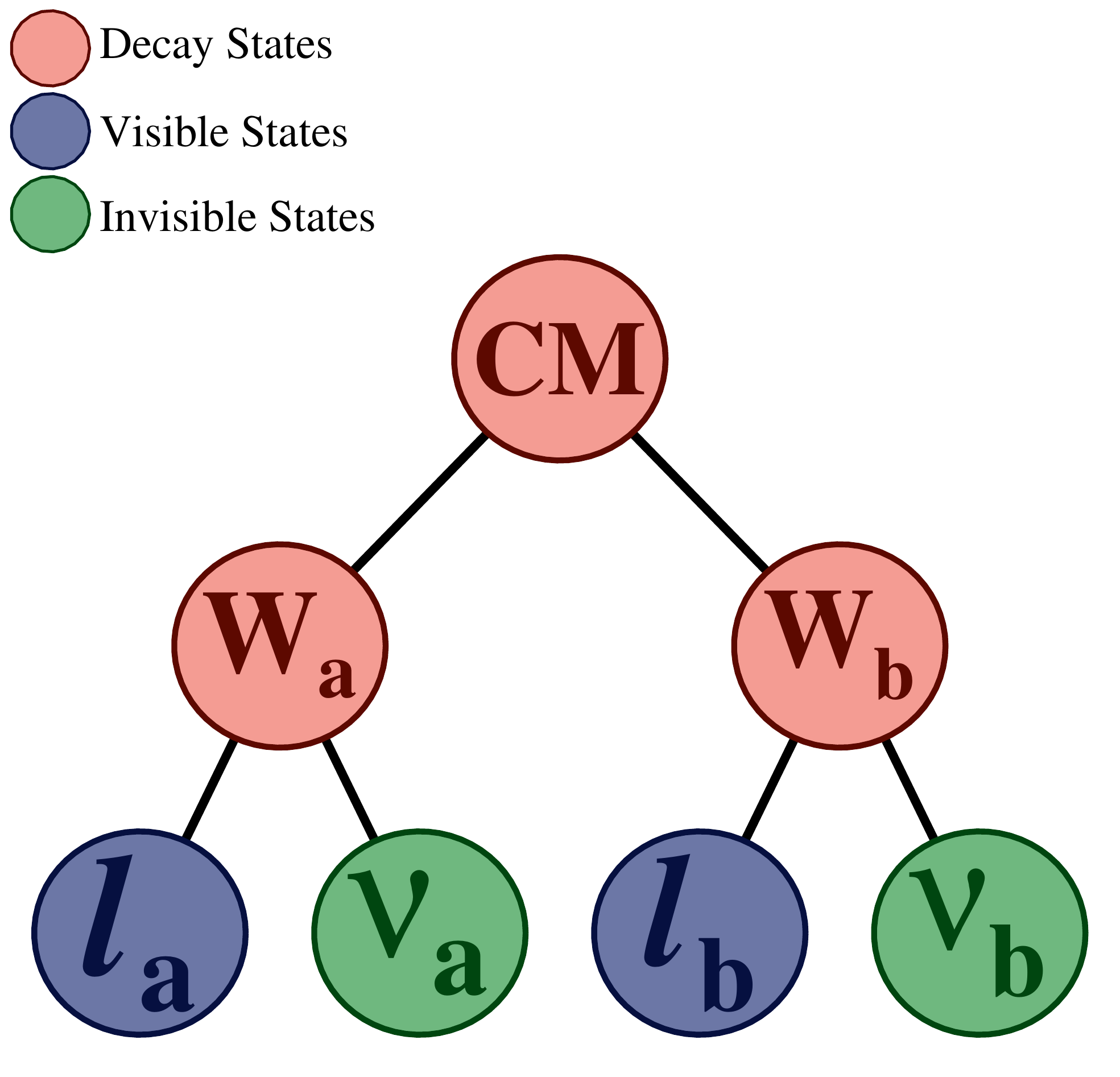}} \hspace{0.5cm}
\subfigure[]{\includegraphics[width=.28\textwidth]{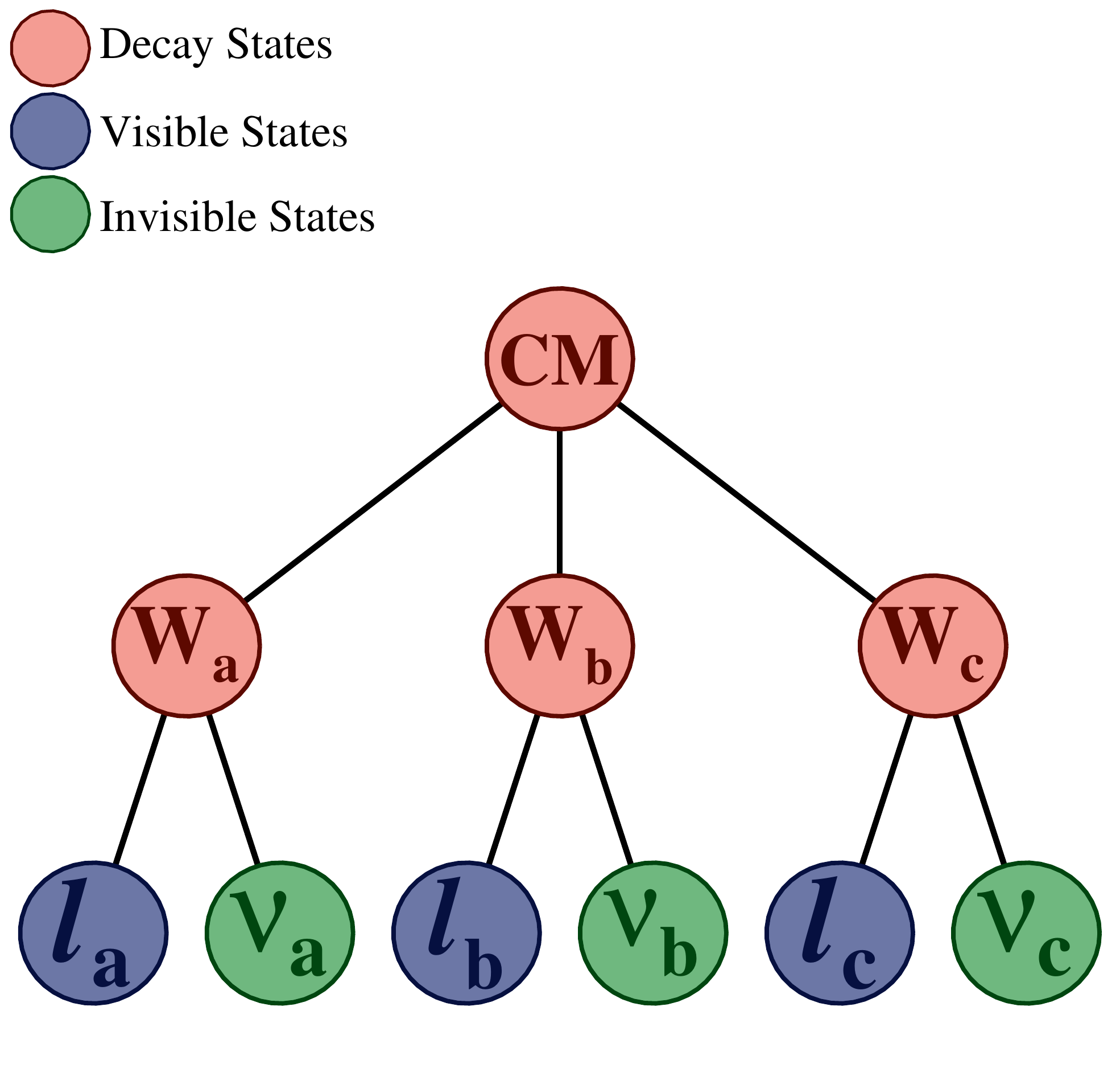}} \hspace{0.5cm}
\subfigure[]{\includegraphics[width=.28\textwidth]{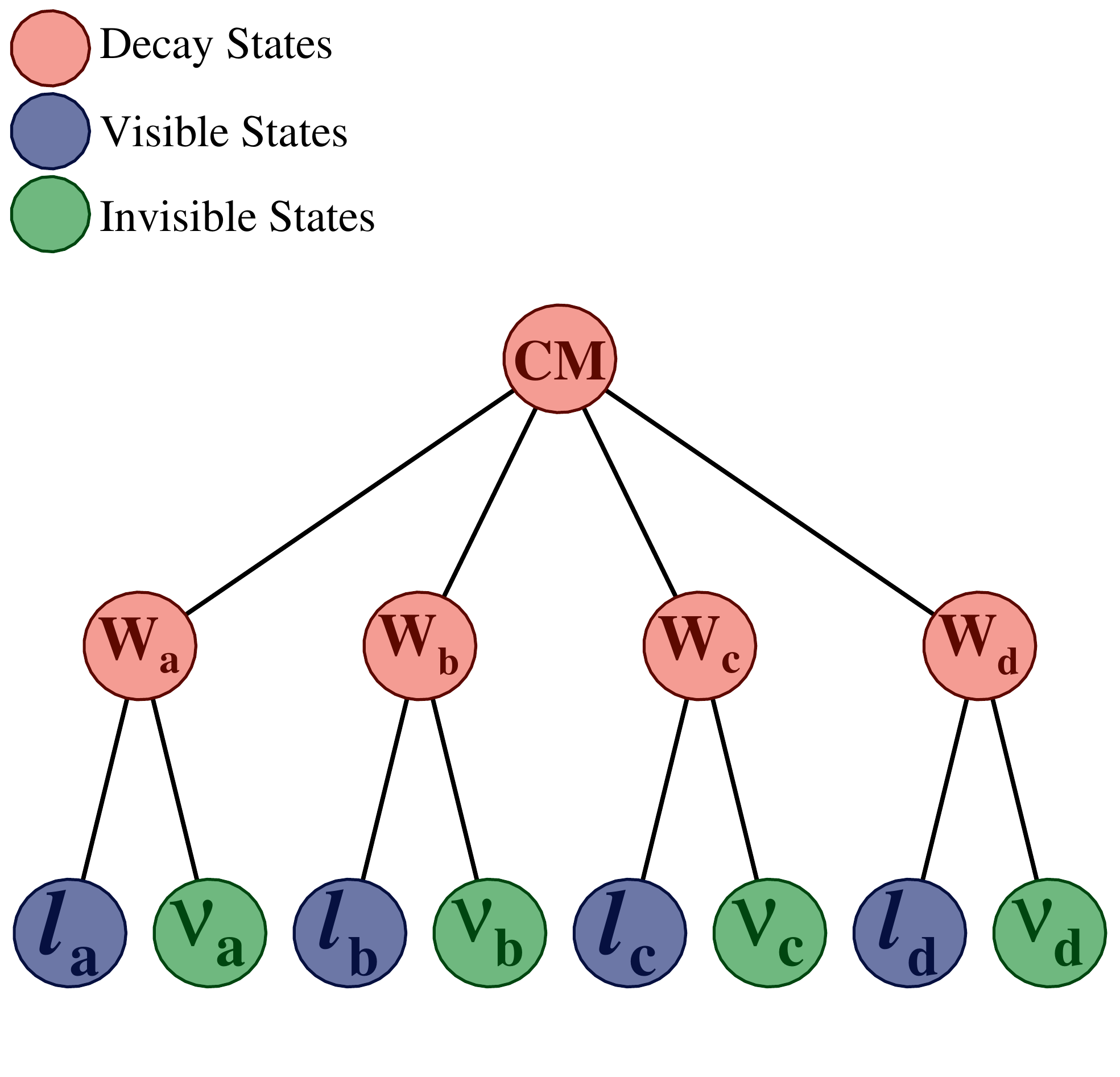}}
\vspace{-0.3cm}
\caption{\label{fig:example_N_Wlnu-decay_NW} Decay trees for $\mathrm{pp} \rightarrow N \geq 2~\times~W(\ell\nu)$ final states with (a) $N=2$, (b) $N=3$, and (c) $N=4$. The $W$ bosons are assumed to be produced non-resonantly, such that there is no phase-space structure beyond the flat $N$-body phase space of the \D{cm} decay and subsequent $W$ decays. While they tend to have similar values in the RJR reconstruction of these events, the $W$ bosons are not explicitly constrained to have equal mass. }
\end{figure}

In some cases, additional complexity is not accompanied by more intermediate structure, with the number of unknowns growing much larger than the number of appropriate constraints. Such a case is the non-resonant production of $N \geq 2$ $W$ bosons at a hadron collider, with each $W$ decaying to a lepton and a neutrino. The $N=2$ case corresponds to the example of $H \rightarrow W(\ell\nu)W(\ell\nu)$ production, described in Section~\ref{subsec:Part2_exampleA}. There, we used the constraint \Mass{\D{$W_a$}}{} = \Mass{\D{$W_b$}}{} and the contra-boost invariant JR~\ref{jr:contra} to determine how to split the invisible system into two separate neutrinos, corresponding to the \D{cm} frame decay in the tree shown in Fig.~\ref{fig:example_N_Wlnu-decay_NW}(a). As we see for the $N > 2$ decay trees in Fig.~\ref{fig:example_N_Wlnu-decay_NW}, the corresponding decay has an increasing number of legs, or velocities relating the \D{cm} frame to its daughter \D{$W_i$} systems' rest frames, and hence a correspondingly larger number of unknowns, with additional neutrino four vectors to choose in the associated JR.
\onecolumngrid

\twocolumngrid

We assume that the $N$ leptons in the final state of each event are identified and reconstructed, with four vectors \pfour{\V{$\ell_i$}}{\lab}, and that the measurement of \met~ in each event can be interpreted as the sum transverse momentum of the total invisible system of neutrinos, $\I{I} = \{ \I{$\nu_{i}$} \}$, with $\pthree{\I{I},T}{\lab} = \met$. With $N$ neutrino four vectors to resolve and only two constraints from the \met~ measurement, there are $4N-2$ values left to choose. 

As the $W$ bosons are expected to have different velocities in the \D{cm} frame when $N \geq 3$, we are unable to introduce a generalization of contra-boost invariance for arbitrary $N$. Applying direct mass constraints between the $W$ bosons can lead to multiple and/or complex solutions for kinematic quantities in events, and cannot be generalized to cases when the intermediate particle masses may be different. The shortcoming of simply choosing an inspired metric and minimizing it w.r.t. each of these unknowns simultaneously is that, with so many d.o.f., the minimization will be able to find small, trivial solutions. In order to resolve the fact that these events have intermediate structure in the form of massive $W$ bosons, we must either factorize the minimization of these unknowns ({\it i.e.} a binary decay tree) or perform the minimization with carefully constructed constraints to retain the structure. The algorithmic approach must also be tractable, in that the mass of the \I{I} system must be chosen as a Lorentz invariant function of the visible lepton four vectors, while the neutrinos' masses must also obey any pre-determined constraints and remain non-tachyonic throughout the space of the minimization.

We adopt an {\it ad-hoc} approach inspired by the expected symmetry between the lepton and neutrino pairs' four vectors due to their common provenance. In the derivation of JR~\ref{jr:invminM2_1}, working in the rest frame of two invisible particles greatly simplified the problem, as their momentum was constrained to be equal and opposite, with a Lorentz invariant value, and only the orientation of this di-neutrino axis left to determine. Generalizing this concept, we choose the relative momentum of the neutrinos in the \D{cm} frame to correspond to that of the leptons in their respective $\V{V} = \{ \V{$\ell_i$} \}$ center-of-mass frame, such that 
\bea
\pthree{\I{$\nu_i$}}{\D{I}} = \mathbf{R}(\alpha,\beta,\gamma)  \pthree{\V{$\ell_i$}}{\D{V}}~,
\eea
where $\mathbf{R}(\alpha,\beta,\gamma)$ is a $3 \times 3$ dimensional rotation matrix described by the three Euler angles $\alpha$, $\beta$, and $\gamma$. With this constraint, we have reduced a potentially $3N-3$ d.o.f. minimization (assuming individual neutrino mass constraints) to 3, and are able to express our choice for \Mass{\I{I}}{} as a compatible Lorentz invariant expression:
\bea
\Mass{\I{I}}{} = \sum_i^{N} |\pthree{\V{$\ell_i$}}{\V{V}}|~,
\eea 
where we have constrained each $\Mass{\I{$\nu_i$}}{} = 0$.

To determine the rotation $\mathbf{R} = \mathbf{R}(\alpha,\beta,\gamma)$, we choose a metric with a linear dependence on the orientation of the neutrinos:
\bea
\label{eqn:fR}
f(\mathbf{R}) &=& \sum_i^N \Mass{\D{$W_i$}}{2} \nonumber \\
&=& \sum_i^N \left( \mass{\V{$\ell_i$}}{2} + 2( \E{\V{$\ell_i$}}{\D{I}} \E{\I{$\nu_i$}}{\D{I}} - \pthree{\V{$\ell_i$}}{\D{I}} \cdot \pthree{\I{$\nu_i$}}{\D{I}} ) \right)  \\
&=& \sum_i^N \left( \mass{\V{$\ell_i$}}{2} + 2  \E{\V{$\ell_i$}}{\D{I}} |\pthree{\V{$\ell_i$}}{\V{V}}| \right) - 2 \sum_i^{N} \pthree{\V{$\ell_i$}}{\D{I}} \cdot \mathbf{R} ~\pthree{\V{$\ell_i$}}{\V{V}}~, \nonumber 
\eea
with only the last term depending on $\mathbf{R}$ and no additional unknowns. The problem of finding $\mathbf{R}$ which minimizes $f(\mathbf{R})$ in Eq.~\ref{eqn:fR} is equivalent to the {\it orthogonal Procrustes problem}~\footnote{{\it Procrustes} was a bandit smith in Greek mythology that adjusted victims to fit his iron bed by stretching or cutting them, similar to the JR~\ref{jr:invminM2}.} in linear algebra. We use the original approach proposed by Sch{\"o}nemann~\cite{Schonemann1966}, to find $\mathbf{R}$ using a closed-formed solution based on the singular value decomposition of the matrix $\mathbf{H} = \mathbf{U}\mathbf{\Lambda}\mathbf{V}^t$, with $\mathbf{H}$ defined as:
\bea
\mathbf{H} = \sum_i^{N} \pthree{\V{$\ell_i$}}{\V{V}} (\pthree{\V{$\ell_i$}}{\D{I}})^t~,
\eea
where $(\pthree{\V{$\ell_i$}}{\D{I}})^t$ is the transpose of the lepton's column three vector. In the singular value decomposition of $\mathbf{H}$, the matrices $\mathbf{U}$ and $\mathbf{V}$ are $3 \times 3$ orthonormal matrices, while $\mathbf{\Lambda}$ is a $3 \times 3$ diagonal matrix with non-negative elements. The matrix $\mathbf{R}$ which minimizes $f(\mathbf{R})$ can be expressed as
\bea
\mathbf{R} = \mathbf{VU}^{t}~.
\eea

These choices effectively minimize the sum of intermediate $W$ masses squared, subject to constraints inspired to minimize this same quantity, anticipating the form of Eq.~\ref{eqn:fR}. The approach can be generalized as a JR:
\begin{jigsaw}[Invisible Minimize Masses$^2$]
\label{jr:invminM2}
If the internal degrees of freedom specifying how an invisible particle, $\I{I} = \{ \I{I$_{i}$} \}$, should split into $N$ particles are unknown, they can be specified by choosing $N$ corresponding visible particles,  $\V{V} = \{ \V{V$_{i}$} \}$, and minimizing the quantity $\sum_i \Mass{\V{V$_{i}$}\I{I$_{i}$}}{2}$, subject to specific constraints. It is assumed that the four vectors of the visible particles are known in the center-of-mass frame, $\D{F} = \{ \V{V}, \I{I}~\}$, as is the four vector of the total \I{I} system, \pfour{\I{I}}{\D{F}}. Furthermore, we assume that the individual invisible particle masses, \Mass{\I{I$_i$}}{} are specified.
The momentum of the invisible particles can be chosen in the di-invisible rest frame, \D{I}, as:
\bea
\pthree{\I{I$_{i}$}}{\D{I}} &=& \hat{p} \left( \mathbf{R}~\pthree{\V{V$_i$}}{\V{V}} \right)~, \\
\eea
where $\hat{p} \geq 0$ is a factor and $\mathbf{R}$ a rotation matrix, scaling and rotating, respectively, the momentum of the visible particles evaluated in the visible center-of-mass frame. The factor $\hat{p}$ is chosen by numerically solving the equation
\bea
\Mass{\I{I}}{} - \sum_i^N \sqrt{ \hat{p}^{~2}|\pthree{\V{V$_i$}}{\V{V}}|^{2} + \Mass{\I{I$_i$}}{2} }~,
\eea
such that \Mass{\I{I}}{} must be chosen to satisfy
\bea
 \Mass{\I{I}}{} \geq \sum_i^N \Mass{\I{I$_i$}}{}~,
\eea
in order to ensure that the invisible particle momenta remain real. Defining the matrix $\mathbf{H}$ as
\bea
\mathbf{H} = \sum_i^{N} \pthree{\V{V$_i$}}{\V{V}} (\pthree{\V{V$_i$}}{\D{I}})^t~,
\eea
we calculate its singular value decomposition, $\mathbf{H} = \mathbf{U}\mathbf{\Lambda}\mathbf{V}^t$, and choose $\mathbf{R}$ as
\bea
\mathbf{R} = \mathbf{VU}^{t}~.
\eea
\end{jigsaw}

We can summarize the analysis strategy for studying non-resonant $N \times W(\ell\nu)$ events as
\begin{enumerate}[noitemsep]
\item Apply the invisible mass JR~\ref{jr:mass}, \\ choosing $\Mass{\I{I}}{} = \sum_i | \pthree{\V{V$_i$}}{\V{V}} |$.
\item Apply the invisible rapidity JR~\ref{jr:rapidity}, choosing \\ \pone{\I{I},z}{\lab} using the collection of leptons, \V{V}.
\item Apply the JR~\ref{jr:invminM2}, specifying the neutrino four vectors in the invisible center-of-mass frame.
\end{enumerate}

\begin{figure}[!htbp]
\centering
\subfigure[]{\includegraphics[width=.28\textwidth]{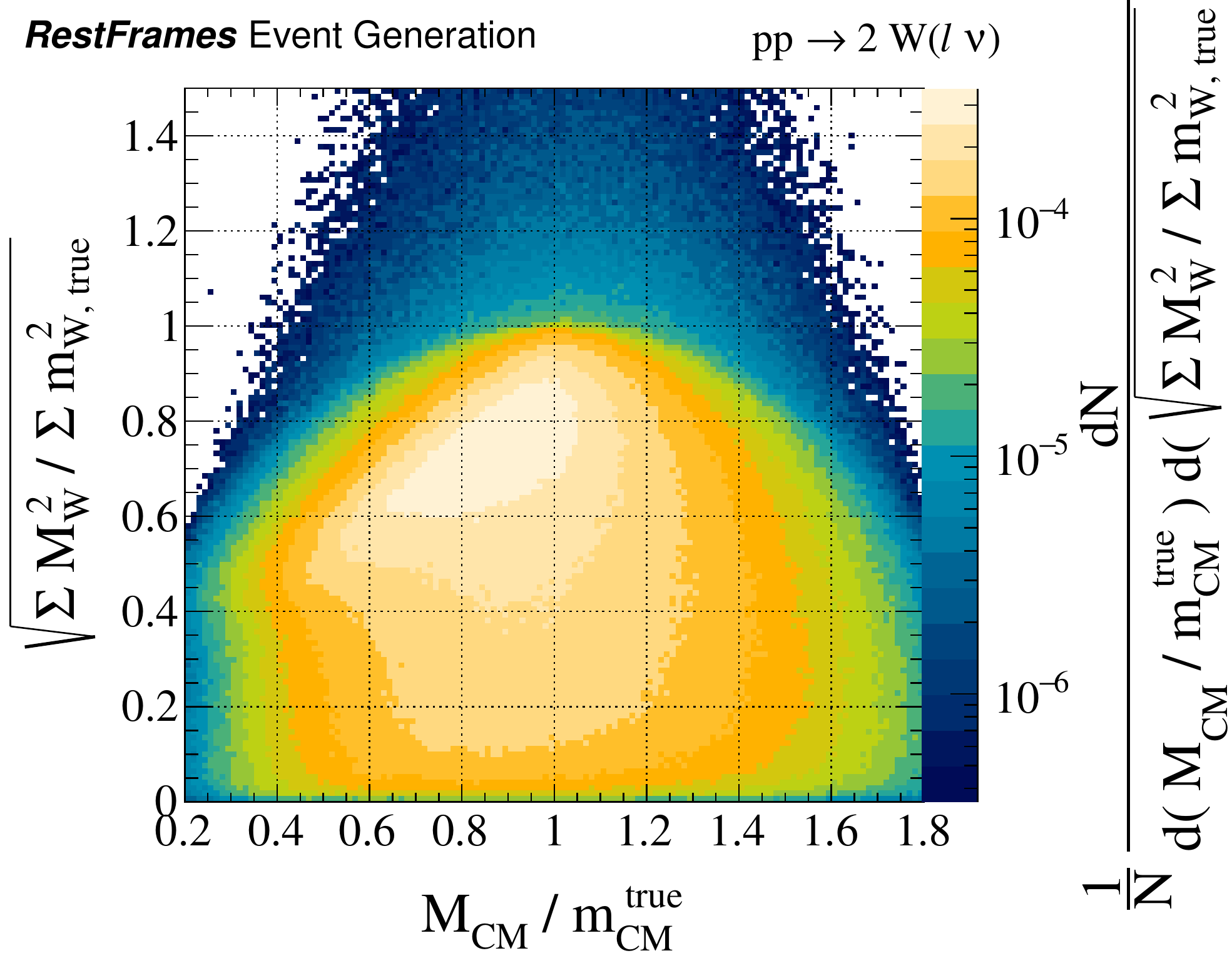}} \hspace{0.5cm}
\subfigure[]{\includegraphics[width=.28\textwidth]{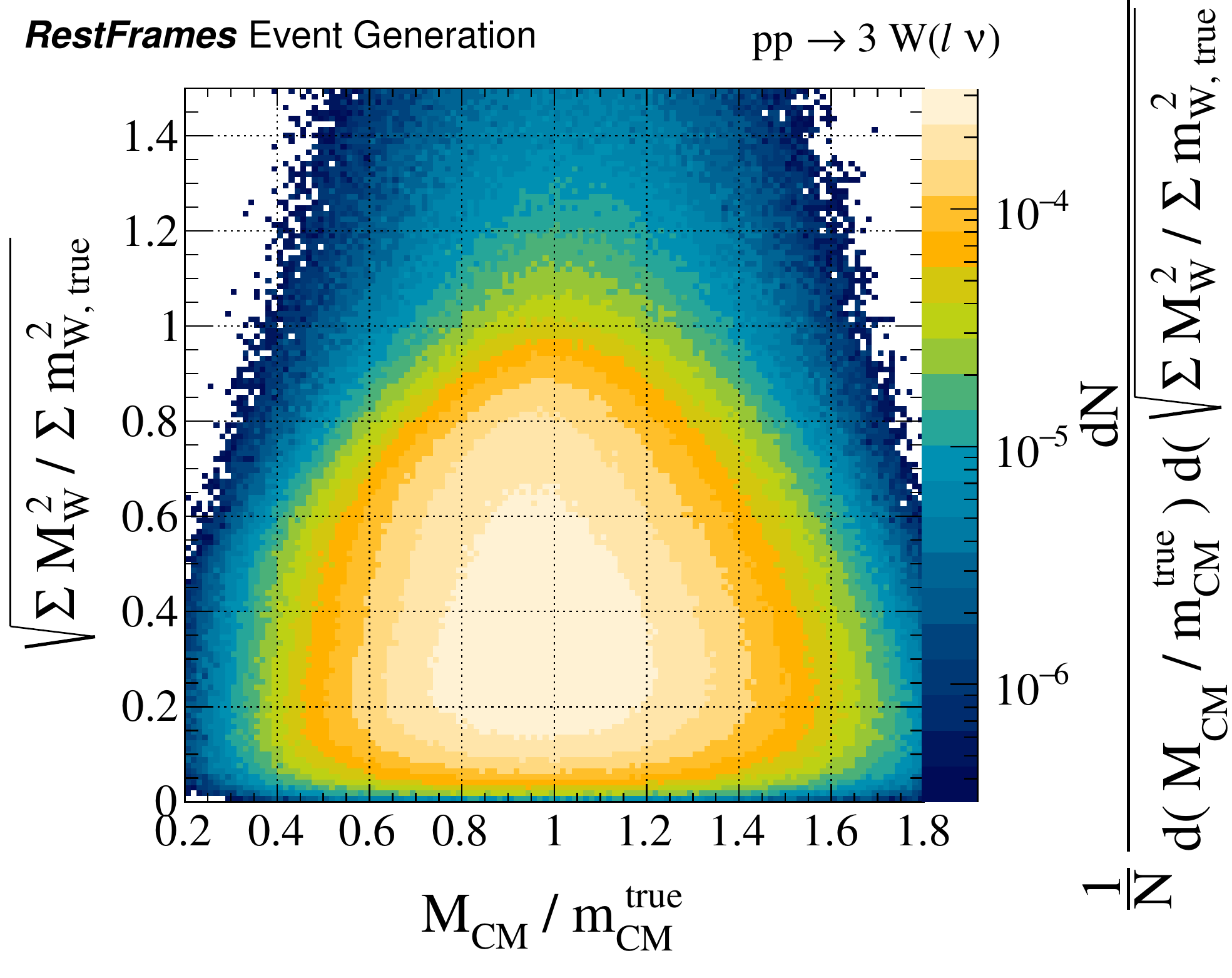}} \hspace{0.5cm}
\subfigure[]{\includegraphics[width=.28\textwidth]{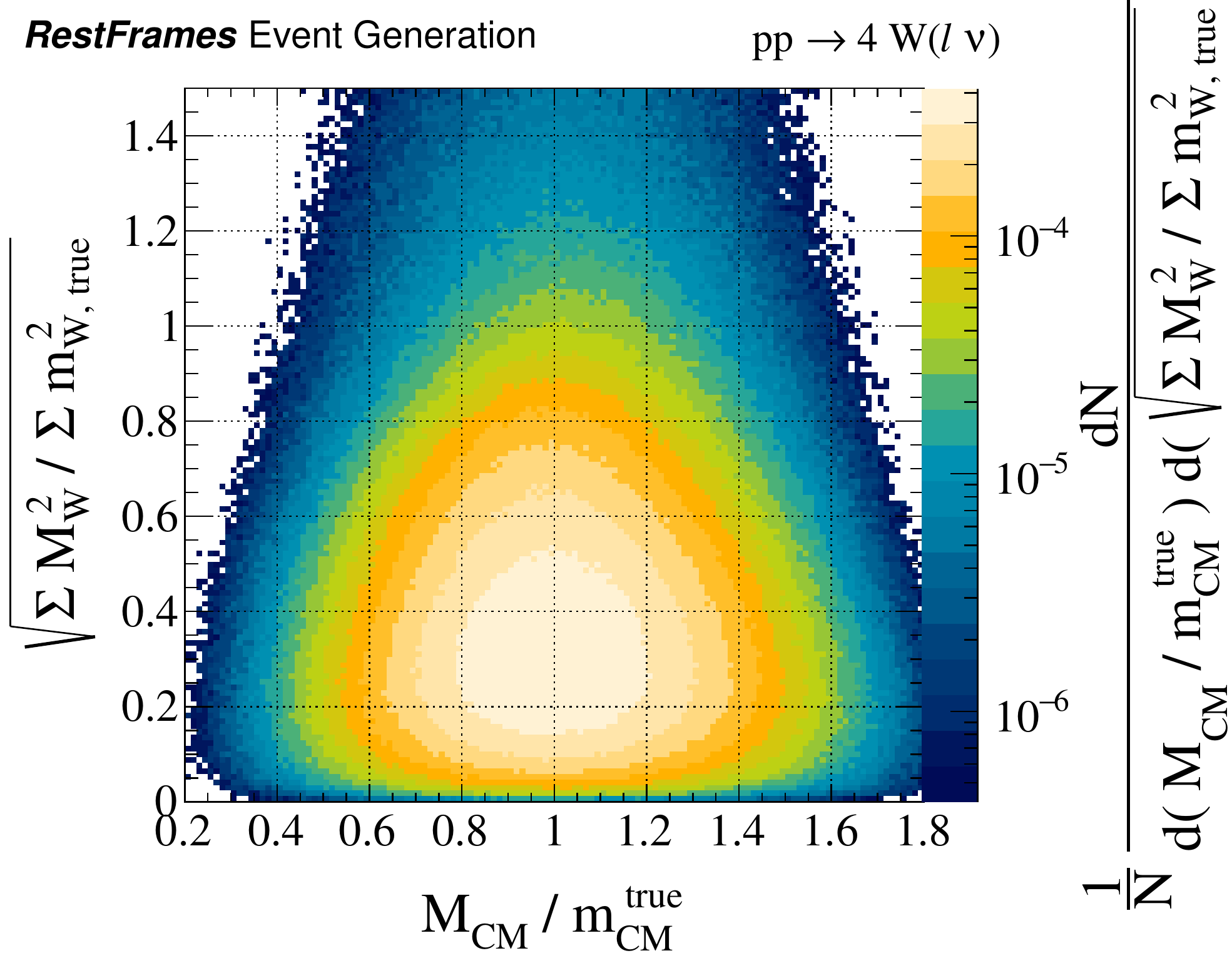}}
\vspace{-0.3cm}
\caption{\label{fig:example_N_Wlnu-mass2D} Distributions of the sum of reconstructed $W$ mass estimators squared, $\sqrt{ (\sum_i \Mass{\D{$W_i$}}{2})/(\sum_i \mass{\D{$W_i$}}{2}) }$, as a function of the estimated invariant mass of the total event, $\Mass{\D{cm}}{}$, for simulated non-resonant $N \times W(\ell\nu)$ events. Both observables are normalized by the true values of the quantities they are estimating. Distributions are shown for (a) $N = 2$, (b) $N = 3$, and (c) $N = 4$.}
\end{figure}

Despite the sparse amount of information available in each event, relative to the number of unknowns, the JR~\ref{jr:invminM2} still allows information about the masses of the individual $W$ bosons, and total event invariant mass, to be inferred.  The sum of $W$ mass estimators squared, which is the quantity we effectively minimized in our analysis of the event, is shown in Fig.~\ref{fig:example_N_Wlnu-mass2D}, as a function of \Mass{\D{cm}}{}, for $N=2,3,4$ $W$ boson events. Normalized by the true quantity, the distribution of this sum exhibits a kinematic edge at one, with resolution slightly degraded for increasing $N$. The total invariant mass of all the $W$ bosons, \Mass{\D{cm}}{}, is estimated with little bias and improving resolution for increasing $N$. This is a result of the guess for $\Mass{\I{I}}{}$ becoming increasingly accurate as the number of visible leptons grows. 
\onecolumngrid

\twocolumngrid

The lack of correlation between the $W$ boson mass and \Mass{\D{cm}}{} estimators recalls Fig.~\ref{fig:example_H_to_WlnuWlnu-MW_v_MH} for $H \rightarrow W^+W^-$ events, Fig.~\ref{fig:example_DiStop_to_hadtopXhadtopX-Ptop_v_MTT} for $\tilde{t}\tilde{t} \rightarrow t_{\rm had} \tilde{\chi}^{0}_{1} t_{\rm had} \tilde{\chi}^{0}_{1} $, and Fig.~\ref{fig:example_X2X2_to_ZllXHggX-PZ_v_MCM} for $\tilde{\chi}_2^0\tilde{\chi}_2^0 \rightarrow Z(\ell\ell) \tilde{\chi}^{0}_{1} h(\gamma\gamma) \tilde{\chi}^{0}_{1}$. The JR~\ref{jr:invminM2} has recovers sensitivity to the mass of the $W$'s, independently of  \Mass{\D{cm}}{}. 

This JR is an important part of the RJR library, as it allows for the analysis of final states with an arbitrary number of invisible particles, with the quality of extracted information dependent on how well they can be paired with visible partners. While it is encouraging that useful information can still be measured in these highly non-resonant cases, additional structure in events can yield much more information, as we see in the following example.


\subsection{$H\rightarrow hh \rightarrow W(\ell\nu)W^*(\ell\nu)W(\ell\nu)W^*(\ell\nu)$}
\label{subsec:Part4_exampleB}

\begin{figure}[htbp]
\centering 
\includegraphics[width=.35\textwidth]{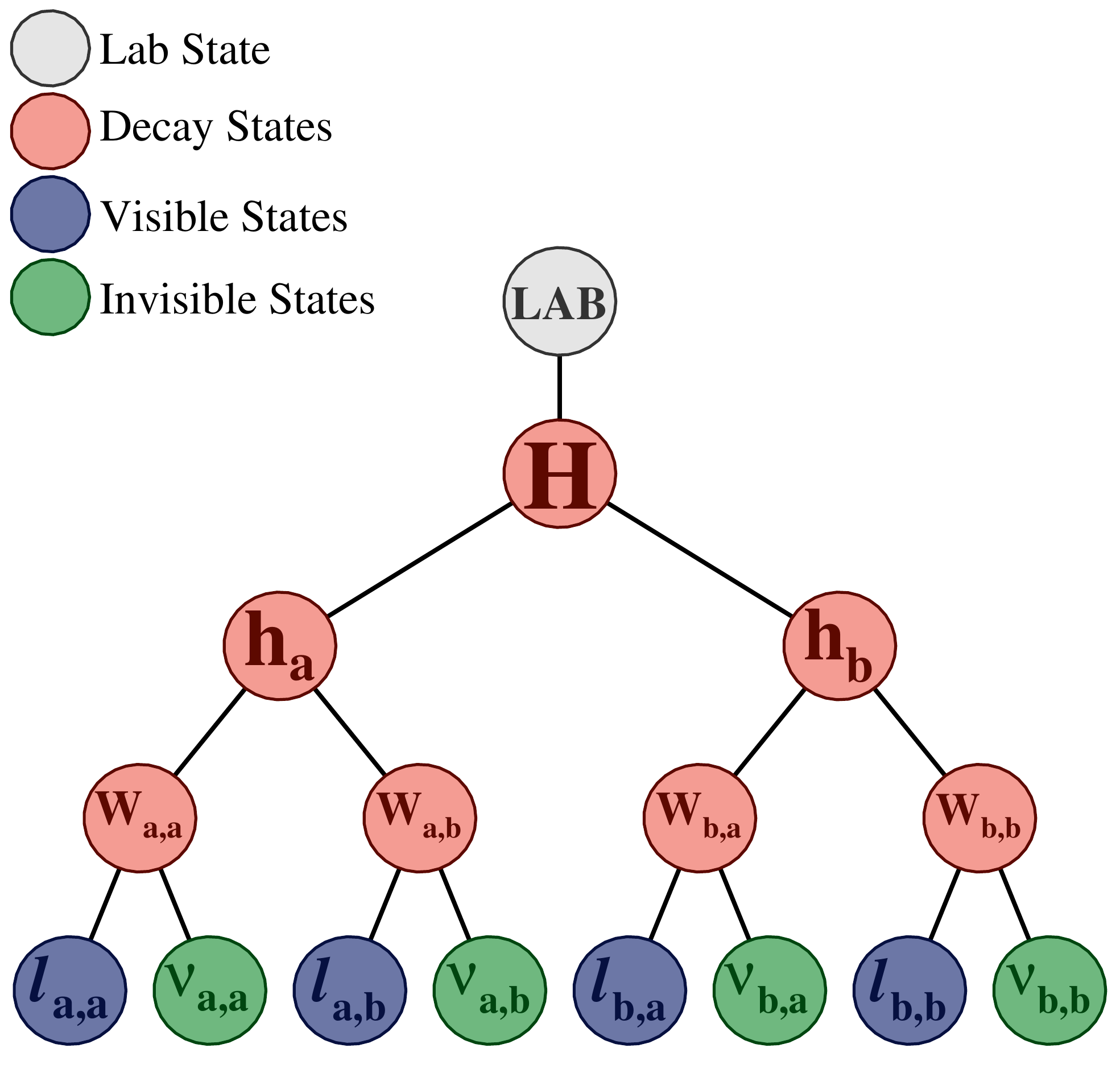}
\vspace{-0.3cm}
\caption{\label{fig:example_H_to_hh_to_4Wlnu-decayTree} A decay tree diagram for heavy, neutral Higgs production decaying to two lighter, neutral Higgs bosons. Each of the lighter Higgs's is assumed to be SM-like and have a mass of $\mass{\D{$h$}}{} = 125$~GeV, further decaying to $W(\ell\nu)W^*(\ell\nu)$. The heavy Higgs mass is chosen as $\mass{\D{$H$}}{} = 750$ GeV. The final state has four visible leptons, with total charge zero.}
\end{figure}

When analyzing events with many invisible particles in the final state, expected symmetries and relations between the intermediate particles possibly appearing in them can be used in the choice of JR's. In this example, we consider the production of a heavy, neutral Higgs boson at a hadron collider, which decays to two, SM-like, Higgs bosons. Each of these lighter Higgs bosons then decays to two $W$ bosons which, in turn, decay to a lepton and neutrino. This leads to four leptons in the final state, $\V{V} = \{ \V{$\ell_{a,a}$}, \V{$\ell_{a,b}$}, \V{$\ell_{b,a}$}, \V{$\ell_{b,b}$} \}$, and likely non-zero measured \met~ associated with four escaping neutrinos, $\I{I} = \{ \I{$\nu_{a,a}$}, \I{$\nu_{a,b}$}, \I{$\nu_{b,a}$}, \I{$\nu_{b,b}$} \}$. The decay tree describing this final state is shown in Fig.~\ref{fig:example_H_to_hh_to_4Wlnu-decayTree}.

As for the non-resonant $4W(\ell\nu)$ example discussed in Section~\ref{subsec:Part4_exampleA}, in order to reconstruct this event we must make choices for all the components of the neutrinos' momentum that we are unable to measure directly. Applying the constraints $\pthree{\I{I},T}{\lab} = \met$ and $\Mass{\I{$\nu_i$}}{} = 0$, there are still 10 under-constrained d.o.f. in each event. While the same JR~\ref{jr:invminM2} can be used to resolve many of these unknowns simultaneously, in this case the expected symmetry between masses and decays of each lighter Higgs boson motivates a different strategy.

In the RJR approach we attempt to factorize the information we hope to extract about each decay step by considering it independently, choosing JR's corresponding to its specific details, and desired constraints. For the decay of the heavy, neutral Higgs boson, $\D{$H$}$, there are several pieces of information we must choose. Firstly, we assume that the total charge of the four leptons in the event is zero, such that there are two lepton/anti-lepton pairs, although not necessarily the same flavor. Defining the two opposite-sign pairs of leptons as $\V{V$_a$} =  \{ \V{$\ell_{a,a}$}, \V{$\ell_{a,b}$} \}$ and $\V{V$_b$} =  \{ \V{$\ell_{b,a}$}, \V{$\ell_{b,b}$} \}$, we choose the assignment which minimizes the quantity $\mass{\V{V$_a$}}{2} + \mass{\V{V$_b$}}{2}$, according to JR~\ref{jr:minM2}. 

When choosing the unknowns describing how the four neutrino system kinematically splits into two pairs in this decay, the expected similarity between the two SM-like Higgs masses can be exploited in the context of a contra-boost invariant JR~\ref{jr:contra}, imposing the constraint $\Mass{\D{$h_a$}}{} = \Mass{\D{$h_b$}}{}$. The application of this JR only describes the momentum of the two neutrino pairs in the approximation of the \D{$H$} rest frame, and uses only information about the total momentum of the pairs \V{V$_a$} and \V{V$_b$}, without resolving that they are each made up of two separate particles. This is a crucial distinction, in that it ensures that the estimators describing each decay are maximally uncorrelated, as they are largely based on different information. 

As explained in Section~\ref{subsec:Part2_exampleC}, the use of the contra-boost invariant JR~\ref{jr:contra} imposes specific constraints on the estimated masses of the individual invisible particles after the pair is split, in this case the two systems of neutrino pairs, $\I{I$_a$} =  \{ \I{$\nu_{a,a}$}, \I{$\nu_{a,b}$} \}$ and $\I{I$_b$} =  \{ \I{$\nu_{b,a}$}, \I{$\nu_{b,b}$} \}$. As the JR forces the two Higgs masses to be equal, any difference between the masses of the two lepton pairs will result in a difference between \Mass{\I{I$_a$}}{} and \Mass{\I{I$_b$}}{}. Since the lepton pairs are coming from the same decays as the neutrinos, the lepton pair masses are good indicators of the corresponding neutrino masses. To prevent any biases in the kinematics of the neutrino system from being introduced by the JR, we choose $\Mass{\I{I$_a$}}{} = 2 | \pthree{\V{$\ell_{b,a/b}$}}{\V{V$_b$}} | \approx \Mass{\V{V$_b$}}{}$ and $\Mass{\I{I$_b$}}{} = 2 | \pthree{\V{$\ell_{a,a/b}$}}{\V{V$_a$}} |$. This corresponds to a choice $\Mass{\I{I}}{} = \mass{\V{V}}{}$.

In the subsequent decays of the Higgs bosons, \D{$h_a$} and \D{$h_b$}, the same decay topology is encountered, with visible and invisible particle pairs splitting in two. As $\mass{\D{$h$}}{} \leq 2\mass{\D{$W$}}{}$, one of the \D{$W$} bosons is generally produced off-shell in each decay, meaning that constraining the two $W$ masses in these decays to be equal would be inaccurate. Instead, a more generic jigsaw minimizing the sum of \D{$W$} masses in each decay, JR~\ref{jr:invminM2}, is used, allowing the individual neutrino mass estimators to be fixed to zero. 

With these choices of JR's, the strategy for analyzing $H\rightarrow hh \rightarrow W(\ell\nu)W^*(\ell\nu)W(\ell\nu)W^*(\ell\nu)$ events at a hadron collider can be summarized as:
\begin{enumerate}[noitemsep]
\item Apply the combinatoric JR~\ref{jr:minM2}, choosing the lepton pairing that minimizes $\mass{\V{V$_a$}}{2}+\mass{\V{V$_b$}}{2}$.
\item Apply the invisible mass JR~\ref{jr:mass}, \\ choosing $\Mass{\I{I}}{} = \mass{\V{V}}{}$.
\item Apply the invisible rapidity JR~\ref{jr:rapidity}, choosing \\ \pone{\I{I},z}{\lab} using the collection of leptons, \V{V}.
\item Apply the contra-boost invariant JR~\ref{jr:contra}, using \\ the constraint $\Mass{\D{$h_a$}}{} = \Mass{\D{$h_b$}}{}$.
\item Apply the JR~\ref{jr:invminM2} for each \D{$h_{a/b}$} decay, \\ minimizing $\Mass{\D{$W_{a/b,a}$}}{2}+\Mass{\D{$W_{a/b,b}$}}{2}$.
\end{enumerate}

In the context of searching for evidence of this phenomenon, the masses and decay angles of the three Higgs bosons are of primary interest. Even though the two SM-like Higgs masses have been constrained to be the same by the contra-boost invariant JR, we can recover independent information about them by using alternative quantities as estimators. The Higgs mass equality required that we set the di-neutrino pair masses equal to that of the di-leptons of the opposite half of the event, a guess that was made for the convenience of the JR. To make the effective mass estimators more independent of this choice, we instead adopt the convention: $M_{h_{a/b}} = 2 \E{\V{V$_{a/b}$}}{\D{$h_{a/b}$}}$, where this equivalence is used only for purposes of data-analysis, and does not change the reconstruction of the event. 

\begin{figure}[!htbp]
\centering 
\subfigure[]{\includegraphics[width=.28\textwidth]{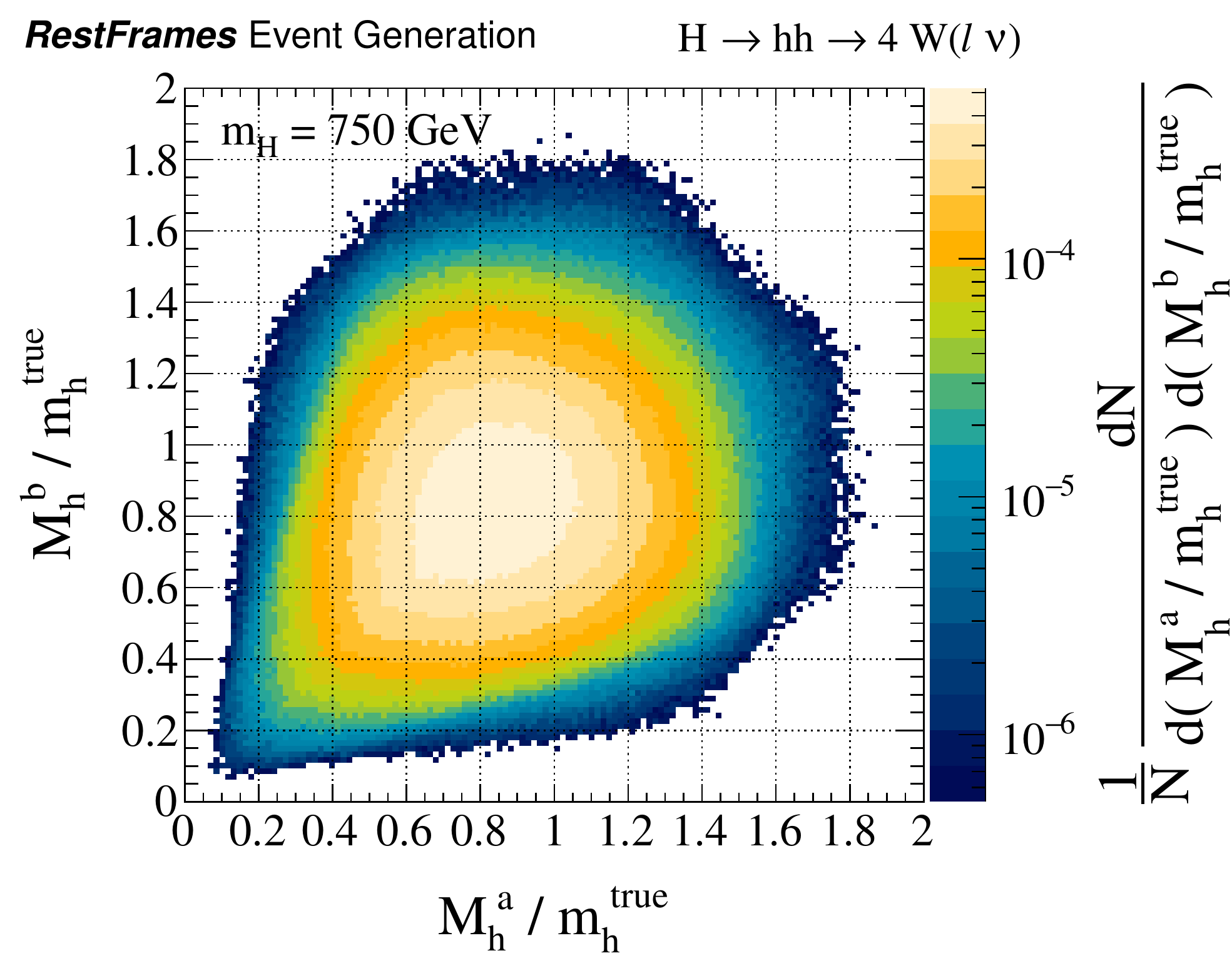}}\hspace{0.5cm}
\subfigure[]{\includegraphics[width=.28\textwidth]{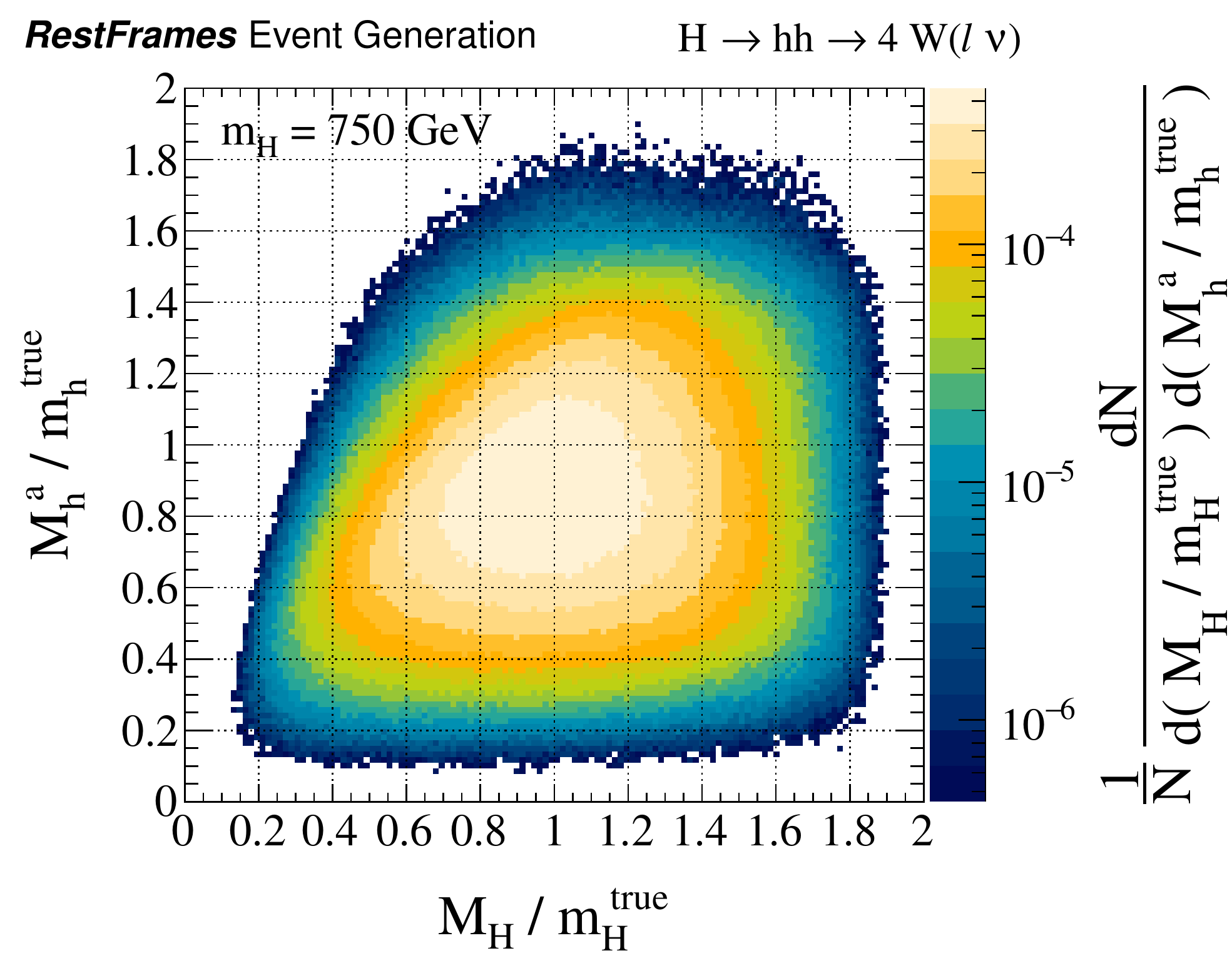}}\hspace{0.5cm}
\subfigure[]{\includegraphics[width=.28\textwidth]{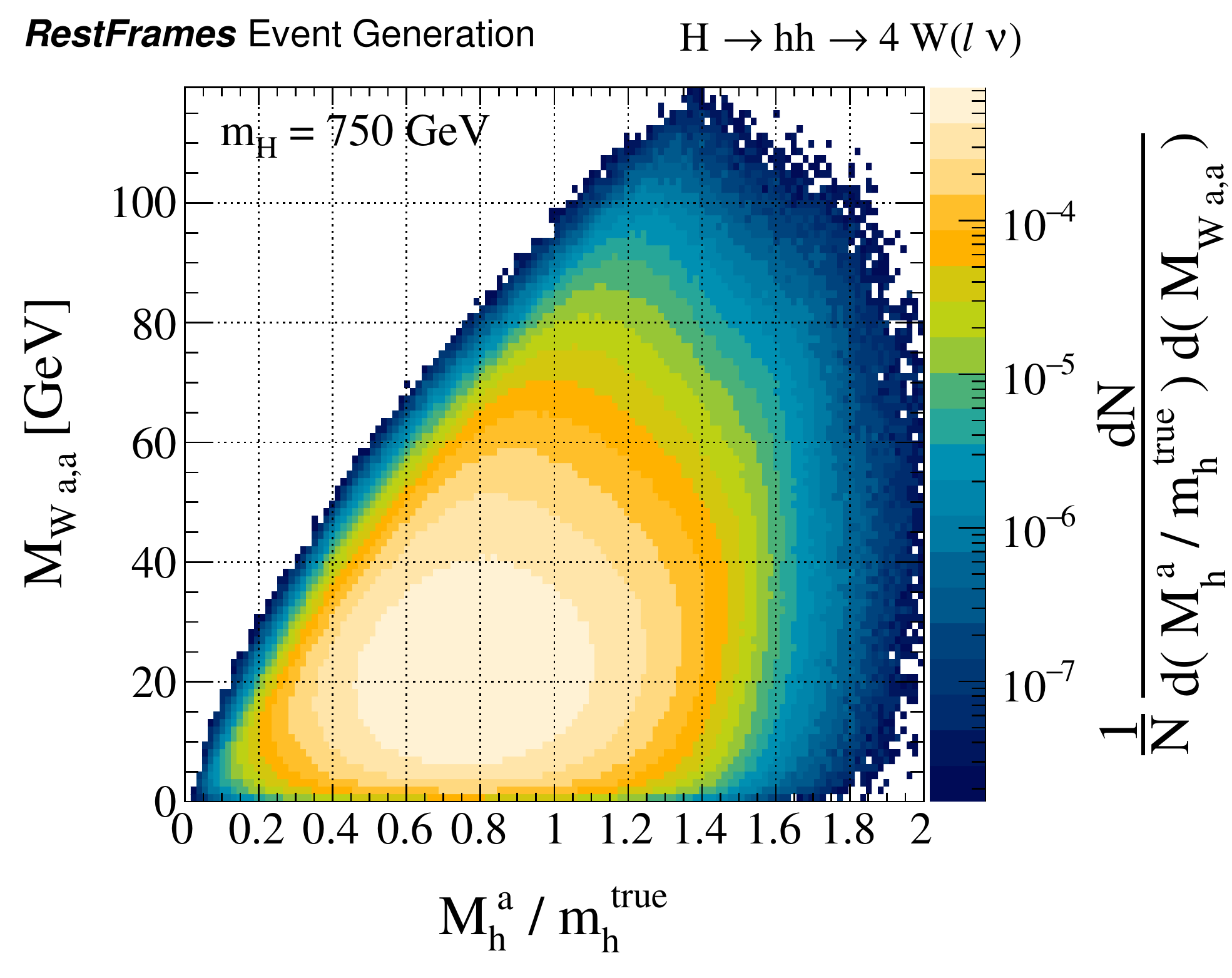}}
\subfigure[]{\includegraphics[width=.28\textwidth]{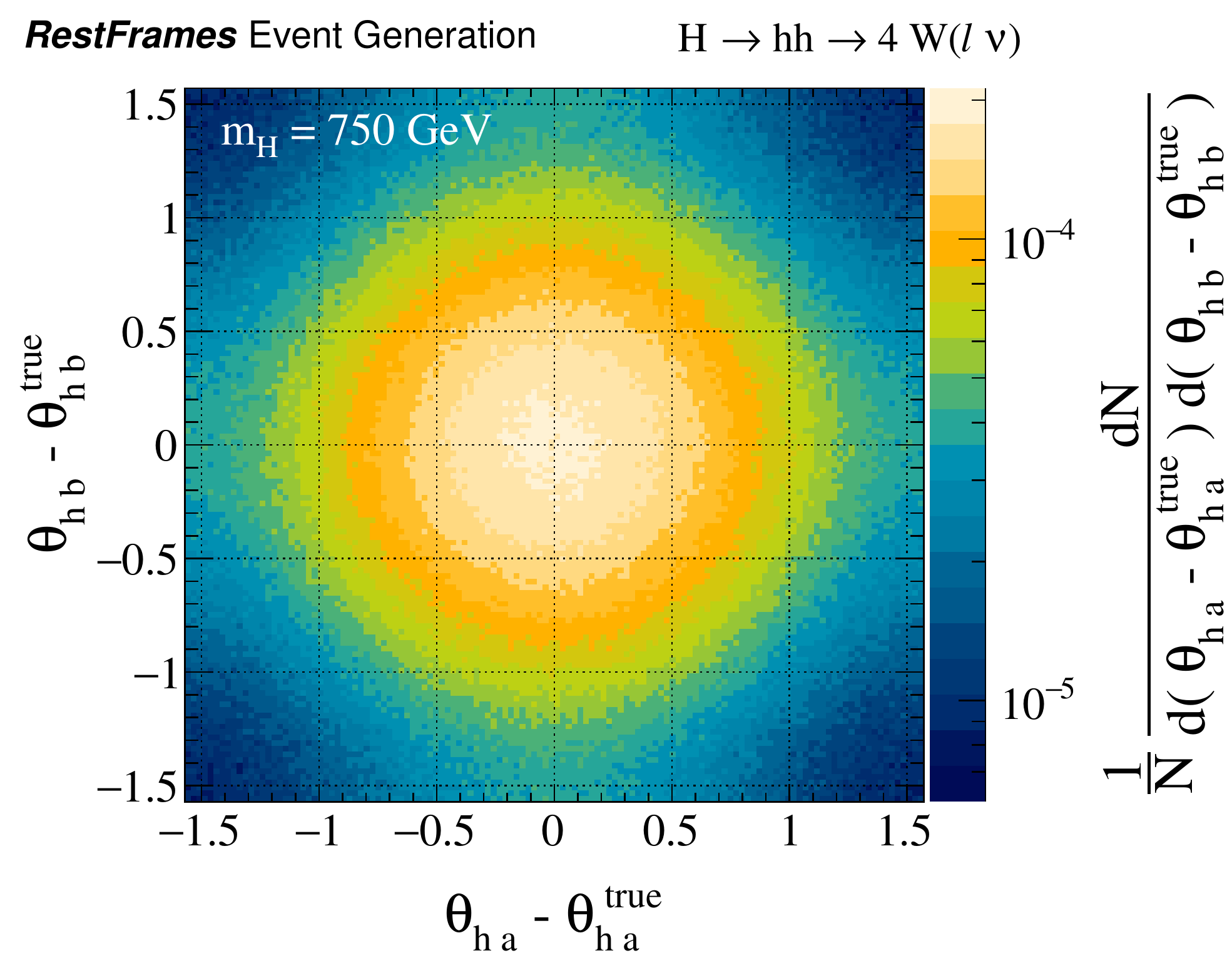}}\hspace{0.5cm}
\subfigure[]{\includegraphics[width=.28\textwidth]{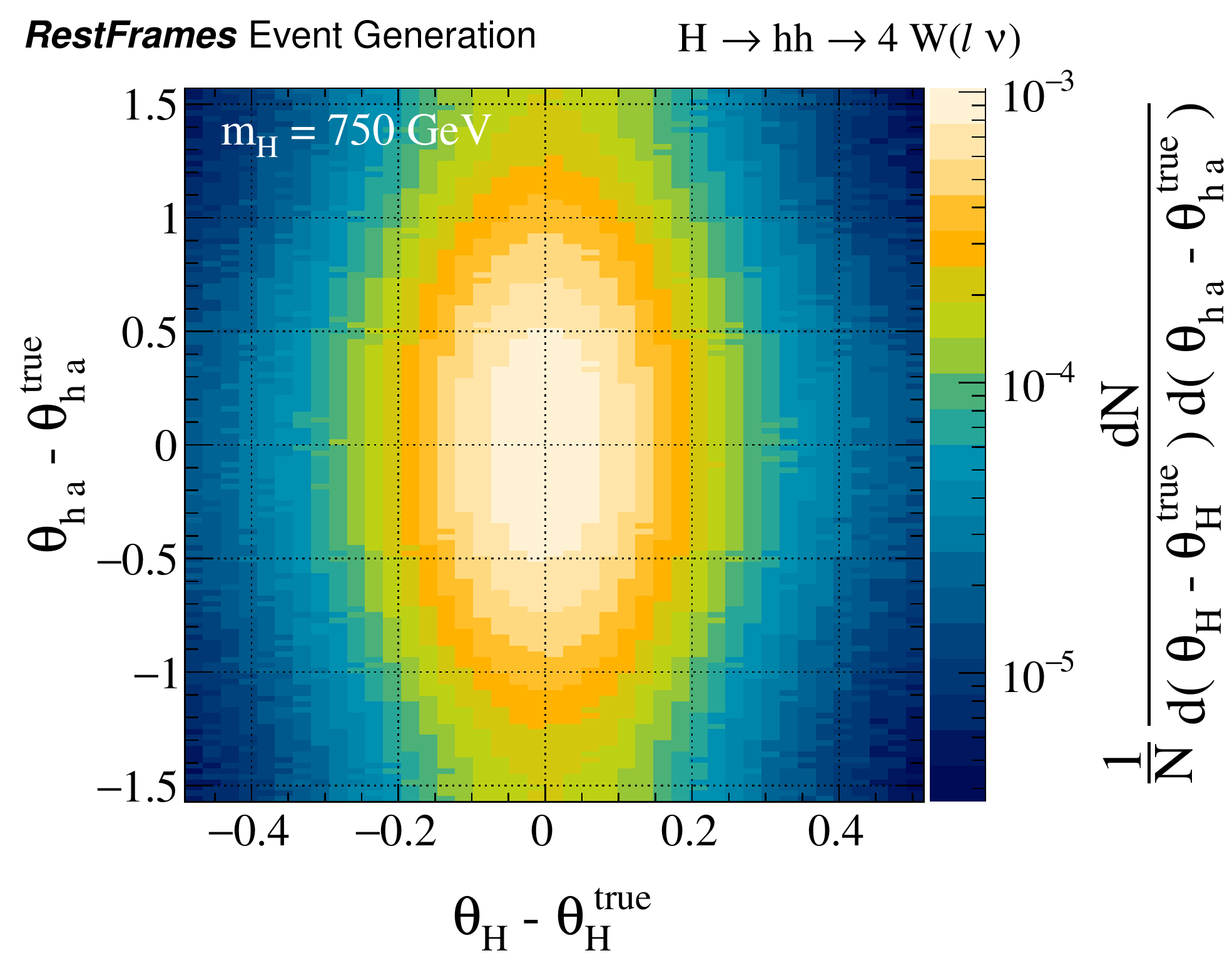}}\hspace{0.5cm}
\subfigure[]{\includegraphics[width=.28\textwidth]{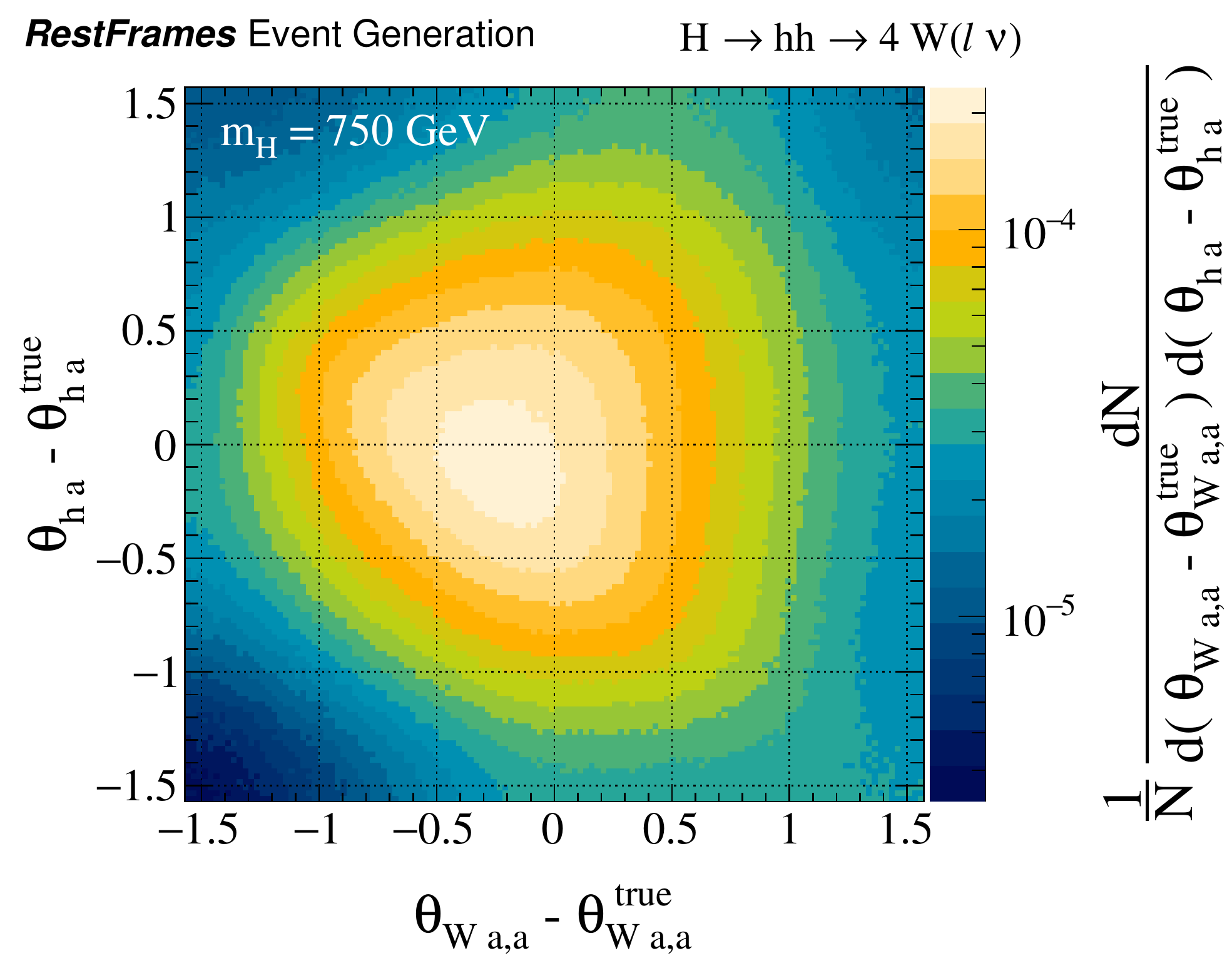}}
\vspace{-0.3cm}
\caption{\label{fig:example_H_to_hh_to_4Wlnu-M2D} Distributions of (a) one SM Higgs mass estimator, as a function of the other, and (b) one SM Higgs mass estimator, as a function of the heavy Higgs mass, (c) the on-shell $W$ mass estimator, $\max(\Mass{\D{$W_{a,a}$}}{},\Mass{\D{$W_{a,b}$}}{})$, as a function of its associated Higgs mass, (d) the reconstructed decay angle of one SM Higgs boson, as a function of the other, and (e) one SM Higgs boson decay angle, as a function of the reconstructed heavy Higgs decay angle, and (f) the on-shell $W$ decay angle, as a function of the corresponding SM Higgs boson decay angle, for simulated $H\rightarrow hh \rightarrow 4W(\ell\nu)$ events. Each observable is normalized appropriately by the true value of the quantity it is estimating, with angles expressed in units radian. }
\end{figure}

The distributions of the mass estimators and reconstructed decay angles of the three Higgs bosons are shown in Fig.~\ref{fig:example_H_to_hh_to_4Wlnu-M2D} for simulated events with $\mass{\D{$H$}}{} = 750$ GeV. The SM-like Higgs masses are almost completely uncorrelated, with both masses showing a slight downward bias relative to the true value and a resolution of $\sim20\%$.  Here, the choice of estimators is contradictory to the reconstructed interpretation of the event resulting from the application of JR's, a strategy which is discussed in more detail in the following example.

The SM-like Higgs mass observables are also independent of the heavy Higgs mass, as seen in Fig.~\ref{fig:example_H_to_hh_to_4Wlnu-M2D}(b), demonstrating that all three masses can be extracted separately, with similar resolution. Similarly, each of the Higgs' reconstructed decay angles can be estimated with excellent precision, independently of each other, with not even small correlations observed in their distributions, as seen in Fig.~\ref{fig:example_H_to_hh_to_4Wlnu-M2D}(d,e). 

There is even more information contained in the approximations of the \D{$W$} rest frames, including masses and decay angles, as can be seen in the distributions of these quantities for the on-shell $W$ in Fig.~\ref{fig:example_H_to_hh_to_4Wlnu-M2D}(c,f). As was the case for the SM-like Higgs decays, these estimators are almost completely uncorrelated with the previous ones, with the distribution of the $W$ mass estimator exhibiting a kinematic edge at the true value, almost identically to $W(\ell\nu)$ pair production in Fig.~\ref{fig:example_N_Wlnu-mass2D} from the previous example. The observables from each decay step behave as if that was the entire decay tree, with little sensitivity to the details of other decays in the event. As additional decays are added, the resolution of some observables degrade, but the accuracy of estimators further up the decay tree improve, as integrating over more degrees of freedom in the event further smooths the kinematics and the applied approximations become better. 

While the ability to accurately measure these quantities is important for studying this type of process, it is equally useful when searching for evidence of it, as SM backgrounds in this final state would have to mimic the targeted process independently in many observable dimensions. The recursive application of JR's at each step in the decay, effectively analyzing each reference frame independently of the others, can be used to derive an appropriate basis of observables for either purpose. 
\onecolumngrid

\twocolumngrid



\subsection{$\tilde{g}\tilde{g}\rightarrow bb\tilde{\chi}^{0}_{1}bb\tilde{\chi}^{0}_{1}$ at a hadron collider}
\label{subsec:Part4_exampleC}

All the previous examples have primarily focused on JR's for resolving the kinematics of invisible particles, with a variety of choices for nearly every decay possibility. In order to be able to analyze every imaginable decay topology with the RJR approach, additional treatments for combinatoric ambiguities, resulting from indistinguishable reconstructed particles appearing in events, are introduced. 

Like the invisible JR~\ref{jr:invminM2} which minimizes the masses squared of potentially many composite particles simultaneously, the combinatoric analogue JR~\ref{jr:minM2} can be generalized to an arbitrarily large number of particles and partitions. But simultaneously choosing many unknowns in a single decay step, rather than factorizing the unknowns into several steps, results in a degradation in resolution of kinematic estimators, as seen in the comparison of analysis strategies for $4W(\ell\nu)$ final states between Sections~\ref{subsec:Part4_exampleA} and~\ref{subsec:Part4_exampleB}. Just as for invisible particle JR's, using combinatoric jigsaws in recursive steps can help resolve intermediate structure in decays.

\begin{figure}[!htbp]
\centering 
\subfigure[]{\includegraphics[width=.238\textwidth]{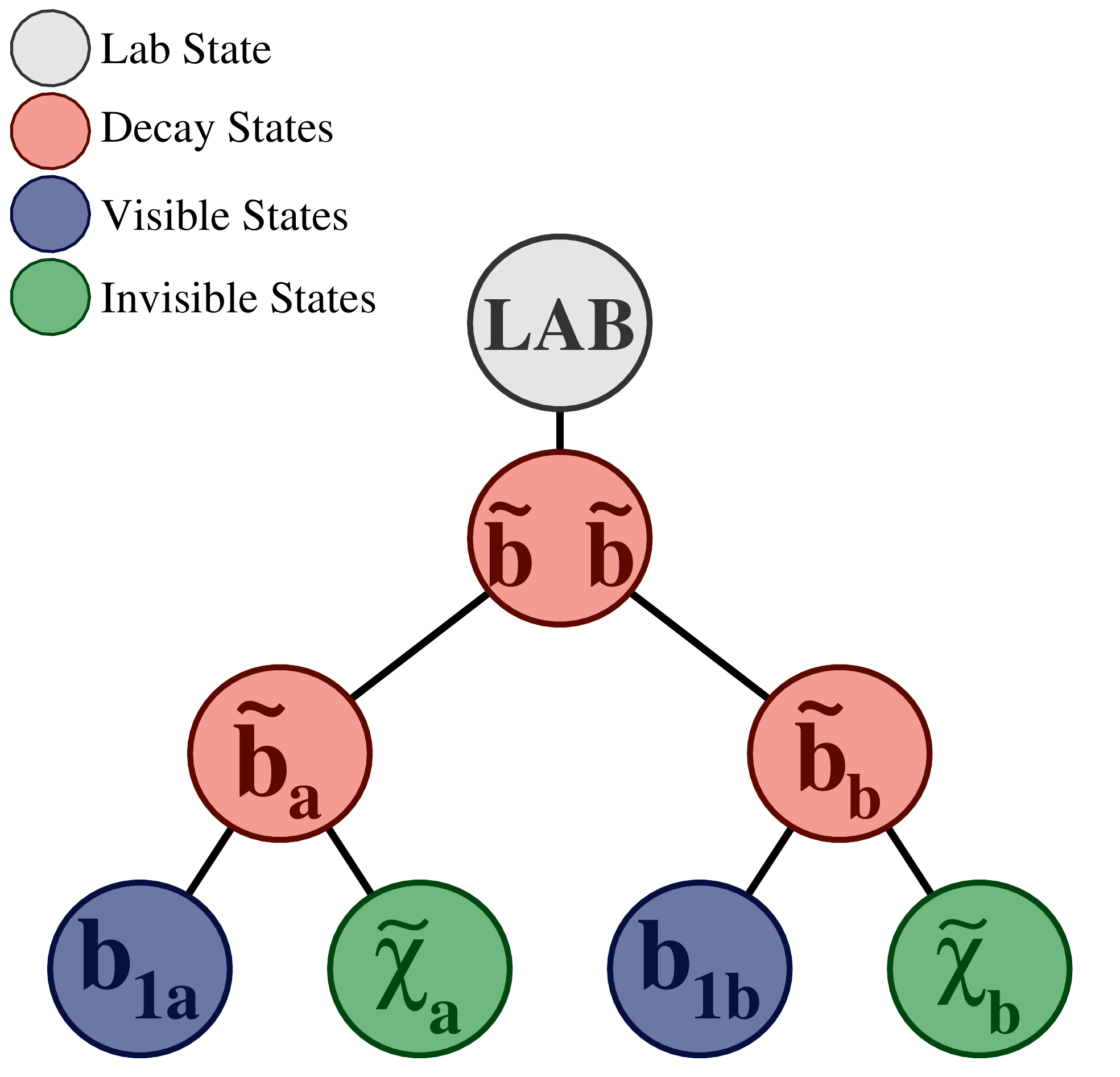}}
\subfigure[]{\includegraphics[width=.238\textwidth]{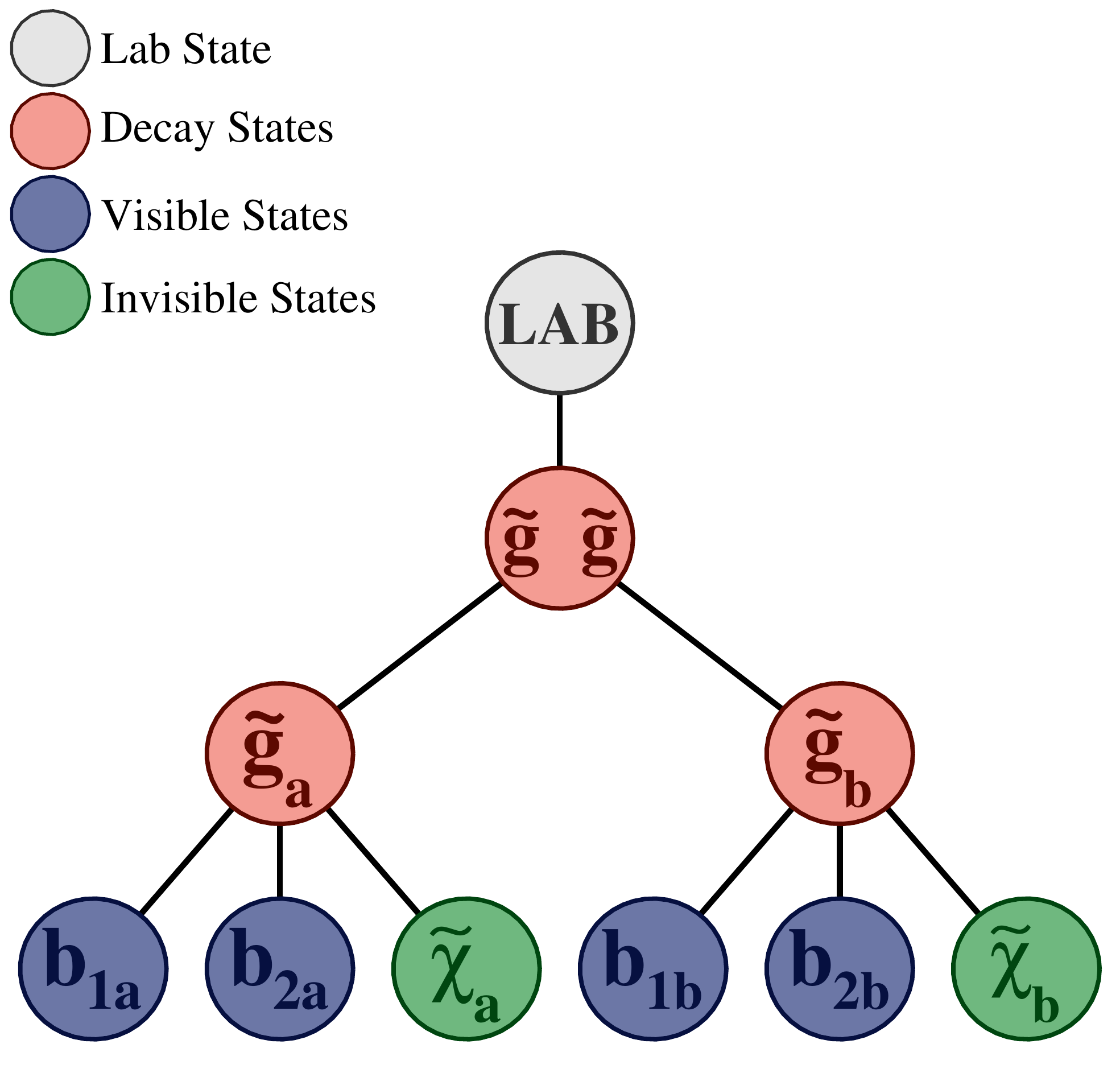}}
\subfigure[]{\includegraphics[width=.238\textwidth]{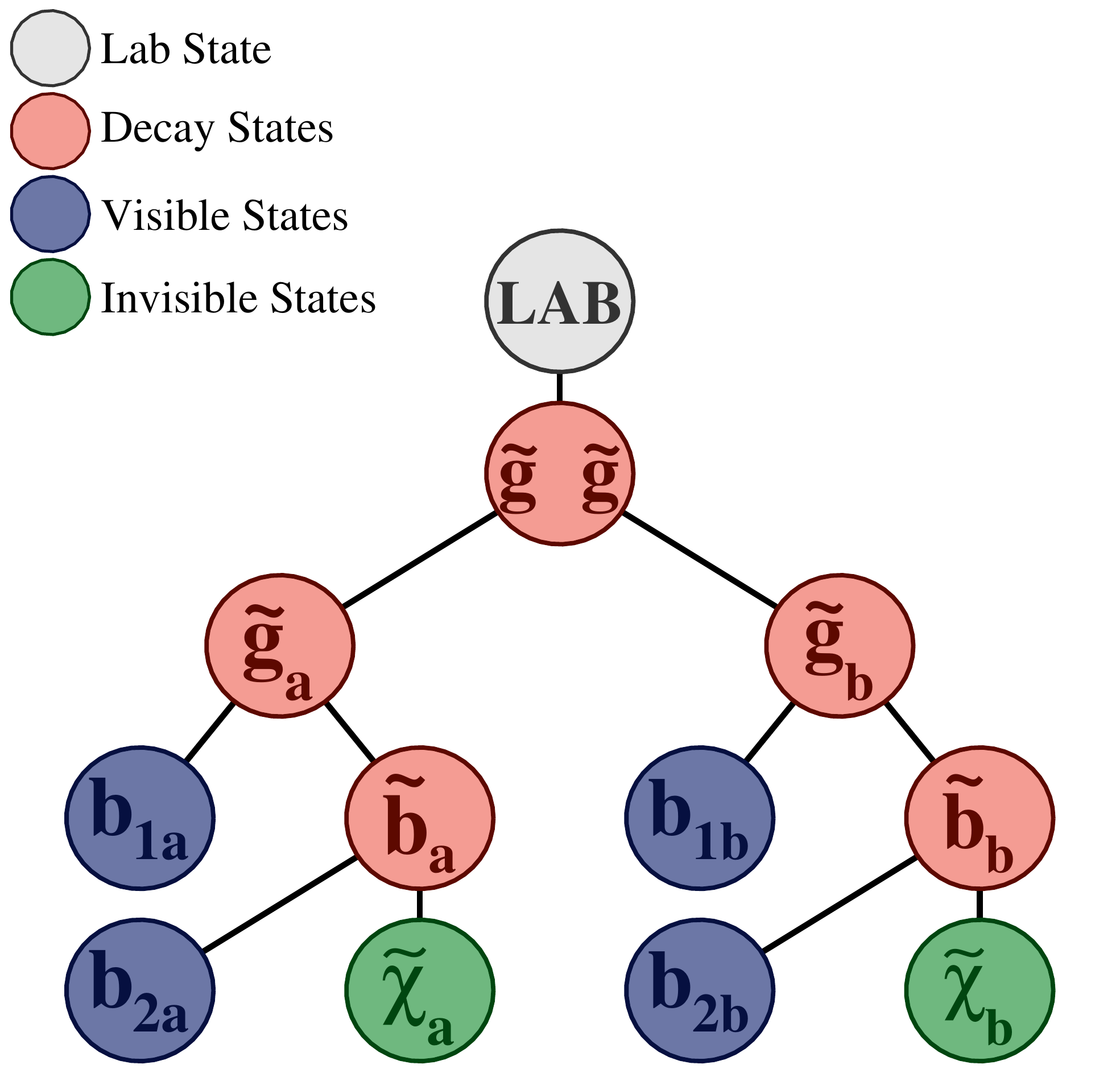}}
\subfigure[]{\includegraphics[width=.238\textwidth]{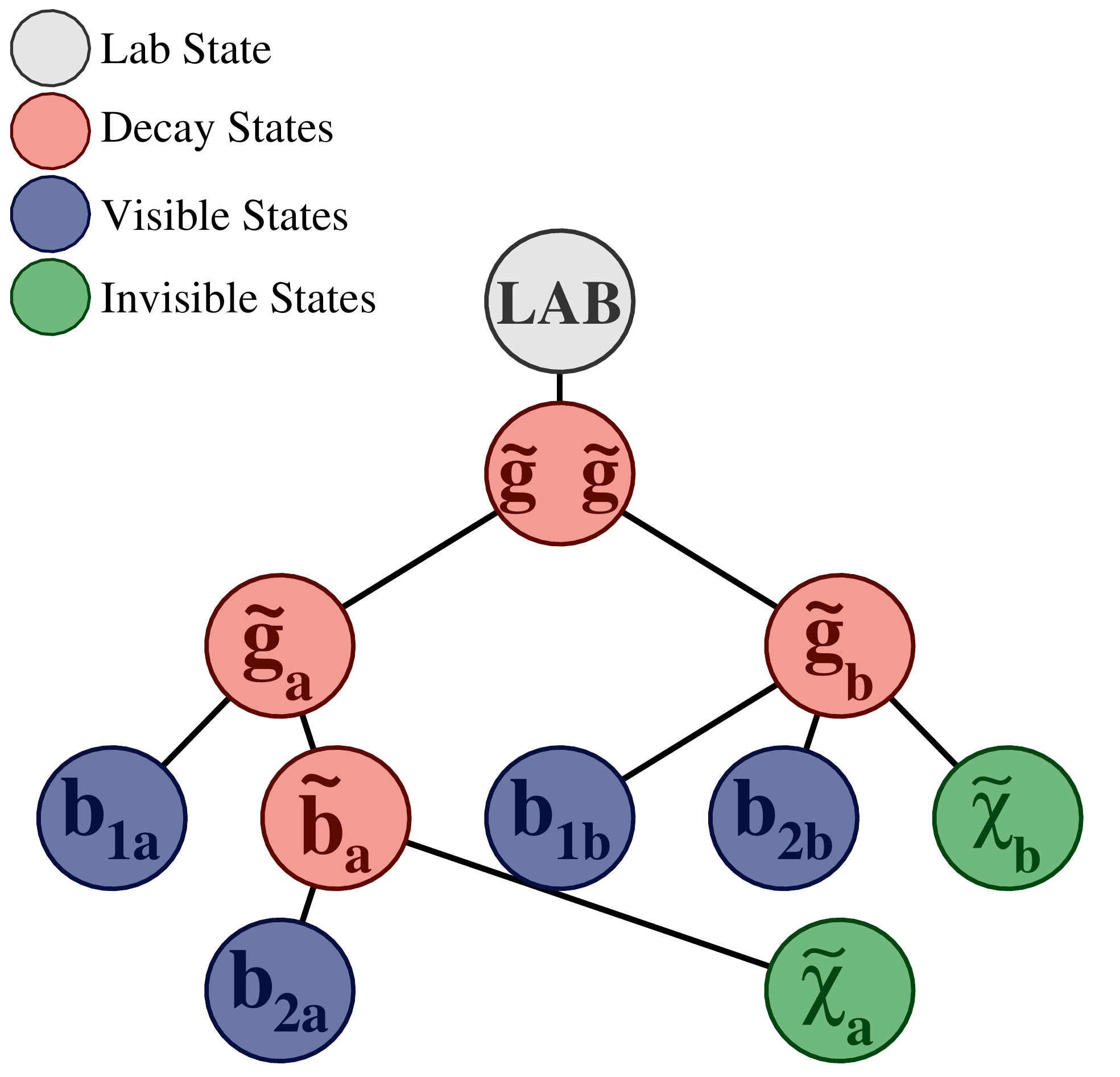}}
\vspace{-0.3cm}
\caption{\label{fig:example_DiGluino_to_bbXbbX-decay} Decay tree diagrams for the pair production of strongly interacting sparticles decaying to final states with $b$-quarks and neutralinos. (a) Two sbottom quarks are produced, each decaying to a $b$-quark and neutralino. (b) Pair produced gluinos each undergo a three-body decay to two $b$-quarks and a neutralino. (c) Two gluinos are produced, each decaying to a $b$-quark and a sbottom quark which, in turn, decays to a $b$-quark and a neutralino. (d) Pair produced gluinos each decay in a different way, one corresponding to the decays in (b), the other the decays in (c). In each of these four scenarios, the mass of the initially produced parent sparticles is 1 TeV, while $\mass{\I{$\tilde{\chi}^{0}_{1}$}}{} = 100$~GeV. When appearing in the decays of gluinos, sbottoms are chosen to have a mass of $\mass{\D{$\tilde{b}$}}{} = 900$~GeV in this example.}
\end{figure}

To demonstrate this idea, we consider the example of gluino and sbottom quark production at a hadron collider, with decays to $b$-quarks and neutralinos. While different $b$-quarks may appear in different places in a decay tree, the reconstructed $b$-tagged jets are indistinguishable, with no direct indication which one is which. We consider four different combinations of gluino and sbottom production and decay, with the processes summarized in Fig.~\ref{fig:example_DiGluino_to_bbXbbX-decay}. The sbottom quarks in these events each decay to a $b$-quark and a neutralino, while two different gluino decays are considered. When the gluino is heavier than the sbottom, it can decay $\tilde{g} \rightarrow b\tilde{b}$, resulting in two $b$-quarks and a neutralino after the sbottom quarks decay. Alternatively, if the squarks are much heavier than the gluino, it can undergo a three-body decay through a virtual sbottom quark to the same final state. The production and decays illustrated in Fig.~\ref{fig:example_DiGluino_to_bbXbbX-decay}(d) contain two gluinos, each decaying in a different way. While kinematically disfavored, this process is included to demonstrate the independent sensitivity of the reconstruction scheme to the two separate decays.

\begin{figure}[htbp]
\centering 
\includegraphics[width=.35\textwidth]{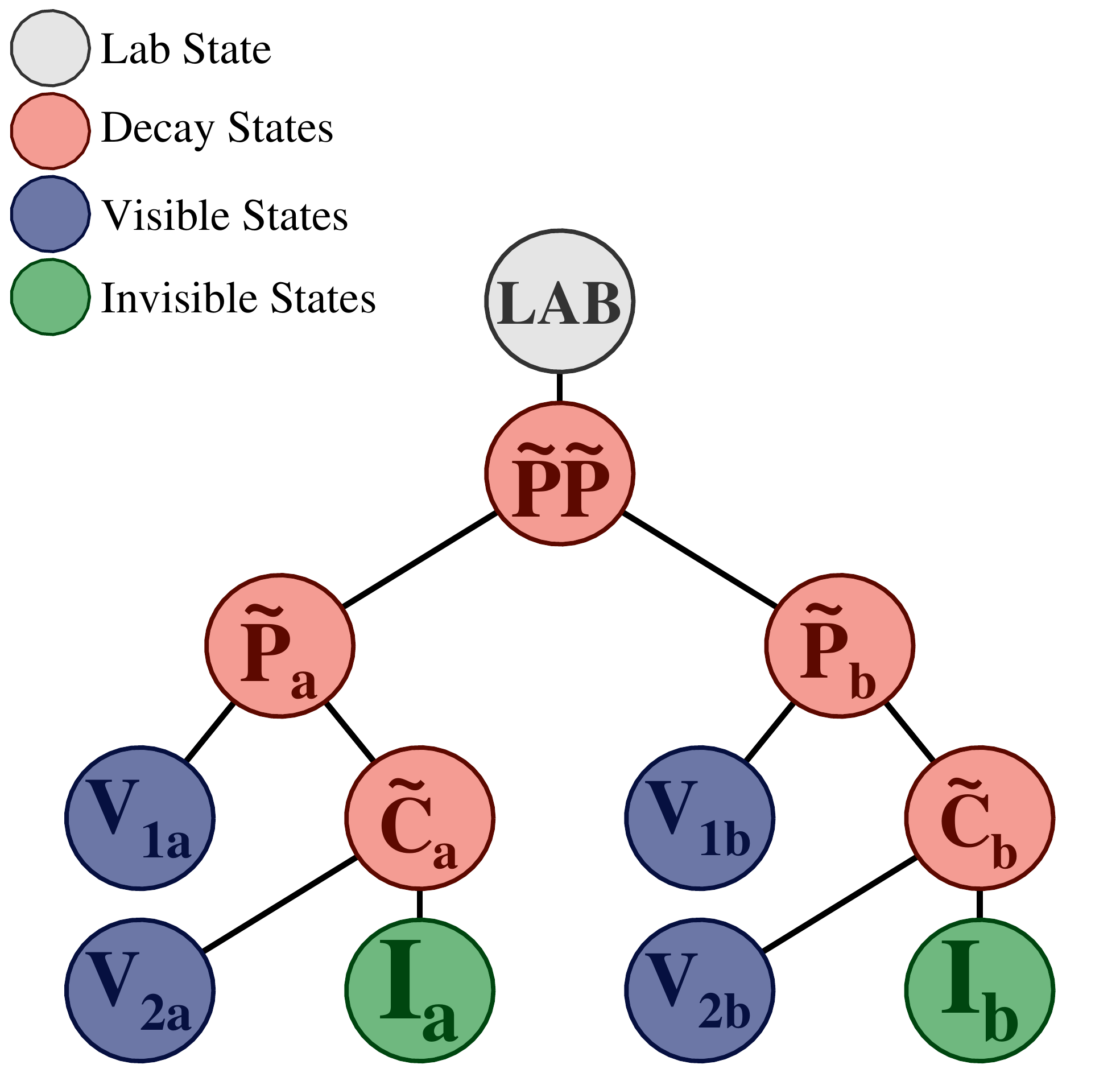}
\vspace{-0.3cm}
\caption{\label{fig:example_DiGluino_to_bbXbbX-decay-ana} Decay tree for analyzing strong sparticle pair production, with the decays described in Fig.~\ref{fig:example_DiGluino_to_bbXbbX-decay}. The intermediate decay states, \D{$\tilde{P}_i$} and \D{$\tilde{C}_i$}, represent the sparticles that may appear in the event, while the visible states, \V{V$_{ij}$}, represent a set of $b$-tagged jets. }
\end{figure}

All of the processes appearing in Fig.~\ref{fig:example_DiGluino_to_bbXbbX-decay} can be analyzed with a single decay tree, shown in Fig.~\ref{fig:example_DiGluino_to_bbXbbX-decay-ana}. The objects \D{$\tilde{P}_{a/b}$} represent the initially produced sparticles, either sbottom quarks or gluinos, while the \D{$\tilde{C}_{a/b}$} are any additional sparticles which might appear in the decays. Neutralinos are represented by the invisible states $\I{I$_{a/b}$}$, with the reconstructed $b$-tagged jets corresponding to the visible states \V{V$_{ij}$}. While the largest number of $b$-tagged jets in the final state is four in the processes explicitly considered in this example, the reconstruction approach adopted here allows for an arbitrarily high number. If, instead, the sbottoms in this example were squarks associated with the light quarks, the visible jets in the final state would be initiated by quarks and gluons, and indistinguishable from any other jets from the underlying event or mis-identified pile-up interactions. To account for larger multiplicities of identical visible particles in the final state, each \V{V$_{ij}$} is interpreted as a set of $b$-tagged jets that can contain a variable number, subject to defined constraints. We require that each \V{V$_{2a/b}$} contain at least one element in each event, while \V{V$_{1a/b}$} are permitted to have none.

\onecolumngrid

\twocolumngrid

Defining $\V{V} = \{ \V{$b_i$} \}$ to be the set of all $b$-tagged jets reconstructed in the event, the combinatoric unknowns are those associated with partitioning this set into these four subsets. Matching the decay tree in Fig.~\ref{fig:example_DiGluino_to_bbXbbX-decay-ana}, the partitioning is done in two steps with separate JR's, the first splitting \V{V} into two sets, $\V{V$_a$} = \{ \V{V$_{1a}$}, \V{V$_{2a}$} \}$ and $\V{V$_b$} = \{ \V{V$_{1b}$}, \V{V$_{2b}$} \}$. While the combinatoric JR~\ref{jr:minM2}, which chooses this partition by minimizing the combination $\mass{\V{V$_a$}}{2}+\mass{\V{V$_b$}}{2}$, can be used, there is a practical limitation in cases where $n_{\V{V}} = |\V{V}|$ is large. Asymptotically, this algorithm scales as $ 2^{n_{\V{V}}-1}n_{\V{V}}\log n_{\V{V}}$, which is computationally taxing. When computational time is limited, it is potentially prohibitive.  To overcome this shortcoming, we introduce an additional combinatoric JR for partitioning a set into two groups which scales as $n_{\V{V}}^3$, with a natural choice of minimized metric. The JR can be defined as follows~\cite{Jackson:2016mfb}:
\begin{jigsaw}[Combinatoric Minimization]
\label{jr:minM}
If there is a set of $n \geq 2$ visible particles, $\V{V} = \{ \V{V$_{1}$},\cdots,\V{V$_{n}$} \}$, we can choose a partition of \V{V} into two subsets, $P_{\V{V}} = \{ S_{\V{V$a$}},  S_{\V{V$b$}} \}$ by effectively minimizing the masses of the two subsets over the space of all valid partitions $P_{\V{V}} \in \mathbb{P}_{\V{V}}$. This rule is applicable when the only requirement on $S_{\V{V$a$}}$ and $S_{\V{V$b$}}$ is that they each contain at least one element.

The partition is chosen by evaluating the four vectors of the visible particles in their mutual center-of-mass frame, and noting that $\pthree{S_{\V{V$a$}}}{\V{V}} = - \pthree{S_{\V{V$b$}}}{\V{V}}$, irrespective of chosen partition. As \mass{\V{V}}{} is also independent of $P_{\V{V}}$, the relation
\bea
\mass{\V{V}}{} = \sqrt{ |\pthree{S_{\V{V$a/b$}}}{\V{V}}|^2 + \mass{\V{V$_a$}}{2} } + \sqrt{ |\pthree{S_{\V{V$a/b$}}}{\V{V}}|^2 + \mass{\V{V$_b$}}{2} }~,
\eea
implies that maximizing the momentum $|\pthree{S_{\V{V$a/b$}}}{\V{V}}|$ is equivalent to simultaneously minimizing \mass{\V{V$_a$}}{} and \mass{\V{V$_b$}}{}. This is accomplished by choosing $P_{\V{V}}$ which maximizes the function:
\bea
f(P_{\V{V}}) = |\pthree{S_{\V{V$a$}}}{\V{V}}| + |\pthree{S_{\V{V$b$}}}{\V{V}}|~,
\eea
which can be done through the determination of the thrust axis in this reference frame with order $4|\V{V}~|^3$ operations.
\end{jigsaw}

The JR~\ref{jr:minM} is used to partition \V{V} into \V{V$_a$} and \V{V$_b$} which, if they contain more than one element, are recursively partitioned into \V{V$_{1a/b}$} and \V{V$_{2a/b}$} using the same JR. In combination with the JR's for resolving the kinematics of the invisible particles in these events, the complete strategy for analyzing these gluino and sbottom quark processes can be summarized as:
\begin{enumerate}[noitemsep]
\item Apply the combinatoric JR~\ref{jr:minM} to partition \\ \V{V} in \V{V$_a$} and \V{V$_b$}.
\item If either \V{V$_a$} or \V{V$_b$} contain more than one element, further partition them into \V{V$_{1a/b}$} and \V{V$_{2a/b}$}\\ using JR~\ref{jr:minM}.
\item Apply the invisible mass JR~\ref{jr:mass}, \\ choosing $\Mass{\I{I}}{2} = \mass{\V{V}}{2} - 4\mass{\V{V$_a$}}{}\mass{\V{V$_b$}}{}$.
\item Apply the invisible rapidity JR~\ref{jr:rapidity}, choosing \\ \pone{\I{I},z}{\lab} using the collection of visible particles, \V{V}.
\item Apply the contra-boost invariant JR~\ref{jr:contra}, \\ using the constraint $\Mass{\D{$\tilde{P}_a$}}{} = \Mass{\D{$\tilde{P}_b$}}{}$.
\end{enumerate}
The recursive partitioning of visible particles into four subsets factorizes the combinatoric uncertainties according to the different decays of Fig.~\ref{fig:example_DiGluino_to_bbXbbX-decay-ana}. As the JR's for the invisible particles have done the same for their associated unknowns, the resulting kinematic estimators can be calculated in each decay frame almost independently. A source of residual correlation between the energies of the visible sets evaluated in each frame is their individual masses. In a sense, the mass of the partitions \V{V$_{a/b}$} include information about the following decays, giving an indication of the number of elements in the sets. Similarly, the individual masses of the smallest sets \V{V$_{ij}$} are sensitive to the composition and number of elements each contains, with very different behavior expected if a set contains one, or more than one, object. 

While the factorization of many uncertainties is ensured by the application of the RJR method, to ensure minimal correlations between observables calculated in different frames we further introduce a family of heuristic variables, \Hvar{$n$}{$m$}{$F$}, defined as
\bea
\label{eqn:Hvar}
\Hvar{$n$}{$m$}{$F$} = \sum_{S_{\V{V$i$}} \in P_{\V{V}}}^{\V{$n$}} |\pthree{S_{\V{V$i$}}}{\D{$F$}}| + \sum_{S_{\I{I$i$}} \in P_{\I{I}}}^{\I{$m$}} |\pthree{S_{\I{I$i$}}}{\D{$F$}}|~,
\eea
where \V{$n$} and \I{$m$} are the number of sets in partitions of all the visible and invisible particles in the event, $P_{\V{V}} = \{ S_{\V{V$i$}} \}$ and $P_{\I{I}} = \{ S_{\I{I$i$}} \}$, respectively. The scalar sum of the momentum of these sets is evaluated in a particular reference frame, \D{$F$}, making each \Hvar{$n$}{$m$}{$F$} an estimate of the mass scale of that frame, at a level of resolution dictated by the sizes of the partitions, which purposefully obfuscate finer event structure.

\begin{figure}[!htbp]
\centering 
\subfigure[]{\includegraphics[width=.238\textwidth]{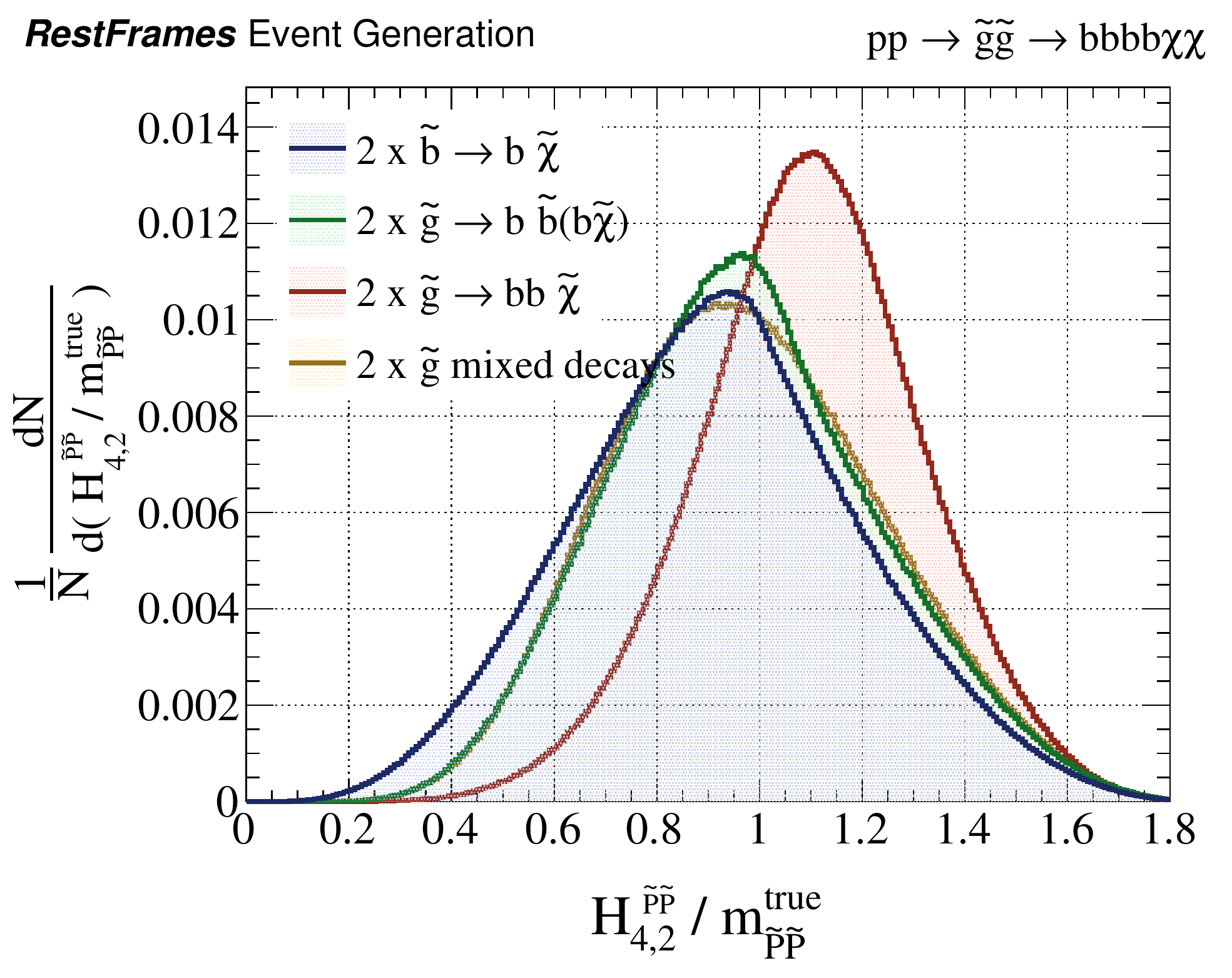}}
\subfigure[]{\includegraphics[width=.238\textwidth]{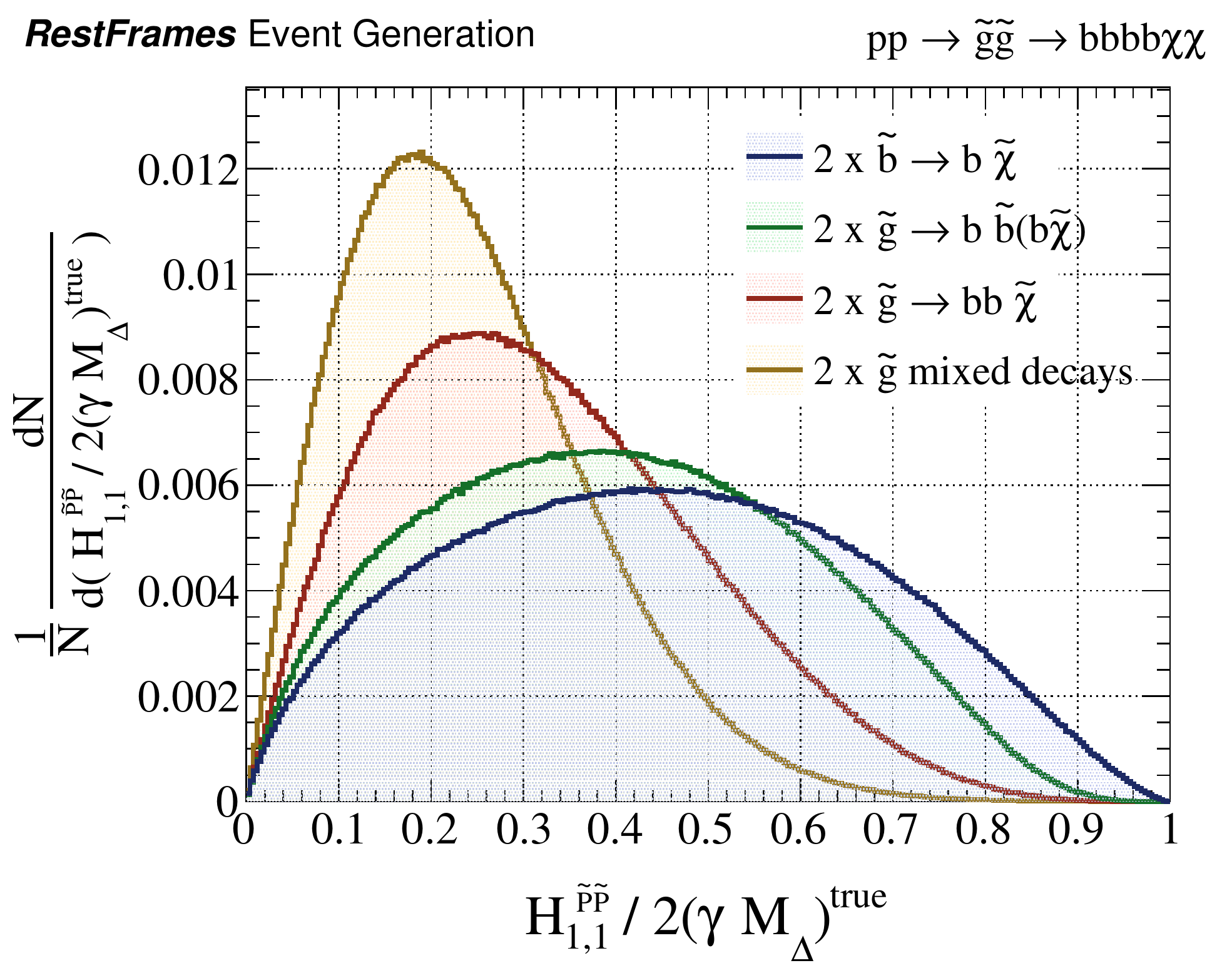}}
\vspace{-0.3cm}
\caption{\label{fig:example_DiGluino_to_bbXbbX-decay-masses} Distributions of (a) the estimator \Hvar{4}{2}{$\tilde{P}\tilde{P}$} and (b) \Hvar{1}{1}{$\tilde{P}\tilde{P}$} for simulated events corresponding to the processes described in Fig.~\ref{fig:example_DiGluino_to_bbXbbX-decay}. Each observable is appropriately normalized by true quantities. }
\end{figure}

For example, an estimator sensitive to the total invariant mass of each of these events, \mass{\D{$\tilde{P}\tilde{P}$}}{}, can be constructed as \Hvar{4}{2}{$\tilde{P}\tilde{P}$}, using the finest partitions of visible and invisible particles in these events considered. The distribution of \Hvar{4}{2}{$\tilde{P}\tilde{P}$} is shown in Fig.~\ref{fig:example_DiGluino_to_bbXbbX-decay-masses}(a) for simulated events of the processes described in Fig.~\ref{fig:example_DiGluino_to_bbXbbX-decay}. For this example, we have chosen to set the masses of the initially produced sparticles, $\D{$\tilde{P}$}$, to 1 TeV, and the masses of the neutralinos to 100 GeV. When they appear in decays of gluinos, the sbottom quark masses are set to 900 GeV. The \Hvar{4}{2}{$\tilde{P}\tilde{P}$} distribution for each of the processes including a sbottom quark scales closely to  \mass{\D{$\tilde{P}\tilde{P}$}}{}, with a slight bias due to the missing masses of the neutralinos themselves, $\sim200$ GeV. For symmetric three-body decays of the gluino, \Hvar{4}{2}{$\tilde{P}\tilde{P}$} is biased to larger values. This is a result of the three-body decay phase-space sometimes giving the neutralinos very little momentum, and the contra-boost invariant JR~\ref{jr:contra} overcompensating in its momentum assignments. 

The estimator \Hvar{1}{1}{$\tilde{P}\tilde{P}$} can be used to examine the same event at a coarser level of resolution, with distributions shown in Fig.~\ref{fig:example_DiGluino_to_bbXbbX-decay-masses}(b). Using a subset of the information going into \Hvar{4}{2}{$\tilde{P}\tilde{P}$}, \Hvar{1}{1}{$\tilde{P}\tilde{P}$} is sensitive to the difference in invisible particle kinematics between the processes, reflecting the fraction of decay phase-space given to the neutralinos, with a kinematic limit at $2\gamma M_{\Delta}$, where $\gamma = \mass{\D{$\tilde{P}\tilde{P}$}}{} / 2 \mass{\D{$\tilde{P}$}}{}$ and $M_{\Delta} = (\mass{\D{$\tilde{P}$}}{2} - \mass{\D{$\tilde{\chi}^0$}}{2})/\mass{\D{$\tilde{P}$}}{}$. When the gluinos decay through an intermediate sbottom quark, the distribution of \Hvar{1}{1}{$\tilde{P}\tilde{P}$} becomes indistinguishable from direct sbottom production in the limit $\mass{\D{$\tilde{b}$}}{} \rightarrow \mass{\D{$\tilde{g}$}}{}$. 

\begin{figure}[!htbp]
\centering
\subfigure[]{\includegraphics[width=.28\textwidth]{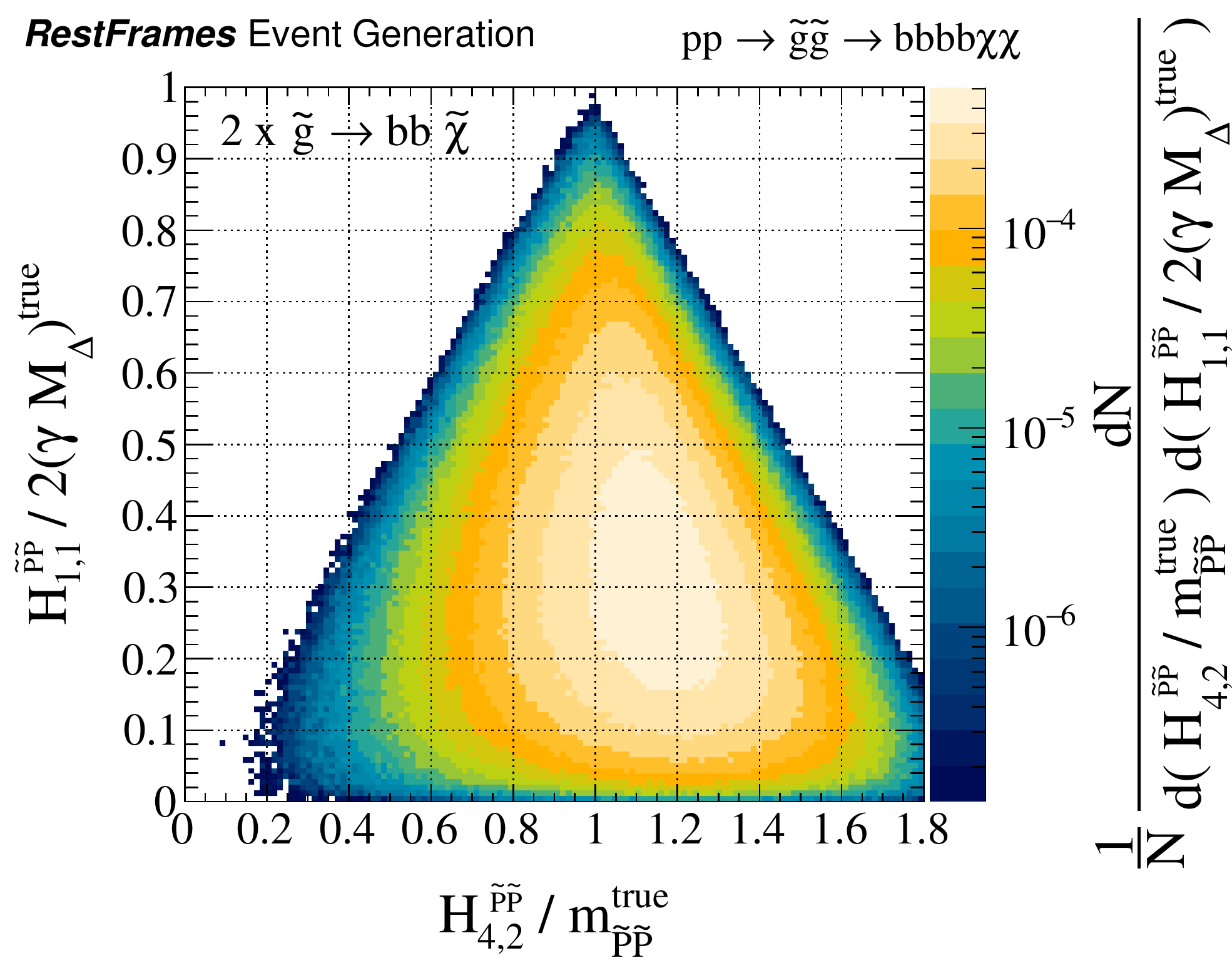}} \hspace{0.5cm}
\subfigure[]{\includegraphics[width=.28\textwidth]{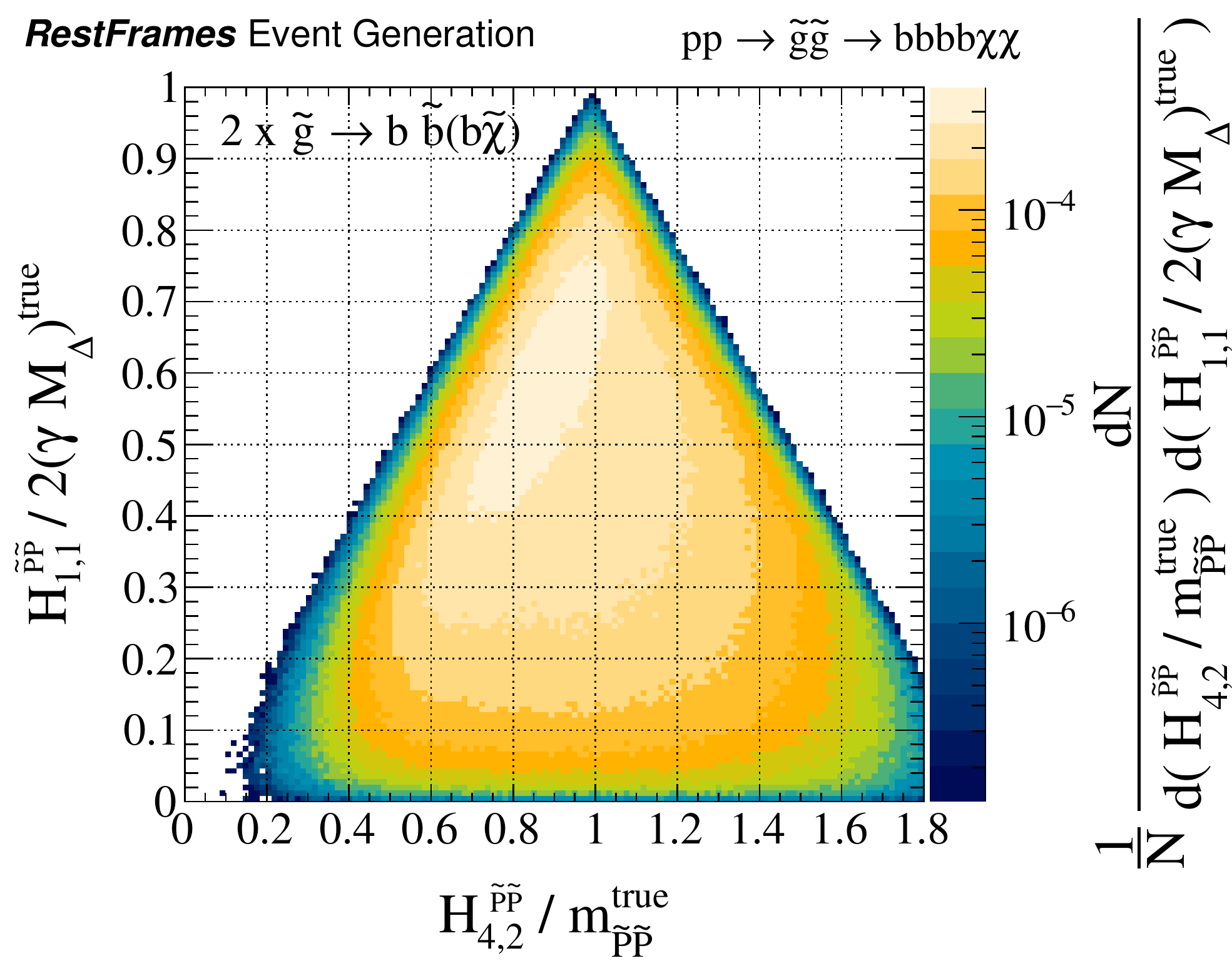}} \hspace{0.5cm}
\subfigure[]{\includegraphics[width=.28\textwidth]{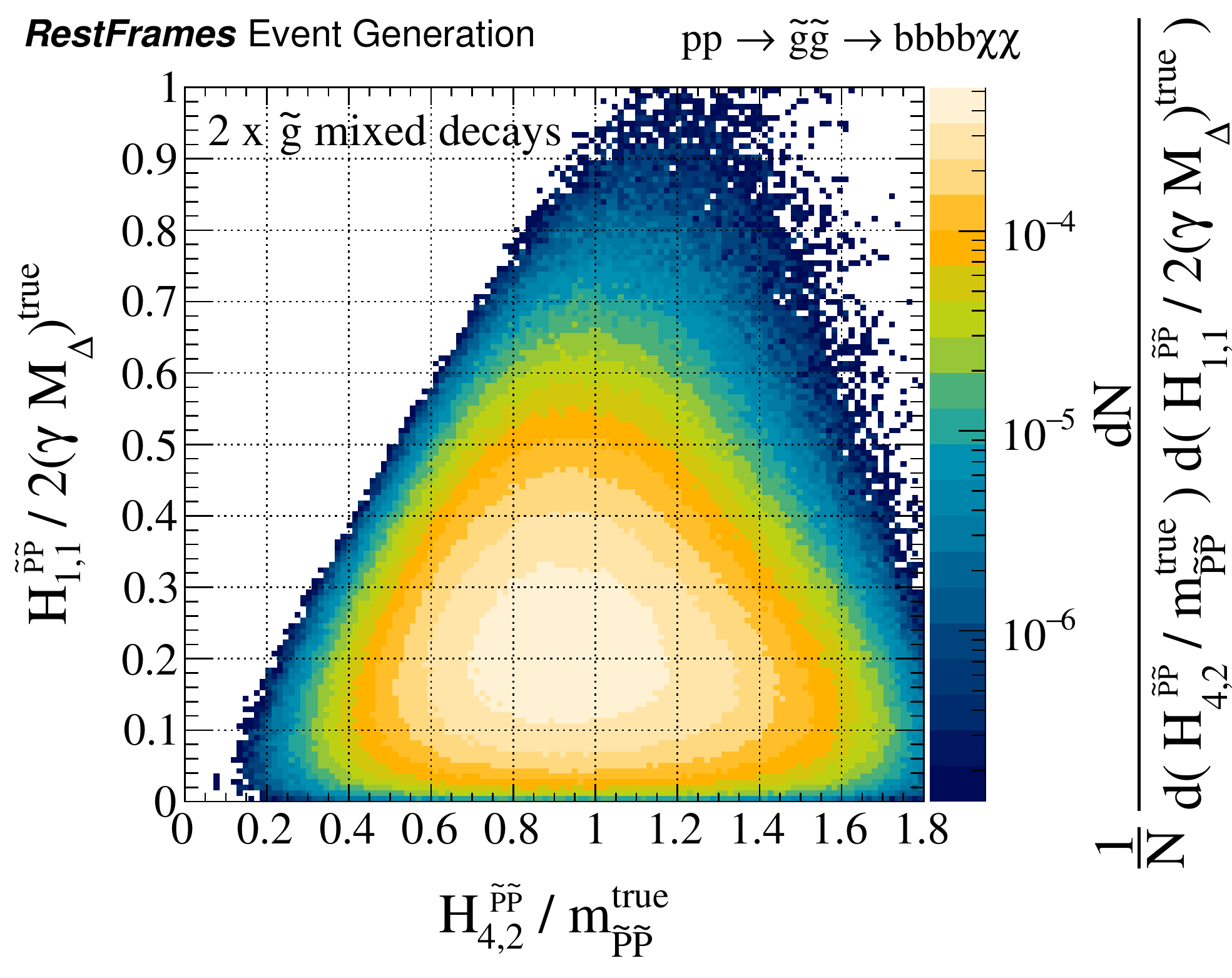}}
\subfigure[]{\includegraphics[width=.28\textwidth]{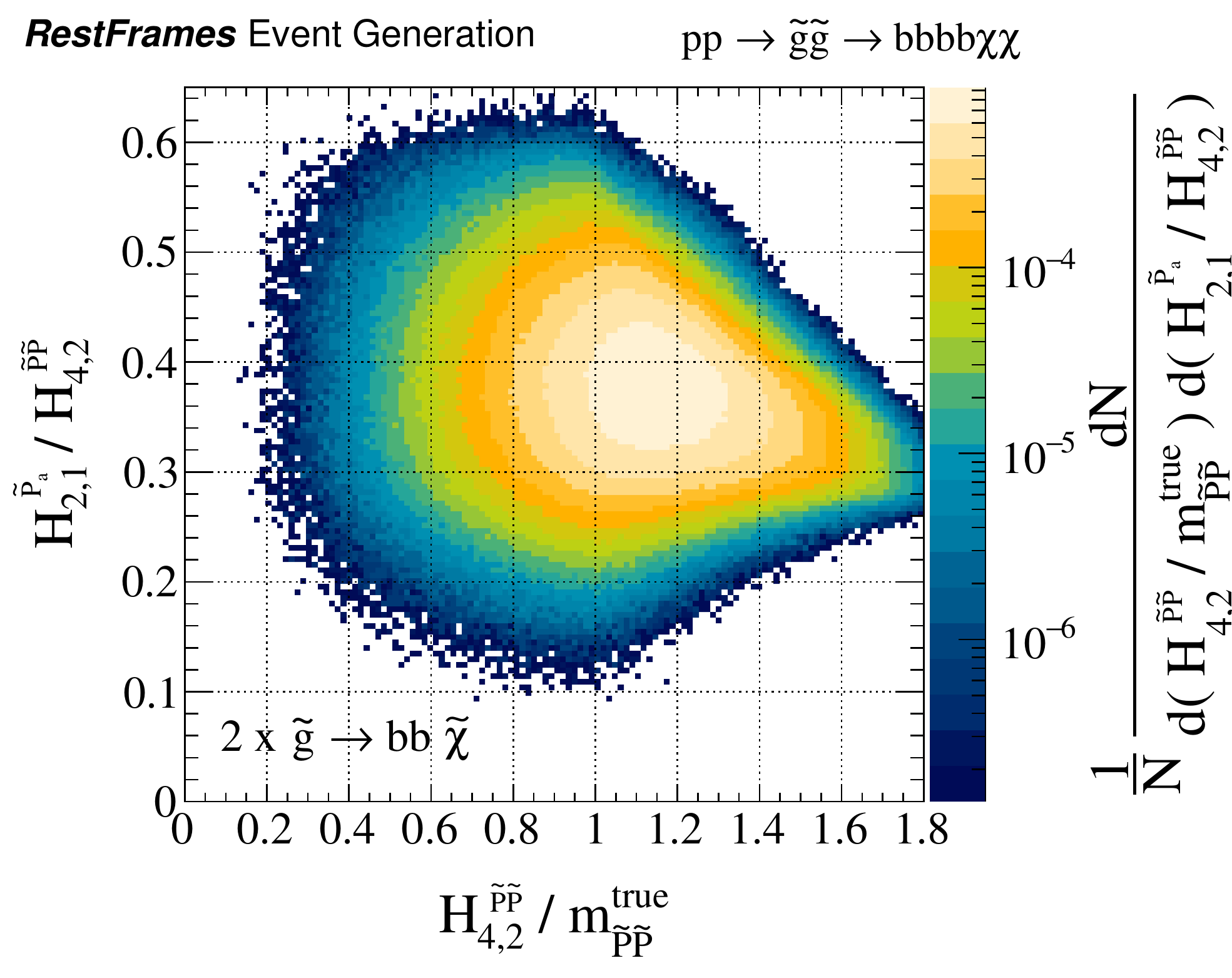}} \hspace{0.5cm}
\subfigure[]{\includegraphics[width=.28\textwidth]{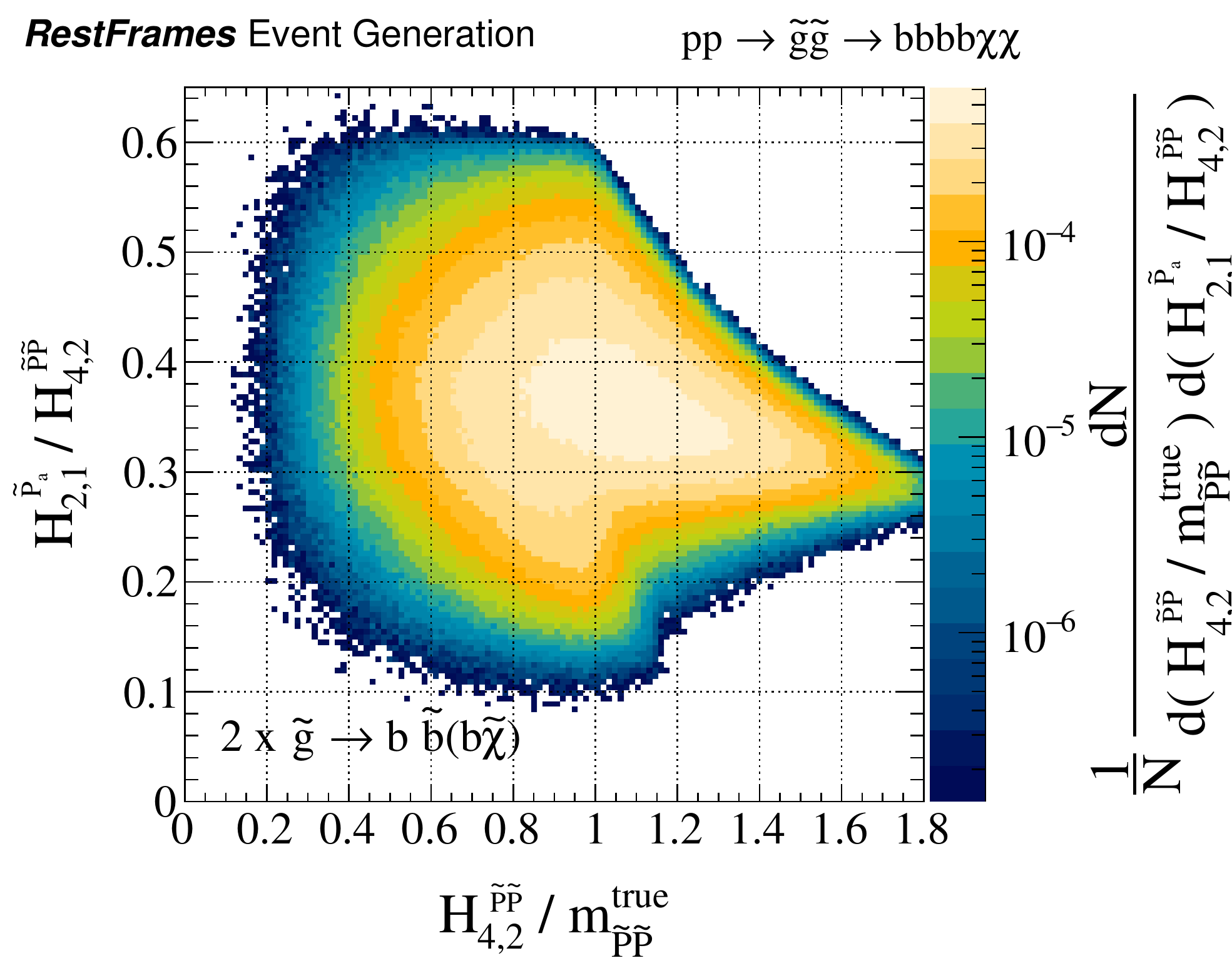}} \hspace{0.5cm}
\subfigure[]{\includegraphics[width=.28\textwidth]{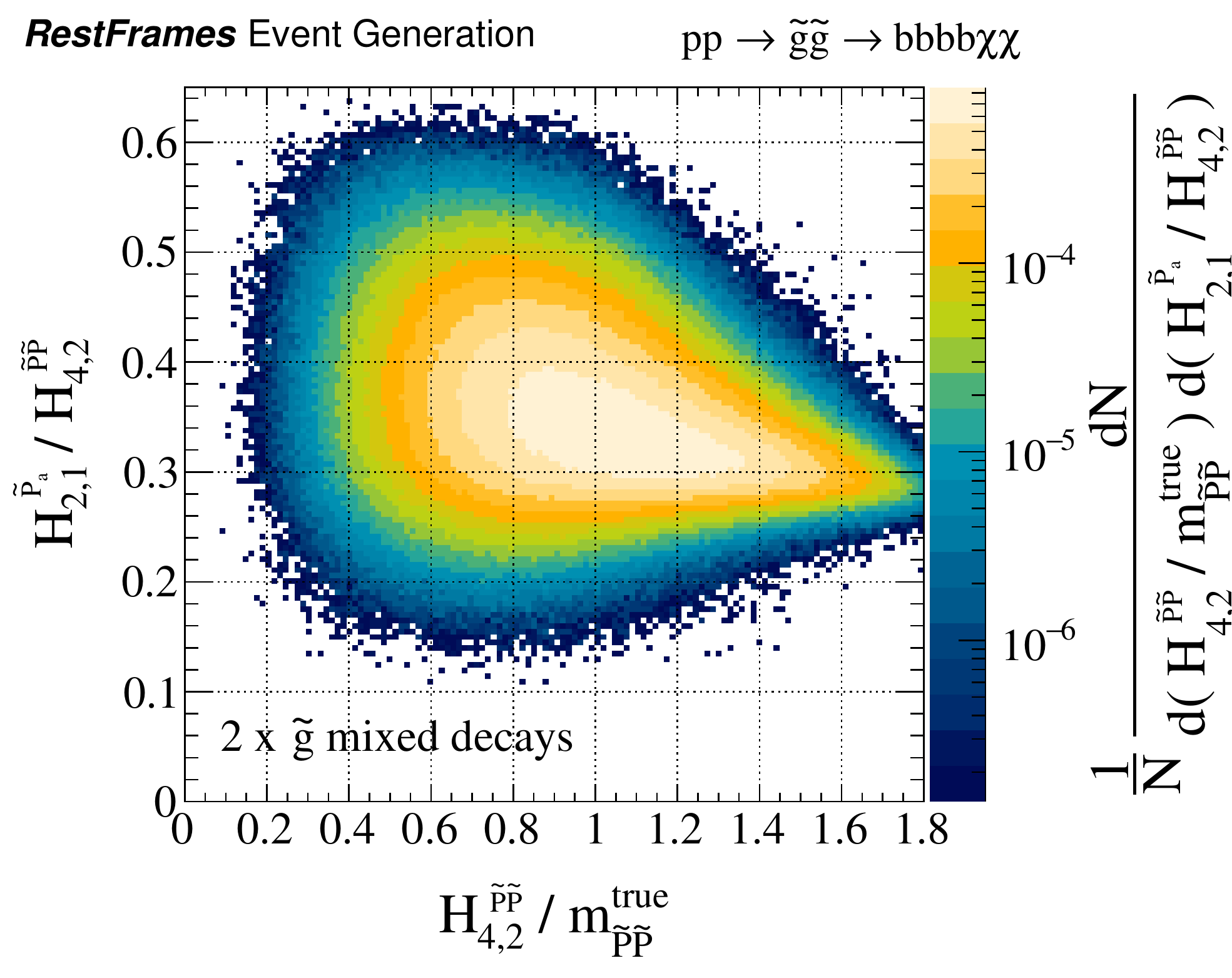}}
\vspace{-0.3cm}
\caption{\label{fig:example_DiGluino_to_bbXbbX-decay-mass2D} Distributions of the estimators (a,b,c) \Hvar{1}{1}{$\tilde{P}\tilde{P}$} and (d,e,f) $\Hvar{2}{1}{$\tilde{P}_a$}/\Hvar{4}{2}{$\tilde{P}\tilde{P}$}$, each as a function of $\Hvar{4}{2}{$\tilde{P}\tilde{P}$}/\mass{\D{$\tilde{P}\tilde{P}$}}{}$, for simulated gluino pair production events corresponding to the decays described in Fig.~\ref{fig:example_DiGluino_to_bbXbbX-decay}. The figures correspond to (a,d) gluino three-body decays, (b,e) gluinos decaying through an intermediate, on-shell sbottom quark, and (c,f) mixed gluino decays. Each observable is normalized, when appropriate, by true quantities. }
\end{figure}

While the estimators \Hvar{4}{2}{$\tilde{P}\tilde{P}$} and \Hvar{1}{1}{$\tilde{P}\tilde{P}$} are constructed from momenta in the same reference frame, the information they contain is largely independent, as demonstrated in Fig.~\ref{fig:example_DiGluino_to_bbXbbX-decay-mass2D}(a,b,c). The residual correlations between the observables are sensitive to the differences in gluino decays, with distinctive behavior for each. 
\clearpage
\onecolumngrid

\twocolumngrid

To better resolve the kinematics of the individual sparticles, the estimators \Hvar{2}{1}{$\tilde{P}_{a}$} and \Hvar{2}{1}{$\tilde{P}_{b}$} can be used, which use partitions of only the visible and invisible particles associated with the hemispheres ``a'' and ``b'', respectively. These observables are sensitive to the masses of these sparticles and, when compared to \Hvar{4}{2}{$\tilde{P}\tilde{P}$}, reveal the presence of the \D{$\tilde{P}$} resonances, as can be seen in Fig.~\ref{fig:example_DiGluino_to_bbXbbX-decay-mass2D}(d,e,f). When these \Hvar{$n$}{$m$}{F} variables are applied in ratio, like $\Hvar{2}{1}{$\tilde{P}_{a/b}$}/\Hvar{4}{2}{$\tilde{P}\tilde{P}$}$, they indicate whether there is a resonant structure between the two different pairs of partitions considered. In the case of the center-of-mass system, \D{$\tilde{P}\tilde{P}$}, ``decaying'' to the individual sparticles, this structure is clearly visible in Fig.~\ref{fig:example_DiGluino_to_bbXbbX-decay-mass2D}, and is estimated with little correlation to the total mass scale, \Hvar{4}{2}{$\tilde{P}\tilde{P}$}. Only small differences in the distributions of these ratios are observed between the different gluino decays considered. 

\begin{figure}[!htbp]
\centering 
\subfigure[]{\includegraphics[width=.238\textwidth]{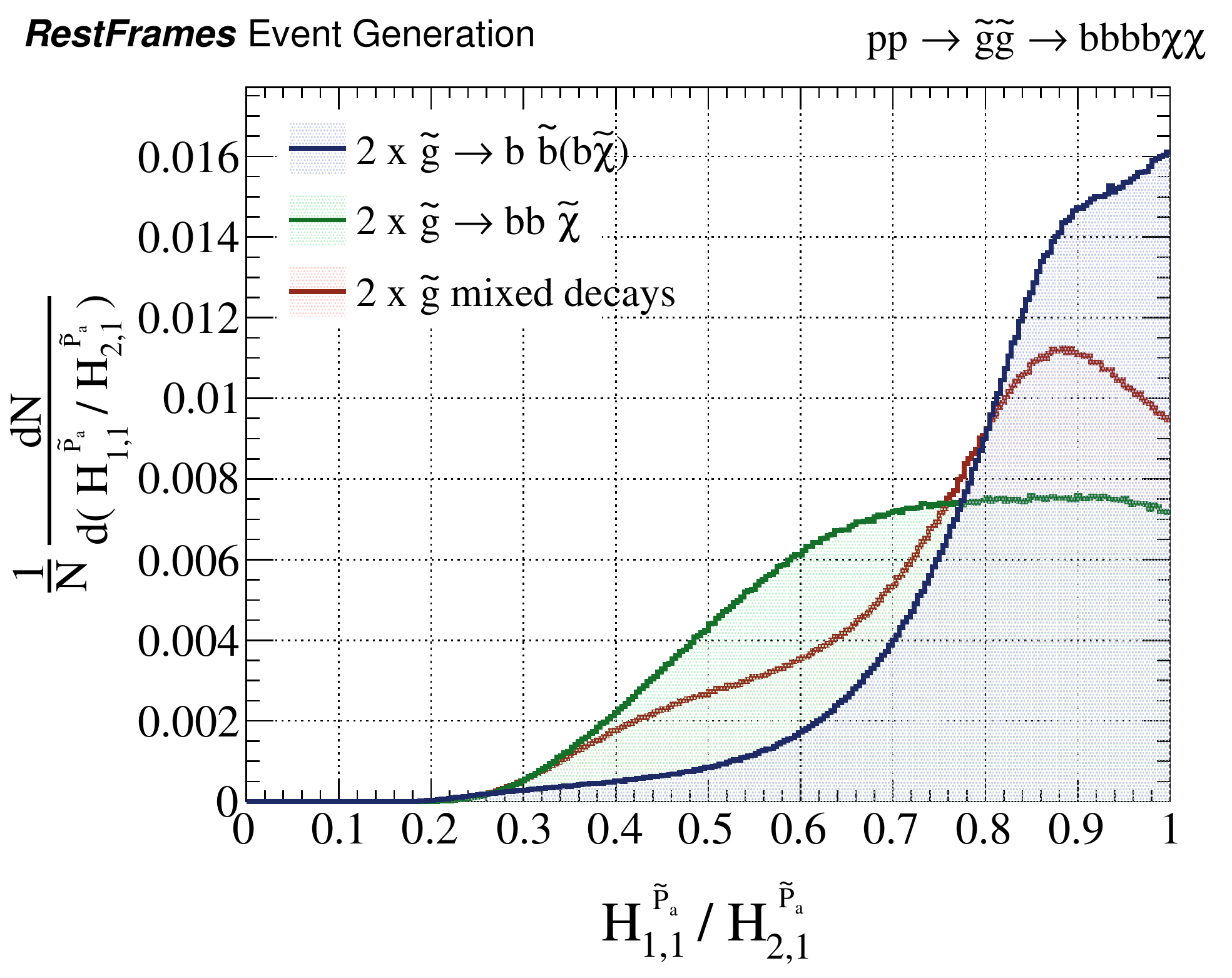}}
\subfigure[]{\includegraphics[width=.238\textwidth]{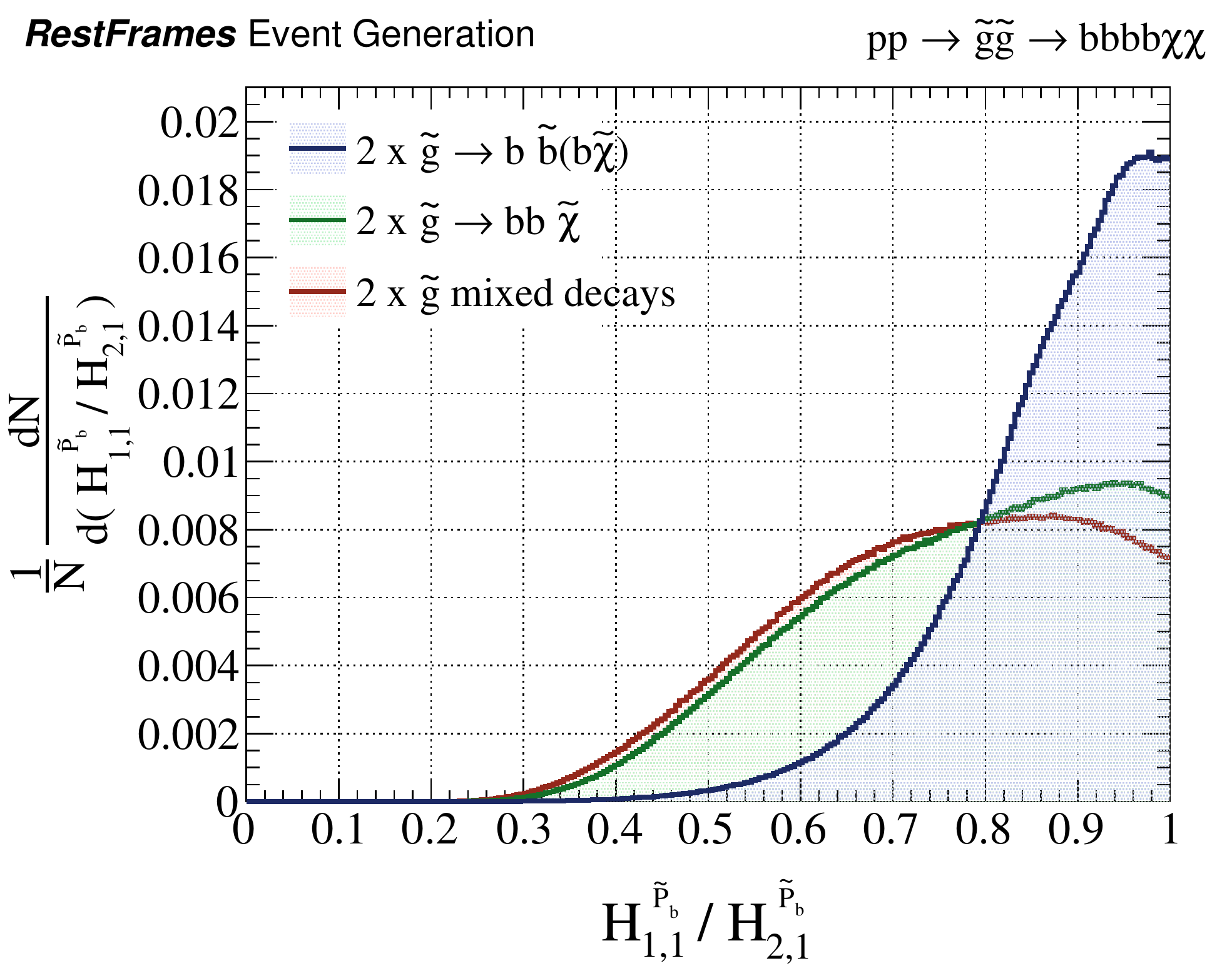}}
\vspace{-0.3cm}
\caption{\label{fig:example_DiGluino_to_bbXbbX-decay-ratios} Distributions of (a) the estimator $\Hvar{1}{1}{$\tilde{P}_a$}/\Hvar{2}{1}{$\tilde{P}_a$}$, and (b) $\Hvar{1}{1}{$\tilde{P}_b$}/\Hvar{2}{1}{$\tilde{P}_b$}$, for simulated gluino pair production events corresponding to the decays described in Fig.~\ref{fig:example_DiGluino_to_bbXbbX-decay}. In the case of events with mixed gluino decays, ``a''  is associated with the intermediate sbottom quark while ``b'' corresponds to the three-body decay. }
\end{figure}

As multiple decays of the gluinos are being considered simultaneously, we take an agnostic approach to the parameterization of observables describing their decays, defining additional estimators $\Hvar{1}{1}{$\tilde{P}_a$}$ and 
$\Hvar{1}{1}{$\tilde{P}_b$}$. These correspond to partial abstractions of the gluino decays, where each visible system \V{V$_{a/b}$} is treated as only one particle, such that, taken in ratio with their respective $\Hvar{2}{1}{$\tilde{P}_{a/b}$}$, they are sensitive to the gluino decay structure. The distributions of these ratios are shown in Fig.~\ref{fig:example_DiGluino_to_bbXbbX-decay-ratios} for each gluino decay, and for each reconstructed hemisphere of the event. We adopt the convention that the hemisphere assigned the highest transverse momentum jet coming from the true gluino ``a'' decay in reconstruction is also assigned the label ``a'' which, in the mixed decay case, corresponds to a decay through an on-shell sbottom quark. The distributions of the ratio $\Hvar{1}{1}{$\tilde{P}_{a/b}$}/\Hvar{2}{1}{$\tilde{P}_{a/b}$}$ have a distinct shape indicative of the type of decay, with the observables corresponding to the different gluino decays in the mixed case each adopting shapes closely resembling those of the symmetric cases. 

\begin{figure}[!htbp]
\centering
\subfigure[]{\includegraphics[width=.28\textwidth]{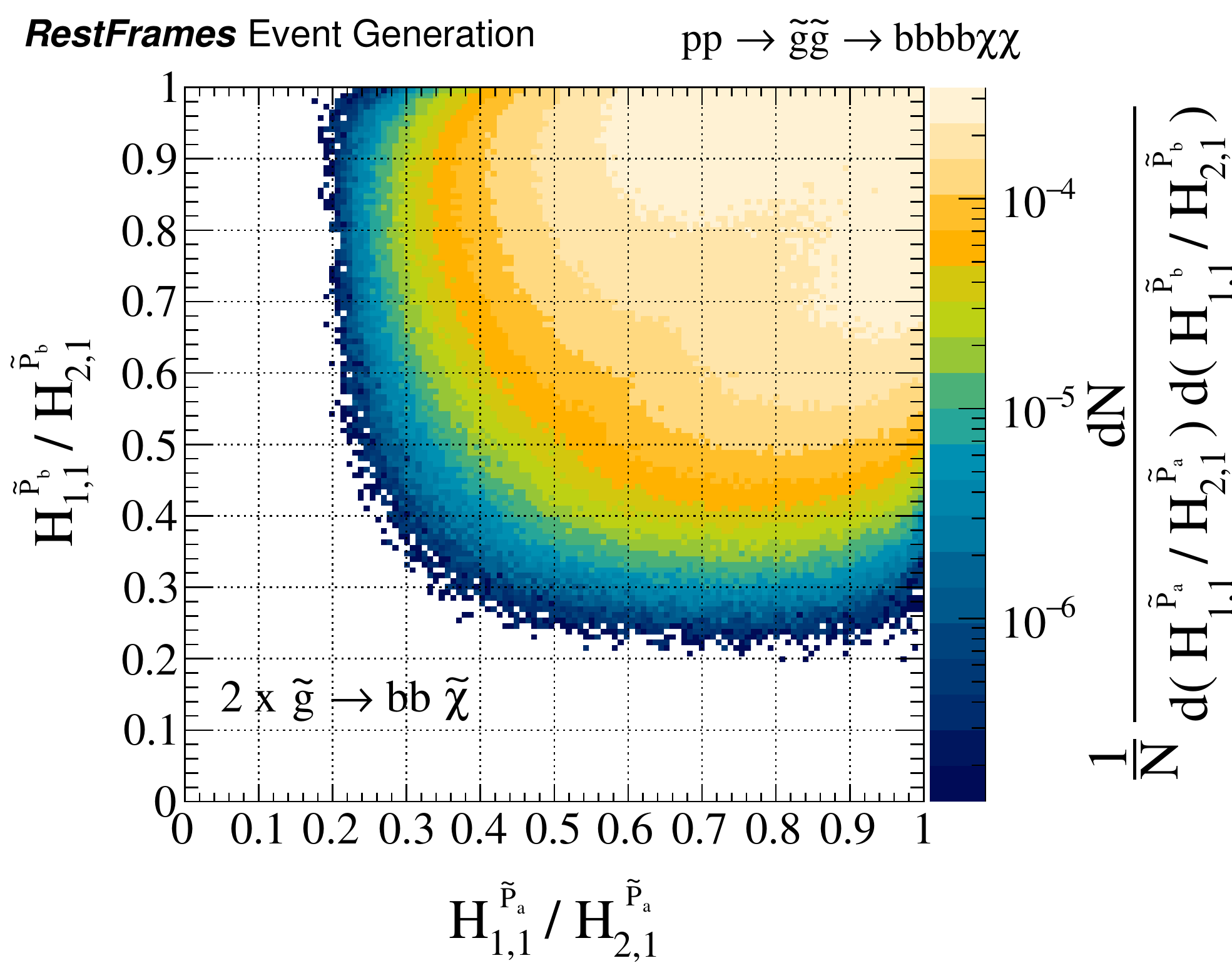}} \hspace{0.5cm}
\subfigure[]{\includegraphics[width=.28\textwidth]{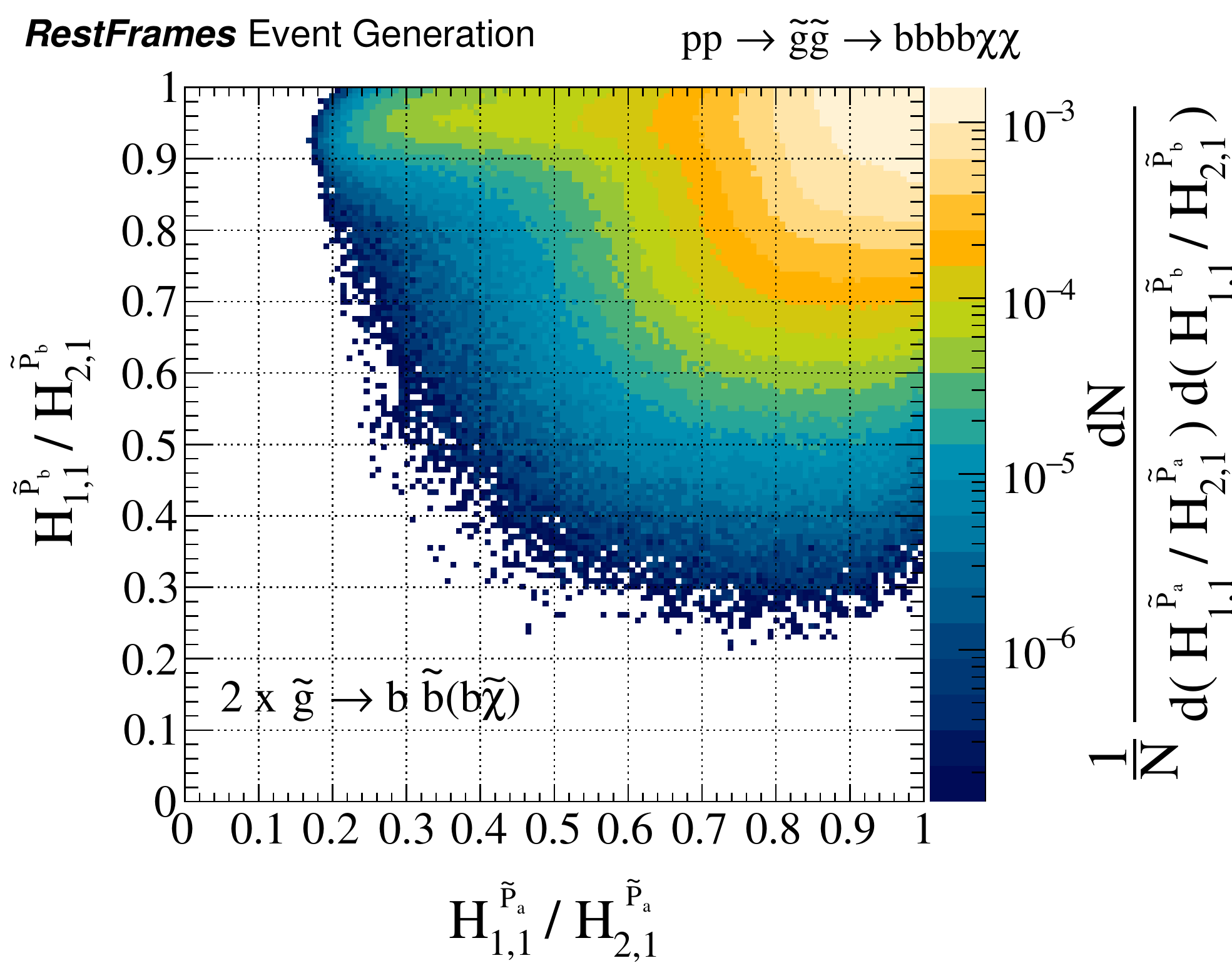}} \hspace{0.5cm}
\subfigure[]{\includegraphics[width=.28\textwidth]{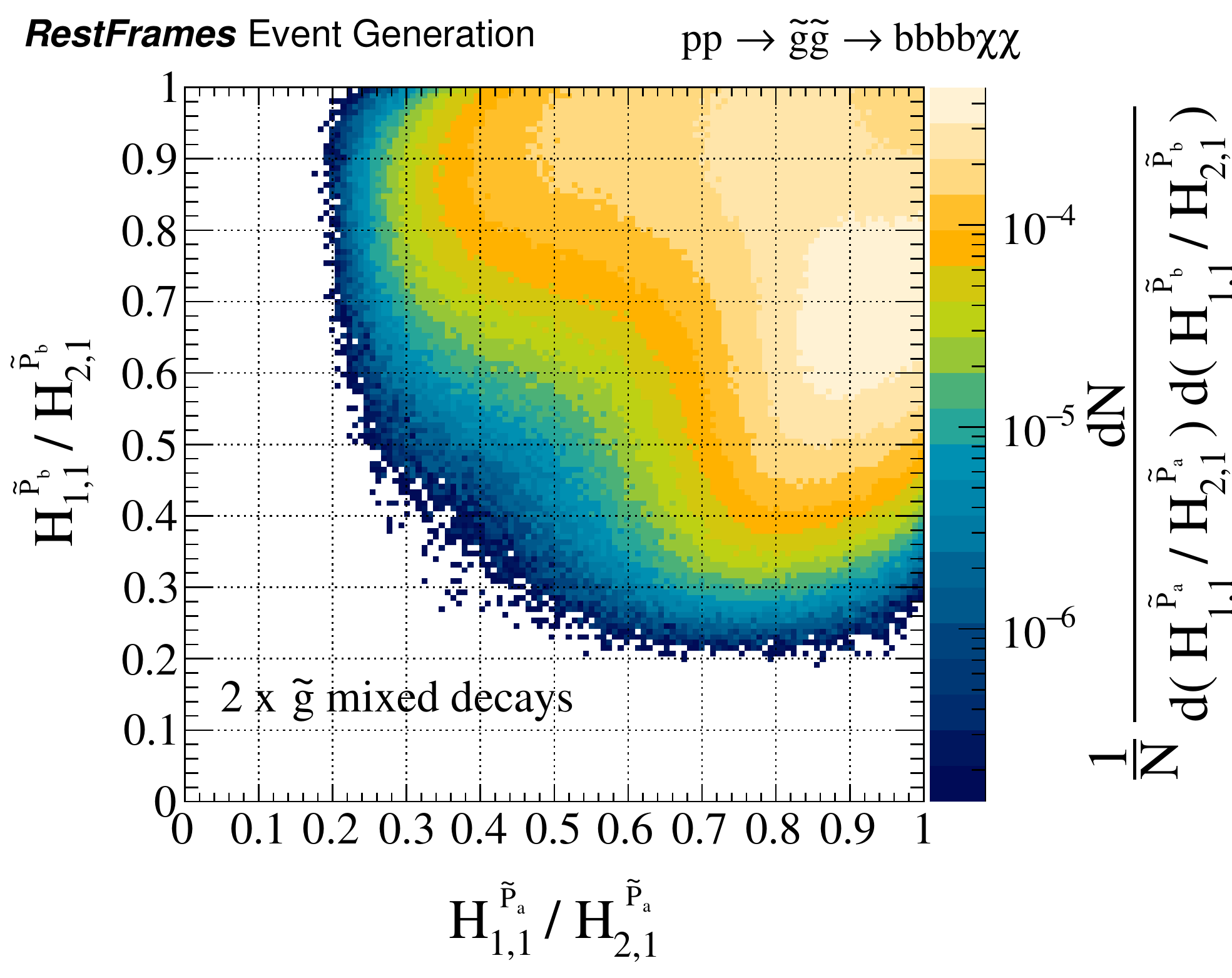}}
\vspace{-0.3cm}
\caption{\label{fig:example_DiGluino_to_bbXbbX-decay-ratio2D} Distributions of $\Hvar{1}{1}{$\tilde{P}_a$}/\Hvar{2}{1}{$\tilde{P}_a$}$, as a function of $\Hvar{1}{1}{$\tilde{P}_b$}/\Hvar{2}{1}{$\tilde{P}_b$}$, for simulated gluino pair production events corresponding to the decays described in  Fig.~\ref{fig:example_DiGluino_to_bbXbbX-decay}. The figures correspond to (a) gluino three-body decays, (b) gluinos decaying through an intermediate, on-shell sbottom quark, and (c) mixed gluino decays. In this case of events with mixed gluino decays, ``a''  is associated with the intermediate sbottom quark while ``b'' corresponds to the three-body decay.  The small lack of symmetry between otherwise symmetric decays observed in the distributions (a,b) is a result of the convention that the true \V{b$_{2a}$} is always associated with the reconstructed ``a'' hemisphere and incorrect combinatoric assignments. }
\end{figure}

The partitioning of information throughout the event reconstruction and observable definition allows for the ratios $\Hvar{1}{1}{$\tilde{P}_{a/b}$}/\Hvar{2}{1}{$\tilde{P}_{a/b}$}$ to be estimated independently for each half of the event, as shown in Fig.~\ref{fig:example_DiGluino_to_bbXbbX-decay-ratio2D}. The small asymmetries in these distributions for the symmetric decays is a result of the convention for assigning the labels ``a'' and ``b'' to the two halves of the event, and appears when the combinatoric assignment of the $b$-tagged jets is incorrect in the reconstruction. Otherwise, the $\Hvar{1}{1}{$\tilde{P}_{a/b}$}/\Hvar{2}{1}{$\tilde{P}_{a/b}$}$ distribution for the $\tilde{g} \rightarrow b\tilde{b}$ decay in the mixed case resembles that of the symmetric decay case, as it independently does for the three-body decay in the opposite event hemisphere.
\onecolumngrid

\twocolumngrid

The basis of observables produced when partitioning the information contained in each reconstructed event into approximately uncorrelated variables through the RJR approach is useful for not just studying these types of decay topologies, but searching for evidence of these phenomena in experimental data. By independently maintaining sensitivity to each decay in the event, background processes must simultaneously fake many different kinematic features in order to be confused with signal. This includes not only the total mass scale of the event, but how energy is shared between the products of each subsequent decay. That these observables are able to distinguish between the different decays of similar sparticles indicates that they are also sensitive to expected differences in SM backgrounds, where the absence of the same resonance structure in the events can be exploited.

\section{Summary}
\label{sec:summary}

As evidenced by the examples discussed in this paper, many of the final state topologies of interest at particle colliders contain both kinematic and combinatoric ambiguities, with missing information resulting from invisible or indistinguishable particles. Recursive Jigsaw Reconstruction is a systematic prescription for overcoming these unknowns, and approximately reconstructing each event in its entirety, resulting in a basis of kinematic estimators sensitive to the masses and decay angles of all the particles appearing in them.

This is accomplished by factorizing all of the unknowns appearing in an event according to which intermediate decays they are related to, and using the library of Jigsaw Rules, interchangeable and configurable algorithms for determining these unknowns, to resolve them while recursively moving through the decay tree describing the event. The JR's for analyzing events with one invisible particle, like $W \rightarrow \ell\nu$ in Section~\ref{subsec:Part1_exampleA}, $t \rightarrow bW(\ell\nu)$ in Section~\ref{subsec:Part1_exampleB}, and $H^{\pm} \rightarrow h(\gamma\gamma)W^{\pm}(\ell\nu)$ in Section~\ref{subsec:Part1_exampleC}, can be combined with more complicated ones in events with multiple invisible particle. 

In events with two invisible particles coming from symmetrically similar decays, an array of different contra-boost invariant JR's can be used to analyze processes like $H \rightarrow W(\ell\nu)W(\ell\nu)$ in Section~\ref{subsec:Part2_exampleA}, $\tilde{t}\tilde{t} \rightarrow t \tilde{\chi}^{0}_1 t \tilde{\chi}^{0}_1$ in Section~\ref{subsec:Part2_exampleB}, and $\tilde{\chi}^{0}_2\tilde{\chi}^0_2 \rightarrow h(\gamma\gamma)\tilde{\chi}^0_1 Z(\ell\ell)\tilde{\chi}^0_1$ in Section~\ref{subsec:Part2_exampleC}. In these cases, both the mass of the total interaction and pair-produced massive particles can be independently determiend with the RJR approach. When there are even more intermediate particles appearing in the final state, like $t\bar{t} \rightarrow bW(\ell\nu)bW(\ell\nu)$ in Sections~\ref{subsec:Part3_exampleA} and~\ref{subsec:Part3_exampleB}, or $\tilde{t}\tilde{t} \rightarrow b\tilde{\chi}^{\pm}(\ell \tilde{\nu})b\tilde{\chi}^{\mp}(\ell\tilde{\nu})$ in Section~\ref{subsec:Part3_exampleC}, estimators sensitive to the masses and decays of these additional particles are also determined. 

There are JR's for arbitrarily complex decays, like non-resonant $N \times W(\ell\nu)$ production in Section~\ref{subsec:Part4_exampleA}. But the real strength of the RJR algorithm is that each of the JR's is designed to only use the momentum of abstractions of the particles in each decay step, where multiple particles that are resolved in later decays are treated as indivisible single particles. This means that many JR's can be combined, recursively, to analyze complicated events like $H \rightarrow hh \rightarrow 4W(\ell\nu)$ in Section~\ref{subsec:Part4_exampleB} with many invisible particles, or $\tilde{g}\tilde{g} \rightarrow bb\tilde{\chi}^0_1 bb\tilde{\chi}^0_1$ in Section~\ref{subsec:Part4_exampleC} with many indistinguishable particles. 

The recursive, factorized, application of JR's when analyzing events yields a complete {\it basis} of kinematic observables, each corresponding to quantities of interest in the event and largely independent of the others. With the library of JR's described in this paper, the RJR algorithm can be used to provide such a basis for {\it any} decay topology imaginable, with the effective resolution of the corresponding estimators limited only by the measurements made in the detector and the imagination of analysts in choosing and assembling the JR pieces.



\begin{acknowledgments}
The authors would like to thank the Harvard Club of Australia for the award of the ``Australia-Harvard Fellowship'' in 2014 and 2015 
which made part of this work possible. PJ is supported by the Australian Research Council Future Fellowship FT130100018.
CR wishes to thank the Harvard Society of Fellows and William F. Milton Fund for their generous support.
\end{acknowledgments}

\bibliography{rj}{}
\bibliographystyle{unsrt}

\end{document}